# Unified Modeling and Experimental Realization of Electrical and Thermal Percolation in Polymer Composites

Navid Sarikhani,[1] Zohreh S. Arabshahi,[2] Abbas Ali Saberi,[3,4,a)] and Alireza Z. Moshfegh[2,5,a)]
[1] *School of Mechanical Engineering, Sharif University of Technology, Tehran 11155-9567, Iran*
[2] *Department of Physics, Sharif University of Technology, Tehran 11155-9161, Iran*
[3] *Department of Physics, University of Tehran, Tehran 14395-547, Iran*
[4] *Max Planck Institute for the Physics of Complex Systems, 01187 Dresden, Germany*
[5] *Institute for Nanoscience and Nanotechnology, Sharif University of Technology, Tehran 14588-8969, Iran*

[a)] Authors to whom correspondence should be addressed: ab.saberi@ut.ac.ir and moshfegh@sharif.edu

Correlations between electrical and thermal conduction in polymer composites are blurred due to the complex contribution of charge and heat carriers at the nanoscale junctions of filler particles. Conflicting reports on the lack or existence of thermal percolation in polymer composites have made it the subject of great controversy for decades. Here, we develop a generalized percolation framework that describes both electrical and thermal conductivity within a remarkably wide range of filler-to-matrix conductivity ratios ($Y_f/Y_m$), covering 20 orders of magnitude. Our unified theory provides a genuine classification of electrical conductivity with typical $Y_f/Y_m \geq 10^{10}$ as insulator–conductor percolation with the standard power-law behavior, and of thermal conductivity with $10^2 \leq Y_f/Y_m \leq 10^4$ as poor–good conductor percolation characterized by two universal critical exponents. Experimental verification of the universal and unified features of our theoretical framework is conducted by constructing a 3D segregated and well-extended network of multiwalled carbon nanotubes in polypropylene as a model polymer matrix under a carefully designed fabrication method. We study the evolution of the electrical and thermal conductivity in our fabricated composites at different loading levels up to 5 vol%. Significantly, we find an ultralow electrical percolation threshold at 0.02 vol% and a record-low thermal percolation threshold at 1.5 vol%. We also apply our theoretical model to a number of 23 independent experimental and numerical datasets reported in the literature, including more than 350 data points, for systems with different microscopic details, and show that all collapse onto our proposed universal scaling function, which depends only on dimensionality.

## I. INTRODUCTION

Polymers, endowed with lightness, durability, and appreciable mechanical properties, in line with their ease of processability and cost-effectiveness, have become an integral part of our modern lifestyle; however, as a matter of fact, "no one is perfect." Most polymers have very low electrical and thermal conductivity, severely limiting their usefulness for many multifunctional device applications. Thus, the fabrication of polymer composites with desired electrical and/or thermal conductivity has been at the forefront of research from the early days of polymer science and still is actively pursued.[1-6] Incorporation of electrically conductive fillers, such as graphite,[7] carbon black,[8] carbon fibers,[9] metal powders,[10] and more recently carbon nanotubes,[11] graphene,[12] and metal nanowires[13] in an insulating polymer matrix usually gives satisfactory results. Above a certain critical filler loading level (percolation threshold), a system-spanning network of connected filler particles emerges throughout the polymer matrix, and the macroscopic electrical conductivity of the composite suddenly increases by several orders of magnitude just within a narrow range of filler loading.[14,15] On the other hand, the addition of various thermally conductive fillers, such as ceramic, metal, or carbon-based fillers, to a polymer matrix has frequently yielded lackluster results, if not disappointing.[16-20] It has been repeatedly reported that the thermal conductivity of polymer composites lacks percolation behavior and increases almost linearly with increasing filler content in accordance with conventional diffusion-type models based on the effective medium theory.[21-24] The small number of reports that exist on observation of thermal percolation in polymer composites, apart from the controversy, are facing two main issues: first, thermal percolation has been observed at relatively high filler loadings (i.e., greater than 15 vol%), and second, no formal percolation model has been provided to describe the observations.[25,26] This is especially surprising, since both conduction of electricity and heat in a medium are described by mathematically equivalent continuum equations and, at least in the case of electrically conductive fillers, it is expected that the established network of connected filler particles which conducts electricity, play a similar role to conduct heat as well. Hence, we begin our discussion by asking the following key questions: (1) Is thermal percolation lacking in polymer composites in all circumstances? (2) Why does, in polymer composites, the thermal conductivity exhibit markedly different behavior in comparison with the electrical conductivity? and, more importantly, (3) Is there any hope to achieve high thermal conductivities at low filler loadings in polymer composites? Or is there a fundamental barrier here?

### A. Thermal percolation controversy

Historically, the general belief has been that the thermal conductivity of macro- or microcomposites cannot be enhanced unless a very high filler loading level (> 50 vol%) is incorporated into the composite.[18,19,27] In this respect, metal fillers with electrons as the dominant heat carrier and non-metal fillers, such as carbon fibers and ceramic powders, with phonons as the main heat carrier, were investigated, and both yielded essentially the same results.[9,10,27] For example, in an influential study in 2002, Mamunya *et al.*[10] added Cu and Ni microparticles into epoxy resin and poly(vinyl chloride) polymer matrices, and while electrical percolation was easily reached in the range of 4–8 vol% filler loading, no thermal percolation was observed even up to a high loading level of 40 vol%. Soon after the discovery of carbon nanotubes (CNT) and especially the realization of their high intrinsic electrical and thermal conductivities, numerous attempts have been made to fabricate CNT–Polymer nanocomposites to exploit the extraordinary physicochemical properties of CNTs.[28,29] In fact, thanks to their very high aspect ratio, CNTs have been very successful in the context of electrical conductivity and have significantly reduced the typical electrical percolation threshold of polymer composites from ~10 vol% in microcomposites to less than 1 vol% in CNT–Polymer nanocomposites.[11,30] However, while the intrinsic thermal conductivity of individual single-walled carbon nanotubes (SWCNT) and multiwalled carbon nanotubes (MWCNT) was measured in the range of 1000–3000 W m$^{-1}$ K$^{-1}$ (along the nanotube axis considering the total cross-sectional area of the tube for heat transport),[31] the thermal conductivity of CNT–Polymer nanocomposites, even at very high nanotube contents, never reached this level and was not even close.[6,16,19,32] For instance, in a highly cited study in 2002, Biercuk *et al.*[33] observed a sharp increase of 5 orders of magnitude in the electrical conductivity of SWCNT–Epoxy nanocomposites at ~0.15 wt% SWCNT loading. However, no thermal percolation was observed in the nanocomposites, and the thermal conductivity enhancement was reported to be only about 2-fold at 1 wt% SWCNT content.

In 2005, Keblinski and co-workers, in a series of follow-up experimental observations[34] and modeling predictions,[35] published their seminal paper[21] with the striking title of "On the Lack of Thermal Percolation in Carbon Nanotube Composites"! The situation became even more complicated when some researchers claimed that carbon nanotube assemblies should be regarded as thermal insulators rather than conductors. The reason was that the nanotube–nanotube junctions were considered to be very inefficient in thermal transport (point contact) because the junctions were thought to suffer from very low thermal conductances.[36,37] Chalopin *et al.*,[36] in 2009, by employing atomistic Green's function calculations, obtained an upper bound of ~5 W m$^{-1}$ K$^{-1}$ for the thermal conductivity of carbon nanotube pellets in sharp contrast with the >1000 W m$^{-1}$ K$^{-1}$ thermal conductivity of individual CNTs. However, the conductor/insulator dilemma was resolved by further theoretical and experimental research that found the thermal conductivity of CNT pellets even above ~200 W m$^{-1}$ K$^{-1}$,[31,38] and Chalopin *et al.*,[39] in an erratum, also accepted huge underestimation in their calculations. Nonetheless, the puzzling low thermal conductivity of CNT–Polymer nanocomposites remained unsolved.

Since the discovery of graphene and demonstrating its remarkable properties, considerable research efforts have been devoted to the experimental realization of thermal percolation in graphene–polymer composites.[40-47] Graphene exhibits an extraordinary in-plane thermal conductivity with often quoted values in the range of 600–4000 W m$^{-1}$ K$^{-1}$ depending on several parameters, such as the number of atomic layers, structural defects, isotope composition, and being suspended or supported.[48,49] However, most studies have failed to observe thermal percolation in graphene–polymer composites.[50-52] In recent years, further research on several other emerging materials, such as expanded graphite,[53] metal nanowires,[54] and hexagonal boron nitride (h-BN),[55] as thermal fillers in polymer composites, has not significantly advanced the field of thermal percolation.

There have been, however, some studies that have reported on the observation of thermal percolation, usually through specialized fabrication techniques, to construct an effective network of filler particles within a polymer matrix for heat conduction. Amongst them, Bonnet *et al.*[56] published one of the first reports on the possibility of thermal percolation in SWCNT–PMMA (Polymethylmethacrylate) nanocomposite films, prepared by solution mixing. They proposed that the electrical and thermal percolation thresholds should coincide since the same network of SWCNTs that conducts electrons could, in principle, conduct phonons. However, as clearly stated by the authors, the accuracy of their data was limited by the accuracy of their experimental setup to measure the thermal conductivity of thin films. At the same time, Yu and coworkers, after some initial failure in the observation of thermal percolation in SWCNT–Epoxy[57] and GNP–Epoxy[58] nanocomposites, reported the evidence of thermal percolation in a composite of SWCNT–GNP–Epoxy.[59] They attributed the superlinear increase of thermal conductivity in this ternary system to a more efficient network formation by combining 1D and 2D nanofillers but they did not provide a model to explain their observation. Also, Shtein and colleagues, in a pioneering and highly influential study,[25] reported the observation of thermal percolation in Graphene–Epoxy nanocomposites prepared by a rather specialized fabrication method (planetary centrifugal mixing technique) for assembling graphene nanoplatelets into a tightly connected network within epoxy. Interestingly, they found the thermal percolation threshold at a significantly higher value (17 vol%) than the electrical percolation threshold (5 vol%). They attempted to interpret their experimental results by the standard power-law percolation model; however, they obtained a non-realistic critical exponent of $t = 0.84$ for thermal percolation, which is far from the well-known values of ~1.3 for 2D networks and ~2.0 for 3D networks [see Eq. (2) and subsequent discussion for details].[14,15] Another important contribution was given by Bandaru *et al.*,[60,61] who reported thermal percolation in nanocomposites of functionalized MWCNTs in reactive ethylene terpolymer (RET), a special polymer that can effectively bond to the COOH groups on the surface of nanotubes, resulting in enhanced CNT–Polymer interactions. They also tried to explain their observations by the standard power-law model but came to a very low critical exponent of $t \approx 0.1$, a sign of inadequate model employment.

The last work we review here is a recent report by a well-known research group led by Balandin, who observed thermal percolation in epoxy composites with graphene and h-BN additives.[26] This group had previously reported the lack of thermal percolation in CNT–Epoxy[62] and Graphene–Epoxy[50] nanocomposites at low and moderate filler loading levels up to 10 vol%. This time, however, by improving their fabrication technique, Kargar, Balandin, and colleagues tackled the high filler loading challenge and captured the thermal percolation threshold at 30 vol% and 23 vol% for graphene and h-BN nanocomposites, respectively.[26] Nevertheless, Balandin and colleagues did not attempt to explain their experimental data by the standard power-law percolation model and preferred to fit the data with the semiempirical Lewis–Nielsen Model. It is worthwhile to note that for the graphene–Epoxy nanocomposite, they obtained the electrical percolation threshold at ~10 vol%, substantially lower than the observed thermal percolation threshold at 30 vol%.

The growing importance of the concept of thermal percolation in the design and fabrication of polymer composites is illustrated in Fig. S2 in the supplementary material by providing statistics on the rapid growth of the annual number of publications in this research field over the past two decades. Table S2 in the supplementary material also provides a broader perspective of thermal percolation in polymer composites by summarizing several key studies in the field. As can be seen from this table, while there are several reports on the lack of thermal percolation in polymer composites with different fillers, such as carbon nanotubes,[33,57,62-71] graphene,[50,53,70-76] and metal particles;[10,77,78] there are also reports of the observation of thermal percolation in polymer composites with various fillers, such as carbon nanotubes,[56,59-61,79-82] graphene,[25,26,82-93] boron nitride,[26,84,94-96] metal particles/nanowires,[54,97-100] metal oxide particles,[101] and ceramic sheets.[102]

### B. Reasoning behind the apparent lack of thermal percolation

Two main reasons are generally given for the lack of thermal percolation in the literature.[21-23,103-107] The first reason is the relatively low thermal conductivity ratio, of the order of 1000, for typical fillers (~100 W m$^{-1}$ K$^{-1}$) and polymer matrices (~0.1 W m$^{-1}$ K$^{-1}$). The corresponding ratio for electrical conductivity in polymer composites is very large, in the range of $10^{14}$–$10^{22}$, for typical conductive fillers ($10^4$–$10^6$ S m$^{-1}$) and matrices (~$10^{-10}$–$10^{-16}$ S m$^{-1}$). This argument is indeed true, and one should not expect a sharp increase in thermal conductivity at the percolation threshold like those seen in electrical conductivity. However, technically speaking, this statement does not preclude the existence of thermal percolation in polymer composites, and even a two-order of magnitude difference in the thermal conductivity of the filler and the matrix is sufficient to expect several times enhancement of the composite thermal conductivity in the vicinity of the percolation threshold, at least based on the existing classical percolation models.[56,104,108]

The second and perhaps the most important reason proposed for the lack of thermal percolation in polymer composites is the high thermal contact resistances between the filler particles. Keblinski and coworkers,[21] in their original article on the lack of thermal percolation, based on molecular dynamics (MD) simulations, argued that the normalized thermal contact resistance per unit area for SWCNT–SWCNT contacts was in the order of the SWCNT–Polymer interfacial thermal resistance, both around ~$10^{-7}$ m$^2$ K W$^{-1}$. They concluded that since the thermal resistances of the CNT–CNT contact and the CNT–Polymer interface were of the same order, considering the very small contact area between nanotubes (point contact), heat has no preference to stay within the network of nanotubes and easily leaks from the walls of nanotubes into the polymer matrix. This favors the dominance of the heat diffusion mechanism, which is in sharp contrast to the picture of thermal percolation within a conductive network (for a comparative discussion of percolation and diffusion processes, see, for example, Refs. 109 and 110). However, a subsequent experiment by Yang *et al.*[111] showed that the thermal contact resistance between multiwalled carbon nanotubes in both crossed and aligned configurations is two orders of magnitude lower than previous estimates, including the MD simulations by Keblinski,[21] and is around $10^{-9}$ m$^2$ K W$^{-1}$. Eventually, in a joint collaboration, Yang, Keblinski, and several other leading researchers of the field,[112] explained this apparent contradiction and found, surprisingly, that the thermal contact resistance normalized to the contact area for CNT–CNT and Graphene–Graphene junctions, linearly depends on the number of layers. Thus, for example, MWCNT–MWCNT junctions have a lower normalized thermal contact resistance than SWCNT–SWCNT junctions, and MWCNTs with thicker walls show better thermal coupling. In light of this new understanding, one may naturally expect that the idea of thermal percolation in polymer composites is revived, at least for fillers such as MWCNTs and multilayer graphene nanosheets.

### C. Our viewpoint and approach

It is evident from the above discussion that recent findings strongly support the existence of thermal percolation in polymer composites. Here, we study the thermal percolation problem from both theoretical and experimental perspectives. From the theoretical point of view, we develop a generalized percolation framework that describes both electrical and thermal conductivity within a remarkably wide range of filler-to-matrix conductivity ratios ($Y_f/Y_m$) covering 20 orders of magnitude. In this respect, a novel phenomenological equation of percolation is proposed as a generalization of the classical power-law percolation model. All the parameters of this equation have a clear and definite physical meaning. It provides a genuine classification of electrical conductivity with $Y_f/Y_m \geq 10^{10}$ as insulator–conductor percolation with the standard power-law behavior, and of thermal conductivity with $10^2 \leq Y_f/Y_m \leq 10^4$ as poor–good conductor percolation characterized by two universal critical exponents. Experimental verification of our unifying theoretical framework is conducted by constructing a 3D segregated and well-extended network of multiwalled carbon nanotubes (MWCNTs) in polypropylene (PP) as a model polymer matrix under a carefully designed fabrication method. Unlike conventional fabrication methods of polymer composites, such as melt mixing, solution mixing, and in situ polymerization,[16,113-115] our novel fabrication method ensures effective wall-to-wall interconnections between MWCNTs at their junctions beyond simple point contacts and without interfacial polymer layer, thus maximizes thermal coupling in the constructed conductive network inside the composites. In a systematic way, we study the evolution of the electrical and thermal conductivity in our fabricated nanocomposites with different loading levels from 0.01 to 10 wt% (equivalent to 0.004 to 4.5 vol%). Significantly, we find an ultralow electrical percolation threshold at 0.02 vol% and a record-low thermal percolation threshold at 1.5 vol% in our MWCNT–PP nanocomposites (Table S2 in the supplementary material). We also apply our theoretical model to a number of 23 independent experimental and numerical datasets reported in the literature, including more than 350 data points, for systems with different microscopic details, and show that all data collapse onto our proposed universal scaling function. We believe that our study sheds light on the underlying physics of thermal percolation and can contribute significantly to resolving the remaining issues in this field toward the realization of its full potential for the fabrication of advanced thermally and/or electrically conductive polymer composites.

## II. THEORY

In this section, before presenting our experimental results, we first highlight some important aspects of percolation theory that are geared toward developing a unified model for electrical and thermal percolation in polymer composites. Specifically, we emphasize the importance of matrix to filler conductivity ratio in percolation behavior. In this regard, we consider three different types of percolation in composites with conductive filler particles [Figs. 1(a)–1(c)] and briefly discuss the governing equation for each type. Then, we propose a master phenomenological equation, which we call the Threshold-Adjusted Generalized Percolation (TAGP) equation, that holds over the whole range of conductivity ratios and thus covers all the percolation types as well as the smooth crossover transition. Without losing generality, in the following discussion, we use the symbol $Y$ as a representation of electrical conductivity ($\sigma$) or thermal conductivity ($k$). The volume percent of filler in composites (which varies between 0 to 100 vol%) is denoted by $\phi$. Matrix and filler are also denoted by "$m$" and "$f$" subscripts, respectively.

### A. Classification

The first type of percolation systems that we consider here, as shown in Fig. 1(a), comprises composites of a conductive matrix with a non-zero conductivity ($Y_m \neq 0$) filled with a superconductive filler ($Y_f \to \infty$). Such a situation is commonly observed in polycrystalline superconductors near their transition temperature.[116-118] In these granular systems, by lowering the temperature, the junctions between superconducting grains randomly reposition from resistive junctions (with ohmic characteristics) to superconducting junctions (with Josephson coupling). When the fraction of superconducting junctions exceeds a specific threshold, called the percolation threshold ($\phi_c$), a sample-spanning superconducting cluster of coupled grains emerges and the whole system becomes a superconductor. The governing equation for this class is of the following power-law form,

$$Y_{eff} \sim (\phi_c - \phi)^{-s}\,, \tag{1}$$

where $s$ is a critical exponent (sometimes called superconductivity or subcritical or pre-percolation exponent) with accepted universal values of 1.30 for 2D systems and 0.73 for 3D systems.[14,15,119] According to Eq. (1) and as can be seen from Fig. 1(a), the effective conductivity of the composite progressively increases by approaching the percolation threshold from the bottom and diverges at this point, indicative of conductor–superconductor transition.

Figure 1(b) shows the second type of percolation in composites, which is more familiar to researchers in the field. In this type, the matrix is considered a perfect insulator ($Y_m = 0$) and the filler is a good conductor (though not a superconductor) with a finite conductivity ($Y_f > 0$). Here, the governing equation is,

$$Y_{eff} \sim (\phi - \phi_c)^{t}\,, \tag{2}$$

where $t$ is another critical exponent (sometimes called conductivity or supercritical or post-percolation exponent) with universal values of 1.30 for 2D and 2.0 for 3D systems.[14,15,120] In this type, starting from the insulating matrix ($\phi = 0$), in the subcritical $\phi < \phi_c$ phase, overlooking the finite-size effects, the conductive filler particles constitute finite isolated clusters (wrapped by the perfect insulating matrix) with no effective contribution in the macroscopic conductivity. However, by reaching the percolation threshold ($\phi_c$), a dramatic change in conductivity (several orders of magnitude) occurs within a very short range of filler loading (a few percent), and the composite undergoes an insulator–conductor transition. At this point, global connectivity emerges and some isolated filler clusters suddenly join together to form a continuous macroscopic network that spans the entire composite. This network, which is called the percolation or infinite cluster, becomes more and more widespread as the filler loading increases and eventually encompasses all isolated clusters of filler particles. Since the electrical conductivity of polymer matrices and fillers typically lie within the range of $10^{-10} - 10^{-16}$ S m$^{-1}$ and $10^{4} - 10^{6}$ S m$^{-1}$, respectively, so the general assumption of conductive filler in an insulating matrix seems quite plausible for the most electrically conductive polymer composites. This has been validated by numerous experiments in the past.[6,11,12]

The third type of percolation problems in composites concerns the mixture of a poor conducting (but not insulating) matrix ($Y_m \neq 0$) and a good conducting (but not superconducting) filler ($Y_f = \text{finite}$). Figure 1(c) shows a schematic illustration of the effective conductivity in this type, which asymptotically approaches the power-law equations of types I and II for $\phi \ll \phi_c$ and $\phi \gg \phi_c$, respectively, and exhibits an intermediate behavior around $\phi_c$. The less familiar type III percolation has been studied by a number of researchers, including Efros,[121] Straley,[122] and Stauffer,[123] especially in analyzing the effective conductivity of metal–dielectric composites under alternating current (AC) at finite frequencies. In such studies with complex AC conductivity, the imaginary component of the electrical conductivity in the dielectric matrix (with capacitive nature) acts as a poor conductor and the real part in the metal filler (with resistive nature) acts as a good conductor.

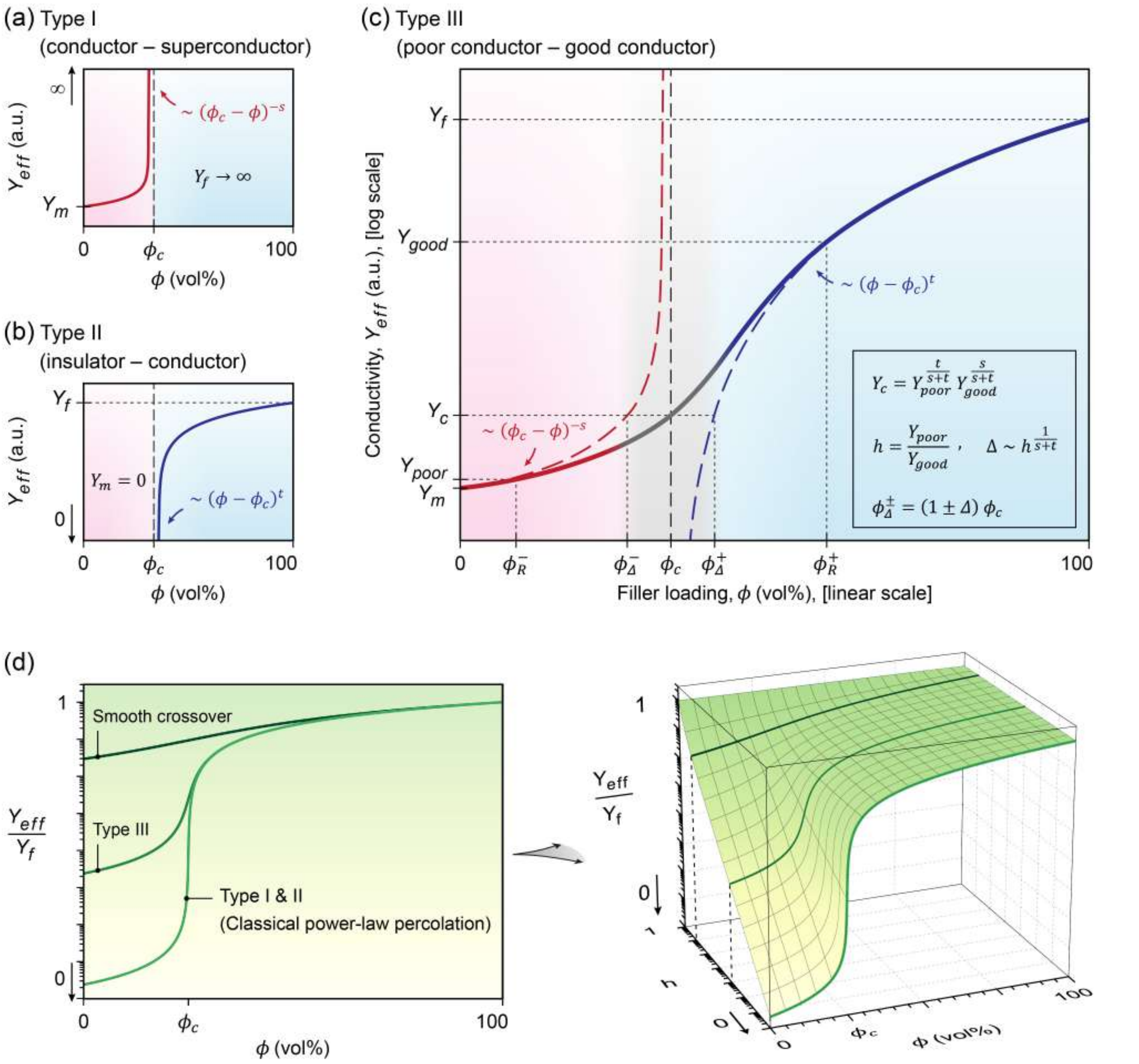

**FIG. 1.** Classification of percolation problems according to the conductivity ratio of the filler to the matrix. (a) Type I, conductor–superconductor. (b) Type II, insulator–conductor. (c) Type III, poor conductor–good conductor. (d)The general form of dimensionless effective conductivity ($Y_{eff}/Y_f$) in terms of filler loading ($\phi$) and conductivity ratio ($h$) based on the TAGP equation.

The effective conductivity of a composite with type III percolation can be expressed by the following scaling relation,[15,121-128]

$$Y_{eff} = Y_{good}\, h^{\frac{t}{s+t}}\, \Psi\left(\epsilon\, h^{\frac{-1}{s+t}}\right), \tag{3}$$

with

$$\epsilon = \frac{\phi - \phi_c}{\phi_c}, \tag{4}$$

$$h = \frac{Y_{poor}}{Y_{good}} \cdot \tag{5}$$

Here, $\epsilon$ denotes for the normalized proximity to the percolation threshold, and $h$ for the conductivity ratio of the matrix to filler. $\Psi(\cdot)$ is a universal scaling function which can be expressed as a three-part piecewise function constructed by series expansion in the integer powers of $h$ for the values of $\phi$ larger or smaller than $\phi_c$ and in the integer powers of $\epsilon$ for the values around $\phi_c$ [see Eq. (S12) in the supplementary material]. The expansions in the leading term give

$$Y_{eff} \sim \begin{cases} Y_{poor}\left(\frac{\phi_c - \phi}{\phi_c}\right)^{-s}, & \phi < \phi_c, \quad h \to 0, \\ Y_{poor}^{\frac{t}{s+t}} Y_{good}^{\frac{s}{s+t}}, & \phi \to \phi_c, \quad 0 < h \ll 1, \\ Y_{good}\left(\frac{\phi - \phi_c}{\phi_c}\right)^{t}, & \phi > \phi_c, \quad h \to 0 \cdot \end{cases} \tag{6}$$

It is important to note that $Y_{poor}$ and $Y_{good}$ depend on the boundary conditions of the region within which the percolation theory can be safely applied, i.e., $[\phi_R^-, \phi_R^+]$ in Fig. 1(c), and thus may or may not be equal to the matrix ($Y_m$) and filler ($Y_f$) conductivity, the point that is also emphasized by other researchers.[129] In particular, the universal properties are characteristic features specific to the critical region close to the percolation threshold (critical point). In this region the macroscopic behavior of the system, regardless of its microscopic details, can be described by scaling laws and universal critical exponents. Away from the critical region, other phenomena such as localization or aggregation of filler particles may come to play and impact the effective conductivity of the composites.[130]

In type III percolation, the transition between the two subcritical $\propto (\phi_c - \phi)^{-s}$ and supercritical $\propto (\phi - \phi_c)^t$ asymptotic behaviors takes place in a so-called smearing region $[\phi_\Delta^-, \phi_\Delta^+]$, where $\phi_\Delta^\pm = (1 \pm \Delta)\,\phi_c$. The effective conductivity of the composite in the subcritical region is mainly affected by the poor conducting matrix, while in the supercritical region, due to the formation of a well-connected and well-extended network of filler particles, the effective conductivity is more influenced by the good conducting filler. But in the smearing region, both the matrix and the filler phase actively contribute to the effective conductivity of the composite, and the percolation threshold is the point at which the contribution of matrix and filler in the conduction process becomes comparable.[128] The parameter $\Delta$, which characterizes the width of the smearing region, scales with the conductivity ratio $h$ as,[125,128]

$$\Delta \sim h^{\frac{1}{s+t}}, \tag{7}$$

which predicts $\Delta \to 0$ for $h \to 0$. This is exactly the case for Types I and II percolation systems, in which the smearing region closes off and shrinks to zero as $h \to 0$ [note the absence of the shaded gray region in Figs. 1(a) and 1(b)].

### B. Generalized percolation equation

In practice, Eq. (6) is oversimplified to fit most experimental data because it assumes constant effective conductivity throughout the smearing region and fails to capture important features related to this transition region. Considering equations with higher-order terms such as Eq. (S12), also introduces additional free parameters in the model that could be the source of artifacts in the analysis. Therefore, it is highly desirable to provide a closed-form equation for effective conductivity that is compatible with the scaling relation [Eq. (3)] and can be easily used to fit experimental data with conventional numerical curve fitting methods. To this aim, we introduce the following phenomenological equation

$$\left(R - \frac{\phi - \phi_c}{\phi_c}\right)\frac{Y_{poor}^{\frac{1}{s}} - Y_{eff}^{\frac{1}{s}}}{Y_{poor}^{\frac{1}{s}} + Y_{eff}^{\frac{1}{s}}} + \left(R + \frac{\phi - \phi_c}{\phi_c}\right)\frac{Y_{good}^{\frac{1}{t}} - Y_{eff}^{\frac{1}{t}}}{Y_{good}^{\frac{1}{t}} + Y_{eff}^{\frac{1}{t}}} = 0, \tag{8}$$

with $R$ being the normalized radius of the critical region, i.e. $|(\phi - \phi_c)/\phi_c| \le R$, over which the percolation process dominates the conduction mechanism. Although the scaling region that also determines the range of validity of the scaling laws is, in principle, assumed to be a small critical region very close to the threshold, here, we propose to widen this region to the extent that the number of free parameters in our model is minimized. As we shall see in the following, our assumption is very well supported by analyzing various observations from numerical simulations and the experimental measurements. To this end, we demand the two amplitudes $Y_{poor}$ and $Y_{good}$ in Eq. (8) to be given by an extrapolation of the effective conductivity $Y_{eff}(\phi)$ to the filler volume fractions $\phi = 0$ and $\phi = 2\phi_c$, respectively, well away from the critical percolation threshold $\phi_c$. The implementation of our proposition gives rise to $Y_{poor} = Y_{eff}(0) = Y_m$ in the sub-critical regime, and $Y_{good} = Y_{eff}(2\phi_c) = Y_{f^*}$ in the supercritical region. Replacing $Y_{poor}$ with $Y_m$ has an extra practical advantage that the value of the matrix conductivity $Y_m$ is either a priori known or easily measurable. Our setting suggests a normalized radius $R = 1$ for the scaling region around the critical point. Consequently, Eq. (8) can be simplified to the following equation, which we call the Threshold-Adjusted Generalized Percolation (TAGP) equation

$$(2\phi_c - \phi)\frac{Y_m^{\frac{1}{s}} - Y_{eff}^{\frac{1}{s}}}{Y_m^{\frac{1}{s}} + Y_{eff}^{\frac{1}{s}}} + \phi\,\frac{Y_{f^*}^{\frac{1}{t}} - Y_{eff}^{\frac{1}{t}}}{Y_{f^*}^{\frac{1}{t}} + Y_{eff}^{\frac{1}{t}}} = 0 \cdot \tag{9}$$

The TAGP equation closely follows the scaling relation Eq. (3) and is highly efficient for numerical calculations. In addition, all the parameters of this equation have a clear and definite physical meaning. Our experimental observations, as well as reanalyzing the data for a wide range of experimental measurements and numerical simulations in the literature that will be discussed in Sec. IV.D, fully confirm the reliability and applicability of the TAGP equation.

It is worth noting that developing phase transition models to regions beyond the critical point is fairly common. As an example, for magnetic systems that have a well-known analogy with the mixed-conductor systems related to our present study,[121,122,124] within the phenomenological Landau theory of the Ising universality class, the critical region around the critical temperature ($T_c$) is considered $|(T-T_c)/T_c| \leq 1$, and the strength of the correlations (coupling) in the system is estimated from the extrapolation of the power-law behavior of the correlation length to $T = 2T_c$.[131] Similarly, in percolation systems, $|(\phi-\phi_c)/\phi_c| \leq 1$ or equivalently $0 \leq \phi \leq 2\phi_c$ has been identified by many researchers as the region where the universal scaling laws of percolation can be safely applied before transitioning from the percolation regime to the effective medium regime.[130,132-136]

It is worthwhile to stress that $Y_{f^*}$ in the TAGP equation is not an artificial parameter at all and, in contrast, it is more accessible parameter than the filler conductivity ($Y_f$) since $Y_{f^*}$ is often one of the main data points in the experimental data collection process. For instance, for many particulate fillers such as carbon nanotubes, it is by no means possible to make a sample with $\phi = 100$ vol% due to the maximum packing fraction limit (about 70 vol% for CNTs[137]) and thus $Y_f$ is beyond the reach of experimental measurements.

The TAGP equation, as its name implies, takes exactly the expected value from the scaling relation at the percolation threshold, i.e. $Y_c = Y_m^{t/(s+t)} Y_{f^*}^{s/(s+t)}$, and from this perspective, it is perfectly adjusted at the threshold point. It can also be easily shown that this equation satisfies the asymptotic behaviors of Eq. (6) at low and high $\phi$ values. Remarkably, the TAGP equation is a generalized equation that holds over the entire range of matrix to filler conductivity ratios ($0 \leq h \leq 1$). Figure 1(d) shows the general form of dimensionless effective conductivity ($Y_{eff}/Y_f$) in terms of filler loading ($\phi$) and conductivity ratio ($h$) based on the TAGP equation. This figure is also presented with more numerical information for 2D and 3D systems in the supplementary material, Sec. S3. It is clear from Fig. 1(d) that the TAGP equation gives rise to the Types I and II percolation models in the limit $h \to 0$. Moreover, in the limit $h \to 1$ where the conductivity of the matrix and the filler are comparable, our TAGP equation predicts a smooth crossover in perfect agreement with experimental observations (see following).

In Sec. S7 of the supplementary material, the problem of determining the exact boundary between Type III percolation and Types I and II percolation from one side and smooth crossover from the other side is considered in details. In addition to the dimensionality of the system and the conductivity ratio, this boundary also depends on the value of the percolation threshold. However, in most practical situations where the percolation threshold is in the range $0.1 \leq \phi_c \leq 10$, when $10 < Y_f/Y_m < 10^8$ we face Type III percolation problems, and thus the necessity to use the TAGP equation. For finite $Y_f$ (i.e., no superconducting filler) and $Y_f/Y_m > 10^8$, again in most practical situations, one can neglect the conductivity of the matrix compared to that of the filler and assume the matrix is an insulator. In this case, one may use TAGP equation interchangeably with the classical power-law percolation.

For the situations where the conductivity of the filler and the matrix are almost of the same order [$Y_f/Y_m \sim O(1)$], percolation is dominated by the diffusion process and a smooth crossover is observed from the conductivity of the matrix to the filler with no singularity. In this regime, our TAGP equation behaves like equations of the effective medium theory (EMT), such as the Maxwell[138-140] and Bruggeman asymmetric[141-143] equations.

There have been proposed other equations in the past which were to some extent compatible with the scaling relation Eq. (3). However, these equations either do not have correct asymptotic behavior at high and low volume fractions,[144-146] or do not give the conductivity around the percolation threshold accurately,[147,148] or are overparametrized giving rise to non-physical fit parameters.[129]

To mention a few, Straley has proposed three implicit parametric equations that must be solved simultaneously to find the conductivity. These equations, in addition to the main physical parameters of the problem ($Y_{poor}$, $Y_{good}$, $\phi_c$, $s$, $t$), contain two additional free parameters and an adjustable function to better match the scaling relation Eq. (3). Although Straley's set of parametric equations has been used to fit some experimental data,[149,150] the large number of involved parameters and its complex mathematical structure for numerical curve fitting have severely limited its widespread applicability and questioned the reliability of the analysis.

Another well-known equation as an alternative to the scaling relation Eq. (3) is the General Effective Media (GEM) equation proposed by McLachlan.[147,148,151] Despite some similarities to our TAGP equation, and obvious physical interpretation of the involved parameters, the GEM equation has an implicit assumption that percolation is valid in the whole range $0 \leq \phi \leq 100$ which is not necessarily true in general. Although the GEM equation respects the asymptotic behavior at limiting volume fractions, emphasis on covering the maximum possible range for $\phi$ causes some critical features to be washed out, and it overestimates the conductivity at the percolation threshold.[152,153]

Similar issue has been addressed for another well-known equation of effective media theory, the Bruggeman Symmetric equation.[144] A detailed comparison of the TAGP equation with some classical conductivity models, including (i) Rule of Mixtures, (ii) Inverse Rule of Mixtures, (iii) Maxwell, (iv) Hashin-Shtrikman, (v) Bruggeman Asymmetric, (vi) Bruggeman Symmetric and (vii) General Effective Media equation (GEM) is presented in the supplementary material, Sec. S4.

An important point to be addressed here is that although it is well known that classical EMT models exhibit a linear behavior between conductivity and filler loading in the dilute limit ($\phi \lesssim 10$ vol%), by increasing filler loading the behavior of conductivity in this these models become nonlinear (more details in Sec. IV.C). Therefore, one way to unambiguously prove the existence of thermal percolation in polymer composites is by achieving a superlinear increase in thermal conductivity with filler loading in the dilute limit, because at this limit the classical EMT models are in the linear regime and are unable to describe nonlinear behaviors. It is to note that the percolation threshold in polymer composites, like many other percolation systems, is a nonuniversal parameter depending on the composite fabrication method, the geometry of filler particles, conductive network topology, and so on.[154-156] Here, we develop a novel fabrication method based on the concept of segregated networks,[114,157,158] which specifically makes it possible to achieve very low percolation thresholds. In this method, the free spaces between compacted polymer particles are used as a scaffold to construct an interconnected and well-extended network of filler particles (space-filling particles). Thus, instead of filling the entire volume of the polymer composite homogeneously, the filler particles are sufficient to fill the available spaces between the polymer particles to form a conductive network (more details in the experimental section, Sec. III). This method, in particular, makes it possible to achieve very low percolation thresholds by adjusting the ratio of the diameter of the polymer particles to the size of the filler particles. To calculate the percolation threshold in composites from the geometry of the filler particles, there is a rigorous theory called Excluded Volume developed by Balberg.[130,159,160] As we will see in Sec. IV.A, electrical percolation in our samples follows the classical power-law percolation equation with an electrical percolation threshold at 0.02 vol%, which this ultralow threshold is well explained by the excluded volume theory. The results of thermal conductivity measurements in our samples show that the superlinear behavior of thermal conductivity versus filler loading also starts from very low loadings (~1 vol%), which cannot be described with any classical EMT models and leaves no doubt that thermal percolation has occurred in the samples [glance ahead briefly to Fig. 6(a)]. However, unlike electrical conductivity, the classical power-law percolation equation cannot satisfactorily describe the thermal conductivity behavior in our sample, and the thermal percolation threshold also no longer follows the excluded volume theory predictions.

In this respect, we argue that thermal percolation in polymer composites in general, and in our samples in particular, indeed belongs to the class of composites of poor and good conductors (Type III percolation characterized by two universal critical exponents $s$ and $t$) and should be modeled by the threshold-adjusted generalized percolation (TAGP) equation. In fact, the thermal conductivity of most polymers is around 0.1 W m$^{-1}$ K$^{-1}$, whereas the thermal conductivity of typical fillers lies in the range 10–1000 W m$^{-1}$ K$^{-1}$.[16,19] Thus, the filler to matrix thermal conductivity ratio in polymer composites may vary between $10^2$ to $10^4$, which definitely falls within the range of applicability of the TAGP equation ($10 < Y_f/Y_m < 10^8$). Note that the corresponding ratio for the electrical conductivity is very large, in the order of $10^{14}$–$10^{22}$, in conventional polymer composites.[6,30] Consequently, the electrical percolation in polymer composites should basically be pursued in the class of insulator–conductor composites (Type II) with the possibility of modeling by the standard power-law percolation. To further emphasize, one of the main reasons for the failure to observe thermal percolation in polymer composites to date has been the use of the inappropriate power-law percolation model in the experimental data analysis. However, we believe that our TAGP equation eliminates all the shortcomings of previous models and is able to properly capture the electrical and thermal percolation within a unified analysis framework.

## III. EXPERIMENT

The fabrication steps of MWCNT–PP nanocomposites (multiwalled carbon nanotubes in polypropylene matrix) and the concept of the segregated conductive network are shown by a simplified schematic in Fig. 2(a). Details are provided in Sec. S1.3 of Materials and Methods in the supplementary material. In brief, MWCNT–PP nanocomposites were fabricated at different carbon nanotube loading levels from 0.01 wt% to 10 wt% (equivalents to 0.004 vol% to 4.5 vol%, according to Table S1 in supplementary material). Initially, the as-received CVD-grown MWCNTs were deagglomerated or so-called unbundled into individual nanotubes. For this purpose, ultrasonication in appropriate solvents was used to prepare solutions of unbundled nanotubes (actually, stable dispersions) with different concentrations from 0.04 mg ml$^{-1}$ to 1.5 mg ml$^{-1}$. Also, primary millimeter-sized polypropylene pellets were powdered into fine microparticles with an average diameter of 50 ± 20 μm. Then, based on the desired weight percent of the nanotubes in the nanocomposite, the MWCNT solution with a specified volume and concentration was added to the container of the polymer powder. The addition process was carried out in one or more steps, each time as much as the retention capacity of the powder microparticles without floating the microparticles in the solution. Thus, the solution permeated all the spaces between the powder microparticles and wetted the surface of all microparticles by capillary action, and then the solvent was evaporated by heating in a vacuum (80–90 °C at ~3 mTorr) to create a uniform coating of MWCNTs on the surface of the polypropylene microparticles.

In the next step, MWCNT-coated PP microparticles were hot compression molded in a carefully selected temperature window ($T_{molding}$ = 168 ± 3 ° C) under relatively low pressure ($P_{molding}$ = 0.5 MPa). The setting of molding temperature and pressure is such that on the one hand, the PP microparticles could get easily softened and deformed. This facilitates their polymer chains to interpenetrate into each other and form a nonporous dense nanocomposite. But on the other hand, in order to maintain the structural interconnectivity of the MWCNT coating on the surface of the PP microparticles during the deformation process, the viscosity of the polymer melt should stay high enough. To do so, the interfacial regions between compressed PP microparticles are used as a scaffold to construct a tightly interconnected and well-extended segregated network of MWCNTs throughout the nanocomposite. After cooling and removing from the mold, the final MWCNT–PP nanocomposites were obtained in which the 3D segregated MWCNT network was firmly secured in its position.

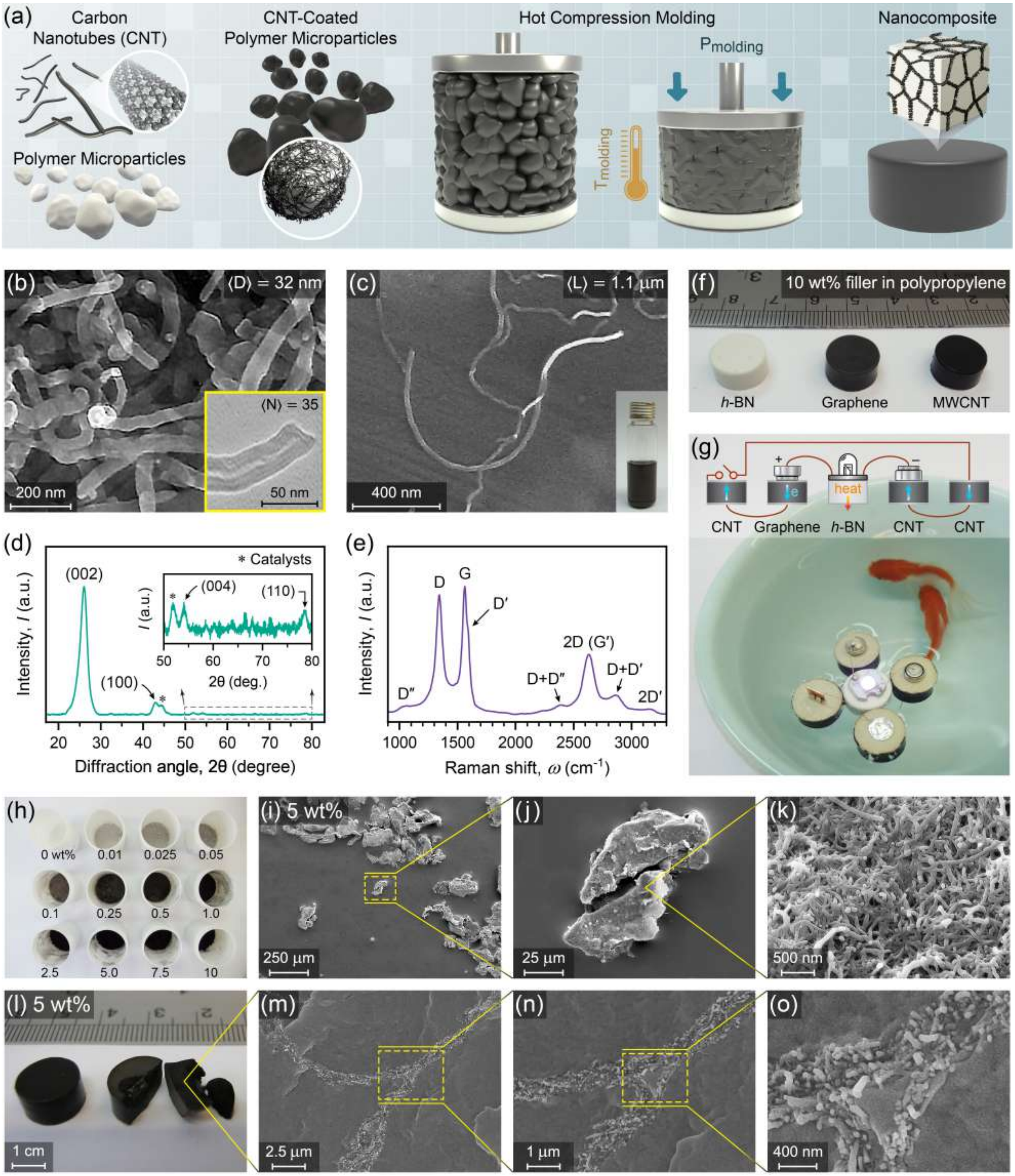

**FIG. 2.** Fabrication and structural characterization of nanocomposites. (a) Simplified schematic of the fabrication steps of MWCNT–PP nanocomposites (multiwalled carbon nanotubes in polypropylene matrix). (b) An example of SEM images of the MWCNTs to determine their average diameter. The inset shows an example of TEM images of an individual MWCNT to determine the average number of walls of the MWCNTs. (c) An example of SEM images of the MWCNTs to determine their average length. The inset shows an exemplary photograph of the prepared MWCNT solutions. (d) X-ray diffraction (XRD) pattern of the MWCNTs. (e) Raman spectrum of the MWCNTs. (f) An exemplary photograph of typical MWCNT–PP, Graphene–PP, and h-BN–PP nanocomposite samples at 10 wt% filler loading. (g) Conceptual demonstration of lightweight, electrically, and thermally conductive polymer composites in practice, using MWCNT–PP and Graphene–PP nanocomposites as part of the wiring of a simple electrical circuit to light a white LED lamp, and an h-BN–PP nanocomposite as a heat sink for the LED in such a way the whole set is floating on water due to low filler loading in the nanocomposites. (see Sec. S9.1 in the supplementary material for more elaboration). (h) PP powders, coated with different MWCNT loading levels from 0.01 to 10 wt% before hot compression molding. (i–k) Representative SEM images from the surface of the 5 wt% MWCNT-coated PP powder particles at different magnifications before molding. (l) Photograph of a typical and a cryofractured 5 wt% MWCNT–PP nanocomposite. (m–o) Representative SEM images from the cryofractured surface of a typical 5 wt% MWCNT–PP nanocomposite showing MWCNT conductive pathways inside it, at three different magnifications.

The main characterizations of our MWCNTs are presented in Figs. 2(b)–2(e). Based on the analysis of several Field Emission Scanning Electron Microscopy (FESEM) images, an example of which is shown in Fig. 2(b), the average diameter of the MWCNTs was determined $D = 32 \pm 6$ nm. Also, the average wall thickness of the MWCNTs was measured $t_{wall} = 12 \pm 2$ nm, based on Transmission Electron Microscopy (TEM) images, [inset of Fig. 2(b)]. Considering the well-known wall-to-wall distance of $d = 0.344$ nm in MWCNTs with a diameter greater than 10 nm,[161,162] the average number of walls in our MWCNTs was obtained $N = 35 \pm 5$. It is also clear from the inset of Fig. 2(b) that the cap termination of our MWCNTs is of the flat cap type, which is more common than the round cap type for such relatively large diameter carbon nanotubes.[163] The average length of the MWCNTs after unbundling by ultrasonication in solvents was measured $L = 1.1 \pm 0.3$ μm, based on the analysis of several SEM images, an example of which is shown in Fig. 2(c). N-Methyl-2-pyrrolidone (NMP) was used as the solvent for the preparation of MWCNT solutions with concentrations up to 0.1 mg ml$^{-1}$, while N-Cyclohexyl-2-pyrrolidone (CHP) was employed as the solvent for higher concentrations [inset of Fig. 2(c)]. Both NMP and CHP are known as excellent solvents for carbon nanotubes.[164] It is worth remarking that in our fabrication method, due to the use of appropriate solvents, there is no need for high ultrasonic powers and times to unbundle the carbon nanotubes. This minimizes the breaking off and length reduction of the nanotubes. For more information on preparation of the MWCNT solutions and related analysis of their stability and quality, see Sec. S1.3 in the supplementary material.

X-ray diffraction (XRD) analysis and Raman spectroscopy were used to evaluate the structural properties and crystalline quality of MWCNTs. Figure 2(d) shows the measured XRD pattern of the MWCNTs. The main (002) peak is clearly visible at $2\theta = 26.0°$ (with Cu Kα radiation and indexed on the basis of the hexagonal graphite crystal structure) that indicates the wall-to-wall distance of $d = 0.342$ nm in accordance with previous reports for large diameter (> 10 nm) MWCNTs.[161,162] This distance is greater than that of the interlayer spacing of crystalline graphite with AB stacking (0.335 nm). This is indicative of the random rotational stacking or so-called turbostratic stacking of the adjacent walls in MWCNTs due to the walls' curvature and geometric constraints.[165-167]

In turbostratic graphite structures, due to the lack of interlayer stacking correlation, two categories of peaks are observed in the XRD spectra. The categories consist of either symmetric or asymmetric (sawtooth shape) interlayer peaks from (00*l*) or (*hk*0) reflections, respectively. General (*hkl*) reflections are not observed in these structures.[165,166,168,169]

In this respect, the symmetric interwall peak of (004) at $2\theta = 54.1°$ and also the asymmetric intrawall peaks of (100) at $2\theta = 43.0°$ and (110) at $2\theta = 78.5°$ were identified. The rest of the weak peaks in the XRD pattern are assigned to the leftover catalyst particles. In particular, the Nickel (111) and Iron (110) peaks at $2\theta \sim 44°$, and the Nickel (200) peak at $2\theta \sim 52°$ can be identified [marked by * in Fig. 2(d)]. Based on the semi-quantitative analysis of these peaks, the amount of catalyst in MWCNTs was estimated to be less than 5 wt%, in agreement with the technical data sheet provided by the supplier.[170]

The Raman spectrum of the MWCNTs using 532 nm laser excitation is shown in Fig. 2(e). The characteristic G peak (related to the graphitic structure of $sp^2$ carbon materials) at $\omega_G \sim 1561$ cm$^{-1}$ and the D peak (activated by structural disorders/defects) at $\omega_D \sim 1344$ cm$^{-1}$ are clearly observed. The intensity ratio of the D-peak to the G-peak is close to 1 ($I_D/I_G \approx 0.9$) which could be a sign of structural disorders in our MWCNTs walls. Such disorders, like Stone–Wales defects, are quite expected in multiwalled carbon nanotubes mass-produced by the CVD method.[171,172] The two small intensity defect-induced peaks D′ and D″ (related to intravalley and intervalley double-resonance Raman processes, respectively) can also be identified at $\omega_{D'} \sim 1591$ cm$^{-1}$ and $\omega_{D''} \sim 1057$ cm$^{-1}$.[173,174] Moreover, the two overtone peaks 2D (G′) at $\omega_{2D} \sim 2633$ cm$^{-1}$ and 2D′ at $\omega_{2D'} \sim 3144$ cm$^{-1}$ , as well as the two combination peaks D+D′ at $\omega_{D+D'} \sim 2861$ cm$^{-1}$ and D+D″ at $\omega_{D+D''} \sim 2393$ cm$^{-1}$ are assigned in the figure.[171,174,175] It is noteworthy that, the 2D (G') peak can be fitted with only a single Lorentzian line shape rather than a doublet as in the Bernal AB-stacked graphite. This implies the turbostratic stacking of the MWCNT walls, which is consistent with our XRD results.[175,176]

In order to demonstrate the efficiency and versatility of our method for fabrication of conductive nanocomposites, in addition to MWCNT–PP nanocomposites, we also fabricated nanocomposites with graphene nanosheets and hexagonal boron nitride (h-BN) nanosheets as fillers in a similar manner. Details of the fabrication of the Graphene–PP and h-BN–PP nanocomposites are provided in Sections S1.4 and S1.5 of the supplementary material. While the h-BN–PP nanocomposites were electrically insulating, the electrical conductivity of the MWCNT–PP nanocomposites was always higher than that of the Graphene–PP nanocomposites at the same weight percentage filler loading. On the other hand, at the same weight percentage of filler loading, h-BN–PP nanocomposites had the highest thermal conductivity, and again MWCNT–PP nanocomposites performed better than Graphene–PP nanocomposites in terms of thermal conductivity. Figure 2(f) provides an example of the fabricated nanocomposites with 10 wt% filler in the polypropylene matrix. Figure 2(g) also demonstrates the concept of lightweight, electrically, and thermally conductive polymer composites in practice. For this purpose, a simple electrical circuit was built consisting of two button batteries (1.5 V), a white LED lamp (with a forward voltage of 2.5 V), four electrically conductive MWCNT–PP and Graphene–PP nanocomposites as part of the wiring of the circuit, and an electrically insulating h-BN–PP nanocomposite as the LED heat sink. The circuit schematic can be seen at the top of the panel. Considering the density of water (1.0 g cm$^{-3}$), polypropylene (0.9 g cm$^{-3}$), and fillers (~2.1 g cm$^{-3}$), the floating of the entire circuit on the water visually demonstrates that the filler loading in our nanocomposites is low. Also, the high brightness of the LED lamp confirms that despite low filler loading in the nanocomposites, they have high electrical conductivity. Moreover, the circuit was left floating on the water for more than an hour. Throughout this period, the luminous intensity of the LED was stable, indicating the good thermal conductivity of the h-BN–PP nanocomposite as a heat sink to transfer the heat generated by the LED to the water (see Sec. S9.1 in the supplementary material for more details). It is necessary to emphasize that, both sides of the MWCNT–PP and Graphene–PP nanocomposites were coated with silver paste to reduce their electrical contact resistance with the connecting wires (note the metallic gray color of the surface of these nanocomposites in the figure). Also, a thin layer of thermal paste was applied between the LED lamp and the h-BN–PP nanocomposite to reduce their thermal contact resistance.

In the following, we mainly focus on electrical and thermal percolation in MWCNT–PP nanocomposites, but there are similar trends in Graphene–PP and h-BN–PP nanocomposites, and our discussion is not limited to a specific filler or matrix. Figure 2(h) shows an example of prepared PP powder samples, coated with different MWCNT loading levels from 0.01 to 10 wt% before hot compression molding. Representative SEM images in Figs. 2(i) and 2(j) reveal that the polymer microparticles had an irregular shape with sizes in the range 50 ± 20 μm. It is also clear from Fig. 2(k) that, for the sample case of PP powder with 5 wt% MWCNT loading, the surface of microparticles was uniformly and densely covered with MWCNTs. To assess the morphology of the formed segregated MWCNT network inside the nanocomposites after hot compression molding, as shown in Fig. 2(l), some of the fabricated samples were cryofractured using liquid nitrogen at -196 °C (well below the glass transition temperature of polypropylene at -20 °C).[177] SEM images in Figs. 2(m)–2(o) show, for instance, typical MWCNT conductive pathways inside a 5 wt% nanocomposite at three different magnifications. The narrow width of the pathways (<1 μm), their sample-spanning feature, and effective wall-to-wall interconnections between MWCNTs at their junctions beyond simple point contacts without interfacial polymer layer are notable in these images. More SEM images from the other powders and nanocomposite samples will be presented in subsequent sections according to the topic under discussion.

Fabrication of MWCNT–PP nanocomposites is usually reported in the literature by one of the three routine methods: (i) Melt mixing[178,179] of MWCNT and PP at high temperatures (>200 °C), (ii) Solution mixing[180,181] of MWCNT and PP in the same solvent, such as xylene, and (iii) In situ polymerization[182,183] of the mixture of MWCNTs and propylene monomers in a batch reactor. These fabrication methods provide a good and homogenous dispersion of MWCNTs in the PP matrix, which leads to improvement in several mechanical properties, such as strength, stiffness, and toughness. However, enhancing electrical and thermal transport properties, particularly thermal conductivity, with these routine methods usually requires high filler loadings. This is while the fabrication of nanocomposites with high filler loadings is rather challenging due to their poor processability and, at the same time, the final products suffer from degraded mechanical properties, brittleness, increased density, as well as high costs. This further implies that electrical and thermal conductivity must be distinguished from mechanical properties, meaning that, instead of a homogenous dispersion of filler, construction of a conducting network of filler particles within the host matrix is required to improve conductivity properties. In this vein, composite fabrication methods based on segregated structures have received increasing attention in recent years.[114,157,184] There also exists a few reports on the fabrication of MWCNT–PP nanocomposites with segregated MWCNT networks;[185-187] however, in the present study, we devise a modified technique that offers several improvements and advantages over the previous studies.

The main distinguishing features of our fabrication technique are: (i) MWCNTs were completely unbundled by ultrasonic-assisted dispersion in excellent solvents of nanotubes, i.e. NMP and CHP, rather than choosing typically employed poor solvents, such as water, ethanol or xylene; (ii) The breaking off and shortening of MWCNTs has been minimized by utilizing an optimized ultrasonic power and time; (iii) A uniform coating of MWCNTs on the PP microparticles has been achieved by employing capillary action and percolation of MWCNT solutions into the PP powder, rather than dry mechanical mixing of MWCNTs and PP microparticles or mixing them in a liquid medium; (iv) The whole process of solvent evaporation has been performed in vacuum to prevent oxidation of nanotubes, polymer powders, and/or the organic solvent; (v) MWCNT-coated PP microparticles have been hot compression molded in an optimized temperature and pressure window to keep the architecture of the conductive network nearly intact during the melting and solidification of the polymer matrix and produce dense nanocomposites without cavities or air bubbles.

## IV. RESULTS AND DISCUSSION

### A. Electrical conductivity of MWCNT–PP nanocomposites

Figure 3(a) shows the electrical conductivity of MWCNT–PP nanocomposites as a function of nanotube loading. The electrical conductivity of neat polypropylene (0 vol% samples) is on the order of $10^{-15}$ S/m.[188,189] While the electrical conductivity of 0.004 vol% and 0.01 vol% samples is also extremely small (< $10^{-12}$ S/m, below our instrumental detection limit), at 0.02 vol% nanotube content, the electrical conductivity suddenly jumps to ~$10^{-7}$ S/m and then follows by a gradual increase with the nanotube loading until it reaches 28.6 S/m at 4.5 vol%. Fitting the experimental data points [orange circles in Fig. 3(a)] with the classical power-law percolation equation [solid blue line in Fig. 3(a)] offers an ultralow percolation threshold at $\phi_c = 0.024 \pm 0.004$ vol% ($\phi_c \approx 0.057$ wt%), where the uncertainty represents one standard error (SE) of the parameter estimate from the regression analysis. This value is among the lowest reported percolation threshold values for MWCNT–PP nanocomposites in the literature[11,185-187] and demonstrates the high efficiency of our fabrication method in the arrangement of nanotubes as a network within the polymer matrix with minimal agglomeration or dead-end branches. Moreover, we find that $t = 1.9 \pm 0.1$ (mean ± SE), in close agreement with the universal value $t = 2.0$ known for 3D percolation systems, which further confirms the formation of a well-extended 3D network of MWCNTs throughout the matrix. We use a normalized version of Eq. (2) (insulator–conductor percolation problems) to fit our experimental data shown in Fig. 3(a):

$$\sigma = \sigma_f \left( \frac{\phi - \phi_c}{100 - \phi_c} \right)^t , \qquad (10)$$

where $\sigma_f$ denotes the value of $\sigma$ at $\phi = 100$ vol%, i.e., the bulk electrical conductivity of filler. The best fit [Eq. (10)] to our experimental data gives $\sigma_f = 18836$ S/m which is in agreement with the technical data sheet value $\sigma_f > 10000$ S/m, provided by the manufacturer of carbon nanotubes.[170] It is to note that, typically, an individual MWCNT has a highly anisotropic electrical conductivity of $\sim 10^6$ S/m along the tube axis and just $\sim 10$ S/m in the radial direction.[190-193] The value for $\sigma_f$ falls indeed between these two limits since both axial and radial conduction effectively contribute to the bulk conductivity of carbon nanotubes. In addition, contact resistance at nanotube–nanotube junctions, including constriction and tunneling/hopping resistance, also plays a pivotal role in the bulk conductivity of MWCNTs.[194-196] Therefore, $\sigma_f$ actually accounts for several complex phenomena, such as anisotropic charge transfer (electrons and holes) within individual nanotubes and quantum tunneling and/or thermally activated hopping charge transfer between nanotubes.[30,197-202] The electrical conductivity of dense MWCNT compacts has been measured to be $\sim 10^3$ S/m, which considering the filling fraction of compacts, yields a total bulk conductivity of $\sim 10^4$ S/m for MWCNTs, in agreement with our finding $\sigma_f = 18836$ S/m.[203-206] Eq. (10) predicts, $\sigma = 0$ for $\phi = \phi_c$, which implies zero electrical conductivity of the nanocomposites at nanotube loadings equal to or lower than $\phi_c$ (Type II, insulator–conductor percolation model). The assumption of PP as a perfect insulating matrix is quite reasonable since the electrical conductivity of neat PP is $\sim 10^{-15}$ S/m, which can be neglected as compared to $\sim 10^4$ S/m for the bulk electrical conductivity of MWCNTs. One may also use TAGP to fit the electrical conductivity data; however, it gives exactly the same results ($t$, $\phi_c$, $\sigma_f$) as in the normalized power-law percolation equation [Eq. (10)]. The reason is that the filler to matrix conductivity ratio in our case is $\sim 10^{19}$ which is well above the $Y_f/Y_m > 10^8$ criterion discussed in Sec. II.B.

A slight deviation from the model is noticeable at the last data point ($\phi = 4.5$ vol% $= 10$ wt%) in the log–linear plot of electrical conductivity versus nanotube loading [Fig. 3(a)]. This deviation becomes more visible in the linear–linear plot, as shown in Fig. 3(b). Careful analysis of the surface of polymer powder particles by SEM, after coating them with carbon nanotubes and before compression molding, reveals that at $\phi = 4.5$ vol% a number of nanotube agglomerates (bundles) are formed on the surface of powder particles whereas the nanotube coating is almost uniform for $\phi \leq 2.2$ vol%. Representative SEM images of the uniform nanotube coating at $\phi = 2.2$ vol% and nanotube bundles at $\phi = 4.5$ vol% are shown in the inset of Fig. 3(b).

In view of our composite preparation procedure, which involves repeated cycles of saturation of PP powder with a solution of MWCNT followed by solvent evaporation, it appears that around $\phi = 4.5$ vol% the MWCNT coating on the surface of powder particles becomes so dense that nanotubes start to agglomerate in the presence of external perturbations, especially the growth and collapse of bubbles during solvent evaporation. However, a close look at our data and SEM images with wider fields of view (Fig. S8 in the supplementary material), indicates that the level of agglomeration is quite low at $\phi = 4.5$ vol% and the nucleation of nanotube bundles is negligible. Nevertheless, due to the onset of carbon nanotube bundling at $\phi = 4.5$ vol%, we did not prepare composites with higher filler loadings. Besides, from the application point of view, conductive composites with high filler loading levels are generally considered less attractive because of their poor processability, increased density, degraded mechanical properties, and high cost. However, it is possible to introduce higher loading of nanotubes into the composites without bundling problems, if necessary, by either slowing down the rate of solvent evaporation or lowering the concentration of MWCNT solution at the cost of increasing the number of saturation–evaporation cycles. One may alternatively prevent the agglomeration by increasing the effective surface area of the polymer scaffold through utilizing powder particles of smaller size. However, this will inevitably shift the percolation threshold to higher values, as for example demonstrated by a recent study on MWCNT–PMMA nanocomposites.[207] Nonetheless, as far as our analysis is concerned, the larger uncertainty in the last data point due to the nanotube bundling has a negligible effect on the estimation of our model parameters because this point is far from the percolation threshold. If we exclude the last data point in our analysis, we find the same $\phi_c = 0.024$ vol% with a slightly modified exponent $t = 2.1$ which is still close to the 3D universality with $t = 2.0$.

Figure 3(c) shows typical SEM images from the surface of polymer particles coated with MWCNTs before compression molding at nanotube loading levels close to the electrical percolation threshold. In agreement with the model, these images also validate the onset of percolation and formation of a system-spanning network of connected carbon nanotubes at $\phi \approx 0.02$ vol%. Furthermore, representative SEM images from the cryofractured samples in Fig. 3(d) demonstrate that a macroscopic connected network of nanotubes emerges around $\phi = 0.02$ vol% and becomes thicker and denser upon increasing the nanotube loading. This 3D conductive network of interconnected MWCNTs is actually constructed from connected patches of MWCNT thin films at the interfacial regions between polymer particles. The whole 3D structure is firmly secured in its position within the polymer matrix through hot compression molding. It can be seen from Fig. 3(d) that, in the vicinity of the electrical percolation threshold, the thickness of the walls in the conductive network (bright pathways in the figure) is less than 100 nm (2 to 4 MWCNT layers).

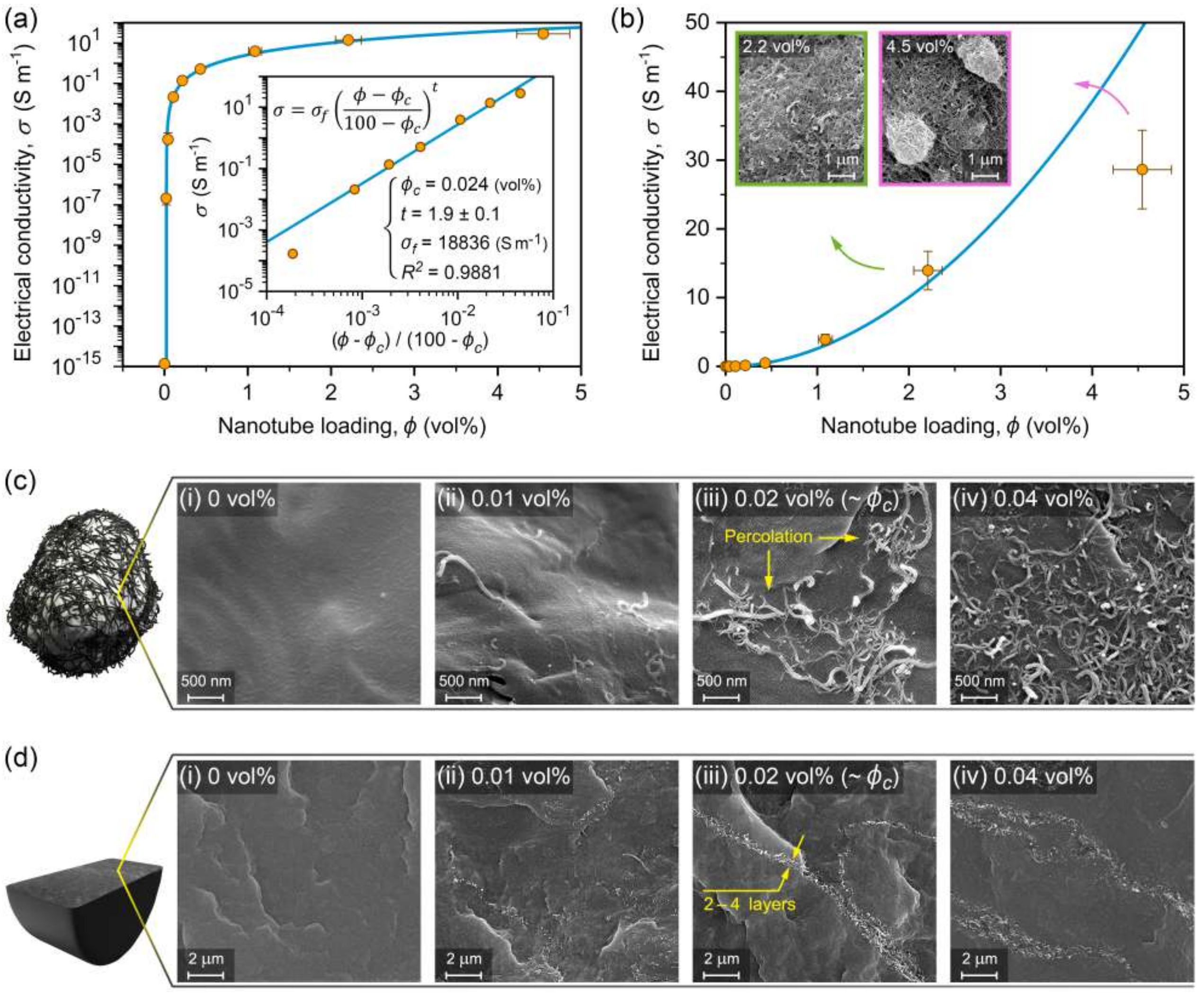

**FIG. 3.** Electrical conductivity of the MWCNT–PP nanocomposites. (a) Electrical conductivity as a function of nanotube loading on a log–linear plot. The inset shows a log–log plot of electrical conductivity versus reduced nanotube loading along with the best-fit parameters obtained from fitting the experimental data points (orange circles) to the normalized power-law percolation model (solid blue line). (b) Electrical conductivity as a function of nanotube loading on a linear–linear plot. The inset shows representative SEM images of the uniform nanotube coating on the surface of polymer powder at $\phi = 2.2$ vol% and nanotube bundles at $\phi = 4.5$ vol%. (c) Typical SEM images from the surface of MWCNT-coated PP powder particles before compression molding. (d) Typical SEM images from the cryofractured surfaces of nanocomposites. The vertical error bars represent one standard deviation (≥ 3 replicates). The horizontal error bars are calculated using the error propagation analysis (see Sec. S1.6 in the supplementary material). If the error bars are not visible in some data points, it is due to that they are smaller than the bullet size. Note that the two illustrations (not to scale) on the left-most side of panels c and d are computer-generated graphics.

As mentioned earlier, the percolation threshold as a nonuniversal quantity may depend on various geometrical details such as the microstructure of the composite. This has been the subject of previous theoretical efforts.[30,114,208-210] One of the most successful and widely accepted theories is the "Soft-Core" model, based on the excluded volume theory.[155,159,210] For composites with a segregated network of spherocylindrical filler particles (with average radius and length, $R_f$ and $L_f$, respectively) in a compressed matrix of spherical polymer particles with average radius $R_m \gg R_f$, the percolation threshold (in volume fraction), is given by,[10,209,210]

$$\varphi_c(\text{vol}) = \frac{(1+s)\left(\frac{4}{3}R_f^3 + R_f^2 L_f\right)}{\frac{32}{3}R_f^3 + 8R_f^2 L_f + R_f L_f^2} \times \left[1 - \left(1 - \frac{nR_f}{R_m}\right)^3\right], \quad (11)$$

where $s$ is a correction factor that accounts for the limited aspect ratio of filler particles (not infinitely long spherocylinders) and $n$ is the average number of filler layers in the thickness direction of conductive pathways between hot compression molded polymer particles. Berhan and Sastry,[209] based on a series of Monte Carlo simulations, developed an empirical equation for the aspect ratio correction factor as $s = c_1\left(R_f/L_f\right)^{c_2}$, which gives $c_1 = 5.231$ and $c_2 = 0.569$ for the case of the Soft-Core model. In addition, the term in the square brackets at the right-hand side of Eq. (11) accounts for the segregated structure of the composite.[10] In fact, contrary to homogenous composites, the available volume for the filler particles in segregated composites is limited to a thin shell around the starting matrix particles. Accordingly, $nR_f = nD_f/2$ can be interpreted as the average thickness of the filler shell on each individual polymer particle before compression molding. Thus, $2nR_f = nD_f$ is actually the thickness of conductive pathways within the composite, after compression molding. Since at the electrical percolation threshold we have $nR_f \ll R_m$, the above-mentioned term in the square brackets is almost equal to $3nR_f/R_m$ which is approximately equal to the ratio of the volume of the filler shell ($4\pi R_m^2 \times nR_f$) to the volume of the polymer particle ($4\pi R_m^3/3$). Eq. (11) can be further simplified by considering the flat caps of our carbon nanotubes and modeling them as cylinders [see TEM image in the inset of Fig. 2(b)].[210] Therefore, for segregated nanotube–polymer composites, Eq. (11) takes the following final form,

$$\phi_c(\text{v\%}) = 100\left[1 + 5.231\left(\frac{R_f}{L_f}\right)^{0.569}\right]\frac{R_f}{8R_f + L_f}\frac{3nR_f}{R_m} \cdot (12)$$

The value of $n$ at the percolation threshold can be estimated from Fig. 3(d)-iii as 3 layers. According to Fig. 2, the average radius and length of nanotubes are $R_f = 16$ nm and $L_f = 1.1\ \mu$m, respectively. The average radius of polymer particles is also $R_m = 25\ \mu$m. With these parameters, both Eqs. (11) and (12) give $\phi_c \approx 0.02$ vol%, in excellent agreement with our experimental observations. This proves that our specifically designed fabrication method provides a nearly perfect segregated network of MWCNTs, close to the theoretical prediction, without unwanted agglomeration or isolation of nanotubes in the polymer matrix. We also tried to calculate the percolation threshold from the Hard-Core@Soft-Shell model;[209] however, assuming a 2 nm tunneling distance[211] with other parameters similar to the Soft-Core model, $\phi_c \approx 0.1$ vol% was obtained. We, therefore, confirm the conclusion made by Balberg[212] that, in the case of nanotube–polymer composites, the Soft-Core model, which neglects tunneling effects and assumes filler particles as penetrable objects, yields more realistic results than the Hard-Core@Soft-Shell model, which albeit accounts for the tunneling transport between filler particles but considers them as rigid and non-flexible objects. The interpenetration of filler particles in the Soft-Core model can be imagined as the local bending of nanotubes to cross each other without any space restrictions.[212]

The last point we would like to discuss here is the range of applicability of percolation equations, for example, Eq. (10), especially at $\phi$ far above the percolation threshold (i.e., $\phi \gg 2\phi_c$). Although critical percolation theory is believed to explain the connectivity transition near the threshold,[123] one may wonder why the power-law percolation equation fits the experimental data very well in a broad range of filler loading. Keblinski and Cleri[132] have tried to address this issue based on the dominant role played by the contact resistances in the total conductivity of the percolating network in off-critical regions. Their argument comes as follows: By increasing the number of filler particles (proportional to the filler loading), the number of contacts between them (proportional to the conductivity) increases quadratically, which can indeed account for the origin of the scaling relation $\sigma \propto (\phi - \phi_c)^2$ valid at filler loading levels much larger than the threshold value. This statement has been recently revisited by Balberg,[130,155,213] in a more rigorous percolation framework which will be discussed in more detail at the end of the following subsection.

### B. Thermal conductivity of MWCNT–PP nanocomposites

Figure 4(a) shows the thermal conductivity of MWCNT–PP nanocomposites as a function of carbon nanotube loading. Unlike electrical conductivity, no abrupt increase of several orders of magnitude is observed here, and thermal conductivity increases smoothly, although still superlinearly, from the value of 0.15 W m$^{-1}$ K$^{-1}$ at 0 vol% nanotube content (neat PP) to 0.55 W m$^{-1}$ K$^{-1}$ at 4.5 vol% (10 wt%) MWCNTs, which is equivalent to 267% enhancement in thermal conductivity. Superlinear increase of thermal conductivity at such a low level of filler loading can be explained neither by the classical effective medium theory nor by the classical power-law percolation, as we will discuss in detail in Sec. IV.C. However, our developed Threshold-Adjusted Generalized Percolation (TAGP) equation can account for such a superlinear behavior, and the experimental data points [orange circles in Fig. 4(a)] are fitted very well to the TAGP equation [solid blue line in Fig. 4(a)]. In the same line of reasoning given for electrical conductivity, the slight deviation from the model at the last data point ($\phi = 4.5$ vol% $= 10$ wt%) can also be attributed to the onset of agglomeration of nanotubes at this loading level. By performing nonlinear curve fitting (details in S1.7. Fitting procedure in the supplementary material), we obtained the thermal percolation threshold of $\phi_c = 1.5 \pm 0.1$ vol%, where the uncertainty represents the 95% confidence interval (CI) of the parameter estimate from the regression analysis. To the best of our knowledge, this is the lowest reported thermal percolation threshold in the literature (see also Table S2 in the supplementary material), indicating the effectiveness of our nanocomposite fabrication method.

The dimensionality-related conductivity critical exponents of thermal percolation were obtained as $s_{th} = 1.3 \pm 0.1$ and $t_{th} = 1.3 \pm 0.2$, both at the 95% confidence level. These values are different from the well-known universal values of $s_{3D} = 0.73$ and $t_{3D} = 2.0$ for 3D percolation systems and are in principle equal to the universal values of $s_{2D} = t_{2D} = 1.3$ for 2D systems.[14,15] This may seem somewhat counterintuitive since the network of nanotubes within the polymer matrix is three-dimensional, and the dimensionality-related critical exponent of electrical conductivity was obtained $t_{el} = 1.9 \pm 0.1$, which also confirms the formation of a well-extended 3D network of nanotubes throughout the matrix. However, observation of nonuniversal values for the critical exponents in composites is common, especially when the percolation equation is used to describe conductivity far beyond the percolation threshold ($|\phi - \phi_c| \gg \phi_c$).[213] In fact, although in random discrete-lattice networks and their continuum counterparts, random composites, the critical behavior of the system does not depend on its microscopic details and depends only on the dimensionality of the system, clustering and spatial correlations under certain conditions can change the critical exponents and the fractal dimensionality of the system.[154,156,214,215] We will discuss this very important observation in more detail at the end of this subsection.

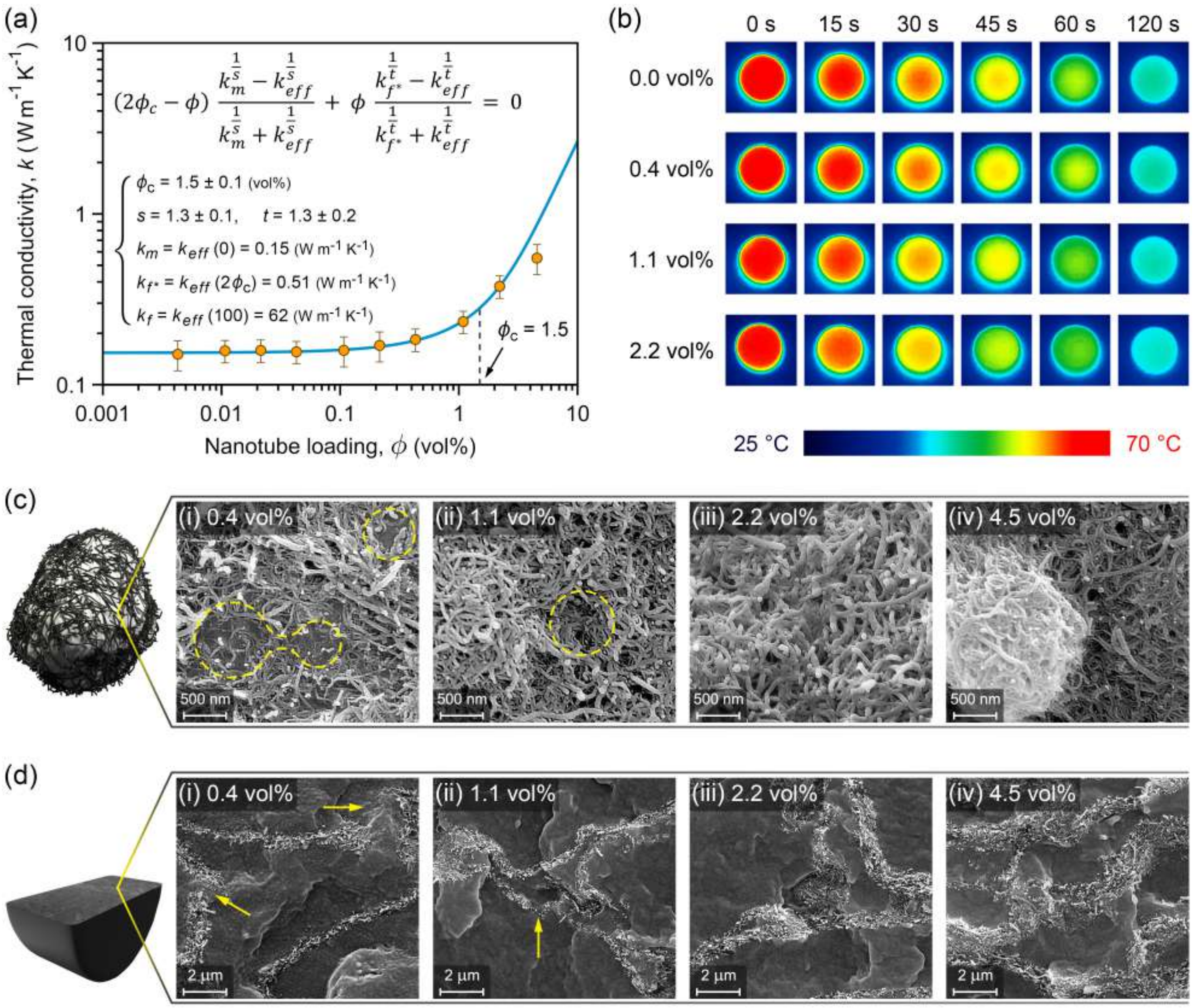

**FIG. 4.** Thermal conductivity of MWCNT–PP nanocomposites. (a) Thermal conductivity as a function of nanotube loading on a log–log plot along with the best-fit parameters obtained from fitting the experimental data points (orange circles) to the Threshold-Adjusted Generalized Percolation (TAGP) Equation (solid blue line). (b) Infrared thermal images of the temperature distribution at the top surface of four preheated samples during natural cooling in ambient air at specific time intervals. (c) Typical SEM images from the surface of MWCNT-coated PP powder particles before compression molding. (d) Typical SEM images from the cryofractured surfaces of nanocomposites. The vertical error bars represent one standard deviation (≥ 3 replicates). The horizontal error bars are also calculated using the error propagation analysis but are not visible due to that they are smaller than the bullet size. (see Sec. S1.6 in the supplementary material). Note that the two illustrations (not to scale) on the left-most side of panels c and d are computer-generated graphics.

Bulk thermal conductivity of matrix and filler were also obtained, $k_m$ = 0.15 W m$^{-1}$ K$^{-1}$ and $k_f$ = 62 W m$^{-1}$ K$^{-1}$, respectively. The obtained value for $k_m$ is in agreement with the reported values of thermal conductivity of isotactic homopolymer polypropylene in the literature.[19,216] As for $k_f$, please note that originally, the parameter that is present in the TAGP equation and is obtained from the fitting procedure is $k_{f^*}$, which is the effective thermal conductivity at $\phi = 2\phi_c$. Nonetheless, there is a direct relation between $k_{f^*}$ and $k_f$, and after determination of all the five parameters of the TAGP equation (i.e., $\phi_c$, $s$, $t$, $k_m$ and $k_{f^*}$), $k_f$ can be immediately calculated from the following equation,

$$k_{f^*} = k_f \left\{ \frac{\left[1 + \left(\frac{k_f}{k_m}\right)^{\frac{1}{s}}\right] - (2\varphi_c - 1)\left[1 - \left(\frac{k_f}{k_m}\right)^{\frac{1}{s}}\right]}{\left[1 + \left(\frac{k_f}{k_m}\right)^{\frac{1}{s}}\right] + (2\varphi_c - 1)\left[1 - \left(\frac{k_f}{k_m}\right)^{\frac{1}{s}}\right]} \right\}^{t} . \quad (13)$$

The value of $k_f$, as expected, is also between the reported axial and radial thermal conductivity of an individual MWCNT.[31,217,218] Similar to the discussion that was given for the interpretation of the $\sigma_f$, as the bulk electrical conductivity of MWCNTs, $k_f$ can also be interpreted as the bulk thermal conductivity of dense MWCNT compacts. Indeed, considering the filling fraction, there is a very good agreement between $k_f$ and the reported bulk thermal conductivity of MWCNT compacts in the literature.[219-223] Moreover, in Sec. S8 in the supplementary material, the correlation of the filler bulk thermal conductivity ($k_f$) with the filler intrinsic thermal conductivity and the filler–filler thermal contact resistance was investigated. In this regard, we used two well-known models of Volkov–Zhigilei[31,38,224] and Zhao *et al.*[24,225] for our MWCNT–PP nanocomposites and found a close agreement between the obtained $k_f$ from TAGP and these two models.

The considerable increase in thermal conductivity around the thermal percolation threshold can also be visualized by the thermography technique. Figure 4(b) shows infrared (IR) thermal images of four samples which were preheated in an oven for 1 hour at 70 °C to ensure the same temperature all over the samples and then immediately transferred onto a heat insulator substrate to naturally cool down at ambient conditions. All samples were short cylinders with the same geometrical dimensions (diameter and height), and the instantaneous temperature of their top surface was recorded every 15 seconds by an infrared thermal imaging camera. It should be mentioned that the emissivity of MWCNT and polypropylene in the IR range is nearly perfect and almost equal to the emissivity of the black body.[226] Accordingly, we set the emissivity equal to 1 in the thermal camera settings for all samples. At the beginning of the cooling process ($t$ = 0 s), all samples were at the initial temperature of 70 °C, as is evident from the left-most column of IR images in Fig. 4(b). Since the samples were insulated from the bottom, they were primarily cooled down by natural convection from their top and lateral surfaces, and with similar dimensionality and boundary conditions, the transient temperature of the samples is governed by their thermal conductivity ($k$) and volumetric heat capacity ($\rho c$) based on the general heat equation [$k\nabla^2 T = \rho c(\partial T/\partial t)$].[227] However, according to direct measurements, density ($\rho$) and specific heat ($c$), and hence their product $\rho c$, is almost the same in all our nanocomposite samples (~1720 kJ $m^{-3}$ $K^{-1}$), which is mainly due to the low filler loading in the samples. Consequently, the temperature response of our samples has a direct correlation with their thermal conductivity. In a sample with higher thermal conductivity, accumulated internal heat is conducted at a higher rate (faster) from the interior of the sample to its outer surfaces, where the heat is dissipated to the surroundings. Taking $t$ = 30 s as an example, the IR images from the top surface of 0 vol% and 0.4 vol% samples are very similar (average temperature of 55.8 °C by image analysis), thus their thermal conductivities should also be almost equal. So, we do not expect appreciable enhancement in thermal conductivity up to 0.4 vol% nanotube loading in harmony with the results of direct thermal conductivity measurement presented in Fig. 4(a). On the other hand, the average temperature of the 1.1 vol% sample is lower than the two previous samples (53.4 °C), which indicates its higher thermal conductivity. More interestingly, the IR images of the 2.2 vol% sample exhibit a nearly uniform temperature distribution all over its top surface (51.4 °C), which is obviously lower than the other samples, visual proof of its considerably higher thermal conductivity. Therefore, our IR thermal images are consistent with the calculated thermal percolation threshold at $\phi_c$ = 1.5 vol% and help visualize a considerable increase in thermal conductivity around this point.

Figure 4(c) shows typical SEM images from the surface of polymer powder particles coated with MWCNTs before compression molding at nanotube loading levels close to the thermal percolation threshold. Careful analysis of these images (and several other images that have not been shown here due to space limitations), has led us to believe that there exists a correlation between the full coverage of the surface of powder particles and the thermal percolation threshold. For powder particles coated with nanotubes at loading levels lower than the thermal percolation threshold ($\phi$ < 1.5 vol%), always some bare regions of polymer without nanotube coating with appreciable area compared to particle area, can be found as indicated for 0.4 vol% and 1.1 vol% powder particles in Fig. 4(c). However, nanotube coating for powder particles with $\phi$ > 1.5 vol% is almost uniform and continuous. As discussed before, far above the thermal percolation threshold, the nanotube coating becomes so dense that nanotubes start to agglomerate, and extra care is needed to avoid the nucleation of nanotube bundles. Representative SEM images from the cryofractured samples around the thermal percolation threshold are provided in Fig. 4(d). Here also, a meaningful relationship between the continuity of nanotube conductive pathways and the thermal percolation threshold can be realized. For 0.4 vol% and 1.1 vol%, as indicated by arrows in the image, there are observable discontinuities in some conductive pathways on the order of their width (i.e., several tens of nanometers), but almost all thick pathways in 2.2 vol% and 4.5 vol% are continuous at least at the level of a hundredth to a thousandth of their width (i.e., below a few nanometers).

An interesting point here is the difference between the thermal percolation threshold ($\phi_{c,th}$ = 1.5 vol%) and the electrical percolation threshold ($\phi_{c,el}$ = 0.024 vol%). This may be surprising, since Fourier's law of heat conduction and Ohm's law of electrical conduction are mathematically equivalent continuum equations (Thermal-Electrical Analogy), and the same network of MWCNTs that conducts electricity can also, in principle, conduct heat.[21] The difference can be understood if one distinguishes between different types of percolation problems. In its very basic definition, the percolation threshold in binary composites is the point at which the share of two constituent phases is equalized in the corresponding transport process.[128] In electrically insulator–conductor systems, the transition from the insulating state to the conducting state occurs in a very narrow range of filler loading around the electrical percolation threshold (electrical smearing region), and practically as soon as a sample-spanning connected network of filler emerges, the composite switches from an insulator to a conductor.

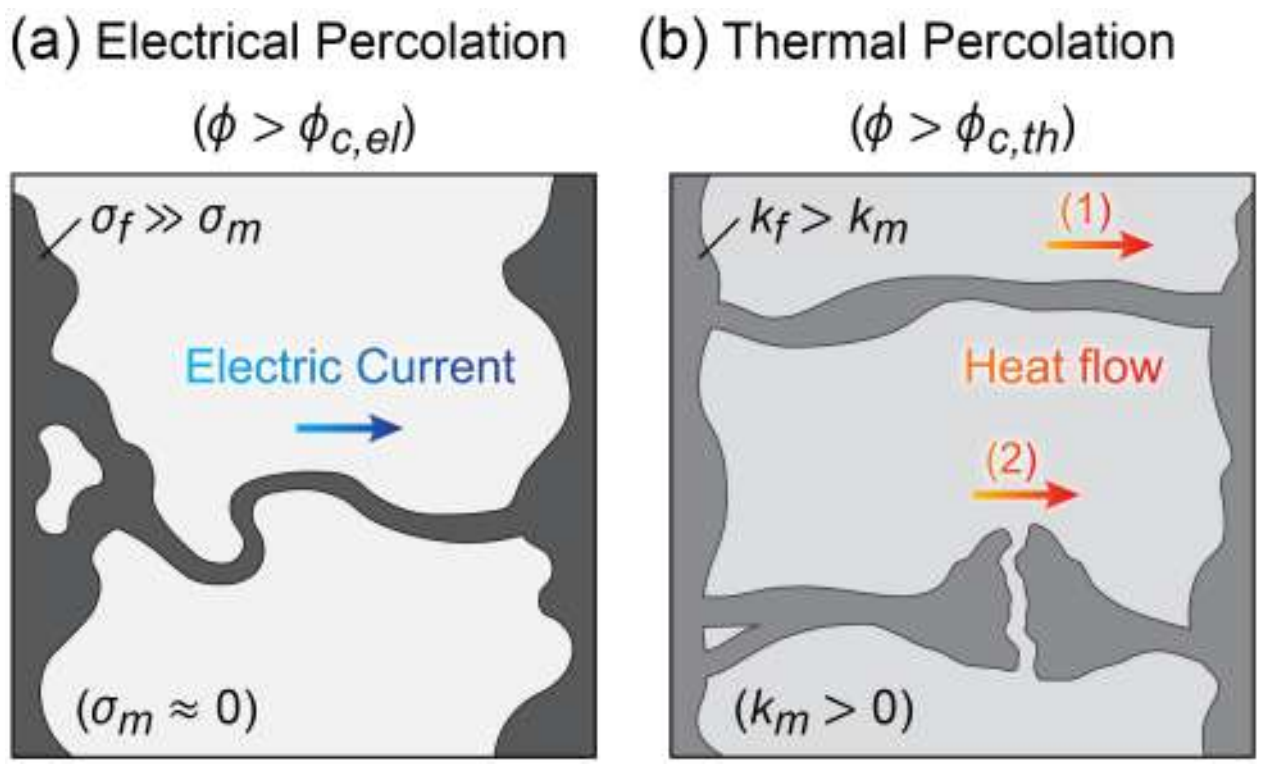

**FIG. 5.** Models of percolation structure and conductive pathways above the (a) electrical, and (b) thermal percolation thresholds. Adapted with permission from Figure 5.13 of Ref. 128. Copyright 2016 Springer.

On the other hand, in a composite of thermally poor and good conductors, both constituent phases can conduct heat, and generally, filler particles (good conductor) are thermally coupled through the matrix (poor conductor) even before that they are physically connected. Thus, both the poor conducting phase and good conducting phase actively contribute to heat transfer in a wide range of filler loading around the thermal percolation threshold (thermal smearing region), and the thermal percolation threshold is just the point at which the share of matrix and filler in thermal conduction becomes equal. Thus, in our nanocomposites, although emerging of a network of filler at the electrical percolation threshold also provides an exclusive pathway for heat transfer through filler particles; however, the heat transfer rate in this sparse network cannot be still competitive to the share of the matrix in total heat transfer rate. Accordingly, the thermal percolation threshold occurs when the network becomes more widespread and denser upon further increase of the nanotube loading. Our calculations based on Eq. (7) indicate that the thermal smearing region spans from 0.56 vol% to 2.44 vol% (1.5×(1 ± 0.62 vol%)), thus in this region, the contribution of both filler and matrix in heat transfer is important. Below 0.56 vol%, the thermal conductivity of the composite is ruled out by the matrix, whereas above the 2.44 vol%, it is governed by the filler. The electrical smearing region, on the contrary, has a very narrow width of $4.4\times10^{-5}$ vol% around $\phi_{c,el}$ = 0.024 vol%, and transition from the insulating state to the conducting state takes place in this narrow region.

In the previous subsection, we demonstrated that the observed electrical percolation threshold ($\phi_{c,el}$ = 0.024 vol%) could be directly related to the microstructure of the composite through the Soft-Core model. However, this model cannot explain the observed thermal percolation threshold at $\phi_{c,th}$ = 1.5 vol%. In fact, the Soft-Core model, which is based on the excluded volume theory, neglects the effect of matrix and assumes two filler particles are connected only when they are in geometric contacts.[155,159,210] While these assumptions are consistent with the situation of electrical percolation, they do not necessarily hold for thermal percolation. According to Fig. 5(a), the percolation structure for electric charge transfer above the electrical percolation threshold comprises narrow continuous conductive pathways of connected filler particles. But, the percolation structure above the thermal percolation threshold may comprise two types of pathways. One type, which is common between electrical and thermal conduction, is narrow continuous pathways completely composed of good conducting phase [pathway 1 in Fig. 5(b)] and the other type, which is peculiar to thermal conduction, is thick discontinuous pathways of good conducting phase with a thin spacer of poor conducting phase [pathway 2 in Fig. 5(b)].[128] The thermal resistance of both types is of the same order, and the gap of poor conducting phase in the second type is compensated by its large cross-sectional area. The maximum allowable thickness of poor conducting phase scales inversely with the ratio of thermal conductivity of good and poor conductors but for typical thermal conductivity ratios of 100 to 1000 is on the order of a hundredth to a thousandth of the width of the pathway. Thus, the thermal percolation threshold does not necessarily require a connected network of filler particles to occur, and models that rely on geometric contacts and neglect the conductivity of the matrix fail to predict it.

In the context of connected filler particles, it should be clarified that two MWCNTs at ~2 nm distance from each other are effectively electrically connected through quantum tunneling, and when they come closer to each other, the physics of charge transfer between them gradually changes to a stronger thermally activated hopping and eventually the band-like electrical conduction.[197-199,201,228] In a similar way, two MWCNTs at ~5 nm distance from each other are effectively thermally connected through Near-Field Infrared Radiation (Super-Planckian Thermal Radiation), and when they come closer to each other, the physics of heat transfer changes to the much more efficient thermal conduction at distances below ~1 nm.[201,228-232] Thus, beyond ~5 nm, two MWCNTs can be practically assumed electrically and thermally disconnected, although they may still be thermally coupled (part of a thermal pathway) through the polymer matrix.

Balberg recently proposed a comprehensive model for nonuniversal behavior in percolation systems.[213] Based on this model, the reasons for observing nonuniversal values for the critical conductivity exponent of 3D systems ($t_{3D}$ = 2.0) can be categorized as (i) long-range interactions (correlations) between filler particles due to, for example, quantum mechanical tunneling and thermally activated hopping, which lead to $2 < t < 10$, (ii) large contact resistances between filler particles, which result in $2 < t < 4$, (iii) reduced dimensionality and formation of a 2D subnetwork within the 3D system, which yields $1.3 < t < 2$, (iv) crossover from percolation to the diffusion-dominated effective medium behavior, which causes $1 < t < 1.3$, and (v) severely constrained structural growth of the conductive network and aggregation (bundling) of the filler particles, which gives rise to $0 < t < 1$. According to Balberg's model, the observation of the smaller-than-universal conductivity critical exponents, which is repeatedly reported in the literature, especially for polymer nanocomposites with the segregated structure, is related to the interplay between the connectivity of the filler particles and the structural variations upon increasing filler content in the composite.[136] Here, connectivity defines as the average number of filler particles that are effectively in contact with a typical filler particle in the composite and is equivalent to the average number of bonds per site ($B$) in classical lattice bond percolation, which itself is given by $B = p_b Z$ where $p_b$ is the bond occupation probability and $Z$ is the lattice coordination number.

Basically, the underlying assumption for describing the critical behavior in continuum media based on $\phi - \phi_c$ is the proportionality of the connectivity and filler content ($B \propto \phi$).[136] This proportionality allows continuum percolation to be topologically mapped to a lattice with a suitable coordination number. Hence, $\phi - \phi_c$ plays the same role in continuum percolation as $p_b$-$p_{bc}$ in the corresponding lattice bond percolation problem.[233] But the relationship between $B$ and $\phi$ is not always linear and depends on the geometry of the filler particles. For a system consisting of long cylinders of length $L$ and radius $r$, as a model for carbon nanotube composites, $B \propto \phi L r^{-1} \langle \sin \theta \rangle$ where $\langle \sin \theta \rangle$ is averaged over all solid angles between connected cylinders in the system.[136] It is clear from this relation, for example, how aggregating nanotubes upon increasing $\phi$ and forming bundles of nanotubes in the composite with an effective radius greater than $r$ (radius of an individual nanotube) can change the dependence of $B$ on $\phi$ from a direct linear relationship to a sublinear form of $B \propto \phi^{\gamma}$, in which $\gamma < 1$. Thus, for the above example, percolation with the universal conductivity critical exponent based on connectivity, $\sigma \propto (B\text{-}B_c)^{t_{3D}}$ is observed as percolation with a smaller-than-universal critical exponent based on filler content $\sigma \propto (\phi\text{-}\phi_c)^t$ where $t \approx \gamma t_{3D}$.

According to the above discussion, the difference between the percolation critical exponents observed for electrical and thermal conductivity in our composites can be attributed to the difference in the nature of the electrical and thermal connectivity between carbon nanotubes.[234,235] Specifically, because MWCNTs are multiple graphene sheets that are concentrically rolled up, at the junction of MWCNTs, electrons and phonons must move perpendicular to the basal plane of constituting graphene sheets, i.e., along the $c$-axis. The electron mean free path along the $c$-axis in layered graphitic materials, such as graphite and MWCNT, is $\Lambda_{c,el} \leq 1$ nm which is comparable to their $c$-axis interlayer separation.[236-239] On the other hand, recent accurate computational and experimental studies indicate a long average phonon mean free path along the $c$-axis in graphitic materials, around $\Lambda_{c,ph} \geq 200$ nm.[112,240-242] From the connectivity point of view, because the coating of carbon nanotubes on our polymer microparticles is of the order of a few tens of nanometers and smaller than $\Lambda_{c,ph}$ (even at the highest loading level of $\phi$ = 4.5 vol%), increasing $\phi$ cannot change connectivity in the direction perpendicular to the surface of the microparticles. In other words, all the nanotubes in the thickness direction of the conductive pathways [Fig. 4(d)] are strongly correlated and their thermal vibrations are coordinated and harmonized. However, due to the very small $\Lambda_{c,el}$ of the electrons compared with the thickness of the nanotube pathways, increasing $\phi$ creates new paths for the electrons to move in the thickness direction of the pathways and increases the electrical connectivity. The dependence of the fractal dimension on the mean free paths of phonons in heat conduction networks of Euclidean dimension D = 3 has already been realized by some researchers and explained by introducing the concept of fractons.[243] Also, similar to our report for thermal conductivity, for the conduction of ions in polymer electrolytes,[244] as well as for the elastic properties of chalcogenide glasses,[245,246] a significant reduction in fractal dimensionality has been reported due to clustering and strong anisotropic spatial correlations.

### C. Comparison with classical models

As mentioned before, neither the classical effective medium theory (EMT) nor the classical power-law percolation cannot explain the superlinear increase of thermal conductivity in our nanocomposites. In Fig. 6(a), the fit of the TAGP equation (solid blue line) with the experimental data points (orange circles) is compared with the major EMT models. The governing equations and underlying assumptions of the EMT models, along with a simplified schematic for each model, are presented in Table S3 in the supplementary material. In general, Parallel (rule of mixtures) and Series (inverse rule of mixtures) models represent the upper and lower bounds of thermal conductivity for composites, respectively.[142,247] For isotropic composites, the upper and lower bounds become more restrictive and are given by Hashin–Shtrikman model.[128,139,248] Not surprisingly, our experimental data points lie between the Hashin–Shtrikman upper and lower bounds. The interesting point here is the linear behavior of all the aforementioned bounds at low filler loading levels, which can be understood from the asymptotic equation of each model as $\varphi \to 0$ given in Fig. 6(a) and is also more elaborated in the supplementary material, Sec. S5.

The most fundamental EMT model is the Maxwell model from which other models are derived.[138-140] This model is analytically obtained for spherical filler particles dispersed in a continuum matrix. However, since the final equation of the Maxwell model is independent of the diameter of the spherical filler particles (i.e., the geometrical shape of inclusions), it is also usually used as the initial estimate for the composites with filler particles of arbitrary shapes.[140] The main assumption in the Maxwell model is that the filler particles must not interact with each other, which restricts this model to only the dilute limit ($\phi \lesssim 10$ vol%).[140,142] The Maxwell model can be extended for composites with moderate and high filler content through the two well-known techniques of Differential Scheme and Self-Consistent Scheme, which lead to the Bruggeman Asymmetric[141-143] and Bruggeman Symmetric[128,141,142] models, respectively. However, as can be seen from Fig. 6(a), the Maxwell and Bruggeman models have an identical linear behavior at low filler loading levels and asymptotically approaches $1 + 3\varphi$ as $\varphi \to 0$ (see Sec. S5 in the supplementary material).

The Maxwell model can be further developed by considering interfacial thermal resistances as well as nonspherical shapes for the filler particles. Considering the effect of interfacial contact resistances between the spherical filler particles and the matrix leads to the so-called Hasselman-Johnson-Benveniste (HJB) model.[247,249,250] Also, considering the filler particles in the shape of an ellipsoid leads to the Fricke model (also known as the Hatta-Taya model).[247,251,252] A more comprehensive model that incorporates both of these models is the Nan model, which its complete equation for randomly oriented spheroidal filler particles in a matrix, considering interfacial thermal resistances, is given in Table S3 in the supplementary material.[247,253] It should be noted, however, that the Nan model also takes simpler forms (sometimes called Nan's simplified models) for carbon nanotube composites in which the aspect ratio (AR) of the nanotubes is assumed to be very high (such as the case for SWCNTs with AR > 1000).[254,255] However, here we use the conventional Nan model for ellipsoidal fillers due to the not too high aspect ratio of MWCNTs used (AR ~ 34). The best fit of the Nan model with our experimental data points is plotted in Fig. 6(a). Like before, the interesting point for the HJB, Fricke, and Nan models, as it can be seen from Fig. 6(a) and is more elaborated in Sec. S5, is almost linear behavior of these models at low filler loading levels and their inadequacy in explaining our observed superlinear increase of thermal conductivity with nanotube loading when $\varphi < 0.05$ ($\phi < 5$ vol%).

Basically, linear behavior at low filler loading levels is an intrinsic feature of EMT models and arises from the condition that the filler particles do not interact with each other in the dilute limit ($\phi \lesssim 10$ vol%). In other words, in the dilute limit of EMT models, the effect of each filler particle on the total thermal conductivity of composite is independent of the effects of other filler particles, and indeed these effects are linearly additive (the superposition principle). Other EMT models that are not discussed here, such as the Every model,[247,256] and models that are developed on the basis of EMT assumptions, such as the Hamilton–Crosser model[257,258] and the Three-Level Homogenization model,[259,260] also show this linear behavior in the dilute limit.

It is worth mentioning that although the nonlinear behavior of thermal conductivity with filler loading is a necessary condition for the proof of thermal percolation, it seems not to be a sufficient condition. As the filler loading level increases beyond the dilute limit, the distorted thermal fields around the filler particles inevitably start to interact with each other, even though the particles are not in direct contact, and thus nonlinear effects come into play. In EMT models, the interaction between filler particles is indirectly accounted for through the interaction of each filler particle with its surrounding effective medium, and conductivity versus filler loading becomes nonlinear at moderate and high loading levels ($\phi \geq 30$ vol%). Figure S3(a) in the supplementary material clearly shows such nonlinear behavior even for non-percolation classical models such as the Maxwell, Bruggeman Asymmetric, and Nan models. These models describe composites in which filler particles are not in direct contact with each other, and even up to high filler loadings, a matrix coating layer completely covers the surface of each filler particle. Therefore, no interconnected network of filler particles is formed in such composites.

Therefore, caution should be taken when attributing thermal percolation to composites in which nonlinear behavior of thermal conductivity is observed at high filler loadings, and other physics than percolation may need to be taken into account. In this regard, our TAGP model can help to distinguish thermal percolation from other nonlinear behaviors accurately. In fact, through the analysis of experimental data using our model, if a percolation threshold can be obtained for the data and the power-law behavior before and after this percolation threshold coincides with the universal critical exponents of percolation ($s$ and $t$), it can be confidently stated that thermal percolation has occurred in the system. We did this in Sec. S10 of the supplementary material for eight well-known and recent reports on thermal percolation and confirmed the observation of thermal percolation in them using the TAGP equation through accurate determination of thermal percolation thresholds and critical exponents. These eight experimental datasets were carefully selected from the literature to cover composites with different fillers and matrices as well as different fabrication methods. As can be seen from Fig. S9 and its accompanying tables in Sec. S10, our model describes all of the selected datasets very well.

To complete the discussion, in addition to classical EMT models, it is also necessary to compare the TAGP equation with classical percolation equations. However, as explained earlier in Sec. II.A, thermal percolation in polymer nanocomposites belongs to a different type of percolation (Type III: poor–good conductor) than electrical percolation (Type II: insulator–conductor), and thus the classical power-law percolation equation for Type II of percolation problems [Eq. (10)], in which the matrix conductivity is neglected, cannot be used for thermal percolation. A number of equations have been proposed to extend the classical power-law percolation equation so that the matrix conductivity and composite microstructure information are considered. The equations proposed by Bonnet *et al.*,[56] Foygel *et al.*,[261] Mamunya *et al.*,[10,262] Huang *et al.*,[263] Zhang *et al.*,[264] and Bandaru and colleagues[60] can be mentioned as the most important extended classical percolation equations. Among these, the Bonnet equation and its variants have received more attention than others, mainly due to their simplicity and appeal to intuition.[25,56,80,94,97] For this reason, here we also adopt a modified version of the Bonnet equation, as a representative of the classical percolation model, and compare it with the TAGP equation. According to the modified Bonnet equation (see Sec. S6 in the supplementary material for details),

$$k_{eff} = k_m + (k_f - k_m)\left(\frac{\phi - \phi_c}{100 - \phi_c}\right)^t, \tag{14}$$

where $\phi_c$ is the percolation threshold (vol%) and $t$ is the percolation critical exponent.

In Fig. 6(b), the TAGP equation is compared with the modified Bonnet equation. As can be seen, the TAGP equation (solid blue line) fits our experimental data (orange circles) better than the modified Bonnet equation (solid black line). In addition, the modified Bonnet equation gives the value of $k_f = 6.7$ W m$^{-1}$ K$^{-1}$ for the bulk thermal conductivity of MWCNTs, which is significantly lower than the results of experimental measurements on dense MWCNT compacts[219-223] and the value obtained from the TAGP equation ($k_f \approx$ 60 W m$^{-1}$ K$^{-1}$). The modified Bonnet equation also gives the value of the thermal percolation threshold equal to $\phi_c =$ 0.09 vol%, which is considerably lower than the value obtained from the TAGP equation ($\phi_c = 1.5$ vol%). Furthermore, the critical exponent is obtained $t = 0.9$ from the Bonnet equation which is distinctly different from the $t =$ 1.3 from the TAGP equation and it does not correspond to any of the two- or three-dimensional universal values of the percolation critical exponent ($t_{2D} = 1.3$ and $t_{3D} = 2.0$). However, for the reasons that will be explained below, we believe that the critical exponent obtained from the modified Bonnet equation cannot be correct.

Based on Balberg's model discussed in Sec. IV.B, the percolation critical exponent value obtained from the modified Bonnet equation, $t = 0.9$, indicates that the development of the conductive network in our composites is limited after the percolation threshold. In other words, after the thermal percolation threshold (here, $\phi_c = 0.09$ vol%), instead of the rapid spatial growth of the backbone and the joining of isolated clusters to it, the filler particles agglomerate and thicken the dead-end branches. But if there really is such a high level of nanotube agglomeration in our nanocomposites, that leads to a critical exponent of thermal percolation of less than one ($t_{th} = 0.9$), then the critical exponent of electrical percolation should not be approximately equal to the universal value ($t_{el} = 1.9 \approx t_{3D}$). Our calculations also show that the removal of the last data point ($\phi = 4.5$ vol% $= 10$ wt%) of thermal conductivity from the fitting process has a negligible effect on the critical exponent of thermal percolation obtained from the modified Bonnet equation. However, if the critical exponent is limited by the nanotube agglomeration, its value strongly depends on the presence or absence of off-threshold points in the fitting process. Moreover, careful examination of SEM images of the surface of polymer powder microparticles coated with carbon nanotubes before compression molding, as well as SEM images of the cryofractured nanocomposites (Figures 3 and 4 in the main text and Fig. S8 in the supplementary material), show that the nanotube agglomeration in our nanocomposites is negligible and only in the last sample with the highest nanotube loading ($\phi = 4.5$ vol% $= 10$ wt%), a small amount of agglomeration is observed. Therefore, it can be concluded that the classical power-law percolation equations, like the classical EMT models, are not able to explain the increase in thermal conductivity with increasing nanotube content in our nanocomposites.

In addition to the classical EMT and percolation models discussed in this section, we also analyzed the experimental thermal conductivity data of our nanocomposites with the models of (i) Lichtenecker Logarithmic Rule of Mixtures (Geometric Mean Model),[73,265] (ii) Agari–Uno,[266,267] (iii) Russell,[268,269] (iv) Cheng–Vachon,[247,270] and (v) Lewis–Nielsen.[26,247,271] Among these models, only the Lewis–Nielsen model fits our data relatively well. However, caution should be taken in applying the Lewis–Nielsen model for thermal percolation problems because this model has been developed basically for composites in which the filler phase remains discontinuous up to the maximum packing fraction ($\phi_m$). In fact, the implicit assumption of the Lewis–Nielsen model is that no system-spanning network of connected filler particles is formed in the composite up to $\phi_m$.

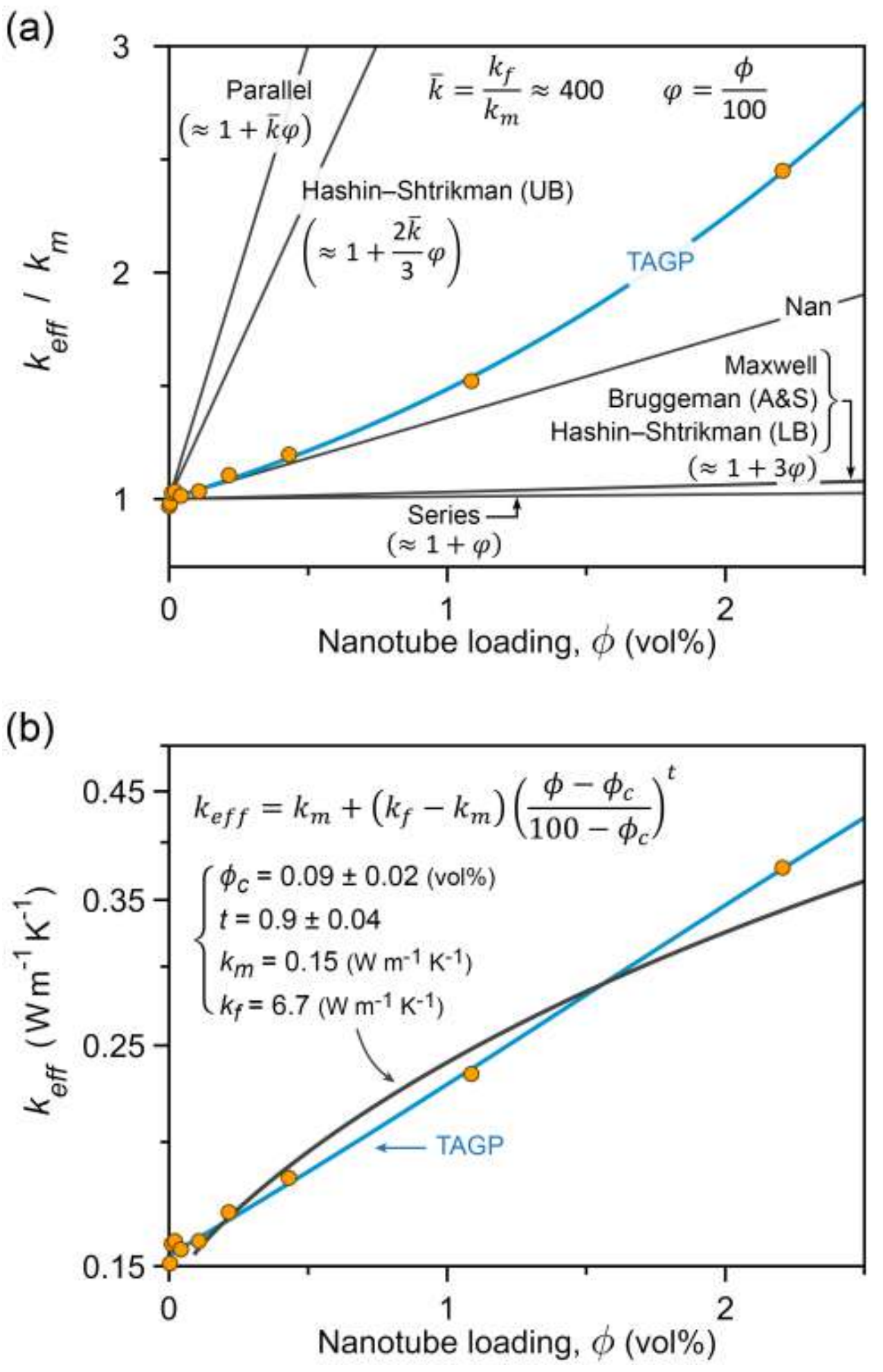

**FIG. 6.** Comparison of the Threshold-Adjusted Generalized Percolation (TAGP) equation with classical models. (a) Comparison of the TAGP equation with the classical effective medium theory (EMT) models. (b) Comparison of the TAGP equation with the modified Bonnet equation as a representative of the classical percolation model. The parameters used in the TAGP equation are the best-fit parameters presented in Fig. 4(a). For Nan's model, the following parameters are used: $a_1 = a_2 = 16$ nm, $a_3 = 550$ nm, $R_I = 8.3 \times 10^{-9}$ m$^2$KW$^{-1}$, $k_m = 0.15$ Wm$^{-1}$K$^{-1}$, and $k_f = 73.8$ Wm$^{-1}$K$^{-1}$.

### D. Electrical and thermal percolation within a unified theory

Let us now unravel the unification character of the TAGP equation in describing the thermal and electrical conductivity of composites within a single theoretical framework. We show that in addition to our thermal conductivity data, the TAGP equation can explain a variety of electrical conductivity measurements in binary mixtures as well. The standard formalism is the so-called Data Collapse technique.[124,272-274] Regardless of the multiple parameters affecting the conduction process in various thermal and electrical systems, close to the percolation threshold (critical point), the macroscopic conductivity of the system (order parameter) is independent of its microscopic details and depends only on the dimensionality of the system.

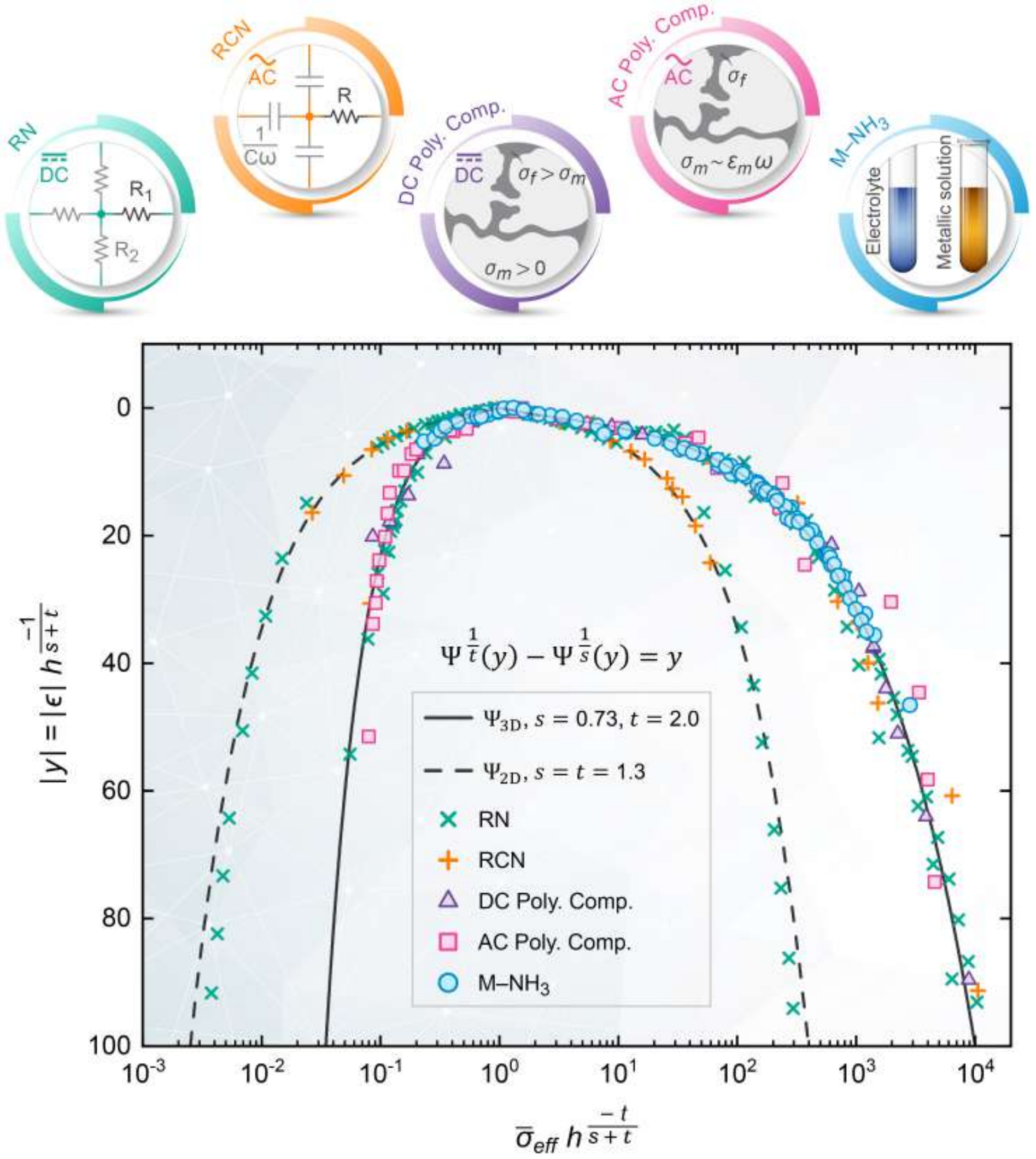

**FIG. 7.** Universal collapse of more than 350 rescaled data points from 23 experimental and numerical datasets collected from the literature (Refs. 244,275-289) onto the proposed scaling function, $\Psi(y)$. The selected data belong to five different general categories of two-component composite systems of good and poor conductors. For each dataset, the normalized proximity to the percolation threshold [$\epsilon = (\phi - \phi_c)/\phi_c$] versus the dimensionless effective conductivity of the composite system ($\bar{\sigma}_{eff} = \sigma_{eff}/\sigma_{good}$) are suitably rescaled with the conductivity ratio ($h = \sigma_{poor}/\sigma_{good}$) by using the universal percolation critical exponents ($s$, $t$). See the supplementary material, Sec. S11, for details. Abbreviations: Resistor Network (RN), Resistor–Capacitor Network (RCN), DC Electrical Conductivity of Conducting Polymer Composites (DC Poly. Comp.), AC Electrical Conductivity of Polymer Composites (AC Poly. Comp.), and Electrical Conductivity of Metal–Ammonia Solutions (M-$NH_3$).

In order to examine our proposition of the unified framework, we have selected five different sets of electrical systems of binary mixtures including two numerical datasets: (i) Resistor Network (RN), and (ii) Resistor–Capacitor Network (RCN), and three experimental datasets: (iii) DC Electrical Conductivity of Composites of Conductive filler in Intrinsically Conducting Polymer, (iv) AC Electrical Conductivity of Polymer Composites, and (v) Electrical Conductivity of Metal–Ammonia Solutions (M-$NH_3$). All these selected systems contain a variable mixture of poor and good conductors (see schematics in Fig. 7) in which a percolation transition in the effective electrical conductivity of the system ($\sigma$) can be observed at a critical fraction of the good conductor ($\phi_c$).

We have extracted more than a total of 350 data points from 23 numerical and experimental datasets reported in the literature.[244,275-289] The details of the selected data points from each data set are given in the supplementary material, Sec. S11. For each extracted dataset we have examined our TAGP equation by leaving the percolation threshold ($\phi_c$) as the "effectively" only fitting parameter. So we have considered the critical exponents as a priori known parameters according to the dimensionality of the systems, i.e., $s = t = 1.3$ for 2D and $s = 0.73$ and $t = 2.0$ for the 3D systems, respectively. Also, in cases where the amount of matrix electrical conductivity ($\sigma_m$) was measured and reported by the authors, we have used this value unchanged. To further check and validate the efficacy of our analysis, we have once considered $\sigma_{f^*}$ as a free parameter to see if our estimate is in agreement with the true values when it is directly compared with the filler electrical conductivity ($\sigma_f$) by using Eq. (13). We have noticed that our estimated results summarized in Sec. S11, are in perfect agreement with those obtained from the filler conductivity. This helped us to evaluate the influence of individual a priori known parameters on the quality of the parameter estimates.

When the percolation threshold ($\phi_c$) for each dataset is determined from our analysis, we make use of the scaling relation Eq. (3) to collapse all rescaled data collected from different physics onto a single universal curve that only depends on the dimensionality. To this aim, we define dimensionless effective electrical conductivity ($\bar{\sigma}_{eff} = \sigma_{eff}/\sigma_{good}$) and the normalized proximity to the percolation threshold [$\epsilon = (\phi - \phi_c)/\phi_c$], and then examine the following rearranged form of Eq. (3)

$$\bar{\sigma}_{eff}\, h^{\frac{-t}{s+t}} = \Psi\left(|\epsilon|\, h^{\frac{-1}{s+t}}\right) \cdot \qquad (15)$$

By inserting Eq. (15) in the TAGP equation [Eq. (9)], and after some manipulation, we will arrive to the following fundamental universal law for $\Psi(\cdot)$

$$\Psi^{\frac{1}{t}}(y) - \Psi^{\frac{1}{s}}(y) = y \cdot \qquad (16)$$

The most remarkable feature of our proposed unified percolation framework is presented in Fig. 7 where all data from different systems perfectly collapse into a universal scaling law $\Psi(\cdot)$ at any given dimensionality. Furthermore, the power series solution of Eq. (16) leads to the following explicit form, which is exact within its radius of convergence ($R_\psi$)

$$\Psi(y) = \sum_{n=0}^{\infty} \frac{(-1)^{n+1} s\, t\, \Gamma\left(n - \frac{n\, t + s\, t}{s+t}\right)}{n!\,(s+t)\,\Gamma\left(1 - \frac{n\, t + s\, t}{s+t}\right)} y^n, \qquad (17)$$

$$R_\Psi = \left(\frac{s}{t}\right)^{\frac{t}{s+t}} + \left(\frac{t}{s}\right)^{\frac{s}{s+t}}, \qquad (18)$$

where $\Gamma(\cdot)$ is the Gamma function.

## V. CONCLUSION AND OUTLOOK

To conclude, we have dealt with the controversial subject of thermal percolation in polymer composites from both theoretical and experimental points of view. From the theoretical standpoint, we have argued that the main reason for not observing a genuine thermal percolation is in fact its masking due to its low conductivity ratio of the filler to the matrix on the order of $10^2$–$10^4$. While in the case of electrical conductivity this ratio is on the order of $10^{14}$–$10^{22}$ which is high enough to capture all aspects of percolation features (such as a rapid jump in conductivity) without controversy. To properly address this long-standing controversy, we have developed a phenomenological percolation theory, i.e., the Threshold-Adjusted Generalized Percolation (TAGP) equation, as a generalization of the classical power-law percolation equation. In TAGP, in addition to the volume percent of the filler at the percolation threshold ($\phi_c$), the thermal conductivity of the filler ($k_f$) and the post-percolation critical exponent ($t$), two other important parameters, i.e., the thermal conductivity of the matrix ($k_m$) and the pre-percolation critical exponent ($s$) are considered. All the parameters of this equation have a clear and definite physical meaning. The TAGP equation provides an accurate picture of the poor–good conductor transition in thermal percolation as a regime between the insulator–conductor transition in electrical percolation on the one hand and the smooth crossover on the other.

From the experimental point of view, it was argued that the effective thermal conductivity of micro- and nano-fillers in polymer composites is highly dependent on the nanocomposite fabrication method and the contact quality between the filler particles due to the very high particle surface to volume ratio. The penetration of the matrix phase between the filler particles or the presence of nanometer gaps between the particles totally suppresses the effective thermal conductivity of filler, thus reducing the conductivity ratio of filler to matrix from a potential value of $10^2$–$10^4$ to less than 10, in which case thermal percolation will no longer be observed. To unambiguously prove the existence of thermal percolation, with a carefully designed fabrication method, nanocomposites of 3D segregated and well-extended network of multiwalled carbon nanotubes (MWCNTs) in polypropylene (PP) were prepared to ensure effective contact between carbon nanotubes. With this novel fabrication method, we demonstrate an ultralow electrical percolation threshold of 0.02 vol% and a record-low thermal percolation threshold of 1.5 vol% in our MWCNT–PP nanocomposites. We believe our present study removes two main obstacles to realizing thermal percolation: (i) By substantially improving the fabrication methods since the previous conventional methods have not been able to create effective thermal contacts between the filler particles, and (ii) By developing a new percolation theory for its realization since previous models have failed to interpret the experimental results for the composites of poor and good conductors.

The TAGP equation has been further validated by analyzing a wide range of conductivity data versus filler loading in the composites reported in the literature, including 8 experimental datasets of thermal conductivity and 23 experimental and numerical datasets of electrical conductivity. Based on this broader perspective, it is possible to have a unified view of electrical percolation and thermal percolation in systems consisting of a binary mixture of a good and a poor conductor. We believe that our study sheds light on the underlying physics of thermal percolation and can contribute significantly to resolving the remaining issues in this field.

As an outlook, our work paves the way for the fabrication of a new generation of highly electrically and thermally conductive polymer composites at unprecedented low filler loading levels with great potential for applications such as electromagnetic interference shielding. We also hope our findings will contribute to the future development of functional polymer composites with independently tunable electrical and thermal conductivities for a multitude of advanced device applications. These applications can range from thermal interface materials, in which high thermal and low electrical conductivity is required, to thermoelectrics, where, on the contrary, low thermal and high electrical conductivity is desired.

Here, we also would like to point out some important and high-priority research directions in the rapidly growing field of thermal percolation. Although the main focus of our present study is on the description of universal features of thermal and electrical percolation within a unified framework, however, our proposed model offers a precise estimation of the percolation threshold that makes it possible to investigate the role of various microstructural properties of the fillers on non-universal properties. For example, the geometry and topology of the filler are of great experimental importance to the optimal design of conductive composites, especially when novel high-aspect-ratio 2D and 1D materials are considered. In this context, several models, including the Soft-Core model [Eq. (12)], have been proposed to relate the electrical percolation threshold ($\phi_{c,el}$) to the microstructural details of insulator–conductor composites (Type II percolation).[159,160,209,261] However, these models are not applicable for thermal percolation in poor–good conductor composites (Type III percolation) due to neglecting matrix conductivity. In fact, to establish a relationship between the thermal percolation threshold ($\phi_{c,th}$) and the microstructural details of the composite (e.g., filler aspect ratio), it is necessary to consider the matrix-to-filler thermal conductivity ratio as one of the main parameters. The TAGP equation can help resolve this issue by providing an accurate assessment of $\phi_{c,th}$ for composites with varying geometrical characteristics and conductivity ratios, through a series of systematic experiments or numerical simulations.

Another important topic from the practical point of view is investigating the effect of filler–matrix interfacial ($R_I$) and filler–filler contact ($R_C$) thermal resistances on thermal percolation. If there is no interconnected network of filler particles in the composite, it is only necessary to consider $R_I$ in the thermal conductivity model. Also, if there is an interconnected network of filler particles, in the case of insulator–conductor composites (Type II percolation), it is sufficient to consider only $R_C$. But in the general case of poor–good conductor composites (Type III percolation), we may encounter cases where the effect of $R_I$ and $R_C$ are both important. In this regard, our percolation model (TAGP) can be advantageous due to the presence of the $k_{f^*}$ parameter in it. In fact, $k_{f^*} = k_{eff}(2\phi_c)$ is the effective thermal conductivity of the composite at filler loading $\phi = 2\phi_c$. At this loading, on the one hand, a well-connected network of filler particles is formed in the composite because $2\phi_c > \phi_c$, thus the effect of $R_C$, can be included in the model. On the other hand, the matrix effect has not been completely eliminated, and the matrix can still play a role in heat conduction because $2\phi_c < 100$ vol%, therefore the $R_I$ effect, can also be considered in the model. Consequently, for any specific geometry of filler particles, the $k_{f^*}$ parameter of our model can be related to $R_I$ and $R_C$, through a series of systematic experiments or numerical simulations. But we should pay attention to the dual role of $R_I$ in thermal percolation, because on the one hand, the reduction of $R_I$ can lead to effective heat transfer in the pathways composed of filler particles with thin matrix spacer layers [pathway 2 in Fig. 5(b)]. But on the other hand, a large decrease in $R_I$ may cause heat leakage from the filler network to the matrix and even prevent thermal percolation.

An interesting research direction that is of great practical importance is the study of thermal percolation in noncuring thermal interface materials, for example, composite thermal pastes consisting of graphene fillers in mineral oil matrices.[88] This category of composites has great potential for heat management in the electronics industry.[46] An in-depth understanding of the mechanism of thermal percolation in these composite systems helps to achieve a better thermal design for them. Also, due to the semi-fluid nature of noncuring thermal pastes, their systematic study may ultimately provide the key to solving another long-lasting unsolved problem, the lack or existence of thermal percolation in nanofluids.[290-294] Currently, state-of-the-art noncuring composites exhibit a certain thermal percolation threshold, although their post-percolation critical exponent ($t$) does not show universal behavior and is less than one.[46,88,295,296] In Sec. S10 of the supplementary material, we have analyzed two sets of experimental data on thermal percolation in noncuring graphene thermal pastes with the TAGP equation and provided remarks for interpreting the model fit results and their relationship to the composite microstructure.[88,295]

Another active research direction in the field is the use of hybrid fillers for independent control of electrical and thermal conductivity, as well as the possibility of benefiting from the synergistic effect of fillers.[25,94,297-301] For example, in binary-filler hybrid composites, by keeping the volume fraction of each filler constant, there will be a percolation threshold for the other filler.[299] So, in these composites, we have a percolation threshold line instead of a percolation threshold point. Careful examination of the shape and location of this line and the critical exponents around it can provide deeper insights into the optimal design of such hybrid composites for fine tuning their properties, and much work remains to be done in this area, both experimentally and theoretically.

## SUPPLEMENTARY MATERIAL

See the supplementary material for a selective Literature Survey, Materials and Methods, comparison of the TAGP equation with Classical Models, application of the TAGP equation for Selected Thermal Percolation Reports, and details of the Data Collapse.

## ACKNOWLEDGMENTS

A.Z.M. would like to acknowledge the financial support from the Iran National Science Foundation (Research Chair Award of Surface and Interface Physics, Grant No. 940009) and the Iran Science Elites Federation (Grant of the Top-100 national science elites). N.S. is deeply indebted to Prof. Mohammad Said Saidi, Prof. Bahar Firoozabadi, and Prof. Siamak Kazemzadeh Hannani for their trust and continuous support that made his contribution to this research possible. In the end, the generous help and valuable suggestions of many colleagues, including Prof. M. Reza Rahimi Tabar, Dr. Morasae Samadi, Dr. Mohammad Qorbani, and Dr. Ali Samieipour, in the various stages of this work, are gratefully acknowledged.

## DATA AVAILABILITY

The data that support the findings of this study are available from the corresponding authors upon reasonable request.

## REFERENCES

1 S. Thomas *et al.*, *Polymer composites, macro- and microcomposites*. (Wiley, 2012).

2 R. K. Gupta, E. Kennel, and K. J. Kim, *Polymer nanocomposites handbook*. (CRC Press, 2009).

3 X. Huang and C. Zhi, *Polymer nanocomposites: Electrical and thermal properties*. (Springer International Publishing, 2016).

4 S. K. Kumar, B. C. Benicewicz, R. A. Vaia, and K. I. Winey, "50th anniversary perspective: Are polymer nanocomposites practical for applications?," *Macromolecules* **50**, 714-731 (2017).

5 I. A. Kinloch *et al.*, "Composites with carbon nanotubes and graphene: An outlook," *Science* **362**, 547-553 (2018).

6 Y. Huang *et al.*, "Tailoring the electrical and thermal conductivity of multi-component and multi-phase polymer composites," *Int. Mater. Rev.* (2019).

7 K. Nagata, H. Iwabuki, and H. Nigo, "Effect of particle size of graphites on electrical conductivity of graphite/polymer composite," *Compos. Interfaces* **6**, 483-495 (1999).

8 I. Balberg, "A comprehensive picture of the electrical phenomena in carbon black-polymer composites," *Carbon* **40**, 139-143 (2002).

9 S. A. Gordeyev *et al.*, "Transport properties of polymer-vapour grown carbon fibre composites," *Phys. B Condens. Matter* **279**, 33-36 (2000).

10 Y. P. Mamunya, V. V. Davydenko, P. Pissis, and E. V. Lebedev, "Electrical and thermal conductivity of polymers filled with metal powders," *Eur. Polym. J.* **38**, 1887-1897 (2002).

11 W. Bauhofer and J. Z. Kovacs, "A review and analysis of electrical percolation in carbon nanotube polymer composites," *Compos. Sci. Technol.* **69**, 1486-1498 (2009).

12 A. J. Marsden *et al.*, "Electrical percolation in graphene-polymer composites," *2D Mater.* **5** (2018).

13 G. A. Gelves, B. Lin, U. Sundararaj, and J. A. Haber, "Low electrical percolation threshold of silver and copper nanowires in polystyrene composites," *Adv. Funct. Mater.* **16**, 2423-2430 (2006).

14 D. Stauffer and A. Aharony, *Introduction to percolation theory*. (Taylor & Francis, 1994).

15 M. Sahimi, *Applications of percolation theory*. (Taylor & Francis, 2003).

16 Z. Han and A. Fina, "Thermal conductivity of carbon nanotubes and their polymer nanocomposites: A review," *Prog. Polym. Sci.* **36**, 914-944 (2011).

17 B. P. Grady, "Thermal conductivity", in *Carbon nanotube-polymer composites: Manufacture, properties, and applications* (Wiley, 2011).

18 N. Burger *et al.*, "Review of thermal conductivity in composites: Mechanisms, parameters and theory," *Prog. Polym. Sci.* **61**, 1-28 (2016).

19 H. Chen *et al.*, "Thermal conductivity of polymer-based composites: Fundamentals and applications," *Prog. Polym. Sci.* **59**, 41-85 (2016).

20 Y. Xu, X. Wang, and Q. Hao, "A mini review on thermally conductive polymers and polymer-based composites," *Compos. Commun.* **24** (2021).

21 N. Shenogina, S. Shenogin, L. Xue, and P. Keblinski, "On the lack of thermal percolation in carbon nanotube composites," *Appl. Phys. Lett.* **87**, 1-3 (2005).

22 X. Xu, J. Chen, J. Zhou, and B. Li, "Thermal conductivity of polymers and their nanocomposites," *Adv. Mater.* **30**, 1705544 (2018).

23 B. Liu *et al.*, "Thermal transport in organic/inorganic composites," *Front. Energy* **12**, 72-86 (2018).

24 C. Huang, X. Qian, and R. Yang, "Thermal conductivity of polymers and polymer nanocomposites," *Mater. Sci. Eng., R* **132**, 1-22 (2018).

25 M. Shtein *et al.*, "Thermally conductive graphene-polymer composites: Size, percolation, and synergy effects," *Chem. Mater.* **27**, 2100-2106 (2015).

26 F. Kargar *et al.*, "Thermal percolation threshold and thermal properties of composites with high loading of graphene and boron nitride fillers," *ACS Appl. Mater. Interfaces* **10**, 37555-37565 (2018).

27 B. Garnier, B. Agoudjil, and A. Boudenne, "Metallic particle-filled polymer microcomposites", in *Polym. Compos.* (Wiley, 2012), Vol. 1.

28 M. Moniruzzaman and K. I. Winey, "Polymer nanocomposites containing carbon nanotubes," *Macromolecules* **39**, 5194-5205 (2006).

29 B. P. Grady, *Carbon nanotube-polymer composites: Manufacture, properties, and applications*. (Wiley, 2011).

30 R. M. Mutiso and K. I. Winey, "Electrical properties of polymer nanocomposites containing rod-like nanofillers," *Prog. Polym. Sci.* **40**, 63-84 (2015).

31 A. M. Marconnet, M. A. Panzer, and K. E. Goodson, "Thermal conduction phenomena in carbon nanotubes and related nanostructured materials," *Rev. Mod. Phys.* **85**, 1295-1326 (2013).

32 K. I. Winey, T. Kashiwagi, and M. Mu, "Improving electrical conductivity and thermal properties of polymers by the addition of carbon nanotubes as fillers," *MRS Bull.* **32**, 348-353 (2007).

33 M. J. Biercuk *et al.*, "Carbon nanotube composites for thermal management," *Appl. Phys. Lett.* **80**, 2767-2769 (2002).

34 S. T. Huxtable *et al.*, "Interfacial heat flow in carbon nanotube suspensions," *Nat. Mater.* **2**, 731-734 (2003).

35 S. Shenogin *et al.*, "Role of thermal boundary resistance on the heat flow in carbon-nanotube composites," *J. Appl. Phys.* **95**, 8136-8144 (2004).

36 Y. Chalopin, S. Volz, and N. Mingo, "Upper bound to the thermal conductivity of carbon nanotube pellets," *J. Appl. Phys.* **105** (2009).

37 R. S. Prasher *et al.*, "Turning carbon nanotubes from exceptional heat conductors into insulators," *Phys. Rev. Lett.* **102** (2009).

38 A. N. Volkov and L. V. Zhigilei, "Scaling laws and mesoscopic modeling of thermal conductivity in carbon nanotube materials," *Phys. Rev. Lett.* **104** (2010).

39 Y. Chalopin, S. Volz, and N. Mingo, "Erratum: Upper bound to the thermal conductivity of carbon nanotube pellets (journal of applied physics (2009) 105 (084301))," *J. Appl. Phys.* **108** (2010).

40 H. Abbasi, M. Antunes, and J. I. Velasco, "Recent advances in carbon-based polymer nanocomposites for electromagnetic interference shielding," *Prog. Mater Sci.* **103**, 319-373 (2019).

41 X. Huang *et al.*, "Thermal conductivity of graphene-based polymer nanocomposites," *Mater. Sci. Eng., R* **142** (2020).

42 A. A. Balandin, "Phononics of graphene and related materials," *ACS Nano* **14**, 5170-5178 (2020).

43 X. Sun *et al.*, "Recent progress in graphene/polymer nanocomposites," *Adv. Mater.* **33** (2021).

44 X. Shen, Q. Zheng, and J. K. Kim, "Rational design of two-dimensional nanofillers for polymer nanocomposites toward multifunctional applications," *Prog. Mater Sci.* **115** (2021).

45 J. Li, X. Liu, Y. Feng, and J. Yin, "Recent progress in polymer/two-dimensional nanosheets composites with novel performances," *Prog. Polym. Sci.*, 101505 (2022).

46 J. S. Lewis *et al.*, "Thermal interface materials with graphene fillers: Review of the state of the art and outlook for future applications," *Nanotechnology* **32** (2021).

47 H. Y. Zhao *et al.*, "Efficient preconstruction of three-dimensional graphene networks for thermally conductive polymer composites," *Nano-Micro Letters* **14** (2022).

48 A. A. Balandin, "Thermal properties of graphene and nanostructured carbon materials," *Nat. Mater.* **10**, 569-581 (2011).

49 E. Pop, V. Varshney, and A. K. Roy, "Thermal properties of graphene: Fundamentals and applications," *MRS Bull.* **37**, 1273-1281 (2012).

50 K. M. F. Shahil and A. A. Balandin, "Graphene-multilayer graphene nanocomposites as highly efficient thermal interface materials," *Nano Lett.* **12**, 861-867 (2012).

51 B. Tang, G. Hu, H. Gao, and L. Hai, "Application of graphene as filler to improve thermal transport property of epoxy resin for thermal interface materials," *Int. J. Heat Mass Transfer* **85**, 420-429 (2015).

52 J. Wang *et al.*, "A multiscale study of the filler-size and temperature dependence of the thermal conductivity of graphene-polymer nanocomposites," *Carbon* **175**, 259-270 (2021).

53 H. S. Kim *et al.*, "Volume control of expanded graphite based on inductively coupled plasma and enhanced thermal conductivity of epoxy composite by formation of the filler network," *Carbon* **119**, 40-46 (2017).

54 S. Bhanushali, P. C. Ghosh, G. P. Simon, and W. Cheng, "Copper nanowire-filled soft elastomer composites for applications as thermal interface materials," *Adv. Mater. Interfaces* **4** (2017).

55 N. Ghahramani, S. A. Seyed Esfahani, M. Mehranpour, and H. Nazockdast, "The effect of filler localization on morphology and thermal conductivity of the polyamide/cyclic olefin copolymer blends filled with boron nitride," *J. Mater. Sci.* **53**, 16146–16159 (2018).

56 P. Bonnet, D. Sireude, B. Garnier, and O. Chauvet, "Thermal properties and percolation in carbon nanotube-polymer composites," *Appl. Phys. Lett.* **91** (2007).

57 A. Yu, M. E. Itkis, E. Bekyarova, and R. C. Haddon, "Effect of single-walled carbon nanotube purity on the thermal conductivity of carbon nanotube-based composites," *Appl. Phys. Lett.* **89** (2006).

58 A. Yu *et al.*, "Graphite nanoplatelet-epoxy composite thermal interface materials," *J. Phys. Chem. C* **111**, 7565-7569 (2007).

59 A. Yu *et al.*, "Enhanced thermal conductivity in a hybrid graphite nanoplatelet - carbon nanotube filler for epoxy composites," *Adv. Mater.* **20**, 4740-4744 (2008).

60 B. W. Kim, S. H. Park, R. S. Kapadia, and P. R. Bandaru, "Evidence of percolation related power law behavior in the thermal conductivity of nanotube/polymer composites," *Appl. Phys. Lett.* **102** (2013).

61 B. W. Kim, S. H. Park, and P. R. Bandaru, "Anomalous decrease of the specific heat capacity at the electrical and thermal conductivity percolation threshold in nanocomposites," *Appl. Phys. Lett.* **105** (2014).

62 R. Gulotty *et al.*, "Effects of functionalization on thermal properties of single-wall and multi-wall carbon nanotube-polymer nanocomposites," *ACS Nano* **7**, 5114-5121 (2013).

63 F. H. Gojny *et al.*, "Evaluation and identification of electrical and thermal conduction mechanisms in carbon nanotube/epoxy composites," *Polymer* **47**, 2036-2045 (2006).

64 A. Moisala, Q. Li, I. A. Kinloch, and A. H. Windle, "Thermal and electrical conductivity of single- and multi-walled carbon nanotube-epoxy composites," *Compos. Sci. Technol.* **66**, 1285-1288 (2006).

65 Y. Mamunya *et al.*, "Electrical and thermophysical behaviour of pvc-MWCNT nanocomposites," *Compos. Sci. Technol.* **68**, 1981-1988 (2008).

66 Y. Yang, M. C. Gupta, J. N. Zalameda, and W. P. Winfree, "Dispersion behaviour, thermal and electrical conductivities of carbon nanotube-polystyrene nanocomposites," *Micro Nano Lett.* **3**, 35-40 (2008).

67 J. Bouchard, A. Cayla, E. Devaux, and C. Campagne, "Electrical and thermal conductivities of multiwalled carbon nanotubes-reinforced high performance polymer nanocomposites," *Compos. Sci. Technol.* **86**, 177-184 (2013).

68 X. Gao *et al.*, "Topological design of inorganic–organic thermoelectric nanocomposites based on "electron–percolation phonon–insulator" concept," *ACS Appl. Energy Mater.* **1**, 2927-2933 (2018).

69 J. Chen, J. Han, and D. Xu, "Thermal and electrical properties of the epoxy nanocomposites reinforced with purified carbon nanotubes," *Mater. Lett.* **246**, 20-23 (2019).

70 B. Krause, P. Rzeczkowski, and P. Pötschke, "Thermal conductivity and electrical resistivity of melt-mixed polypropylene composites containing mixtures of carbon-based fillers," *Polymers* **11** (2019).

71 G. Zhao *et al.*, "Flame synthesis of carbon nanotubes on glass fibre fabrics and their enhancement in electrical and thermal properties of glass fibre/epoxy composites," *Composites, Part B* **198** (2020).

72 J. U. Jang *et al.*, "Electrically and thermally conductive carbon fibre fabric reinforced polymer composites based on nanocarbons and an in-situ polymerizable cyclic oligoester," *Sci. Rep.* **8** (2018).

73 O. Maruzhenko *et al.*, "Improving the thermal and electrical properties of polymer composites by ordered distribution of carbon micro- and nanofillers," *Int. J. Heat Mass Transfer* **138**, 75-84 (2019).

74 M. Cierpisz, J. McPhedran, Y. He, and A. Edrisy, "Characterization of graphene-filled fluoropolymer coatings for condensing heat exchangers," *J. Compos. Mater.* **55**, 4305-4320 (2021).

75 S. Yang *et al.*, "The fabrication of polyethylene/graphite nanoplatelets composites for thermal management and electromagnetic interference shielding application," *J. Mater. Sci.* **57**, 1084–1097 (2022).

76 A. Mirabedini *et al.*, "Scalable production and thermoelectrical modeling of infusible functional graphene/epoxy nanomaterials for engineering applications," *Industrial & Engineering Chemistry Research* **61**, 5141-5157 (2022).
77 A. I. Misiura, Y. P. Mamunya, and M. P. Kulish, "Metal-filled epoxy composites: Mechanical properties and electrical/thermal conductivity," *J. Macromol. Sci., Part B: Phys.* **59**, 121-136 (2020).
78 C. Muhammed Ajmal *et al.*, "In-situ reduced non-oxidized copper nanoparticles in nanocomposites with extraordinary high electrical and thermal conductivity," *Mater. Today* **48**, 59-71 (2021).
79 R. Haggenmueller *et al.*, "Single wall carbon nanotube/polyethylene nanocomposites: Thermal and electrical conductivity," *Macromolecules* **40**, 2417-2421 (2007).
80 S. Y. Kwon *et al.*, "A large increase in the thermal conductivity of carbon nanotube/polymer composites produced by percolation phenomena," *Carbon* **55**, 285-290 (2013).
81 A. Shayganpour *et al.*, "Stacked-cup carbon nanotube flexible paper based on soy lecithin and natural rubber," *Nanomaterials* **9** (2019).
82 G. Shachar-Michaely *et al.*, "Mixed dimensionality: Highly robust and multifunctional carbon-based composites," *Carbon* **176**, 339-348 (2021).
83 J. Huang *et al.*, "Massive enhancement in the thermal conductivity of polymer composites by trapping graphene at the interface of a polymer blend," *Compos. Sci. Technol.* **129**, 160-165 (2016).
84 D. An *et al.*, "A polymer-based thermal management material with enhanced thermal conductivity by introducing three-dimensional networks and covalent bond connections," *Carbon* **155**, 258-267 (2019).
85 F. Kargar *et al.*, "Dual-functional graphene composites for electromagnetic shielding and thermal management," *Adv. Electron. Mater.* **5** (2019).
86 Z. Wu *et al.*, "Synergistic effect of aligned graphene nanosheets in graphene foam for high-performance thermally conductive composites," *Adv. Mater.* **31**, e1900199 (2019).
87 Z. Barani *et al.*, "Multifunctional graphene composites for electromagnetic shielding and thermal management at elevated temperatures," *Adv. Electron. Mater.* **6** (2020).
88 S. Naghibi *et al.*, "Noncuring graphene thermal interface materials for advanced electronics," *Adv. Electron. Mater.* **6** (2020).
89 K. M. Burzynski *et al.*, "Graphite nanocomposite substrates for improved performance of flexible, high-power algan/gan electronic devices," *ACS Appl. Electron. Mater.* **3**, 1228-1235 (2021).
90 W. Dai *et al.*, "Multiscale structural modulation of anisotropic graphene framework for polymer composites achieving highly efficient thermal energy management," *Adv. Sci.* **8**, 2003734 (2021).
91 J.-u. Jang *et al.*, "Thermal percolation behavior in thermal conductivity of polymer nanocomposite with lateral size of graphene nanoplatelet," *Polymers* **14**, 323 (2022).
92 J. U. Jang *et al.*, "Enhanced thermal conductivity of graphene nanoplatelet filled polymer composite based on thermal percolation behavior," *Compos. Commun.* **31** (2022).
93 S. Shi *et al.*, "3D printed polylactic acid/graphene nanocomposites with tailored multifunctionality towards superior thermal management and high-efficient electromagnetic interference shielding," *Chem. Eng. J.* **450** (2022).
94 M. Shtein, R. Nadiv, M. Buzaglo, and O. Regev, "Graphene-based hybrid composites for efficient thermal management of electronic devices," *ACS Appl. Mater. Interfaces* **7**, 23725-23730 (2015).
95 A. Gurijala *et al.*, "Castable and printable dielectric composites exhibiting high thermal conductivity via percolation-enabled phonon transport," *Matter* **2**, 1015-1024 (2020).
96 S. H. Ryu *et al.*, "Quasi-isotropic thermal conduction in percolation networks: Using the pore-filling effect to enhance thermal conductivity in polymer nanocomposites," *ACS Appl. Polym. Mater.* **3**, 1293-1305 (2021).
97 B. Shi *et al.*, "Thermal percolation in composite materials with electrically conductive fillers," *Appl. Phys. Lett.* **113** (2018).
98 D. Suh *et al.*, "Significantly enhanced phonon mean free path and thermal conductivity by percolation of silver nanoflowers," *Phys. Chem. Chem. Phys.* **21**, 2453-2462 (2019).
99 H. Bark, M. W. M. Tan, G. Thangavel, and P. S. Lee, "Deformable high loading liquid metal nanoparticles composites for thermal energy management," *Adv. Energy Mater.* **11** (2021).
100 I. Y. Forero-Sandoval *et al.*, "Percolation threshold of the thermal, electrical and optical properties of carbonyl-iron microcomposites," *Appl. Compos. Mater.* **28**, 447-463 (2021).
101 B. S. Chang *et al.*, "Thermal percolation in well-defined nanocomposite thin films," *ACS Appl. Mater. Interfaces* **14**, 14579-14587 (2022).
102 M. C. Vu *et al.*, "High thermal conductivity enhancement of polymer composites with vertically aligned silicon carbide sheet scaffolds," *ACS Appl. Mater. Interfaces* **12**, 23388-23398 (2020).
103 S. Kumar, M. A. Alam, and J. Y. Murthy, "Effect of percolation on thermal transport in nanotube composites," *Appl. Phys. Lett.* **90** (2007).
104 W. Tian and R. Yang, "Effect of interface scattering on phonon thermal conductivity percolation in random nanowire composites," *Appl. Phys. Lett.* **90** (2007).
105 A. Khoubani, T. M. Evans, and T. S. Yun, "Thermal percolation in mixtures of monodisperse spheres," *Granular Matter* **22** (2020).
106 S. Chen, Q. Liu, L. Gorbatikh, and D. Seveno, "Does thermal percolation exist in graphene-reinforced polymer composites? A molecular dynamics answer," *J. Phys. Chem. C* **125**, 1018-1028 (2021).
107 I. Y. Forero-Sandoval *et al.*, "Electrical and thermal percolation in two-phase materials: A perspective," *J. Appl. Phys.* **131** (2022).
108 J. P. Clerc *et al.*, "La percolation," **8**, 3-105 (1983).
109 V. K. S. Shante and S. Kirkpatrick, "An introduction to percolation theory," *Adv. Phys.* **20**, 325-357 (1971).
110 H. L. Frisch and J. M. Hammersley, "Percolation processes and related topics," *J. Soc. Indust. Appl. Math.* **11**, 894-918 (1963).
111 J. Yang *et al.*, "Contact thermal resistance between individual multiwall carbon nanotubes," *Appl. Phys. Lett.* **96** (2010).
112 J. Yang *et al.*, "Phonon transport through point contacts between graphitic nanomaterials," *Phys. Rev. Lett.* **112**, 205901 (2014).
113 H. Deng *et al.*, "Progress on the morphological control of conductive network in conductive polymer composites and the use as electroactive multifunctional materials," *Prog. Polym. Sci.* **39**, 627-655 (2014).
114 H. Pang, L. Xu, D. X. Yan, and Z. M. Li, "Conductive polymer composites with segregated structures," *Prog. Polym. Sci.* **39**, 1908-1933 (2014).
115 S. N. Leung, "Thermally conductive polymer composites and nanocomposites: Processing-structure-property relationships," *Composites, Part B* **150**, 78-92 (2018).
116 G. Deutscher, O. Entin-Wohlman, S. Fishman, and Y. Shapira, "Percolation description of granular superconductors," *Phys. Rev. B* **21**, 5041-5047 (1980).

117 H. J. Herrmann, B. Derrida, and J. Vannimenus, “Superconductivity exponents in two- and three-dimensional percolation,” *Phys. Rev. B* **30**, 4080-4082 (1984).
118 M. Sahimi, “Finite-size scaling calculation of conductivity of three-dimensional conductor-superconductor networks at percolation threshold,” *J. Phys. C: Solid State Phys.* **17**, L355-L358 (1984).
119 C. W. Nan, Y. Shen, and J. Ma, “Physical properties of composites near percolation,” *Annu. Rev. Mater. Res.* **40**, 131-151 (2010).
120 N. A. Mohd Radzuan, A. B. Sulong, and J. Sahari, “A review of electrical conductivity models for conductive polymer composite,” *Int. J. Hydrogen Energy* **42**, 9262-9273 (2017).
121 A. L. Efros and B. I. Shklovskii, “Critical behaviour of conductivity and dielectric constant near the metal-non-metal transition threshold,” *Physica Status Solidi (b)* **76**, 475-485 (1976).
122 J. P. Straley, “Critical phenomena in resistor networks,” *J. Phys. C: Solid State Phys.* **9**, 783-795 (1976).
123 D. Stauffer, “Scaling theory of percolation clusters,” *Phys. Rep.* **54**, 1-74 (1979).
124 D. C. Hong, H. E. Stanley, A. Coniglio, and A. Bunde, “Random-walk approach to the two-component random-conductor mixture: Perturbing away from the perfect random resistor network and random superconducting-network limits,” *Phys. Rev. B* **33**, 4564-4573 (1986).
125 J. P. Clerc, G. Giraud, J. M. Laugier, and J. M. Luck, “The electrical conductivity of binary disordered systems, percolation clusters, fractals and related models,” *Adv. Phys.* **39**, 191-309 (1990).
126 D. J. Bergman and D. Stroud, "Physical properties of macroscopically inhomogeneous media", in *Solid state physics - advances in research and applications*, edited by Henry Ehrenreich and David Turnbull (Academic Press, 1992), Vol. 46, pp. 147-269.
127 C. W. Nan, “Physics of inhomogeneous inorganic materials,” *Prog. Mater Sci.* **37**, 1-116 (1993).
128 A. A. Snarskii *et al.*, *Transport processes in macroscopically disordered media: From mean field theory to percolation*. (Springer New York, 2016).
129 J. P. Straley, “Cooperative phenomena in resistor networks and inhomogeneous conductors,” **40**, 118-127 (1978).
130 I. Balberg, “The physical fundamentals of the electrical conductivity in nanotube-based composites,” *J. Appl. Phys.* **128** (2020).
131 N. Goldenfeld, *Lectures on phase transitions and the renormalization group*. (CRC Press, 2018).
132 P. Keblinski and F. Cleri, “Contact resistance in percolating networks,” *Phys. Rev. B* **69**, 184201 (2004).
133 J. Li and S.-L. Zhang, “Conductivity exponents in stick percolation,” *Phys. Rev. E* **81**, 021120 (2010).
134 M. Zezelj and I. Stanković, “From percolating to dense random stick networks: Conductivity model investigation,” *Phys. Rev. B* **86** (2012).
135 R. M. Mutiso and K. I. Winey, “Electrical percolation in quasi-two-dimensional metal nanowire networks for transparent conductors,” *Phys. Rev. E* **88** (2013).
136 I. Balberg, D. Azulay, Y. Goldstein, and J. Jedrzejewski, “Possible origin of the smaller-than-universal percolation-conductivity exponent in the continuum,” *Phys. Rev. E* **93** (2016).
137 J. Hone, M. Whitney, C. Piskoti, and A. Zettl, “Thermal conductivity of single-walled carbon nanotubes,” *Phys. Rev. B* **59**, R2514-R2516 (1999).
138 J. C. Maxwell, *A treatise on electricity and magnetism (vol 1)*. (Dover Publications, 1954).
139 T. C. Choy, *Effective medium theory: Principles and applications*. (OUP Oxford, 2015).
140 V. A. Markel, “Introduction to the Maxwell Garnett approximation: Tutorial,” *J. Opt. Soc. Am. A* **33**, 1244-1256 (2016).
141 D. A. G. Bruggeman, “Berechnung verschiedener physikalischer konstanten von heterogenen substanzen. I. Dielektrizitätskonstanten und leitfähigkeiten der mischkörper aus isotropen substanzen,” *Ann Phys Leipzig* **416**, 636-664 (1935).
142 R. Pal, *Electromagnetic, mechanical, and transport properties of composite materials*. (Taylor & Francis, 2014).
143 P. Cosenza *et al.*, “Effective medium theories for modelling the relationships between electromagnetic properties and hydrological variable in geomaterials: A review,” *Near Surface Geophysics* **7**, 563-578 (2009).
144 S. Kirkpatrick, “Percolation and conduction,” *Rev. Mod. Phys.* **45**, 574-588 (1973).
145 A. Davidson and M. Tinkham, “Phenomenological equations for the electrical conductivity of microscopically inhomogeneous materials,” *Phys. Rev. B* **13**, 3261-3267 (1976).
146 C. G. Granqvist and O. Hunderi, “Conductivity of inhomogeneous materials: Effective-medium theory with dipole-dipole interaction,” *Phys. Rev. B* **18**, 1554-1561 (1978).
147 D. S. McLachlan, M. Blaszkiewicz, and R. E. Newnham, “Electrical resistivity of composites,” *J. Am. Ceram. Soc.* **73**, 2187-2203 (1990).
148 D. S. McLachlan, “Evaluating the microstructure of conductor-insulator composites using effective media and percolation theories,” *MRS Proceedings* **411**, 309 (1995).
149 J. P. Straley, “Thermoelectric properties of inhomogeneous materials,” *J. Phys. D: Appl. Phys.* **14**, 2101-2105 (1981).
150 O. Levy and D. J. Bergman, “Scaling behaviour of the thermopower in a two-component composite near a percolation threshold,” *J. Phys. A: Math. Gen.* **25**, 1875-1884 (1992).
151 D. S. McLachlan, “The percolation exponents for electrical and thermal conductivities and the permittivity and permeability of binary composites,” *Phys. B Condens. Matter* **606**, 412658 (2021).
152 M. A. J. Michels, “Scaling relations and the general effective-medium equation for isolator-conductor mixtures,” *J. Phys.: Condens. Matter* **4**, 3961-3966 (1992).
153 D. McLachlan, W. Heiss, C. Chiteme, and J. Wu, “Analytic scaling functions applicable to dispersion measurements in percolative metal-insulator systems,” *Phys. Rev. B* **58**, 13558-13564 (1998).
154 M. Sahimi, *Heterogeneous materials i: Linear transport and optical properties*. (Springer New York, 2006).
155 I. Balberg, "Principles of the theory of continuum percolation", in *Encyclopedia of complexity and systems science*, edited by Robert A. Meyers (Springer, 2020), pp. 1-61.
156 A. A. Saberi, "Application of percolation theory to statistical topographies", in *Encyclopedia of complexity and systems science*, edited by Robert A. Meyers (Springer, 2020), pp. 1-19.
157 Z. Ju *et al.*, “Unveiling the dimensionality effect of conductive fillers in thick battery electrodes for high-energy storage systems,” *Appl. Phys. Rev.* **7**, 041405 (2020).
158 Y. Oh and M. F. Islam, “Preformed nanoporous carbon nanotube scaffold-based multifunctional polymer composites,” *ACS Nano* **9**, 4103-4110 (2015).
159 I. Balberg, C. H. Anderson, S. Alexander, and N. Wagner, “Excluded volume and its relation to the onset of percolation,” *Phys. Rev. B* **30**, 3933-3943 (1984).

160 G. Ambrosetti *et al.*, "Solution of the tunneling-percolation problem in the nanocomposite regime," *Phys. Rev. B* **81** (2010).
161 C. H. Kiang *et al.*, "Size effects in carbon nanotubes," *Phys. Rev. Lett.* **81**, 1869-1872 (1998).
162 O. V. Kharissova and B. I. Kharisov, "Variations of interlayer spacing in carbon nanotubes," *RSC Adv.* **4**, 30807-30815 (2014).
163 S. Dresselhaus, G. Dresselhaus, and P. C. Eklund, *Science of fullerenes and carbon nanotubes: Their properties and applications*. (Elsevier Science, 1996).
164 S. D. Bergin *et al.*, "Multicomponent solubility parameters for single-walled carbon nanotube-solvent mixtures," *ACS Nano* **3**, 2340-2350 (2009).
165 O. Zhou *et al.*, "Defects in carbon nanostructures," *Science* **263**, 1744-1747 (1994).
166 D. Reznik, C. Olk, D. Neumann, and J. Copley, "X-ray powder diffraction from carbon nanotubes and nanoparticles," *Phys. Rev. B* **52**, 116 (1995).
167 Y. Maniwa *et al.*, "Multiwalled carbon nanotubes grown in hydrogen atmosphere: An x-ray diffraction study," *Phys. Rev. B* **64**, 731051-731054 (2001).
168 Z. Q. Li *et al.*, "X-ray diffraction patterns of graphite and turbostratic carbon," *Carbon* **45**, 1686-1695 (2007).
169 R. Mitsuyama *et al.*, "Chirality fingerprinting and geometrical determination of single-walled carbon nanotubes: Analysis of fine structure of x-ray diffraction pattern," *Carbon* **75**, 299-306 (2014).
170 Technical data sheet: MWCNTs (>95%, od: 30-50 nm). https://www.us-nano.com/inc/sdetail/249 (accessed February 10, 2022).
171 J. B. Wu *et al.*, "Raman spectroscopy of graphene-based materials and its applications in related devices," *Chem. Soc. Rev.* **47**, 1822-1873 (2018).
172 Á. Kukovecz, G. Kozma, and Z. Kónya, "Multi-walled carbon nanotubes", in *Springer handbook of nanomaterials* (Springer, 2013), pp. 147-188.
173 J.-B. Wu, M.-L. Lin, and P.-H. Tan, "Raman spectroscopy of monolayer and multilayer graphenes", in *Raman spectroscopy of two-dimensional materials*, edited by Ping-Heng Tan (Springer, 2019), pp. 1-27.
174 A. Jorio and R. Saito, "Raman spectroscopy for carbon nanotube applications," *J. Appl. Phys.* **129**, 021102 (2021).
175 A. C. Ferrari and D. M. Basko, "Raman spectroscopy as a versatile tool for studying the properties of graphene," *Nat. Nanotechnol.* **8**, 235-246 (2013).
176 A. Jorio, M. S. Dresselhaus, R. Saito, and G. Dresselhaus, *Raman spectroscopy in graphene related systems*. (Wiley, 2011).
177 B. Ellis and R. Smith, *Polymers: A property database*. (CRC Press/Taylor & Francis Group, 2008).
178 S. H. Lee, E. Cho, S. H. Jeon, and J. R. Youn, "Rheological and electrical properties of polypropylene composites containing functionalized multi-walled carbon nanotubes and compatibilizers," *Carbon* **45**, 2810-2822 (2007).
179 P. Pötschke, F. Mothes, B. Krause, and B. Voit, "Melt-mixed pp/MWCNT composites: Influence of CNT incorporation strategy and matrix viscosity on filler dispersion and electrical resistivity," *Polymers* **11** (2019).
180 C. C. Chu *et al.*, "Electrical conductivity and thermal stability of polypropylene containing well-dispersed multi-walled carbon nanotubes disentangled with exfoliated nanoplatelets," *Carbon* **50**, 4711-4721 (2012).
181 J. Wang *et al.*, "More dominant shear flow effect assisted by added carbon nanotubes on crystallization kinetics of isotactic polypropylene in nanocomposites," *ACS Appl. Mater. Interfaces* **7**, 1364-1375 (2015).
182 A. Funck and W. Kaminsky, "Polypropylene carbon nanotube composites by in situ polymerization," *Compos. Sci. Technol.* **67**, 906-915 (2007).
183 A. A. Koval'chuk *et al.*, "Synthesis and properties of polypropylene/multiwall carbon nanotube composites," *Macromolecules* **41**, 3149-3156 (2008).
184 S. H. Park *et al.*, "High areal capacity battery electrodes enabled by segregated nanotube networks," *Nat. Energy* (2019).
185 X. Zhang *et al.*, "Electrically conductive polypropylene nanocomposites with negative permittivity at low carbon nanotube loading levels," *ACS Appl. Mater. Interfaces* **7**, 6125-6138 (2015).
186 H. Y. Wu *et al.*, "Simultaneously improved electromagnetic interference shielding and mechanical performance of segregated carbon nanotube/polypropylene composite via solid phase molding," *Compos. Sci. Technol.* **156**, 87-94 (2018).
187 H. Y. Wu *et al.*, "Injection molded segregated carbon nanotube/polypropylene composite for efficient electromagnetic interference shielding," *Ind. Eng. Chem. Res.* **57**, 12378-12385 (2018).
188 K. A. Imran, J. Lou, and K. N. Shivakumar, "Enhancement of electrical and thermal conductivity of polypropylene by graphene nanoplatelets," *J. Appl. Polym. Sci.* **135** (2018).
189 J. E. Mark, *Polymer data handbook*. (Oxford University Press, USA, 2009).
190 H. Dai, E. W. Wong, and C. M. Lieber, "Probing electrical transport in nanomaterials: Conductivity of individual carbon nanotubes," *Science* **272**, 523-526 (1996).
191 T. W. Ebbesen *et al.*, "Electrical conductivity of individual carbon nanotubes," *Nature* **382**, 54-56 (1996).
192 B. Bourlon *et al.*, "Determination of the intershell conductance in multiwalled carbon nanotubes," *Phys. Rev. Lett.* **93**, 176806-176801-176806-176804 (2004).
193 A. Stetter, J. Vancea, and C. H. Back, "Determination of the intershell conductance in a multiwall carbon nanotube," *Appl. Phys. Lett.* **93** (2008).
194 B. P. Grady, "Electrical properties", in *Carbon nanotube-polymer composites: Manufacture, properties, and applications* (Wiley, 2011).
195 N. F. Zorn and J. Zaumseil, "Charge transport in semiconducting carbon nanotube networks," *Appl. Phys. Rev.* **8**, 041318 (2021).
196 R. Hanus *et al.*, "Thermal transport in defective and disordered materials," *Appl. Phys. Rev.* **8**, 031311 (2021).
197 T. Hu and B. I. Shklovskii, "Theory of hopping conductivity of a suspension of nanowires in an insulator," *Phys. Rev. B* **74** (2006).
198 A. B. Kaiser and V. Skakalova, "Electronic conduction in polymers, carbon nanotubes and graphene," *Chem. Soc. Rev.* **40**, 3786-3801 (2011).
199 P. Pipinys and A. J. C. E. J. o. P. Kiveris, "Variable range hopping and/or phonon-assisted tunneling mechanism of electronic transport in polymers and carbon nanotubes," *Cent. Eur. J. Phys.* **10**, 271-281 (2012).
200 L. He and S. C. Tjong, "Zener tunneling in polymer nanocomposites with carbonaceous fillers", in *Nanocrystalline materials: Their synthesis-structure-property relationships and applications* (Elsevier, 2014), pp. 377-406.
201 S. Gong, Z. H. Zhu, and Z. Li, "Electron tunnelling and hopping effects on the temperature coefficient of resistance of carbon nanotube/polymer nanocomposites," *Phys. Chem. Chem. Phys.* **19**, 5113-5120 (2017).

202 R. Hull *et al.*, "Stochasticity in materials structure, properties, and processing - a review," *Appl. Phys. Rev.* **5**, 011302 (2018).
203 T. W. Ebbesen and P. M. Ajayan, "Large-scale synthesis of carbon nanotubes," *Nature* **358**, 220-222 (1992).
204 D. J. Yang *et al.*, "Thermal conductivity of multiwalled carbon nanotubes," *Phys. Rev. B* **66**, 1654401-1654406 (2002).
205 B. Marinho *et al.*, "Electrical conductivity of compacts of graphene, multi-wall carbon nanotubes, carbon black, and graphite powder," *Powder Technol.* **221**, 351-358 (2012).
206 M. Ghislandi *et al.*, "Electrical conductivities of carbon powder nanofillers and their latex-based polymer composites," *Composites, Part A* **53**, 145-151 (2013).
207 S. H. Ryu *et al.*, "The effect of polymer particle size on three-dimensional percolation in core-shell networks of pmma/MWCNTs nanocomposites: Properties and mathematical percolation model," *Compos. Sci. Technol.* **165**, 1-8 (2018).
208 F. J. J. o. M. S. Lux, "Models proposed to explain the electrical conductivity of mixtures made of conductive and insulating materials," *J. Mater. Sci.* **28**, 285-301 (1993).
209 L. Berhan and A. M. Sastry, "Modeling percolation in high-aspect-ratio fiber systems. I. Soft-core versus hard-core models," *Phys. Rev. E* **75** (2007).
210 R. M. Mutiso and K. I. Winey, "Electrical conductivity of polymer nanocomposites", in *Polymer science: A comprehensive reference* (Elsevier, 2012), Vol. 7, pp. 327-344.
211 C. Li, E. T. Thostenson, and T. W. Chou, "Dominant role of tunneling resistance in the electrical conductivity of carbon nanotube-based composites," *Appl. Phys. Lett.* **91** (2007).
212 I. Balberg, "The importance of bendability in the percolation behavior of carbon nanotube and graphene-polymer composites," *J. Appl. Phys.* **112** (2012).
213 I. Balberg, "Unified model for pseudononuniversal behavior of the electrical conductivity in percolation systems," *Phys. Rev. Lett.* **119** (2017).
214 A. A. Saberi, "Recent advances in percolation theory and its applications," *Phys. Rep.* **578**, 1-32 (2015).
215 M. Li *et al.*, "Percolation on complex networks: Theory and application," *Phys. Rep.* **907**, 1-68 (2021).
216 H. F. Mark, *Encyclopedia of polymer science and technology*. (Wiley-Interscience, 2007).
217 M. Fujii *et al.*, "Measuring the thermal conductivity of a single carbon nanotube," *Phys. Rev. Lett.* **95**, 065502 (2005).
218 P. Kim, L. Shi, A. Majumdar, and P. L. McEuen, "Thermal transport measurements of individual multiwalled nanotubes," *Phys. Rev. Lett.* **87**, 2155021-2155024 (2001).
219 P. Wang, R. Xiang, and S. Maruyama, "Thermal conductivity of carbon nanotubes and assemblies," *Advances in Heat Transfer* **50** (2018).
220 X. J. Hu *et al.*, "3-omega measurements of vertically oriented carbon nanotubes on silicon," *J. Heat Transfer* **128**, 1109-1113 (2006).
221 Y. Xu, Y. Zhang, E. Suhir, and X. Wang, "Thermal properties of carbon nanotube array used for integrated circuit cooling," *J. Appl. Phys.* **100** (2006).
222 Y. Son *et al.*, "Thermal resistance of the native interface between vertically aligned multiwalled carbon nanotube arrays and their $SiO_2$ /Si substrate," *J. Appl. Phys.* **103** (2008).
223 M. B. Jakubinek *et al.*, "Thermal and electrical conductivity of array-spun multi-walled carbon nanotube yarns," *Carbon* **50**, 244-248 (2012).
224 A. N. Volkov and L. V. Zhigilei, "Heat conduction in carbon nanotube materials: Strong effect of intrinsic thermal conductivity of carbon nanotubes," *Appl. Phys. Lett.* **101** (2012).
225 X. Zhao *et al.*, "Thermal conductivity model for nanofiber networks," *J. Appl. Phys.* **123** (2018).
226 V. Vavilov and D. Burleigh, *Infrared thermography and thermal nondestructive testing*. (Springer, 2020).
227 T. L. Bergman, A. S. Lavine, F. P. Incropera, and D. P. DeWitt, *Fundamentals of heat and mass transfer*. (Wiley, 2020).
228 K. Kloppstech *et al.*, "Giant heat transfer in the crossover regime between conduction and radiation," *Nat. Commun.* **8** (2017).
229 A. Fiorino *et al.*, "Giant enhancement in radiative heat transfer in sub-30 nm gaps of plane parallel surfaces," *Nano Lett.* **18**, 3711-3715 (2018).
230 J. Yang *et al.*, "Observing of the super-planckian near-field thermal radiation between graphene sheets," *Nat. Commun.* **9** (2018).
231 K. Kim *et al.*, "Radiative heat transfer in the extreme near field," *Nature* **528**, 387-391 (2015).
232 V. Chiloyan, J. Garg, K. Esfarjani, and G. Chen, "Transition from near-field thermal radiation to phonon heat conduction at sub-nanometre gaps," *Nat. Commun.* **6**, 6755 (2015).
233 A. Malekan, S. Saber, and A. A. Saberi, "Exact finite-size scaling for the random-matrix representation of bond percolation on square lattice," *Chaos* **32**, 023112 (2022).
234 D. G. Cahill *et al.*, "Nanoscale thermal transport," *J. Appl. Phys.* **93**, 793-818 (2003).
235 D. G. Cahill *et al.*, "Nanoscale thermal transport. Ii. 2003-2012," *Appl. Phys. Rev.* **1**, 011305 (2014).
236 I. L. Spain, "The electronic properties of graphite", in *Chemistry and physics of carbon, a series of advances*, edited by P. L. Walker and Peter A. Thrower (Marcel Dekker, Inc., 1973), Vol. 8, pp. 87-94.
237 M. S. Dresselhaus and G. Dresselhaus, "Intercalation compounds of graphite," *Adv. Phys.* **51**, 1-186 (2002).
238 D. D. L. Chung, "Review graphite," *J. Mater. Sci.* **37**, 1475-1489 (2002).
239 D. D. L. Chung, "Graphite", in *Carbon materials: Science and applications* (World Scientific Publishing Company PTE Limited, 2019), pp. 21-83.
240 Z. Wei *et al.*, "Phonon mean free path of graphite along the c - axis," *Appl. Phys. Lett.* **104** (2014).
241 Q. Fu *et al.*, "Experimental evidence of very long intrinsic phonon mean free path along the c -axis of graphite," *Appl. Phys. Lett.* **106** (2015).
242 H. Zhang, X. Chen, Y. D. Jho, and A. J. Minnich, "Temperature-dependent mean free path spectra of thermal phonons along the c-axis of graphite," *Nano Lett.* **16**, 1643-1649 (2016).
243 T. Nakayama, K. Yakubo, and R. L. Orbach, "Dynamical properties of fractal networks: Scaling, numerical simulations, and physical realizations," *Rev. Mod. Phys.* **66**, 381-443 (1994).
244 J. T. Gostick and A. Z. Weber, "Resistor-network modeling of ionic conduction in polymer electrolytes," *Electrochim. Acta* **179**, 137–145 (2015).
245 K. Tanaka, "Structural phase transitions in chalcogenide glasses," *Phys. Rev. B* **39**, 1270-1279 (1989).
246 R. Zallen, *The physics of amorphous solids*. (Wiley, 2008).
247 W. Lin, "Modeling of thermal conductivity of polymer nanocomposites", in *Modeling and prediction of polymer nanocomposite properties* (Wiley, 2013), pp. 169-200.
248 Z. Hashin and S. Shtrikman, "A variational approach to the theory of the effective magnetic permeability of multiphase materials," *J. Appl. Phys.* **33**, 3125-3131 (1962).
249 D. P. H. Hasselman and L. F. Johnson, "Effective thermal conductivity of composites with interfacial thermal barrier resistance," *J. Compos. Mater.* **21**, 508-515 (1987).

250 Y. Benveniste, “Effective thermal conductivity of composites with a thermal contact resistance between the constituents: Nondilute case,” *J. Appl. Phys.* **61**, 2840-2843 (1987).

251 H. Fricke, “A mathematical treatment of the electric conductivity and capacity of disperse systems i. The electric conductivity of a suspension of homogeneous spheroids,” *Phys. Rev.* **24**, 575-587 (1924).

252 H. Hatta and M. Taya, “Effective thermal conductivity of a misoriented short fiber composite,” *J. Appl. Phys.* **58**, 2478-2486 (1985).

253 C. W. Nan, R. Birringer, D. R. Clarke, and H. Gleiter, “Effective thermal conductivity of particulate composites with interfacial thermal resistance,” *J. Appl. Phys.* **81**, 6692-6699 (1997).

254 C. W. Nan, Z. Shi, and Y. Lin, “A simple model for thermal conductivity of carbon nanotube-based composites,” *Chem. Phys. Lett.* **375**, 666-669 (2003).

255 C. W. Nan, G. Liu, Y. Lin, and M. Li, “Interface effect on thermal conductivity of carbon nanotube composites,” *Appl. Phys. Lett.* **85**, 3549-3551 (2004).

256 A. G. Every, Y. Tzou, D. P. H. Hasselman, and R. Raj, “The effect of particle size on the thermal conductivity of zns/diamond composites,” *Acta Metallurgica Et Materialia* **40**, 123-129 (1992).

257 R. L. Hamilton and O. K. Crosser, “Thermal conductivity of heterogeneous two-component systems,” *Industrial & Engineering Chemistry Fundamentals* **1**, 187-191 (1962).

258 X. Yang *et al.*, “A review on thermally conductive polymeric composites: Classification, measurement, model and equations, mechanism and fabrication methods,” *Adv. Compos. Hybrid Mater.* **1**, 207-230 (2018).

259 R. Prasher *et al.*, “Effect of aggregation on thermal conduction in colloidal nanofluids,” *Appl. Phys. Lett.* **89** (2006).

260 P. Keblinski, "Modeling of heat transport in polymers and their nanocomposites", in *Handbook of materials modeling: Applications: Current and emerging materials*, edited by Wanda Andreoni and Sidney Yip (Springer, 2018), pp. 1-23.

261 M. Foygel *et al.*, “Theoretical and computational studies of carbon nanotube composites and suspensions: Electrical and thermal conductivity,” *Phys. Rev. B* **71** (2005).

262 E. P. Mamunya, V. V. Davidenko, and E. V. Lebedev, “Effect of polymer-filler interface interactions on percolation conductivity of thermoplastics filled with carbon black,” *Compos. Interfaces* **4**, 169-176 (1997).

263 J. Huang *et al.*, “Effective thermal conductivity of epoxy matrix filled with poly(ethyleneimine) functionalized carbon nanotubes,” *Compos. Sci. Technol.* **95**, 16-20 (2014).

264 G. Zhang *et al.*, “A percolation model of thermal conductivity for filled polymer composites,” *J. Compos. Mater.* **44**, 963-970 (2010).

265 K. Lichtenecker, “The electrical output resistance of artificial and natural aggregates - the influence of the composition of "building blocks" in flat lattice arrays,” *Physikalische Zeitschrift* **25**, 193-204 (1924).

266 Y. Agari and T. Uno, “Estimation on thermal conductivities of filled polymers,” *J. Appl. Polym. Sci.* **32**, 5705-5712 (1986).

267 D. M. Bigg, "Thermal conductivity of heterophase polymer compositions", in *Thermal and electrical conductivity of polymer materials* (Springer, 1995), pp. 1-30.

268 H. W. Russell, “Principles of heat flow in porous insulators,” *J. Am. Ceram. Soc.* **18**, 1-5 (1935).

269 R. C. Progelhof, J. L. Throne, and R. R. Ruetsch, “Methods for predicting the thermal conductivity of composite systems: A review,” *Polymer Engineering & Science* **16**, 615-625 (1976).

270 S. C. Cheng and R. I. Vachon, “The prediction of the thermal conductivity of two and three phase solid heterogeneous mixtures,” *Int. J. Heat Mass Transfer* **12**, 249-264 (1969).

271 T. B. Lewis and L. E. Nielsen, “Dynamic mechanical properties of particulate-filled composites,” *J. Appl. Polym. Sci.* **14**, 1449-1471 (1970).

272 H. E. Stanley, *Introduction to phase transitions and critical phenomena*. (Oxford University Press, 1987).

273 S. H. E. Rahbari, A. A. Saberi, H. Park, and J. Vollmer, “Characterizing rare fluctuations in soft particulate flows,” *Nat. Commun.* **8** (2017).

274 J. Fan *et al.*, “Universal gap scaling in percolation,” *Nat. Phys.* **16**, 455-461 (2020).

275 S. Hahne and U. Schindewolf, “Temperature and pressure dependence of the nonmetal-metal transition in sodium–ammonia solutions (electrical conductivity and pressure–volume–temperature data up to 150°C and 1000 bars),” *J. Phys. Chem.* **79**, 2922–2928 (1975).

276 I. Webman, J. Jortner, and M. H. Cohen, “Numerical simulation of electrical conductivity in microscopically inhomogeneous materials,” *Phys. Rev. B* **11**, 2885–2892 (1975).

277 J. Jortner and M. H. Cohen, “Metal-nonmetal transition in metal-ammonia solutions,” *Phys. Rev. B* **13**, 1548–1568 (1976).

278 M. Hirasawa, Y. Nakamura, and M. Shimoji, “Electrical conductivity and thermoelectric power of concentrated lithium-ammonia solutions,” *Ber. Bunsen-Ges./PCCP* **82**, 815–818 (1978).

279 S. Sunde, “Calculation of conductivity and polarization resistance of composite SOFC electrodes from random resistor networks,” *J. Electrochem. Soc.* **142**, L50–L52 (1995).

280 G. J. Lee, K. D. Suh, and S. S. Im, “Study of electrical phenomena in carbon black–filled HDPE composite,” *Polym. Eng. Sci.* **38**, 471–477 (1998).

281 Zdekamsk, V. Kueslek, and J. Paek, “AC conductivity of carbon fiber-polymer matrix composites at the percolation threshold,” *Polym. Compos.* **23**, 95–103 (2002).

282 S. Ju, T. Y. Cai, and Z. Y. Li, “Percolative magnetotransport and enhanced intergranular magnetoresistance in a correlated resistor network,” *Phys. Rev. B* **72** (2005).

283 E. Kymakis and G. A. J. Amaratunga, “Electrical properties of single-wall carbon nanotube-polymer composite films,” *J. Appl. Phys.* **99**, 084302 (2006).

284 T. B. Murtanto, S. Natori, J. Nakamura, and A. Natori, “AC conductivity and dielectric constant of conductor-insulator composites,” *Phys. Rev. B* **74**, 115206 (2006).

285 I. Singh *et al.*, “Optical and electrical characterization of conducting polymer-single walled carbon nanotube composite films,” *Carbon* **46**, 1141–1144 (2008).

286 J. Zhang, M. Mine, D. Zhu, and M. Matsuo, “Electrical and dielectric behaviors and their origins in the three-dimensional polyvinyl alcohol/MWCNT composites with low percolation threshold,” *Carbon* **47**, 1311–1320 (2009).

287 M. Abu-Abdeen, A. S. Ayesh, and A. A. Al Jaafari, “Physical characterizations of semi-conducting conjugated polymer-CNTs nanocomposites,” *J. Polym. Res.* **19**, 9839 (2012).

288 J. Sun *et al.*, “Parallel algorithm for the effective electromagnetic properties of heterogeneous materials on 3D RC network model,” in *Proceedings of the 10$^{th}$ International Symposium on Antennas, Propagation, and EM Theory, ISAPE2012*, Xi'an, China, 2012 (IEEE, Piscataway, NJ), pp. 1214–1218.

289 E. Persky *et al.*, “Non-universal current flow near the metal-insulator transition in an oxide interface,” *Nat. Commun.* **12**, 3311 (2021).

290 P. Keblinski, J. A. Eastman, and D. G. Cahill, "Nanofluids for thermal transport," *Mater. Today* **8**, 36-44 (2005).
291 P. Keblinski, R. Prasher, and J. Eapen, "Thermal conductance of nanofluids: Is the controversy over?," *J. Nanopart. Res.* **10**, 1089-1097 (2008).
292 J. J. Wang, R. T. Zheng, J. W. Gao, and G. Chen, "Heat conduction mechanisms in nanofluids and suspensions," *Nano Today* **7**, 124-136 (2012).
293 R. Zheng *et al.*, "Thermal percolation in stable graphite suspensions," *Nano Lett.* **12**, 188-192 (2012).
294 L. Ma *et al.*, "Viscosity and thermal conductivity of stable graphite suspensions near percolation," *Nano Lett.* **15**, 127-133 (2015).
295 S. Sudhindra, F. Kargar, and A. A. Balandin, "Noncured graphene thermal interface materials for high-power electronics: Minimizing the thermal contact resistance," *Nanomaterials* **11** (2021).
296 S. Sudhindra *et al.*, "Specifics of thermal transport in graphene composites: Effect of lateral dimensions of graphene fillers," *ACS Appl. Mater. Interfaces* **13**, 53073-53082 (2021).
297 J. S. Lewis *et al.*, "Thermal and electrical conductivity control in hybrid composites with graphene and boron nitride fillers," *Mater. Res. Express* **6** (2019).
298 X. Zhang *et al.*, "Preparation of highly thermally conductive but electrically insulating composites by constructing a segregated double network in polymer composites," *Compos. Sci. Technol.* **175**, 135-142 (2019).
299 Z. Barani *et al.*, "Thermal properties of the binary-filler hybrid composites with graphene and copper nanoparticles," *Adv. Funct. Mater.* **30**, 1904008 (2020).
300 J. Ren *et al.*, "Enhanced thermal conductivity of epoxy composites by introducing graphene@boron nitride nanosheets hybrid nanoparticles," *Mater. Des.* **191** (2020).
301 Z. Xu *et al.*, "Enhanced thermal conductivity and electrically insulating of polymer composites," *J. Mater. Sci.* **56**, 4225-4238 (2021).

Supplementary Material

# Unified Modeling and Experimental Realization of Electrical and Thermal Percolation in Polymer Composites

Navid Sarikhani,[1] Zohreh S. Arabshahi,[2] Abbas Ali Saberi,[3,4,a)] and Alireza Z. Moshfegh [2,5,a)]

[1)] *School of Mechanical Engineering, Sharif University of Technology, Tehran 11155-9567, Iran*
[2)] *Department of Physics, Sharif University of Technology, Tehran 11155-9161, Iran*
[3)] *Department of Physics, University of Tehran, Tehran 14395-547, Iran*
[4)] *Max Planck Institute for the Physics of Complex Systems, 01187 Dresden, Germany*
[5)] *Institute for Nanoscience and Nanotechnology, Sharif University of Technology, Tehran 14588-8969, Iran*

[a)] Authors to whom correspondence should be addressed: ab.saberi@ut.ac.ir and moshfegh@sharif.edu

## Table of Contents

## S1 Materials and Methods

### S1.1 Materials

The isotactic homopolymer Polypropylene (PP) supplied by Marun Petrochemical Complex, Iran (Grade: C30S) with a melt flow rate (MFR) of 6 g/10 min (ISO 1133, 230 °C/2.16 kg) and density of 0.9 g/cm$^3$ was used as the polymer matrix for the fabrication of the composites. The Chemical Vapor Deposition (CVD) grown Multiwalled Carbon Nanotubes (MWCNTs) purchased from US Research Nanomaterials Inc., USA (Product No. US4309) with carbon nanotube purity of > 95 wt%, the nominal outer diameter of 20–30 nm, inner diameter of 5–10 nm, length of 10–30 μm, specific surface area of > 110 m$^2$/g, and true density of ~2.1 g/cm$^3$, were used (after deagglomeration) as the filler for the fabrication of the MWCNT–PP composites. Graphite fine powder extra pure purchased from Merck (Product No. 1042062500, 99.5%, < 50 μm) and hexagonal Boron Nitride (h-BN) powder purchased from Alfa Aesar (Product No. 11078, 99.5%, < 10 μm), were used (after exfoliation) as the filler for the fabrication of the Graphene–PP and h-BN–PP composites, respectively. Silver paste from Asahi Chemical Research Laboratory, Japan (Product No. SW1100-1, 80 mΩ/□) was applied for making electrical contacts. Deionized (DI) water with a resistivity of 18.3 MΩ cm was used during all preparation procedures. All other chemicals and solvents used in this work were obtained from commercial sources (Sigma-Aldrich and Merck) in analytical grade and used as received without any further purification.

### S1.2 Instrumentation and characterization

The morphology of samples was investigated with Field Emission Scanning Electron Microscopy (FESEM, TESCAN MIRA$_3$ LMU) at an accelerating voltage of 15.0 kV and Transmission Electron Microscopy (TEM, Zeiss EM900) at an accelerating voltage of 200 kV. For FESEM analysis, before imaging, powder and nanocomposite samples were coated with a 10 nm Au layer by DC sputtering (DSR1 Desk Sputter Coater, Nanostructured Coatings Co.). The X-ray diffraction (XRD) patterns were obtained by a PANalytical X-ray diffractometer (X'Pert PRO MPD) with Cu $K_\alpha$ irradiation ($\lambda_{K\alpha_1}$=1.5406 Å, $\lambda_{K\alpha_2}$=1.5444 Å, $I_{K\alpha_2}/I_{K\alpha_1}$ =0.5) at a generator voltage of 40 kV. The PANalytical X'Pert HighScore Plus software was used for data analysis. Raman spectra were acquired using a Raman Microscope (Teksan, TAKRAM SRM) with a 532 nm laser at 100 mW laser power and the spectral resolution of 6 cm$^{-1}$. An analytical balance (AND GR-202) equipped with a density determination kit (AND AD-1653) was used to measure the density of nanocomposite samples. Ultrasonication was carried out in a bath sonicator (Elmasonic, P60H) with an effective power of 180 W at 37 kHz ultrasonic frequency. Centrifugation was performed by a universal centrifuge (Pole Ideal Tajhiz, PIT-320) at 6000 rpm for 1 h. Ultraviolet-visible (UV-vis) absorption spectra were recorded with a UV-vis spectrophotometer (Jasco V-530). A hot plate magnetic stirrer (IKA, RCT basic) equipped with an accurate temperature sensor (IKA, ETS-D4 fuzzy) was employed for solvent evaporation. The electrical conductivity (volume resistivity) of the nanocomposites was measured in a two-probe configuration after applying silver paste onto the both sides of the samples to reduce the contact resistance by using a digital multimeter (Sanwa PC5000) and/or digital insulation tester (Mastech, MS5205). The thermal conductivity and thermal diffusivity of the nanocomposites were measured based on the transient plane heat source (hot disc) method according to the ISO 22007-2:2015 standard by using a Kapton-insulated nickel sensor with a 4 mm diameter. Infrared (IR) thermal images were recorded by an IR thermal imaging camera (Olip Systems, ThermoCam P200).

## S1.3 Fabrication of MWCNT–PP nanocomposites

Polymer nanocomposites were fabricated with multiwalled carbon nanotube (MWCNT) filler in polypropylene (PP) matrix at different loading levels from 0.01 wt% to 10 wt% according to Table S1. The nanocomposites were in the form of cylinders of 10.7 mm diameter and 5.0–5.3 mm height (depending on the filler loading). To check the reproducibility of the measurements and to more accurately estimate the experimental errors/uncertainties, a number of replicate samples (up to 15 samples) were also fabricated. In addition, to assess the size-effect, several other nanocomposites with larger dimensions were fabricated that the results of their electrical and thermal measurements were not significantly different from the samples presented in Table S1.

To calculate the filler loading in terms of volume percent, $\phi$ (vol%), based on the filler loading in weight percent, $\phi$ (wt%), the following equation was used,

$$\phi(\text{vol\%}) = 100 \times \frac{\rho_m\, \phi(\text{wt\%})}{\rho_m\, \phi(\text{wt\%}) + \rho_f\, [100 - \phi(\text{wt\%})]} \quad \text{(S1)}$$

where $\rho_m$ and $\rho_f$ are the density of the matrix and filler, respectively.

**Table S1** Specifications and fabrication process parameters of the MWCNT–PP nanocomposite samples

| Sample | $\phi$ (wt%) | $\phi$ (vol%)[a] | $m_{PP}$ (mg) | $m_{MWCNT}$ (mg) | Solvent | $C_{Solution}$ (mg/ml) | $V_{Solution}$ (ml)[b] |
|---|---|---|---|---|---|---|---|
| 1 | 0 | 0 | 500 | 0 | – | – | – |
| 2 | 0.01 | 0.004 | 500 | 0.050 | NMP | 0.04 | 1.25 |
| 3 | 0.025 | 0.01 | 500 | 0.125 | NMP | 0.1 | 1.25 |
| 4 | 0.05 | 0.02 | 500 | 0.250 | NMP | 0.1 | 2.50 |
| 5 | 0.1 | 0.04 | 500 | 0.501 | CHP | 0.4 | 1.25 |
| 6 | 0.25 | 0.1 | 500 | 1.25 | CHP | 1.0 | 1.25 |
| 7 | 0.5 | 0.2 | 500 | 2.51 | CHP | 1.0 | 2.51 |
| 8 | 1.0 | 0.4 | 500 | 5.05 | CHP | 1.0 | 5.05 |
| 9 | 2.5 | 1.1 | 500 | 12.82 | CHP | 1.5 | 8.55 |
| 10 | 5.0 | 2.2 | 500 | 26.32 | CHP | 1.5 | 17.54 |
| 11 | 7.5 | 3.4 | 500 | 40.54 | CHP | 1.5 | 27.03 |
| 12 | 10 | 4.5 | 500 | 55.56 | CHP | 1.5 | 37.04 |

[a] $\phi$ (vol%) is calculated from $\phi$ (wt%) based on Eq. (S1) considering $\rho_{PP} = 0.9$ (g/cm$^3$) and $\rho_{MWCNT} = 2.1$ (g/cm$^3$).

[b] $V_{Solution} = m_{MWCNT}/C_{Solution}$

As shown in Fig. S1, our MWCNT–PP nanocomposite fabrication method consisted of four main steps, which are described in the following subsections.

<u>MWCNT Solutions Preparation</u>:

To make the best use of the exceptional electrical/thermal conductivity of carbon nanotubes in composites, it is necessary to first unbundle or deagglomerate them. Here we used the strategy of ultrasonic-assisted dispersion in liquid media for this purpose. In this regard, two organic solvents N-Methyl-2-pyrrolidone (NMP) and N-Cyclohexyl-2-pyrrolidone (CHP), which are known as excellent solvents for carbon nanotubes, were chosen. The maximum dispersibility (concentration after centrifugation) of carbon nanotubes in NMP and CHP was reported to be ~0.1 mg/ml and ~3.5 mg/ml, respectively,[1] which we also independently tested and validated for our MWCNTs. Accordingly, NMP solvent, which evaporates at relatively lower temperatures, was used to prepare dispersions with a concentration of less than or equal to 0.1 mg/ml, and CHP solvent was used to prepare more concentrated dispersions [Fig. S1(a)]. In a typical solution-preparation process, based on Table S1, an exact amount of nanotube was first weighed according to the required concentration and added to 10 ml of the appropriate solvent in a round-bottom test tube. The mixture was then placed in a bath sonicator and 10 min ultrasonicated for every 1 mg of MWCNTs (frequency of 37 kHz and effective power of 180 W). This ultrasonication procedure was selected through a Design of Experiment (DoE) study to effectively deagglomerate MWCNTs and prepare a stable dispersion on the one hand and to minimize the reduction in the length of MWCNTs during ultrasonication on the other hand. The stability of the dispersions and the absence of MWCNT bundles in them was confirmed by centrifugation of the dispersions at 6000 rpm for 1 h and making sure that their concentration did not change before and after centrifugation, (measured using UV–vis spectroscopy through the Beer–Lambert law). Also, the reduction in the length of MWCNTs with the ultrasonication time was monitored by SEM images and it was concluded that the ultrasonication procedure used effectively deagglomerates the nanotubes while keeping their length at an appropriate level (L = 1.1 μm on average).

Polypropylene (PP) Powder Preparation:

In the segregated network method for the fabrication of polymer composites, the free spaces between the compacted polymer particles are used as a scaffold to construct a segregated network of filler particles. For MWCNT–PP nanocomposites, the smaller the PP particle size, the finer the conductive segregated network. However, at the same time, the PP particle size must be large enough so that the MWCNTs can be well dispersed on the surface of the particles and do not agglomerate during the formation of the segregated network. Thus, according to the average length of 1.1 μm for the unbundled MWCNTs in the previous step, the average size of 50 μm was selected to prepare PP powder. To this end, the initial millimeter-sized PP pellets were powdered using a pulverizing machine at -70 °C (below the glass transition temperature of polypropylene at about -20 °C).[2] The obtained powder was then screened by laboratory sieves and the powder particles with an average diameter of 50 ± 20 μm were used to fabricate the nanocomposites [Fig. S1(b)]. It should be noted that nanocomposites with powders of different diameters from 100 μm to 300 μm were also made, but in this study only the results of 50 μm powder have been reported, which gives a finer and more uniform network of nanotubes in nanocomposites.

MWCNT–Coated PP Microparticles Preparation:

Uniform coating of polymer powder microparticles with filler particles is of great importance in the segregated network method. For this purpose, some researchers have combined polymer and filler particles by mechanical mixing, while others have mixed polymer and filler particles in a common solvent.[3] Here we present for the first time a method based on capillary action which provides a uniform coating of MWCNTs on the surface of PP microparticles. To do this, as shown in Fig. S1(c), the MWCNT solution prepared in the first step was added to a vacuum flask containing the PP powder prepared in the second step so that no part of the powder remained unwetted but at the same time the powder was not saturated with the solution. For a flask containing 500 mg of PP powder with an average diameter of 50 μm, we obtained the retention capacity of the MWCNT solution 1.2–1.3 ml. In fact, by adding this amount of MWCNT solution to PP powder, the solution percolates through all the spaces between the powder microparticles due to capillary action, so that the surface of all particles is wetted with the MWCNT solution without the particles floating in the solution.

The whole evaporation process was performed in a silicone oil bath under a vacuum of ~3 mTorr in order to evaporate the organic solvent faster and prevent the solvent, MWCNTs, and PP microparticles from oxidizing.

The oil bath temperature was set to 80 °C for NMP evaporation and to 90 °C for CHP evaporation. Under these conditions, each cycle of addition and evaporation of the MWCNT solution took about 1 min. It is important to note that according to Table S1, to achieve high loading levels of MWCNT in nanocomposites, it is necessary to add the solution to the PP powder several times (1.2-1.3 ml each time) without saturating the powder. For example, to fabricate sample 8 (1 wt%), it is necessary to add 5.05 ml of MWCNT solution in CHP at a concentration of 1.0 mg/ml to 500 mg of PP powder in four cycles and evaporate the CHP solvent in each cycle. Note that to recycle the organic solvent and also to prevent the solvent vapors from entering the vacuum pump and contaminating the pump oil, it is better to use a cold trap (~0 °C) before the pump inlet.

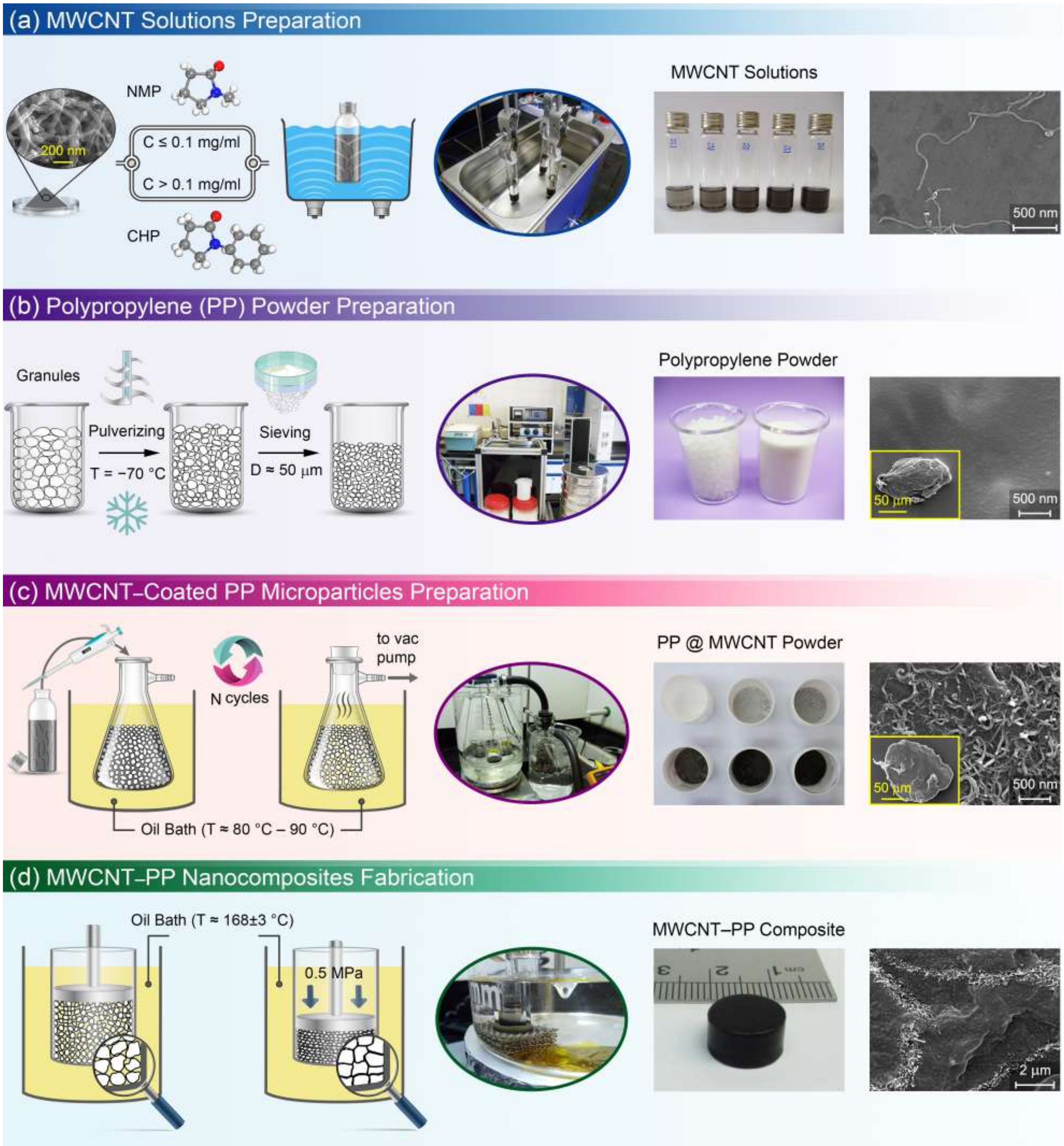

**FIG. S1.** Fabrication steps of MWCNT–PP nanocomposites

MWCNT–PP Nanocomposites Fabrication:

To construct a segregated network of carbon nanotubes, MWCNT-coated PP powder microparticles must be compressed in the mold under controlled temperature and pressure. The choice of temperature and pressure should be such that the PP microparticles soften, deform, and lock together under the molding conditions and form a dense nanocomposite without pores by closing the open spaces between them. On the other hand, the temperature and pressure should not be so high that the structure of the MWCNT coating on the PP microparticles is destroyed and loses its continuity and connectivity. In this way, the MWCNT coatings bond together during controlled compression molding of PP microparticles, forming a well-extended and tightly interconnected segregated network throughout the composite.

In line with this goal, through a Design of Experiment (DoE) study, the temperature window of 168 ± 3 °C and the pressure of 0.5 MPa was selected for the fabrication of nanocomposites. In order to more precisely control the temperature, the compression molding of PP@MWCNT powder microparticles was conducted in the silicone oil bath. During the compression molding process, the piston moved continuously down the mold, and at the same time, the air in the spaces between the PP microparticles escaped from the distance between the piston and the mold body (clearance = 100 μm) until the piston stopped moving and the microparticles completely merged together [Fig. S1(d)]. Typically, the time to fabricate a sample with a diameter of 10.7 mm and a height of 5.2 mm in these molding conditions was about 5 min. After the compression molding was completed, the mold was removed from the oil bath without removing pressure from the sample and it was allowed to naturally cool down to room temperature, and then the pressure was released. SEM images from the cryofractured surfaces of nanocomposite samples (using liquid nitrogen at -196 °C) confirmed the formation of a macroscopic connected network of MWCNTs and the absence of cavities or air bubbles in the samples. It is worthwhile to note that during the cooling and solidification process, the nanocomposites underwent thermal contraction (the thermal expansion coefficient of PP is ~ 0.0001 $K^{-1}$),[2] which facilitated their removal from the mold. In addition, this thermal contraction helped to better connectivity of the nanotubes to each other and make more effective contact between them inside the composite.

## S1.4 Fabrication of Graphene–PP nanocomposites

The fabrication steps of our Graphene–PP nanocomposites are the same as the fabrication steps of our MWCNT–PP nanocomposites, but with two differences in the Solution Preparation and Coating of PP Microparticles steps. As the first difference, in the preparation step of graphene solutions, the well-known Liquid Phase Exfoliation (LPE) method in a mixed solvent of isopropyl alcohol (IPA) and water with 40 vol% IPA and 60 vol% water was used.[4,5] For example, to obtain a 5 ml graphene solution, 50 mg graphite powder was first added to a mixture of 4 ml IPA and 6 ml water (initial concentration of 5 mg/ml). The mixture was then placed in a bath sonicator for 1 h (frequency of 37 kHz and effective power of 180 W). The resulting black dispersion was centrifuged for 15 min at 500 rpm and the 5 ml supernatant was collected. The exact concentration of the collected dispersion was determined 1.5 mg/ml by UV–vis spectroscopy through the Beer–Lambert law. The second difference between the fabrication steps of the Graphene–PP and MWCNT–PP nanocomposites is in the preparation step of the graphene-coated PP microparticles where due to the rapid evaporation of the IPA/water mixture, there was no need to use a vacuum. In practice, each cycle of addition of a 1.2–1.3 ml graphene solution to a 500 mg PP powder (see Table S1 for instructions) and its complete evaporation in an oil bath (~100 °C) at ambient pressure took less than 1 min.

### S1.5 Fabrication of h-BN–PP nanocomposites

The fabrication steps of our h-BN–PP nanocomposites are the same as the fabrication steps of our MWCNT–PP nanocomposites, but with two differences in the Solution Preparation and Coating of PP Microparticles steps. As the first difference, in the preparation step of h-BN solutions, the well-established method of Milling-Assisted Liquid Phase Exfoliation with some modifications was used.[6-8] Briefly, to obtain 5 ml of h-BN solution, 150 mg h-BN powder was first added to 9 g urea and grounded in mortar and pestle for 1 h. The resulting powder was then dissolved in water and centrifuged at 9000 rpm to precipitate h-BN and separate it from water-soluble urea. The h-BN precipitate was again dispersed several times in water and precipitated by centrifugation to completely remove the urea. After the washing process was completed, the obtained h-BN powder was redispersed in 25 ml of IPA and placed in a bath sonicator (frequency 37 kHz and effective power of 180 W) for 2 hours for Liquid Phase Exfoliation. The resulting milky dispersion was then centrifuged for 15 min at 1000 rpm and the 5 ml supernatant was collected. The exact concentration of the collected dispersion was determined 1.0 mg/ml by UV–vis spectroscopy through the Beer–Lambert law. The second difference between the fabrication steps of the h-BN–PP and MWCNT–PP nanocomposites is in the preparation step of the h-BN-coated PP microparticles where due to the rapid evaporation of the IPA solvent, there was no need to use a vacuum. In practice, each cycle of addition of a 1.2–1.3 ml h-BN solution to a 500 mg PP powder (see Table S1 for instructions) and its complete evaporation in an oil bath (~80 °C) at ambient pressure took less than 1 min.

### S1.6 Error analysis

Error analysis can be divided into two general categories of (i) uncertainties in direct measurements and (ii) uncertainties in indirect measurements.[9,10] The vertical error bars in Figures 3 and 4 belong to the first category and represent one standard deviation of multiple (≥ 10) electrical or thermal conductivity direct measurements on separately prepared samples (≥ 3) for each data point. The outcomes of measurements on replicate samples were combined by the Weighted Averages technique[9] to obtain the mean and the random component of uncertainty. The manufacturer's stated instrumental accuracy (for electrical or thermal conductivity measurement) then was combined with the random component of uncertainty using the sum in quadrature to compute the total uncertainty for each data point. The final results are presented as (mean ± total uncertainty) in the vertical error bars. Note that in logarithmic scale, as intended, error bars do not look symmetric around the mean. The horizontal error bars in Figures (3) and (4), on the other hand, belong to the second category (uncertainties in indirect measurements) and were calculated by the Error Propagation analysis,[9,11] on the basis of standard deviation, considering various sources of uncertainty, mainly: (i) determination of the concentration of MWCNT solutions, (ii) measurement of the MWCNT solution volume at each cycle of saturation and evaporation, and (iii) uncertainty in equipment, such as balances, micropipettes and UV–vis spectrophotometers.

### S1.7 Fitting procedure

Fitting electrical and thermal conductivity data to the selected models was performed and cross-checked in multiple statistical software packages and in each case various iteration algorithms, initial conditions and constraints on parameters were considered.

For electrical conductivity, the best-fit parameters in Fig. 3a were obtained by nonlinear curve fitting of the experimental electrical conductivity data to the normalized power-law percolation model (Eq. 10) in the Total Least-Squares sense with the weighted Orthogonal Distance Regression (ODR) algorithm, without any constraints on the three independent parameters of the model, *i.e.* $\phi_c$, $t$ and $\sigma_f$. Here, the OriginPro 2018 software package (OriginLab Corporation, Northampton, MA, USA) was used for the final implementation of the curve fitting. We chose the ODR nonlinear curve fitting strategy instead of the conventional linear least-squares fitting (by taking the logarithm of Eq. 10 and linearization), because the later method only includes those data points that lie beyond the percolation threshold ($\phi > \phi_c$) and one of our data points at $\phi = 0.0214$ vol% was very close the actual percolation threshold. Accordingly, if we wanted to include this point in the curve fitting process, we inevitably imposed the $\phi_c < 0.0214$ vol% constraint on our problem and if we omitted this point, we lost one of our key data points. On the other hand, the ODR method which belongs to the category of generalized least-squares regression models (Total Least-Squares subcategory),[12] is more flexible in dealing with the data points that are very close to the percolation threshold because in addition to the uncertainty in the dependent variable (here $\sigma$) it also takes into account the uncertainty in the independent variable (here $\phi$). Thus, by employing the ODR algorithm, we were able to include the important data point at $\phi = 0.0214$ vol% in the fitting process without imposing any redundant constraints on $\phi_c$. Since the uncertainties in the measured nanotube loading and electrical conductivity of each sample ($\phi_i$,$\sigma_i$) are different from other samples, it is necessary to assign weights to each $\phi_i$ and $\sigma_i$ in the curve fitting process. In this respect, the most common choice is the reciprocal square of the corresponding standard deviation (inverse-variance) of the measured value.[9] Based on our error analysis, which indicated that the uncertainties (standard deviation) in $\phi_i$ and $\sigma_i$ were directly proportional to their values, we considered the weights of $w_{\phi_i} = 1/\phi_i^2$ and $w_{\sigma_i} = 1/\sigma_i^2$ in our nonlinear fitting calculations (weighted ODR algorithm). It should also be noted that in the apparently unweighted linear least-squares fitting of the electrical conductivity data to the power-law percolation model through the change-of-variable of $y_i = \log \sigma_i$, implicitly less weight is given to less precise data (higher $\sigma_i$). We obtained $\phi_c = 0.0243$ vol% and $t = 1.9$ by the weighted ODR algorithm whereas the conventional linear least-squares fitting yielded $\phi_c = 0.0208$ vol% and $t = 2.2$. Although the two sets of parameters are close to each other, we reported the ODR results in the main text and Fig. 3 which we believe are more accurate and realistic.

For thermal conductivity, the best-fit parameters in Fig. 4a were obtained by nonlinear curve fitting of the experimental thermal conductivity data to the TAGP equation (Eq. 9) in the both NonLinear and Total Least-Squares senses. Nonlinear least-square curve fitting was implemented using the lsqnonlin solver in MATLAB R2021a software package. For Total least-square curve fitting, ODR algorithm in OriginPro 2018 was used. The results of both methods were essentially the same within the accuracy of numerical calculations, but in any case, the results of MATLAB calculations have always been reported for thermal conductivity.

## S2 Literature Survey

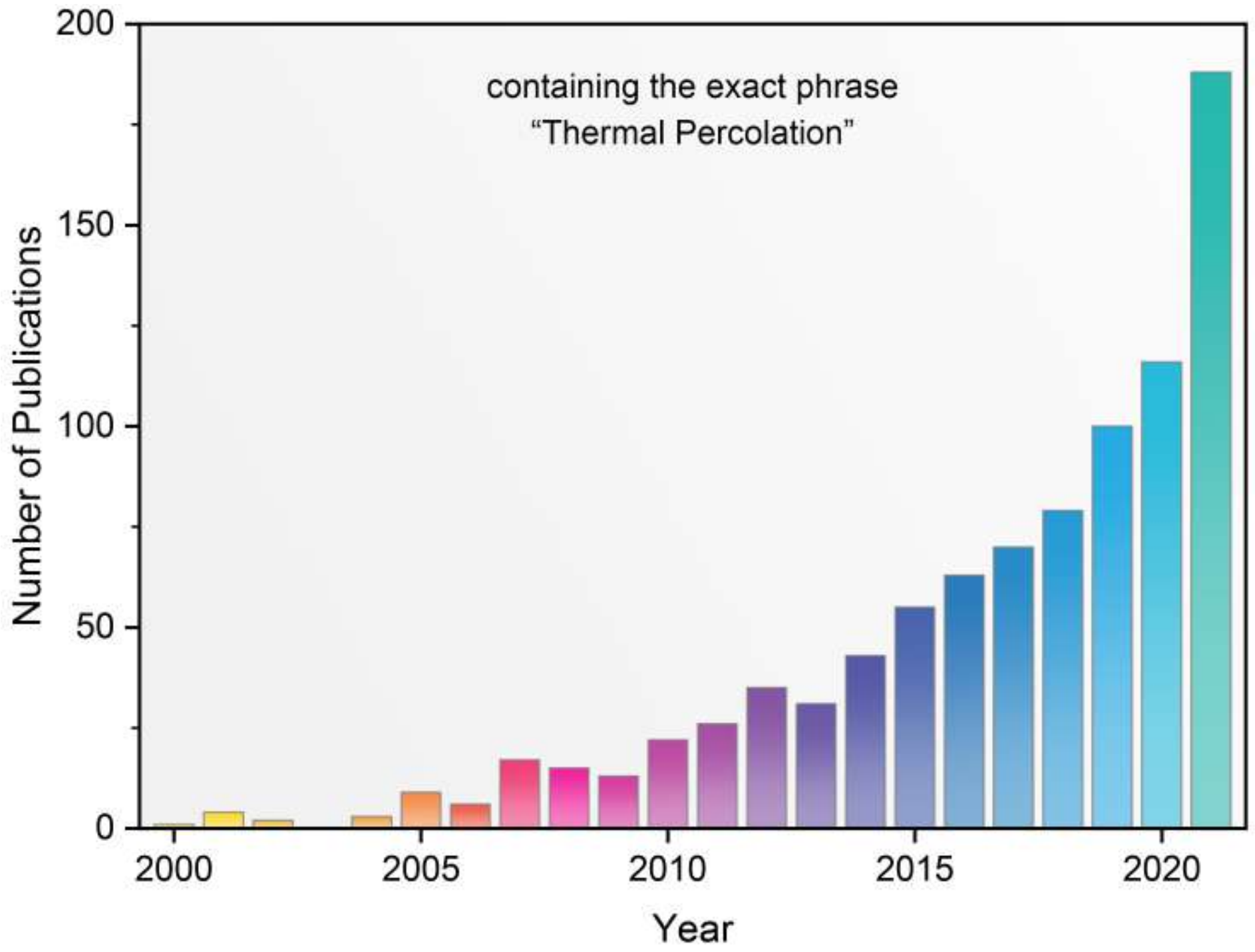

**FIG. S2.** The growing importance of the concept of percolation in the research field of heat transfer, reflected in the increasing annual number of publications containing the exact phrase "Thermal Percolation" in their Title, Abstract, Keywords, or Main Text, from 2000 to 2021, based on Web of Science, Scopus and Google Scholar (extracted and compiled: January 2022)

**Table S2** A chronological summary of the key studies in the field of polymer composites that have measured the variation of both electrical conductivity ($\sigma$) and thermal conductivity ($k$) as a function of filler loading ($\phi$) and reported observation (✓) or lack (✗) of thermal percolation[a]

| | | | | Electrical Percolation | | Thermal Percolation | | | |
|---|---|---|---|---|---|---|---|---|---|
| # | Year | Composite system | Fabrication method | $\phi_{c,el}$ (vol%) | $t_{el}$ | $\phi_{c,th}$ (vol%) | $t_{th}$ | Remarks | Ref. |
| 1 | 2002 | (Cu- or Ni-MP) in (Epoxy or PVC) | Shear/Powder mixing | 4–8 | > 2.4 | ✗ – | – | Modified logarithmic rule of mixtures model for thermal | 13 |
| 2 | 2002 | SWCNT in Epoxy | Sonication & Mixing | > 0.05 | – | ✗ – | – | – | 14 |
| 3 | 2006 | CNT in Epoxy | Mill mixing | 0.05 | – | ✗ – | – | Hatta–Taya model for thermal, Single, double or multiwalled CNTs | 15 |
| 4 | 2006 | CNT in Epoxy | Shear mixing | 0.001 | – | ✗ – | – | SWCNT or MWCNT as fillers | 16 |
| 5 | 2006 | SWCNT in Epoxy | Shear mixing | > 0.04 | ~2.0 | ✗ – | – | – | 17 |
| 6 | 2007 | SWCNT in PMMA | Solution mixing | 0.3 | 2.25 | ✓ 0.3 | 2.25 | Thin films, Limited by experimental accuracy, Non-universal exponents | 18 |
| 7 | 2007 | SWCNT in PE | Solution mixing | 0.3 | – | ✓ 0.3 | 1.57 | Limited by experimental accuracy, Non-universal $t_{th}$ | 19 |
| 8 | 2008 | MWCNT in PVC | Mechanical mixing | 0.05 | 3.3 | ✗ – | – | Lichtenecker logarithmic rule of mixtures model for thermal | 20 |
| 9 | 2008 | MWCNT in PS | Solution mixing | 1 | – | ✗ – | – | Thin films | 21 |
| 10 | 2008 | SWCNT–GNP in Epoxy | Shear mixing | – | – | ✓ – | – | No fit to any percolation models | 22 |
| 11 | 2012 | Graphene in Epoxy | Shear mixing | – | – | ✗ – | – | Maxwell-Garnett for thermal | 23 |
| 12 | 2013 | CNT in Epoxy | Shear mixing | 0.5 | – | ✗ – | – | SWCNT or MWCNT as fillers | 24 |
| 13 | 2013 | MWCNT in (PES or Phenoxy) | Melt mixing | > 0.3 | ~2.0 | ✗ – | – | – | 25 |
| 14 | 2013 | MWCNT in RET | Solution mixing | 2.3 | 5.7 | ✓ 2.2 | 0.1 | Non-universal $t_{th}$, No thermal percolation in *cross-plane* | 26 |
| 15 | 2013 | MWCNT in PDMS | Solution mixing | 0.04 | 1.5 | ✓ 0.03 | 1.2 | Non-universal exponents, thin films (1 mm), long-CNT filler (~0.1 mm) | 27 |
| 16 | 2014 | MWCNT in RET | Solution mixing | 2 | 2 | ✓ 2 | 0.3 | Non-universal $t_{th}$ | 28 |
| 17 | 2015 | Graphene in Epoxy | Ball milling & Mixing | 5 | 0.91 | ✓ 17 | 0.84 | Thin films, Non-universal $t_{th}$ | 29 |
| 18 | 2015 | hBN in Epoxy | Ball milling & Mixing | – | – | ✓ 17 | 0.84 | Thin films, Non-universal $t_{th}$ | 30 |
| 19 | 2016 | Graphene in (PCL or PLA) | Solution mixing | 0.11 | – | ✓ 0.11 | – | No fit to any percolation models | 31 |

**Table S2** (continued)

| # | Year | Composite system | Fabrication method | Electrical Percolation | | Thermal Percolation | | Remarks | Ref. |
|---|---|---|---|---|---|---|---|---|---|
| | | | | $\phi_{c,el}$ (vol%) | $t_{el}$ | $\phi_{c,th}$ (vol%) | $t_{th}$ | | |
| 20 | 2017 | Expanded graphite in Epoxy | Ultrasonic mixing | – | – | ✗ – | – | Plasma treated graphite, Essentially linear variation of $k$ with $\phi$ | 32 |
| 21 | 2017 | Cu-NW in Silicone | Sponge filling | ≳ 0.6 | – | ✓ ≳ 0.6 | – | Electron dominant heat transport, No fit to any percolation models | 33 |
| 22 | 2018 | MWCNT in PVDF | Ultrasonic mixing | 0.096 | 2 | ✗ – | – | Every model for thermal | 34 |
| 23 | 2018 | Ag-NW in PVDF | Solution mixing | 2.25 | 2 | ✓ 2.25 | 2 | Electron dominant heat transport, Wiedemann-Franz law, thin films | 35 |
| 24 | 2018 | Graphene in Epoxy | Shear & Mechanical mixing | 10 | – | ✓ 30 | – | No fit to any percolation models for thermal, Lewis–Nielsen model | 36 |
| 25 | 2018 | hBN in Epoxy | Shear & Mechanical mixing | – | – | ✓ 23 | – | No fit to any percolation models for thermal, Lewis–Nielsen model | 36 |
| 26 | 2019 | Graphene in UHMWPE | Segregated network | 0.21 | 3.0 | ✗ – | – | Lichtenecker logarithmic rule of mixtures model for thermal | 37 |
| 27 | 2019 | (GNP, CF, CNT, CB or G) in PP | Melt mixing | ≥ 2.5 | – | ✗ – | – | Hatta–Taya and Xue models for thermal | 38 |
| 28 | 2019 | Graphene/CF in pCBT | Shear mixing & Hot pressing | ~0.7 | – | ✗ – | – | – | 39 |
| 29 | 2019 | SWCNT in Epoxy | Shear mixing | ~0.6 | – | ✗ – | – | – | 40 |
| 30 | 2019 | Ag nanoflower in PU | Solution Mixing | 4 | – | ✓ 4 | – | Electron dominant heat transport, Wiedemann-Franz law, Thin films | 41 |
| 31 | 2019 | 3D hBN–rGO in NR | Infiltration & Vulcanization | – | – | ✓ 2.3 | 0.35 | Non-universal $t_{th}$ | 42 |
| 32 | 2019 | Graphene/GrF in NR | Pouring & Vulcanization | 6.2 | – | ✓ 6.2 | – | No fit to any percolation models, Highly anisotropic | 43 |
| 33 | 2019 | SCCNT in (Lecithin and NR) | Solution Mixing | ~2.7 | – | ✓ ~6.8 | – | No fit to any percolation models, thin films | 44 |
| 34 | 2019 | Graphene in Epoxy | Shear & Mechanical mixing | ~12 | – | ✓ 22 | – | No fit to any percolation models | 45 |
| 35 | 2020 | MWCNT/GFF in Epoxy | Infiltration into laminates & Pressing | 3.76 | ~0.5 | ✗ – | – | Highly anisotropic | 46 |
| 36 | 2020 | Ni-MP in Epoxy | Shear mixing | 4.0 | 2.9 | ✗ – | – | Lichtenecker logarithmic rule of mixtures model for thermal | 47 |
| 37 | 2020 | Nanodiamond in PP | Melt mixing | – | – | ✗ – | – | 2–35 wt% nanodiamond loading | 48 |
| 38 | 2020 | Magnetized hBN in Epoxy | Solution assisted shear mixing | – | – | ✓ 23.18 | 3 | Non-universal $t_{th}$, Highly anisotropic | 49 |
| 39 | 2020 | Graphene in Epoxy | Shear & Mechanical mixing | 2.5 | 4.28 | ✓ 11.4 | – | No fit to any percolation models for thermal | 50 |

**Table S2** (continued)

| # | Year | Composite system | Fabrication method | Electrical Percolation $\phi_{c,el}$ (vol%) | Electrical Percolation $t_{el}$ | Thermal Percolation $\phi_{c,th}$ (vol%) | Thermal Percolation $t_{th}$ | Remarks | Ref. |
|---|---|---|---|---|---|---|---|---|---|
| 40 | 2020 | Aligned SiC in Epoxy | Infiltration | ~2 | – | ✓ 1.78 | 0.46 | Non-universal $t_{th}$, Highly anisotropic | 51 |
| 41 | 2020 | Graphene in Mineral oil | Solution mixing | – | – | ✓ 1.9 | 0.32 | Non-universal $t_{th}$ | 52 |
| 42 | 2021 | Cu-NP in Epoxy | Sintering & Infiltration | 16 | 1.45 | ✗ – | – | Metallic filler, linear variation of $k$ with $\phi$ | 53 |
| 43 | 2021 | Graphene in PFA | Powder mixing & Electrostatic spraying | ≤ 5 | – | ✗ – | – | Thin films | 54 |
| 44 | 2021 | 3D DAGF in Epoxy | Infiltration | – | – | ✓ 0.67 | 0.97 | Non-universal $t_{th}$, No thermal percolation in *in-plane* | 55 |
| 45 | 2021 | EGaIn in PDMS | Shear mixing | – | – | ✓ ~40 | – | No fit to any percolation models for thermal, Lewis–Nielsen model | 56 |
| 46 | 2021 | hBN in PMMA | Solution mixing | – | – | ✓ 25 | 0.78 | Non-universal $t_{th}$ | 57 |
| 47 | 2021 | MWCNT–Graphene in Epoxy | Shear mixing | 0 | – | ✓ 8 | – | Hybrid filler system with constant $\phi_{MWCNT} \approx 0.08$ vol% | 58 |
| 48 | 2021 | Carbonyl-iron MP in Polyester | Sonication & Mixing | 46 | – | ✓ 38 | – | Zhang extended percolation model for thermal | 59 |
| 49 | 2021 | GNP in PDMS | Shear & Mechanical mixing | ~7 | – | ✓ ~4 | – | No fit to any percolation models | 60 |
| 50 | 2022 | GNP in PE | Melt mixing | 6.4 | 2.98 | ✗ – | – | Agari model for thermal | 61 |
| 51 | 2022 | GNP in Epoxy | Sonication & Mixing | 0.5 | – | ✗ – | – | – | 62 |
| 52 | 2022 | GNP in pCBT | Shear mixing & Hot pressing | – | – | ✓ ~6.2 | 1.01 | Non-universal $t_{th}$ | 63 |
| 53 | 2022 | $Fe_3O_4$-NP in PS-b-P4VP(PDP) | Solution mixing | – | – | ✓ 9 | – | No fit to any percolation models | 64 |
| 54 | 2022 | GNP in PA6 | Melt mixing | – | – | ✓ 18.5 | 0.84 | Thin films, Non-universal $t_{th}$ | 65 |
| 55 | 2022 | Graphene in PLA | Solution mixing | 1.72 | 2.26 | ✓ 2.27 | – | No fit to any percolation models | 66 |
| **56** | **2022** | **MWCNT in PP** | **Segregated Network** | **0.02** | **1.9** | ✓ **1.5** | **1.3** | **TAGP Equation** | **This work** |

[a] $\phi_{c,el}$ and $\phi_{c,th}$ are the electrical and thermal percolation thresholds, and $t_{el}$ and $t_{th}$ are the electrical and thermal critical exponents, respectively as reported by the authors. These parameters are usually obtained from the fit of experimental data to the classical power-law percolation equation, *i.e.* Eq. (2) for electrical and Eq. (9) for thermal percolation, see the main text of the paper for the equations. However, some authors have determined $\phi_{c,el}$ and $\phi_{c,th}$ only on the basis of a visual examination of the plot of the conductivity as a function of filler loading and have not reported any percolation exponents.

**Abbreviations and acronyms in the alphabetical order (see next page):**

**CB**: Carbon black, **CF**: Carbon fiber, **CNT**: Carbon nanotube, **DAGF**: Dual assembled graphene framework, **DWCNT**: Double-walled carbon nanotube, **EGaIn**: Eutectic gallium–indium, **G**: Graphite, **GFF**: Glass fiber fabric, **GNP**: Graphite nanoplatelet, **GrF**: Graphene foam, **hBN**: Hexagonal boron nitride, **MP**: Microparticle, **MWCNT**: Multiwalled carbon nanotube, **NDS**: Nanodiamond soot, **NP**: Nanoparticle, **NR**: Natural rubber, **NW**: Nanowire, **PA6**: Polyamide 6, **pCBT**: Polymerized cyclic butylene terephthalate, **PCL**: Poly(ε-caprolactone), **PDMS**: Polydimethylsiloxane, **PE**: Polyethylene, **PES**: Polyethersulfone, **PFA**: Perfluoroalkoxy, **PLA**: Poly(lactic acid), **PMMA**: Poly(methyl methacrylate), **PS**: Polystyrene, **PS-b-P4VP(PDP)**: Polystyrene-block-poly(4-vinylpyridine), PS-b-P4VP, and 3-pentadecylphenol, PDP, **PU**: Polyurethane, **PVC**: Poly(vinyl chloride), **PVDF**: Poly(vinylidene fluoride), **RET**: Reactive ethylene terpolymer, **rGO**: Reduced graphene oxide, **SCCNT**: Stacked-cup carbon nanotube, **SWCNT**: Single-walled carbon nanotube, **UHMWPE**: Ultra-high molecular weight polyethylene

## S3 Scaling Theory Analysis of the TAGP Equation

The effective conductivity of a composite of good and poor conductors with the intrinsic conductivities of $k_f$ and $k_m$, respectively, in analogy with the scaling theory of phase transitions, can be expressed as,[67-72]

$$\bar{k}_{eff} = |\epsilon|^t \, \Phi_{\pm}\left(\frac{h}{|\epsilon|^{s+t}}\right) \tag{S2}$$

or alternatively

$$\bar{k}_{eff} = h^{\frac{t}{s+t}} \, \Psi\left(\frac{\epsilon}{h^{\frac{1}{s+t}}}\right) \tag{S3}$$

where

$$\bar{k}_{eff} = \frac{k_{eff}}{k_{good}} \tag{S4}$$

$$\epsilon = \frac{\varphi - \varphi_c}{\varphi_c} \tag{S5}$$

$$h = \frac{k_{poor}}{k_{good}} \tag{S6}$$

$$\Phi_+(x) = A_+ + B_+ x + C_+ x^2 + \cdots \tag{S7}$$

$$\Phi_-(x) = B_- x + C_- x^2 + \cdots \tag{S8}$$

$$\Psi(y) = \Psi_0 + \Psi_1 y + \Psi_2 y^2 + \cdots \tag{S9}$$

accordingly

$$\bar{k}_{eff} = \begin{cases} (-\epsilon)^t \, \Phi_-\left(\dfrac{h}{(-\epsilon)^{s+t}}\right), & \epsilon < 0, \quad h \to 0 \\ h^{\frac{t}{s+t}} \, \Psi\left(\dfrac{\epsilon}{h^{\frac{1}{s+t}}}\right), & \epsilon \to 0, \quad 0 < h \ll 1 \\ \epsilon^t \, \Phi_+\left(\dfrac{h}{\epsilon^{s+t}}\right), & \epsilon > 0, \quad h \to 0 \end{cases} \tag{S10}$$

or

$$\bar{k}_{eff} = \begin{cases} (-\epsilon)^t \left\{ B_- \frac{h}{(-\epsilon)^{s+t}} + C_- \left[ \frac{h}{(-\epsilon)^{s+t}} \right]^2 + \cdots \right\}, & \epsilon < 0, \quad h \to 0 \\ h^{\frac{t}{s+t}} \left[ \Psi_0 + \Psi_1 \frac{\epsilon}{h^{\frac{1}{s+t}}} + \Psi_2 \left( \frac{\epsilon}{h^{\frac{1}{s+t}}} \right)^2 + \cdots \right], & \epsilon \to 0, \quad 0 < h \ll 1 \\ \epsilon^t \left[ A_+ + B_+ \frac{h}{\epsilon^{s+t}} + C_+ \left( \frac{h}{\epsilon^{s+t}} \right)^2 + \cdots \right], & \epsilon > 0, \quad h \to 0 \end{cases} \tag{S11}$$

which yields

$$\bar{k}_{eff} = \begin{cases} (-\epsilon)^t \left\{ B_- \frac{h}{(-\epsilon)^{s+t}} + C_- \left[ \frac{h}{(-\epsilon)^{s+t}} \right]^2 + \cdots \right\}, & \epsilon < 0, \quad h \to 0 \\ h^{\frac{t}{s+t}} \left[ \Psi_0 + \Psi_1 \frac{\epsilon}{h^{\frac{1}{s+t}}} + \Psi_2 \left( \frac{\epsilon}{h^{\frac{1}{s+t}}} \right)^2 + \cdots \right], & \epsilon \to 0, \quad 0 < h \ll 1 \\ \epsilon^t \left[ A_+ + B_+ \frac{h}{\epsilon^{s+t}} + C_+ \left( \frac{h}{\epsilon^{s+t}} \right)^2 + \cdots \right], & \epsilon > 0, \quad h \to 0 \end{cases} \tag{S12}$$

By keeping only, the leading terms

$$\bar{k}_{eff} = \begin{cases} B_- \, h \, (-\epsilon)^{-s} + O(h^2), & \epsilon < 0, \quad h \to 0 \\ \Psi_0 \, h^{\frac{t}{s+t}} + O(\epsilon), & \epsilon \to 0, \quad 0 < h \ll 1 \\ A_+ \, \epsilon^t + O(h), & \epsilon > 0, \quad h \to 0 \end{cases} \tag{S13}$$

or finally

$$k_{eff} \sim \begin{cases} k_{poor} \left( \frac{\varphi_c - \varphi}{\varphi_c} \right)^{-s}, & \varphi < \varphi_c, \quad h \to 0 \\ k_{poor}^{\frac{t}{s+t}} \, k_{good}^{\frac{s}{s+t}}, & \varphi \to \varphi_c, \quad 0 < h \ll 1 \\ k_{good} \left( \frac{\varphi - \varphi_c}{\varphi_c} \right)^t, & \varphi > \varphi_c, \quad h \to 0 \end{cases} \tag{S14}$$

## Bruggeman (Symmetric) Effective Medium Approximation (D = 2)

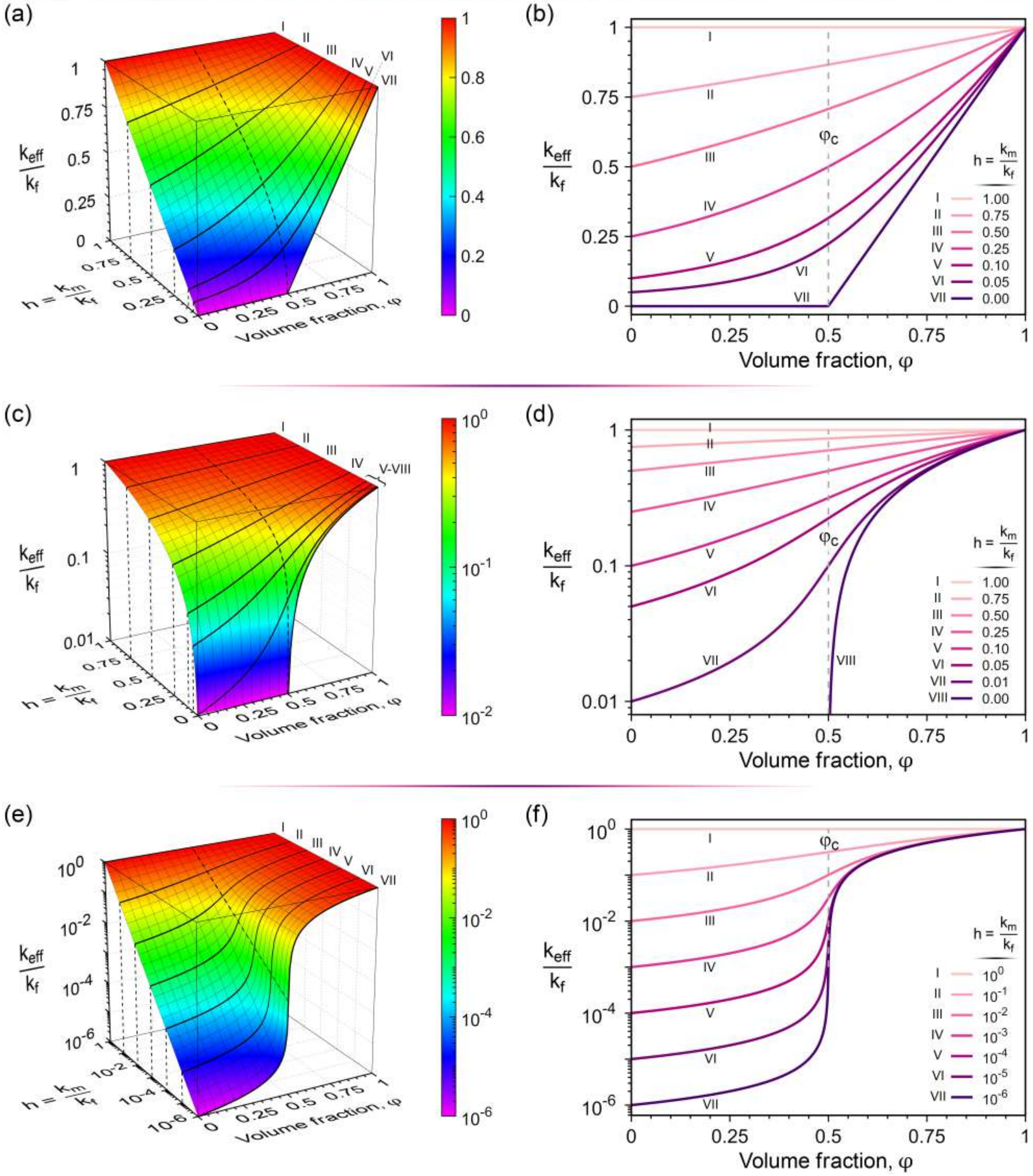

## Bruggeman (Symmetric) Effective Medium Approximation (D = 3)

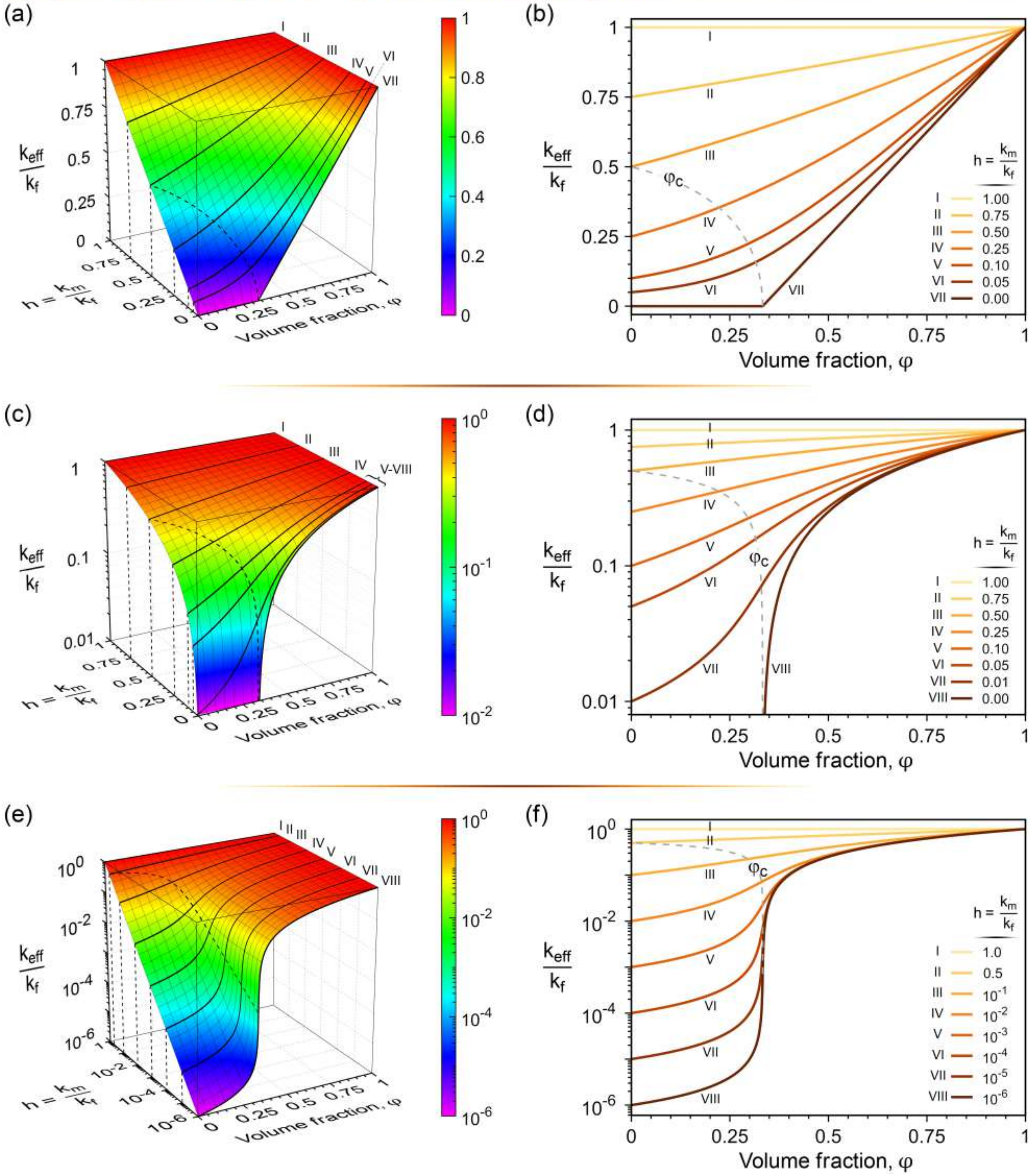

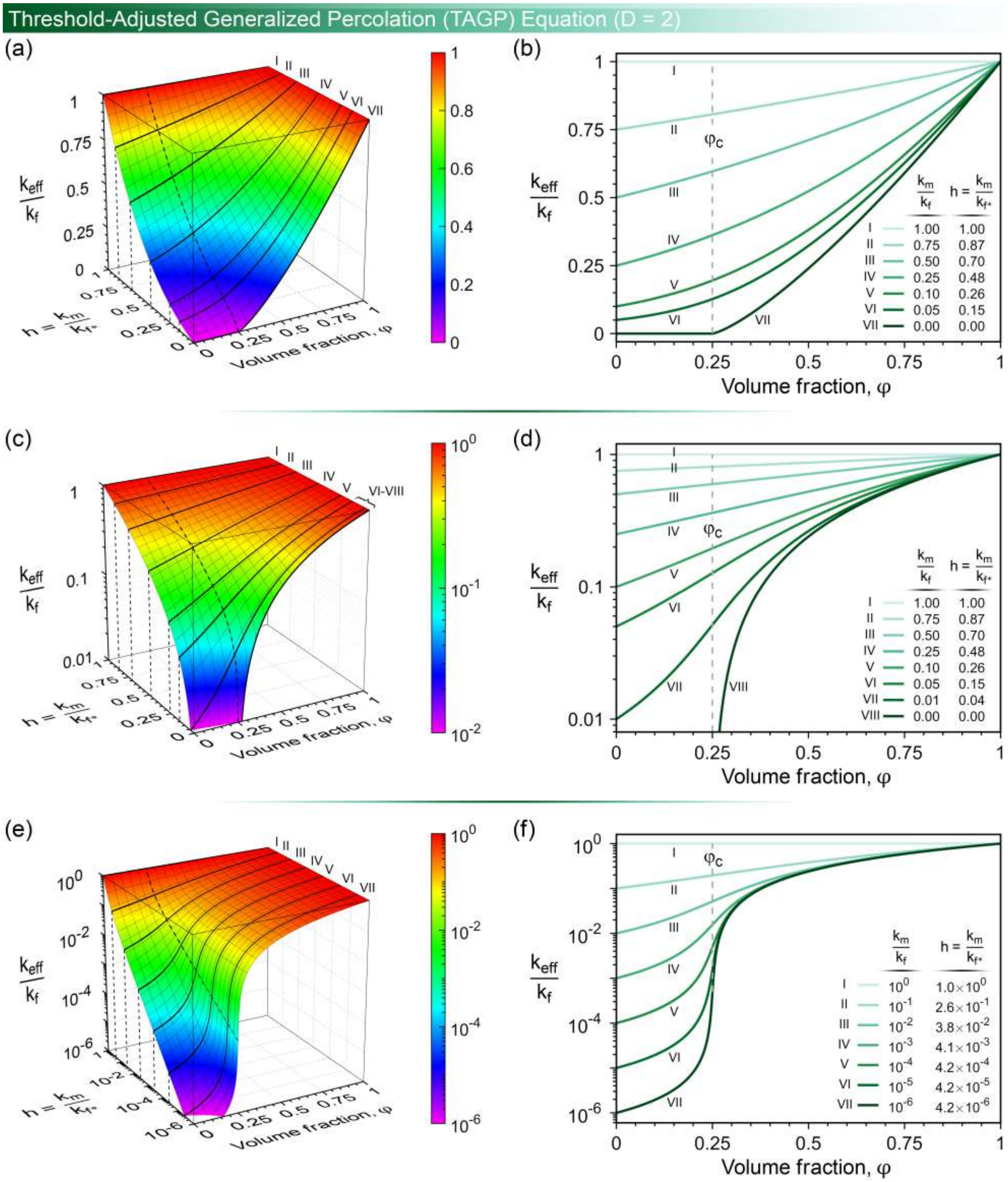
Threshold-Adjusted Generalized Percolation (TAGP) Equation (D = 2)
(a)
(b)
(c)
(d)
(e)
(f)
keff / kf
h = km / kf*
Volume fraction, φ
φc
km / kf
I 1.00 1.00
II 0.75 0.87
III 0.50 0.70
IV 0.25 0.48
V 0.10 0.26
VI 0.05 0.15
VII 0.00 0.00
I 1.00 1.00
II 0.75 0.87
III 0.50 0.70
IV 0.25 0.48
V 0.10 0.26
VI 0.05 0.15
VII 0.01 0.04
VIII 0.00 0.00
VI-VIII
I 10^0 1.0×10^0
II 10^-1 2.6×10^-1
III 10^-2 3.8×10^-2
IV 10^-3 4.1×10^-3
V 10^-4 4.2×10^-4
VI 10^-5 4.2×10^-5
VII 10^-6 4.2×10^-6

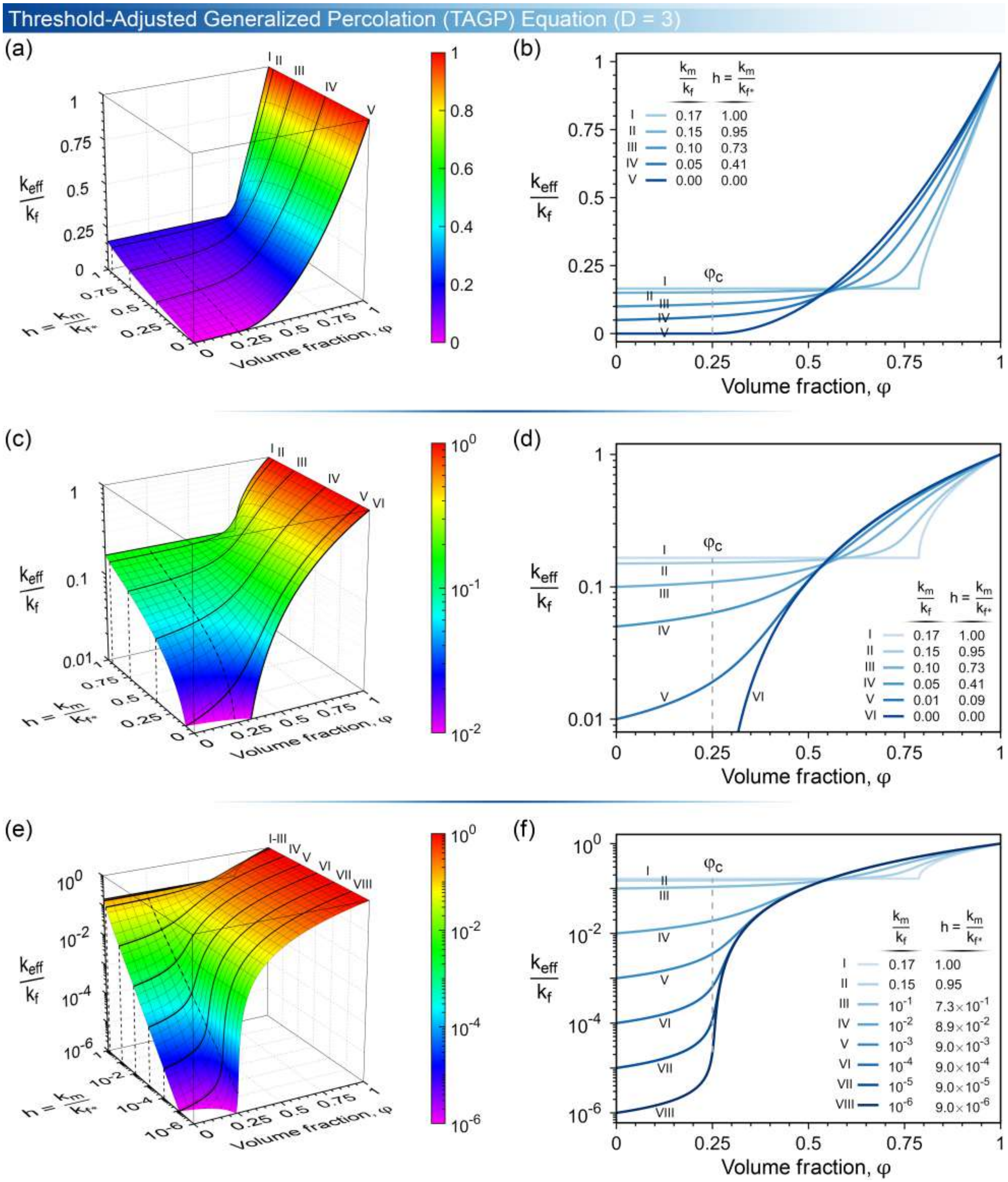
Threshold-Adjusted Generalized Percolation (TAGP) Equation (D = 3)
(a)
(b)
(c)
(d)
(e)
(f)
k_eff / k_f
h = k_m / k_f*
Volume fraction, φ
φ_c
k_m / k_f
I 0.17 1.00
II 0.15 0.95
III 0.10 0.73
IV 0.05 0.41
V 0.00 0.00
I 0.17 1.00
II 0.15 0.95
III 0.10 0.73
IV 0.05 0.41
V 0.01 0.09
VI 0.00 0.00
I 0.17 1.00
II 0.15 0.95
III 10^-1 7.3×10^-1
IV 10^-2 8.9×10^-2
V 10^-3 9.0×10^-3
VI 10^-4 9.0×10^-4
VII 10^-5 9.0×10^-5
VIII 10^-6 9.0×10^-6

## S4 Comparison of the TAGP Equation with the Major Classical Conductivity Models

**Table S3** Summary of the major classical conductivity models for composites [a]

| Model | Equation | Visualization and Remarks ($k_m$: ▭, $k_f$: ■, $k_{eff}$: ▩) | Ref. |
|---|---|---|---|
| **Rule of Mixtures**<br>– Parallel Model<br>– Voigt Model<br>– Wiener Upper Bound<br>– Arithmetic Mean Model | $k_{eff} = (1-\varphi)k_m + \varphi k_f$ | 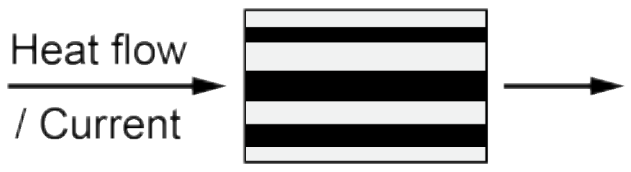<br>Upper bound for anisotropic composites with aligned filler particles, Perfect thermal contact ($R_I = 0$) | 73,74 |
| **Inverse Rule of Mixtures**<br>– Series Model<br>– Reuss Model<br>– Wiener Lower Bound<br>– Harmonic Mean Model | $\frac{1}{k_{eff}} = \frac{1-\varphi}{k_m} + \frac{\varphi}{k_f}$ | 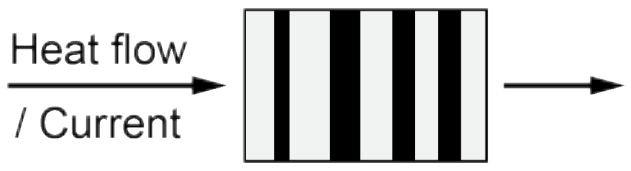<br>Lower bound for anisotropic composites with aligned filler particles, Perfect thermal contact ($R_I = 0$) | 73,74 |
| **Maxwell**<br>– Maxwell Garnett<br>– Maxwell–Eucken<br>– Maxwell–Wagner<br>– Clausius–Mossotti | $\frac{k_{eff}-k_m}{k_{eff}+2k_m} = \varphi\,\frac{k_f-k_m}{k_f+2k_m}$<br>$\longrightarrow k_{eff} = k_m\frac{k_f+2k_m+2\varphi(k_f-k_m)}{k_f+2k_m-\varphi(k_f-k_m)}$ | 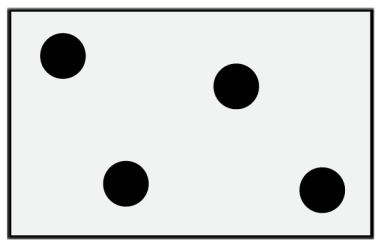<br>Sphere, $R_I = 0$<br>Isotropic composite, Low concentrations ($0 \le \varphi \lesssim 0.1$), Non-interacting spheres of the same size as filler, Perfect thermal contact ($R_I = 0$) | 75-77 |
| **Hasselman–Johnson–Benveniste** | $\frac{k_{eff}-k_m}{k_{eff}+2k_m} = \varphi\,\frac{k_f^c-k_m}{k_f^c+2k_m}$<br>$\longrightarrow k_{eff} = k_m\frac{k_f^c+2k_m+2\varphi(k_f^c-k_m)}{k_f^c+2k_m-\varphi(k_f^c-k_m)}$<br>where<br>$D$: diameter of spherical inclusions<br>$R_I$: interfacial thermal resistance [b]<br>$k_f^c = \frac{k_f}{1+\frac{2R_I k_f}{D}}$ | 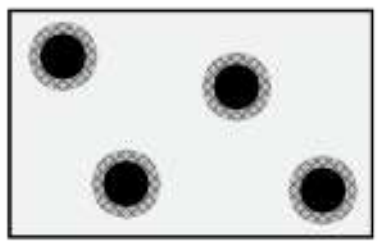<br>Sphere, $R_I \neq 0$<br>Isotropic composite, Medium concentrations ($0 \le \varphi < \pi/6 \approx 0.52$), Non-interacting spheres of the same size as filler, Imperfect thermal contact ($R_I \neq 0$) | 73,78,79 |
| **Fricke**<br>– Hatta–Taya | $\frac{k_{eff}-k_m}{k_{eff}+Ak_m} = \varphi\,\frac{k_f-k_m}{k_f+Ak_m}$<br>$\longrightarrow k_{eff} = k_m\frac{k_f+Ak_m+A\varphi(k_f-k_m)}{k_f+Ak_m-\varphi(k_f-k_m)}$<br>where<br>$L_{ii}$: depolarization factor [c]<br>$A = 2 - \frac{(1-3L_{11})(1-3L_{33})(k_f-k_m)}{k_m+3L_{11}L_{33}(k_f-k_m)}$ | 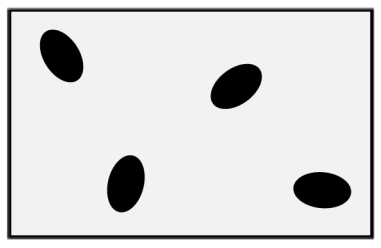<br>Spheroid, $R_I = 0$<br>Isotropic composite, Medium concentrations ($0 \le \varphi \lesssim 0.4$), Non-interacting randomly oriented spheroids as filler, Perfect thermal contact ($R_I = 0$) | 73,80,81 |

**Table S3** (continued)

| Model Name(s) | Formula | Visualization and Remarks<br>($k_m$: , $k_f$: , $k_{eff}$: ) | Ref. |
|---|---|---|---|
| **Nan** | $\frac{k_{eff}-k_m}{k_{eff}+Ak_m}=\varphi\,\frac{k_f^c-k_m}{k_f^c+Ak_m}$<br>$\rightarrow k_{eff}=k_m\frac{k_f^c+Ak_m+A\varphi\left(k_f^c-k_m\right)}{k_f^c+Ak_m-\varphi\left(k_f^c-k_m\right)}$<br>where<br>$a_i$: semi-axes of spheroidal inclusions ($a_1=a_2\neq a_3$)<br>$R_I$: interfacial thermal resistance [b]<br>$L_{ii}$: depolarization factors [c]<br>$\frac{1}{r_{ii}}=\left(\frac{2}{a_1}+\frac{1}{a_3}\right)L_{ii}\rightarrow r_{ii}=\frac{a_1a_3}{(a_1+2a_3)L_{ii}}$<br>$k_{ii}^c=\frac{k_f}{1+\frac{R_I\,k_f}{r_{ii}}}$<br>$\beta_{ii}=\frac{k_{ii}^c-k_m}{k_m+L_{ii}(k_{ii}^c-k_m)}$<br>$M=2\beta_{11}(1-L_{11})+\beta_{33}(1-L_{33})$<br>$N=2\beta_{11}L_{11}+\beta_{33}L_{33}$<br>$A=\frac{M}{N}$<br>$k_f^c=\frac{3+M}{3-N}k_m$ | 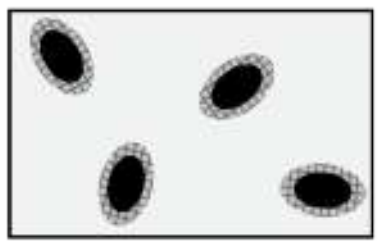<br>Spheroid, $R_I\neq 0$<br>Isotropic composite, Medium concentrations ($0\leq\varphi\lesssim 0.4$), Non-interacting randomly oriented spheroids as filler, Imperfect thermal contact ($R_I\neq 0$) | 73,82 |
| **Hashin–Shtrikman Upper Bound** | $k_{eff}=k_f+\frac{1-\varphi}{\frac{1}{k_m-k_f}+\frac{\varphi}{3k_f}}$ | 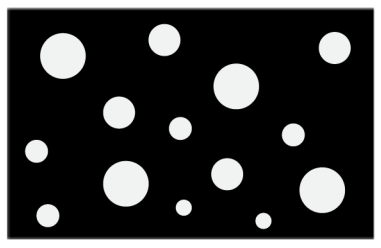<br>Upper bound for isotropic composites, Non-interacting spheres of variable sizes as filler with the conductivity of $k_m$ embedded in a continuum of $k_f$, Perfect thermal contact ($R_I=0$) | 69,76,83 |
| **Hashin–Shtrikman Lower Bound** [d] | $k_{eff}=k_m+\frac{\varphi}{\frac{1}{k_f-k_m}+\frac{1-\varphi}{3k_m}}$ | 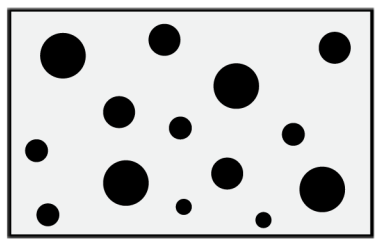<br>Lower bound for isotropic composites, Non-interacting spheres of variable sizes as filler with the conductivity of $k_f$ embedded in a continuum of $k_m$, Perfect thermal contact ($R_I=0$) | 69,76,83 |

**Table S3** (continued)

| Model Name(s) | Formula | Visualization and Remarks<br>($k_m$: light grey box, $k_f$: black box, $k_{eff}$: dark grey box) | Ref. |
|---|---|---|---|
| **Bruggeman Asymmetric**<br>– Bruggeman–Hanai | $(1-\varphi)=\left(\frac{k_m}{k_{eff}}\right)^{\frac{1}{3}}\left(\frac{k_f-k_{eff}}{k_f-k_m}\right)$ | 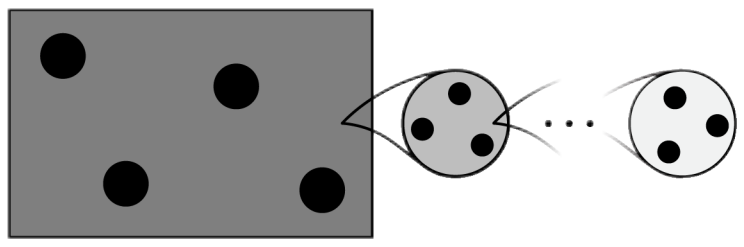<br>Differential Scheme, $R_I=0$<br>Isotropic composite, Step-by-step addition of small amounts of non-interacting spheres as filler with the conductivity of $k_f$ into a continuum with variable effective conductivity, starting from $k_m$ and ending to $k_{eff}$, Perfect thermal contact ($R_I=0$) | 74,84,85 |
| **Bruggeman Symmetric**<br>– Bruggeman Effective Medium Approximation (EMA)<br>– Bruggeman Self-Consistent Field Theory<br>– Bruggeman–Landauer | $(1-\varphi)\,\frac{k_m-k_{eff}}{k_m+2k_{eff}}+\varphi\,\frac{k_f-k_{eff}}{k_f+2k_{eff}}=0$<br>$\longrightarrow k_{eff}=\frac{1}{4}\left(\kappa+\sqrt{\kappa^2+8k_mk_f}\right)$<br>where $\kappa=(3\varphi-1)k_f+(2-3\varphi)k_m$ | 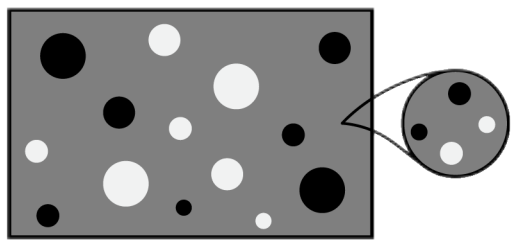<br>Self-Consistent Scheme, $R_I=0$<br>Isotropic composite, Two types of non-interacting spheres as filler with the conductivities of $k_m$ and $k_f$ embedded in a continuum of $k_{eff}$, Perfect thermal contact ($R_I=0$), Gives a fixed percolation threshold at $\varphi_c=\frac{1}{3}$ | 69,74,85 |
| **mGEM**<br>– modified General Effective Media Equation<br>– modified McLachlan Equation | $(1-\varphi)\,\frac{k_m^{\frac{1}{s}}-k_{eff}^{\frac{1}{s}}}{k_m^{\frac{1}{s}}+A\,k_{eff}^{\frac{1}{s}}}+\varphi\,\frac{k_f^{\frac{1}{t}}-k_{eff}^{\frac{1}{t}}}{k_f^{\frac{1}{t}}+A\,k_{eff}^{\frac{1}{t}}}=0$<br>where $A=\frac{1-\varphi_c}{\varphi_c}$ | | 86,87 |
| **TAGP**<br>– Threshold-Adjusted Generalized Percolation Equation | $(2\varphi_c-\varphi)\,\frac{k_m^{\frac{1}{s}}-k_{eff}^{\frac{1}{s}}}{k_m^{\frac{1}{s}}+k_{eff}^{\frac{1}{s}}}+\varphi\,\frac{k_{f^*}^{\frac{1}{t}}-k_{eff}^{\frac{1}{t}}}{k_{f^*}^{\frac{1}{t}}+k_{eff}^{\frac{1}{t}}}=0$ | | This work |

[a] Filler volume fraction ($\varphi$), filler volume fraction at the percolation threshold ($\varphi_c$), composite effective conductivity ($k_{eff}$), matrix bulk conductivity ($k_m$), filler bulk conductivity ($k_f$), composite effective conductivity at $\varphi=2\varphi_c$ ($k_{f^*}$), pre-percolation critical exponent ($s$) where $s_{2D}=1.3$ and $s_{3D}=0.73$, post-percolation critical exponent ($t$) where $t_{2D}=1.3$ and $t_{3D}=2.0$.

[b] Specific interfacial thermal resistance between the filler and the matrix ($R_I$) is defined as:

$$R_I=\frac{\Delta T_i}{\dot{q}_i/A_i}\left(\frac{m^2K}{W}\right)$$

where $\Delta T_i$ $(K)$ is the temperature discontinuity at the interface of filler particles and the matrix, $\dot{q}_i$ $(W)$ is the heat flow across the interface and $A_i$ $(m^2)$ is the nominal area of the interface. It depends mainly on material properties (Kapitza resistance) and the surface roughness of filler particles as well as the pressure and temperature at the interface.[88,89]

[c] Depolarization factors ($L_{ii}$) of a general ellipsoid depend only on the shape of the ellipsoid and are defined as:[81,90]

$$L_{ii} = \frac{a_1\, a_2\, a_3}{2} \int_0^{\infty} \frac{ds}{(a_i^2 + s)\sqrt{(a_1^2 + s)(a_2^2 + s)(a_3^2 + s)}}\ , \qquad \text{i = 1, 2, 3}$$

where $a_1$, $a_2$ and $a_3$ are the lengths of the principal semi-axes of the ellipsoid (volume $= \frac{4\pi}{3} a_1 a_2 a_3$).

For a spheroid ($a_1 = a_2$), depolarization factors have closed-form expressions in terms of elementary functions as follows:

– Prolate (elongated) spheroid ($a_1 = a_2 < a_3$):

$$L_{11} = L_{22} = \frac{p^2}{2(p^2-1)} - \frac{p}{2(p^2-1)^{3/2}} \cosh^{-1} p\,, \quad \text{where } p = \frac{a_3}{a_1} > 1$$

$$L_{33} = 1 - 2L_{11}$$

– Oblate (flattened) spheroid ($a_1 = a_2 > a_3$):

$$L_{11} = L_{22} = \frac{p^2}{2(p^2-1)} + \frac{p}{2(1-p^2)^{3/2}} \cos^{-1} p\,, \quad \text{where } p = \frac{a_3}{a_1} < 1$$

$$L_{33} = 1 - 2L_{11}$$

Three special cases that commonly encountered in nanocomposites, are:

– Spheres ($a_1 = a_2 = a_3 = D/2$) $\rightarrow$ $L_{11} = L_{22} = L_{33} = 1/3$

– Disks ($a_1 = a_2,\ a_3 = 0$) $\rightarrow$ $L_{11} = L_{22} = 0,\ L_{33} = 1$ , *e.g.* graphene nanosheets

– Needles ($a_1 = a_2,\ a_3 \rightarrow \infty$) $\rightarrow$ $L_{11} = L_{22} = 1/2\,,\ L_{33} = 0$ , *e.g.* long carbon nanotubes

[d] Note that the equation of the Hashin–Shtrikman lower bound model is identical to the equation of the Maxwell model. Nevertheless, due to the historical importance of these models, they both are included in the table. See also Ref. 91 for an interesting commentary on who should be given the credit for the Maxwell model.

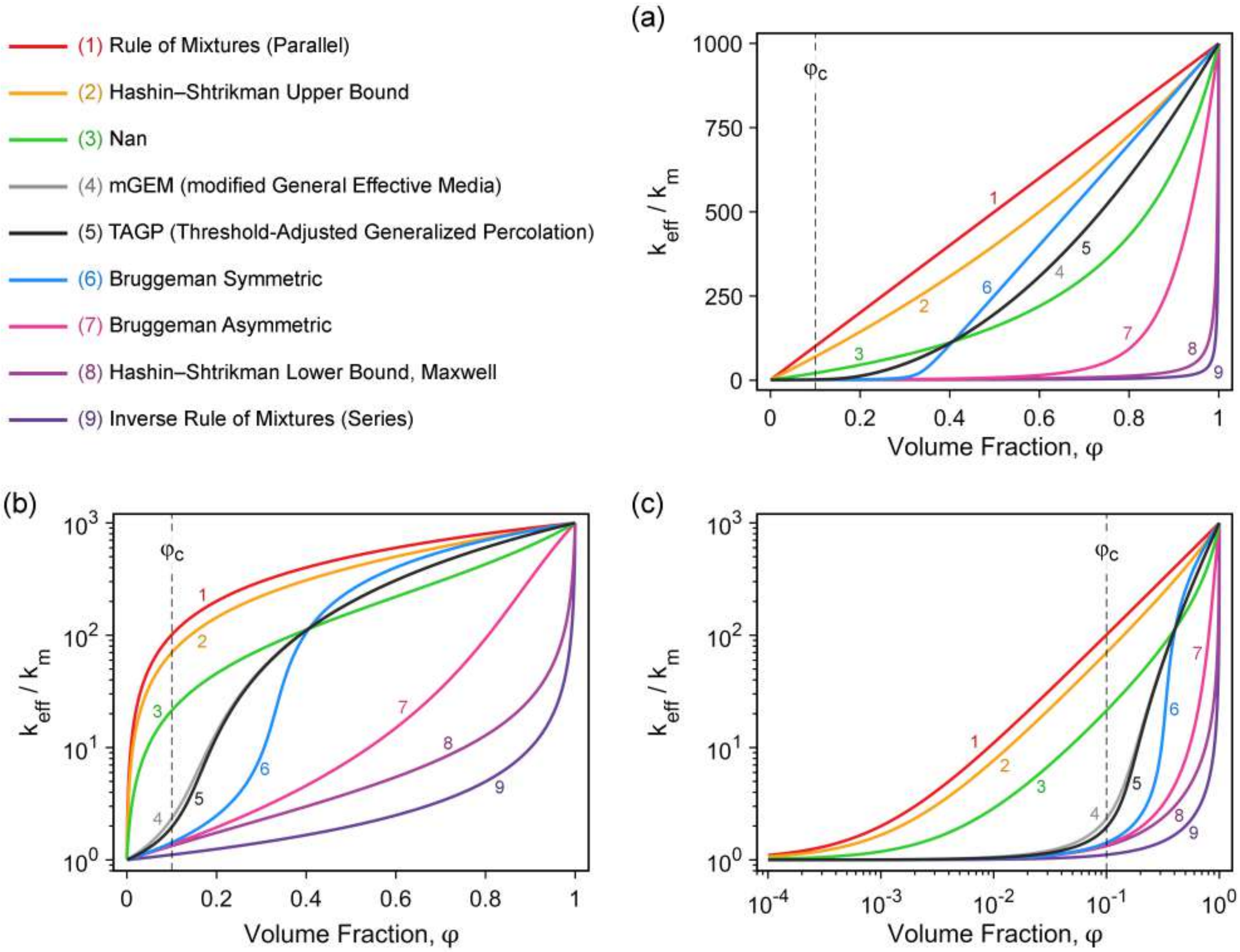

**FIG. S3.** Comparison of the TAGP with the major classical conductivity models by using the following parameters: $k_{eff}(0) = k_m = 0.1$ Wm$^{-1}$K$^{-1}$, $k_{eff}(1) = k_f = 100$ Wm$^{-1}$K$^{-1}$, $\varphi_c = 0.1$ vol, $s = 0.73$, $t = 2$, $a_1 = a_2 = 10$ nm, $a_3 = 500$ nm, $R_I = 10^{-8}$ m$^2$KW$^{-1}$. For the Nan Model, where the interfacial resistance is not included in $k_f$ and is explicitly taken into account, $k_f = 1740.7$ Wm$^{-1}$K$^{-1}$ is used which results $k_{eff,Nan}(1) = k_f^c = 100$ Wm$^{-1}$K$^{-1}$.

## S5 Series Expansion of the Major Classical Conductivity Models

(For the precise definition of the parameters involved in each model, please refer to Table S3)

❖ **Rule of Mixtures (Parallel):**

$$k_{eff} = (1-\varphi)k_m + \varphi k_f \qquad \text{(S15)}$$

$$\rightarrow \frac{k_{eff}}{k_m} = 1 + (\bar{k}-1)\varphi \qquad \bar{k} = \frac{k_f}{k_m} \quad \text{(S16)}$$

$$\Rightarrow \boxed{\boldsymbol{\frac{k_{eff}}{k_m} \approx 1 + \bar{k}\varphi}} \qquad \boldsymbol{\bar{k} \rightarrow \infty} \quad \textbf{(S17)}$$

❖ **Inverse Rule of Mixtures (Series):**

$$\frac{1}{k_{eff}} = \frac{1-\varphi}{k_m} + \frac{\varphi}{k_f} \qquad \text{(S18)}$$

$$\rightarrow \frac{k_{eff}}{k_m} = 1 + \frac{\bar{k}-1}{\bar{k}}\varphi + \left(\frac{\bar{k}-1}{\bar{k}}\right)^2 \varphi^2 + \mathcal{O}(\varphi^3) \qquad \varphi \rightarrow 0\,,\ \bar{k} = \frac{k_f}{k_m} \quad \text{(S19)}$$

$$\Rightarrow \boxed{\boldsymbol{\frac{k_{eff}}{k_m} \approx 1 + \varphi}} \qquad \boldsymbol{\varphi \rightarrow 0\,,\ \bar{k} \rightarrow \infty} \quad \textbf{(S20)}$$

❖ **Maxwell:**

$$k_{eff} = k_m \frac{k_f + 2k_m + 2\varphi(k_f - k_m)}{k_f + 2k_m - \varphi(k_f - k_m)} \qquad \text{(S21)}$$

$$\rightarrow \frac{k_{eff}}{k_m} = 1 + 3\frac{\bar{k}-1}{\bar{k}+2}\varphi + 3\left(\frac{\bar{k}-1}{\bar{k}+2}\right)^2 \varphi^2 + \mathcal{O}(\varphi^3) \qquad \varphi \rightarrow 0\,,\ \bar{k} = \frac{k_f}{k_m} \quad \text{(S22)}$$

$$\Rightarrow \boxed{\boldsymbol{\frac{k_{eff}}{k_m} \approx 1 + 3\,\varphi}} \qquad \boldsymbol{\varphi \rightarrow 0\,,\ \bar{k} \rightarrow \infty} \quad \textbf{(S23)}$$

❖ **Hasselman–Johnson–Benveniste:**

$$k_{eff} = k_m \frac{k_f^c + 2k_m + 2\varphi(k_f^c - k_m)}{k_f^c + 2k_m - \varphi(k_f^c - k_m)} \qquad k_f^c = \frac{k_f}{1 + \frac{2\,R_I\,k_f}{D}} \tag{S24}$$

$$\rightarrow \frac{k_{eff}}{k_m} = 1 + 3\frac{\overline{k^c} - 1}{\overline{k^c} + 2}\varphi + 3\left(\frac{\overline{k^c} - 1}{\overline{k^c} + 2}\right)^2 \varphi^2 + \mathcal{O}(\varphi^3) \qquad \varphi \to 0\,,\ \overline{k^c} = \frac{k_f^c}{k_m}\,,\ \ \bar{k} = \frac{k_f}{k_m} \tag{S25}$$

$$\rightarrow \frac{k_{eff}}{k_m} \approx 1 + 3\ \frac{Nu - 1}{Nu + 2}\varphi \qquad \varphi \to 0\,,\ k_f \to \infty\,,\ Nu = \frac{D}{2R_I k_m} \tag{S26}$$

$$\Rightarrow \boxed{\boldsymbol{\frac{k_{eff}}{k_m} \approx 1 + 3\,\varphi}} \qquad \boldsymbol{\varphi \to 0\,,\ \bar{k} \to \infty\,,\ Nu \to \infty} \tag{S27}$$

❖ **Fricke:**

$$k_{eff} = k_m \frac{k_f + Ak_m + A\varphi(k_f - k_m)}{k_f + Ak_m - \varphi(k_f - k_m)} \tag{S28}$$

$$\rightarrow \frac{k_{eff}}{k_m} = 1 + (A+1)\frac{\bar{k} - 1}{\bar{k} + A}\varphi + (A+1)\left(\frac{\bar{k} - 1}{\bar{k} + A}\right)^2 \varphi^2 + \mathcal{O}(\varphi^3) \qquad \varphi \to 0\,,\ \bar{k} = \frac{k_f}{k_m} \tag{S29}$$

$$\rightarrow \frac{k_{eff}}{k_m} \approx 1 + \frac{1}{3}\frac{(\bar{k} - 1)(\bar{k} + 5)}{(\bar{k} + 1)}\varphi \qquad \varphi \to 0\,,\ a_3 \to \infty \text{ (long fibers)} \tag{S30}$$

$$\Rightarrow \boxed{\boldsymbol{\frac{k_{eff}}{k_m} \approx 1 + \frac{\bar{k}}{3}\varphi}} \qquad \boldsymbol{\varphi \to 0\,,\ a_3 \to \infty\,,\ \bar{k} \to \infty} \tag{S31}$$

❖ **Nan:**

$$k_{eff} = k_m \frac{k_f^c + Ak_m + A\varphi(k_f^c - k_m)}{k_f^c + Ak_m - \varphi(k_f^c - k_m)} \qquad A = \frac{M}{N}\,,\ \ k_f^c = \frac{3 + M}{3 - N}k_m \tag{S32}$$

$$\rightarrow \frac{k_{eff}}{k_m} = 1 + (A+1)\frac{\overline{k^c} - 1}{\overline{k^c} + A}\varphi + (A+1)\left(\frac{\overline{k^c} - 1}{\overline{k^c} + A}\right)^2 \varphi^2 + \mathcal{O}(\varphi^3) \qquad \varphi \to 0\,,\ \overline{k^c} = \frac{k_f^c}{k_m} \tag{S33}$$

$$\rightarrow \frac{k_{eff}}{k_m} \approx 1 + \frac{1}{3}\frac{(Nu+1)\bar{k}^2 + (4Nu - 5)\bar{k} - 5Nu}{(Nu+1)\bar{k} + Nu}\varphi \qquad \begin{gathered}\varphi \to 0\,,\ a_3 \to \infty \text{ (long fibers)} \\ Nu = \frac{a_1}{R_I k_m}\,,\ \ \bar{k} = \frac{k_f}{k_m}\end{gathered} \tag{S34}$$

$$\Rightarrow \boxed{\boldsymbol{\frac{k_{eff}}{k_m} \approx 1 + \frac{\bar{k}}{3}\varphi}} \qquad \boldsymbol{\varphi \to 0\,,\ a_3 \to \infty\,,\ k_f \to \infty} \tag{S35}$$

❖ **Hashin–Shtrikman Upper Bound:**

$$k_{eff} = k_f + \frac{1-\varphi}{\frac{1}{k_m - k_f} + \frac{\varphi}{3k_f}} \tag{S36}$$

$$\rightarrow \frac{k_{eff}}{k_m} = 1 + (2\bar{k}+1)\frac{\bar{k}-1}{3\bar{k}}\varphi + (2\bar{k}+1)\left(\frac{\bar{k}-1}{3\bar{k}}\right)^2 \varphi^2 + \mathcal{O}(\varphi^3) \qquad \varphi \to 0\,,\ \bar{k} = \frac{k_f}{k_m} \tag{S37}$$

$$\Rightarrow \boxed{\boldsymbol{\frac{k_{eff}}{k_m} \approx 1 + \frac{2\bar{k}}{3}\,\varphi}} \qquad \boldsymbol{\varphi \to 0\,,\ \bar{k} \to \infty} \tag{S38}$$

❖ **Hashin–Shtrikman Lower Bound:**

$$k_{eff} = k_m + \frac{\varphi}{\frac{1}{k_f - k_m} + \frac{1-\varphi}{3k_m}} \tag{S39}$$

$$\rightarrow \frac{k_{eff}}{k_m} = 1 + 3\frac{\bar{k}-1}{\bar{k}+2}\varphi + 3\left(\frac{\bar{k}-1}{\bar{k}+2}\right)^2 \varphi^2 + \mathcal{O}(\varphi^3) \qquad \varphi \to 0\,,\ \bar{k} = \frac{k_f}{k_m} \tag{S40}$$

$$\Rightarrow \boxed{\boldsymbol{\frac{k_{eff}}{k_m} \approx 1 + 3\,\varphi}} \qquad \boldsymbol{\varphi \to 0\,,\ \bar{k} \to \infty} \tag{S41}$$

❖ **Bruggeman Asymmetric:**

$$(1-\varphi) = \left(\frac{k_m}{k_{eff}}\right)^{\frac{1}{3}} \left(\frac{k_f - k_{eff}}{k_f - k_m}\right) \tag{S42}$$

$$\Rightarrow \boxed{\boldsymbol{\frac{k_{eff}}{k_m} \approx 1 + 3\,\varphi}} \qquad \boldsymbol{\varphi \to 0\,,\ \bar{k} \to \infty} \tag{S43}$$

❖ **Bruggeman Symmetric:**

$$(1-\varphi)\ \frac{k_m - k_{eff}}{k_m + 2k_{eff}} + \varphi\ \frac{k_f - k_{eff}}{k_f + 2k_{eff}} = 0 \tag{S44}$$

$$\rightarrow \frac{k_{eff}}{k_m} = 1 + 3\frac{\bar{k}-1}{\bar{k}+2}\varphi + 9\left(\frac{\bar{k}}{\bar{k}+2}\right)\left(\frac{\bar{k}-1}{\bar{k}+2}\right)^2 \varphi^2 + \mathcal{O}(\varphi^3) \qquad \varphi \to 0\,,\ \bar{k} = \frac{k_f}{k_m} \tag{S45}$$

$$\Rightarrow \boxed{\boldsymbol{\frac{k_{eff}}{k_m} \approx 1 + 3\,\varphi}} \qquad \boldsymbol{\varphi \to 0\,,\ \bar{k} \to \infty} \tag{S46}$$

## S6 Extension of the Classical Power-Law Percolation Equation

The normalized classical power-law percolation equation for the effective thermal conductivity of composites ($k_{eff}$) as a function of filler volume fraction ($\varphi$) is given by:

$$k_{eff} = k_f \left(\frac{\varphi - \varphi_c}{1 - \varphi_c}\right)^t \tag{S47}$$

where the pre-exponential factor $k_f$ is the bulk thermal conductivity of filler, $t$ is the percolation critical exponent, and $\varphi_c$ is the percolation threshold. However, this model ignores the thermal conductivity of the matrix ($k_m$), while in polymer nanocomposites, the thermal conductivity of the matrix, unlike electrical conductivity, cannot be neglected.

A number of equations have been proposed to extend the classical power-law percolation equation to include the effect of the finite (non-zero) thermal conductivity of the matrix in polymer composites. One of the commonly used extended percolation equations is the equation proposed by Bonnet and colleagues.[18] Inspired by the parallel model in which the both filler and matrix phases are assumed continuous, the Bonnet equation considers the contribution of the matrix to the effective conductivity of the composite as $(1 - \varphi)k_m$, which is then summed with the filler contribution calculated from the normalized classical power-law percolation equation (the top rows of Figures S4a and S4b). Accordingly, the Bonnet equation is written as:

$$k_{eff} = (1 - \varphi)k_m + k_f \left(\frac{\varphi - \varphi_c}{1 - \varphi_c}\right)^t \tag{S48}$$

But if we rearrange the equation of the parallel model, we find that the effective thermal conductivity can be interpreted in yet another way by this model:

$$k_{eff} = (1 - \varphi)k_m + \varphi k_f = k_m + (k_f - k_m)\varphi \tag{S49}$$

According to this second interpretation, the effective conductivity of the composite can be considered as the superposition of a hypothetical phase with the conductivity of $(k_f - k_m)$ instead of the actual filler phase, on a continuous matrix phase ($k_m$) that fills the entire composite space (the bottom row of Figure S4a).

Consequently, while maintaining the spirit of the Bonnet equation, the modified Bonnet equation can be formulated as follows (the bottom row of Figure S4b):

$$k_{eff} = k_m + (k_f - k_m)\left(\frac{\varphi - \varphi_c}{1 - \varphi_c}\right)^t \tag{S50}$$

The modified Bonnet equation, like the original Bonnet equation, predicts a power-law growth in thermal conductivity with filler content, proportional to $(\varphi - \varphi_c)^t$, after the percolation threshold. We also have $k_{eff}(\varphi)\big|_{\varphi=1} = k_f$ in both models. But the modified Bonnet equation has the advantage over the original Bonnet equation that it does not calculate the thermal conductivity of the composite at the onset of percolation ($\varphi = \varphi_c$) less than the thermal conductivity of the matrix.

It should be emphasized that when $\varphi_c$ is small, as is the case in our nanocomposites with $\varphi_c \lesssim 0.01$, the difference between the two models is practically insignificant, and fitting the experimental data with either of the equations yield the same results in terms of the fitting parameters (*i.e.* $k_m$, $k_f$, $t$, and $\varphi_c$).

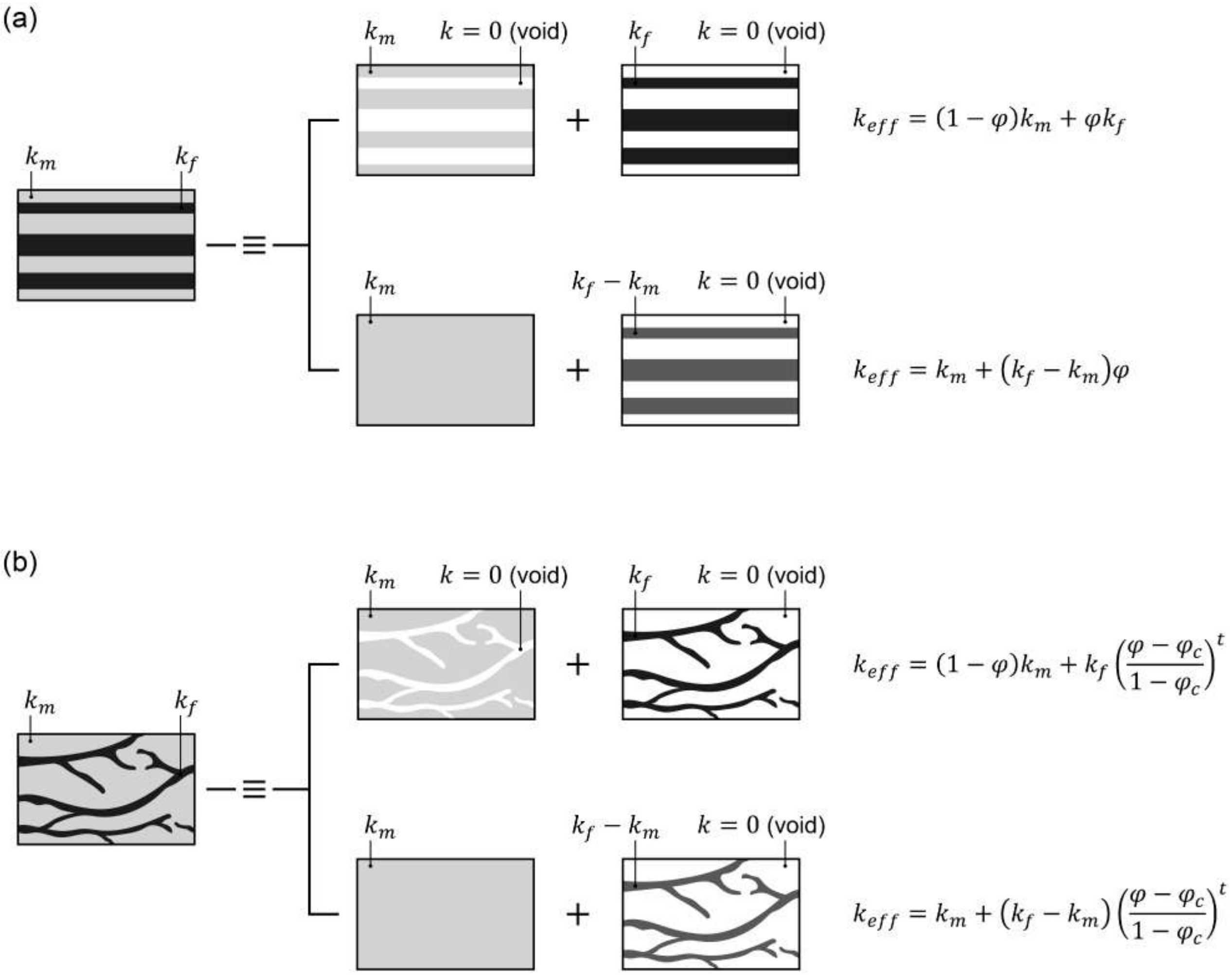

**FIG. S4.** (a) Two possible interpretations of the Parallel Model (Rule of Mixtures), (b) Two possible extensions of the Classical Power-Law Percolation equation for composites to include the effect of the finite (non-zero) conductivity of the matrix in analogy with the Parallel Model.

## S7 Definition of Good and Poor Conductors

### S7.1 Upper bound of conductivity ratio

There exists an upper bound for the conductivity ratio of a composite of good and poor conductors, above which one can neglect the conductivity of the poor conductor. Above this bound, the conductivity model simplifies to Type II of percolation in composites, that is a good conductor incorporated into an insulating matrix, with a well-known standard power-law percolation behavior. In Figure S5, we compare the TAGP with the normalized power-law percolation equation and analyze the characteristic width of the smearing region as a criterion for distinguishing between Type II and Type III percolation models. The following equations are used to construct Figure S5.

$$\text{TAGP: } \frac{(2\varphi_c - \varphi)\left(Y_m^{\frac{1}{s}} - Y_{eff}^{\frac{1}{s}}\right)}{Y_m^{\frac{1}{s}} + Y_{eff}^{\frac{1}{s}}} + \frac{\varphi\left(Y_{f^*}^{\frac{1}{t}} - Y_{eff}^{\frac{1}{t}}\right)}{Y_{f^*}^{\frac{1}{t}} + Y_{eff}^{\frac{1}{t}}} = 0 \tag{S51}$$

$$Y_{f^*} = Y_f \left\{ \frac{\left[1 + \left(\frac{Y_f}{Y_m}\right)^{\frac{1}{s}}\right] - (2\varphi_c - 1)\left[1 - \left(\frac{Y_f}{Y_m}\right)^{\frac{1}{s}}\right]}{\left[1 + \left(\frac{Y_f}{Y_m}\right)^{\frac{1}{s}}\right] + (2\varphi_c - 1)\left[1 - \left(\frac{Y_f}{Y_m}\right)^{\frac{1}{s}}\right]} \right\}^t \tag{S52}$$

$$\text{Normalized Percolation Equation: } Y_{eff} = Y_f \left(\frac{\varphi - \varphi_c}{1 - \varphi_c}\right)^t \tag{S53}$$

$$\text{Characteristic Width of the Smearing Region: } \Delta = \left(\frac{Y_m}{Y_{f^*}}\right)^{\frac{1}{s+t}} \tag{S54}$$

As shown in Figure S5a, at a fixed percolation threshold of 0.01 (1 vol%), the TAGP closely match the normalized power-law percolation equation in volume fractions above the percolation threshold for filler to matrix conductivity ratios equal to or above $10^8$. Below this value the difference between the two equations is noticeable and by further reducing the conductivity ratio this difference becomes more significant. Figure S5b indicates the importance of the percolation threshold in the matching of the TAGP and the classical percolation equation. At a fixed conductivity ratio of $Y_f/Y_m = 10^6$, the TAGP deviates from the classical percolation equation at $\varphi_c = 0.01$ (1 vol%), whereas the matching is excellent at $\varphi_c = 0.1$ (10 vol%). To cast these observations in a quantitative framework, we adopt the characteristic width of the smearing region as a criterion for distinguishing between (i) situations that the classical percolation equation works (*i.e.* situations that the poor conductor can be regarded as an insulator) and (ii) situations that one should use the TAGP (*i.e.* situations that the conductivity of the poor conductor cannot be neglected). It can be inferred from Figure S5c that, in 3D systems, above the conductivity ratio of $10^9$, $\Delta$ is smaller than 0.1 for all practical situations that the percolation threshold is in the range of 0.001 to 0.1 (0.1 to 10 vol%). The actual width of the smearing region is $2\phi_c\Delta$ (see Fig. 1 in the main text and be noted that we use the symbol $\varphi$ for the filler volume fraction which varies between 0 and 1 and the symbol $\phi$ for the filler volume percent which varies between 0 and 100). Thus, filler to matrix conductivity ratios of $10^9$ and above guarantee the validity of classical percolation equation and the approximation of the poor conductor as an insulator ($2\phi_c\Delta < 0.1$ vol%). One should be cautious when using classical percolation equation for $Y_f/Y_m < 10^9$. In particular, for $Y_f/Y_m < 10^4$ the deviation of the classical percolation equation from the TAGP becomes significant ($\Delta > 0.1$) and the conductivity of the poor conductor cannot be neglected anymore. For 2D systems, $Y_f/Y_m > 10^7$ guarantees the validity of the classical percolation equation in all practical situations ($0.001 < \varphi_c < 0.1$) and again for $Y_f/Y_m < 10^4$ serious disparity between the classical power-law percolation equation and the TAGP is observed.

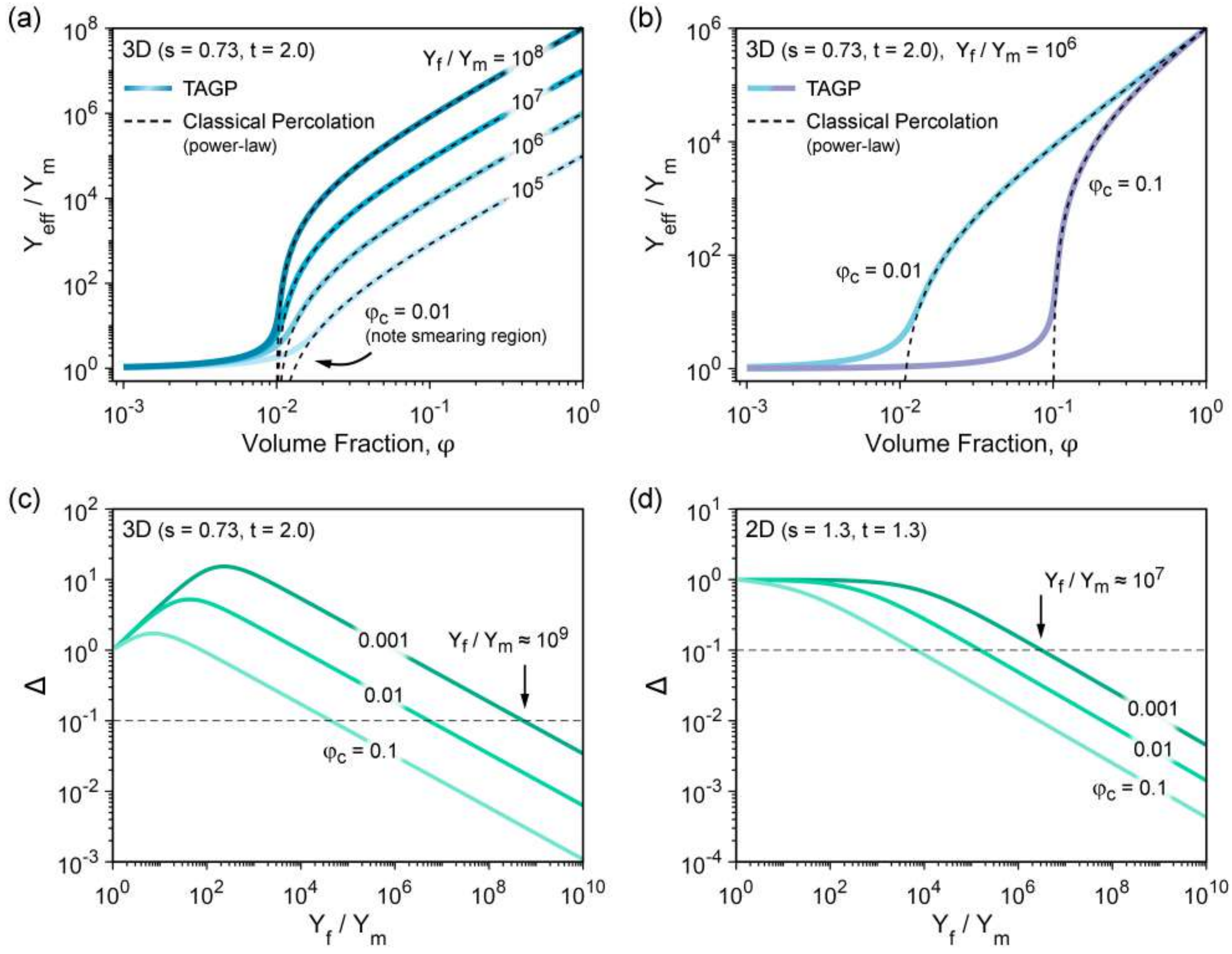

**FIG S5.** (a) Comparison of the TAGP with the classical power-law percolation model at different filler to matrix conductivity ratios ($Y_f/Y_m$) for a fixed percolation threshold of $\varphi_c = 0.01$ ($\phi_c = 1$ vol%), (b) Comparison of the TAGP with the classical power-law percolation model at two different percolation thresholds of 0.01 (1 vol%) and 0.1 (10 vol%) for a fixed filler to matrix conductivity ratio of $10^6$, (c and d) The characteristic width of the smearing region ($\Delta$) as a function of filler to matrix conductivity ratio at three different percolation thresholds of 0.001 (0.1 vol%), 0.01 (1 vol%) and 0.1 (10 vol%) for 3D and 2D systems.

## S7.2 Lower Bound of Conductivity Ratio

The main characteristic of percolation in composites with a large conductivity contrast between the filler and the matrix is an abrupt increase in the conductivity at the percolation threshold. In the composites of good and poor conductors this increase is slightly broadened, but still remains sharp, and occurs in a small transition region, called smearing region. However, when the conductivity of the filler is of the same order of the conductivity of the matrix, the width of the smearing region becomes increasingly large and no abrupt or sharp increase in the conductivity of the composite takes place. Accordingly, there is a lower bound for filler to matrix conductivity ratios below which the picture of percolation becomes blur and only a smooth transition from the conductivity of the matrix to the filler is observed. This later case can be well described in the framework of effective medium theory (EMT) and its specifically developed models, such as the Maxwell, Hashin–Shtrikman, Bruggeman symmetric and Bruggeman Asymmetric models.

Figure S6 shows the overall behavior of the TAGP at the low conductivity ratios of the filler to the matrix. It is evident from Figure S6a that, in 2D systems, at a fixed percolation threshold ($\varphi_c = 0.1$ or $\phi_c = 10$ vol%, here), by decreasing the filler to matrix conductivity ratio ($Y_f/Y_m$) from $10^8$ to $10^2$, the width of the smearing region increases and the variation of the effective conductivity becomes smoother. By further decreasing of the conductivity ratio to 10 and below 10 (Figure S6b), only a smooth transition of the effective conductivity from the conductivity of the matrix to the filler is observed without any noticeable smearing region or percolation behavior.

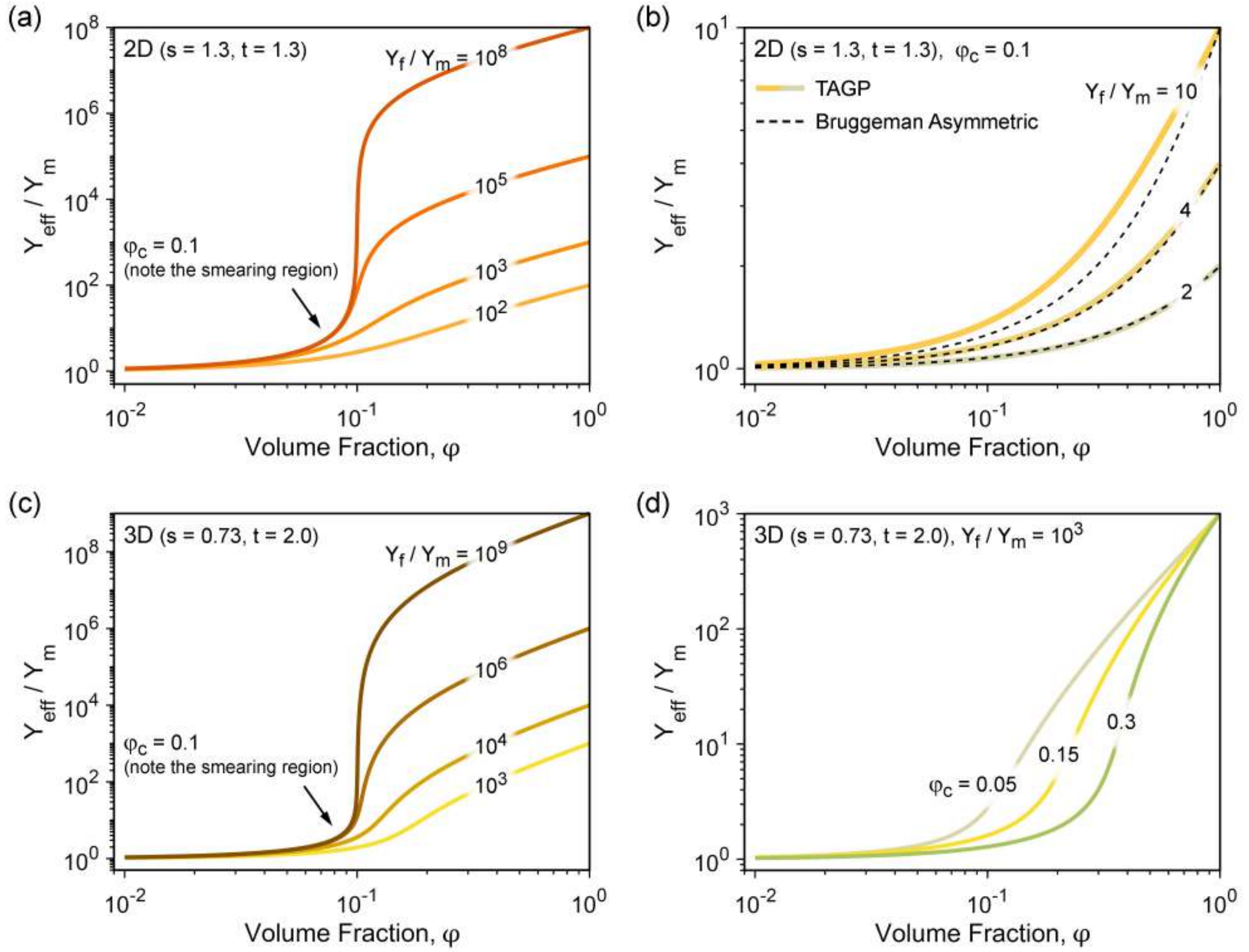

**FIG S6.** (a) Effective conductivity of a 2D composite system as calculated by the TAGP at different filler to matrix conductivity ratios ($Y_f/Y_m$) for a fixed percolation threshold of $\varphi_c = 0.1$ ($\phi_c = 10$ vol%), (b) Comparison of the TAGP with the asymmetric Bruggeman equation in a 2D composite system at low conductivity ratios of the filler to the matrix for a fixed percolation threshold of 0.1 (10 vol%), (c) Effective conductivity of a 3D composite system as calculated by the TAGP at different filler to matrix conductivity ratios for a fixed percolation threshold of 0.1 (10 vol%), (d) Effective conductivity of a 3D composite system as calculated by the TAGP at three different percolation thresholds of 0.05 (5 vol%), 0.15 (15 vol%) and 0.3 (30 vol%) for a fixed filler to matrix conductivity ratio of $10^3$.

In fact, when the conductivities of the filler and the matrix are of the same order ($Y_f/Y_m < 10$), there are no significant differences between the TAGP and EMT equations, as shown in Figure S6b, which compares, for example, the TAGP and the Bruggeman asymmetric equation at this regime. The trend is similar for 3D systems as depicted in Figure S6c. By decreasing the conductivity ratio from $10^9$ to $10^3$, the variation of composite effective conductivity with filler volume fraction becomes more gradual in 3D systems. Figure S6d, illustrates the effect of percolation threshold on the effective conductivity of composites. At a fixed conductivity ratio of $Y_f/Y_m = 10^3$, the effective conductivity exhibits a smoother transition at lower percolation thresholds.

The TAGP shows an interesting behavior when the filler to matrix conductivity ratio approaches unity. In 2D systems, similar to other EMT equations, the TAGP reduces to the Maxwell equation as $Y_f/Y_m \longrightarrow 1$. However, in 3D systems, at any given percolation threshold values there exists a minimum for $Y_f/Y_m$, below which the TAGP has no solution. This minimum can be approximated from $[(1-\varphi_c)/\varphi_c]^t$. At this stage, we cannot unambiguously interpret the physical meaning of this mathematical peculiarity of the TAGP in 3D systems but we think it deserves a more detailed study and careful experimental assessment. Nevertheless, the general conclusion of our above discussion is straightforward that when the conductivity of the filler and matrix are of the same order ($Y_f/Y_m < 10$), there are no significant differences between the percolation and diffusion processes and the TAGP becomes comparable to EMT equations, such as the Maxwell and the Bruggeman asymmetric equations.

## S8 Microstructural modeling of bulk thermal conductivity of filler

The bulk thermal conductivity of filler determined from the TAGP model can be related to the geometrical details and intrinsic thermal conductivity of the filler particles and the thermal contact resistance between them. In this section, we investigate this relationship for our MWCNT–PP nanocomposites using thermal models developed for fiber networks. In this regard, we use the two well-known models of Volkov–Zhigilei[92-95] and Zhao *et al.*[96,97] with the following general form,

$$k_f = \frac{k_f^{\infty}}{1 + c_1 \langle N_c \rangle Bi} \tag{S55}$$

<u>Input parameters</u>:

| | |
|---|---|
| $L_f$ | Average length of filler particles |
| $D_f$ | Average diameter of filler particles |
| $k_{f,int}$ | Intrinsic thermal conductivity of individual filler particles |
| $h$ | Thermal contact conductance of filler–filler contacts |

<u>Output parameter</u>:

| | |
|---|---|
| $k_f$ | Bulk thermal conductivity of dense filler compacts ($\varphi \approx 1$) |

<u>Intermediate parameters and calculations for 3D random nanofiber networks</u>:

| | | Volkov–Zhigilei[93] | Zhao *et al.*[96] |
|---|---|---|---|
| $c_1$ | Constant | $\frac{1}{12} \approx 0.0833$ | $\frac{0.1852\pi}{4} \approx 0.1455$ |
| $c_2$ | Constant | $\frac{2}{9\pi} \approx 0.0707$ | $\frac{4 \times 0.1852}{\pi(2+\pi)} \approx 0.0459$ |
| $A_f$ | Cross-sectional area | $\frac{\pi D_f^2}{4}$ | $\frac{\pi D_f^2}{4}$ |
| $\langle N_c \rangle$ | Average contact number per nanofiber | $\frac{2L_f}{D_f}$ | $\frac{4}{2+\pi}\frac{L_f}{D_f} \approx 0.7780\frac{L_f}{D_f}$ |
| $k_f^{\infty}$ | The value of $k_f$ assuming $k_{f,int} \to \infty$ | $c_2 \frac{hL_f^2}{D_f^3}$ | $c_2 \frac{hL_f^2}{D_f^3}$ |
| $Bi$ | Biot Number | $\frac{hL_f}{k_{f,int}A_f}$ | $\frac{hL_f}{k_{f,int}A_f}$ |

For our MWCNTs, $L_f$=1.1×$10^{-6}$ m and $D_f$=3.2×$10^{-8}$ m. The intrinsic thermal conductivity of our CVD grown MWCNTs can also be estimated to be about 1000 W $m^{-1}K^{-1}$.[94,98-100] In addition, the thermal contact conductance between our MWCNTs with an average diameter of 32 nm can be considered as $h \approx 2.0 \times 10^{-8}$ W $K^{-1}$ (specific thermal contact resistance of $R_C \approx 10^{-9}$ $m^2K\,W^{-1}$).[94,101] Using these input parameters, the bulk thermal conductivity of dense MWCNT compacts ($\varphi \approx 1$) is obtained as 45.2 W $m^{-1}K^{-1}$ from the Volkov–Zhigilei model and 30.1 W $m^{-1}K^{-1}$ from the model of Zhao *et al.*, which, considering the approximations involved, is in fair agreement with $k_f \approx 62$ W $m^{-1}K^{-1}$ obtained from the analysis of our experimental data with the TAGP model.

As discussed in the Introduction section of the main text, the specific thermal contact resistance between MWCNTs ($R_C \approx 10^{-9}$ $m^2K\,W^{-1}$) is two orders of magnitude smaller than the specific thermal contact resistance between SWCNTs ($R_C \approx 10^{-7}$ $m^2K\,W^{-1}$) and the specific interfacial thermal resistance between carbon nanotubes and the matrix ($R_I \approx 10^{-7}$ $m^2K\,W^{-1}$). Interestingly, if $R_C \approx 10^{-7}$ $m^2K\,W^{-1}$ is used instead of $R_C \approx 10^{-9}$ $m^2K\,W^{-1}$ for MWCNTs in the models of Volkov–Zhigilei and Zhao *et al.*, then $k_f \approx 0.6$ W $m^{-1}K^{-1}$ is obtained from both models, which illustrates the significant effect of contact resistances on the bulk thermal conductivity of filler. Therefore, $k_f \approx 62$ W $m^{-1}K^{-1}$ determined from the TAGP model for bulk thermal conductivity of filler in our samples confirms the low thermal contact resistance between MWCNTs in the fabricated composites and the good thermal coupling of nanotubes.

In the rest of this section, more details of the models of Volkov–Zhigilei and Zhao *et al.* are provided.

The model of Volkov–Zhigilei:[93]

$$k_{f,net} = \frac{k_{f,net}^{\infty}}{1 + \frac{\langle N_c \rangle}{12} Bi} \tag{S56}$$

Input parameters:

| | |
|---|---|
| $L_f$ | Average length of filler particles |
| $D_f$ | Average diameter of filler particles |
| $k_{f,int}$ | Intrinsic thermal conductivity of individual filler particles |
| $\varphi$ | Volume fraction |
| $h$ | Thermal contact conductance of filler–filler contacts |

Output parameter:

| | |
|---|---|
| $k_{f,net}$ | Effective thermal conductivity of the network |

Intermediate parameters and calculations for 3D random networks:

| | |
|---|---|
| $R_f = \frac{D_f}{2}$ | Average radius of filler particles |
| $A_f = \frac{\pi D_f^2}{4}$ | Cross-sectional area |
| $n_V = \frac{\varphi}{\pi R_f^2 L_f}$ | Volume number density of filler particles |
| $\bar{n}_V = n_V R_f L_f^2$ | Density parameter |
| $\langle N_c \rangle = \pi \bar{n}_V$ | Average contact number per nanofiber |
| $Bi = \frac{h L_f}{k_{f,int} A_f}$ | Biot Number |
| $k_{f,net}^{\infty} = \frac{\pi h \bar{n}_V^2}{36 R_f}$ | Effective thermal conductivity of the network assuming infinite intrinsic filler thermal conductivity |

The model of Zhao *et al.*:[96]

$$k_{f,net} = k_{f,int} n_s A_f \frac{Bi\langle N_c \rangle}{\frac{2\langle|\cos\theta|\rangle L_f}{\langle H \rangle} + Bi\langle N_c \rangle} \langle|\cos\theta|\rangle \quad \text{(S57)}$$

Input parameters:

| | |
|---|---|
| $L_f$ | Average length of filler particles |
| $D_f$ | Average diameter of filler particles |
| $k_{f,int}$ | Intrinsic thermal conductivity of individual filler particles |
| $\varphi$ | Volume fraction |
| $h$ | Thermal contact conductance of filler–filler contacts |

Output parameter:

| | |
|---|---|
| $k_{f,net}$ | Effective thermal conductivity of the network |

Intermediate parameters and calculations for 3D random networks:

| | |
|---|---|
| $r = \frac{L_f}{D_f}$ | Aspect ratio |
| $A_f = \frac{\pi D_f^2}{4}$ | Cross-sectional area |
| $\langle H \rangle = 0.1852 L_f$ | Average center-to-center distance of the two connected nanofibers in the z direction |
| $\langle|\cos\theta|\rangle = \frac{2}{\pi}$ | $\theta$ is the angle between a nanofiber and the direction of macroscopic heat transfer |
| $n_s = \frac{\varphi}{2A_f}$ | Areal number density of nanofibers through the total cross-sectional area of the network |
| $\langle N_c \rangle = \frac{4r\varphi}{2+\pi\varphi}$ | Average contact number per nanofiber |
| $Bi = \frac{hL_f}{k_{f,int} A_f}$ | Biot Number |

## S9 Additional Figures and Complementary Information

### S9.1 Lightweight, electrically, and thermally conductive polymer composites

Figure S7 demonstrates in more detail the concept of lightweight, electrically, and thermally conductive polymer composites mentioned in the main text. As shown in Fig. S7a, a simple electrical circuit was built consisting of two button batteries (1.5 V), a white LED lamp (with a forward voltage of 2.5 V), four electrically conductive MWCNT–PP and Graphene–PP nanocomposites as part of the wiring of the circuit, and an electrically insulating h-BN–PP nanocomposite as the LED heat sink.

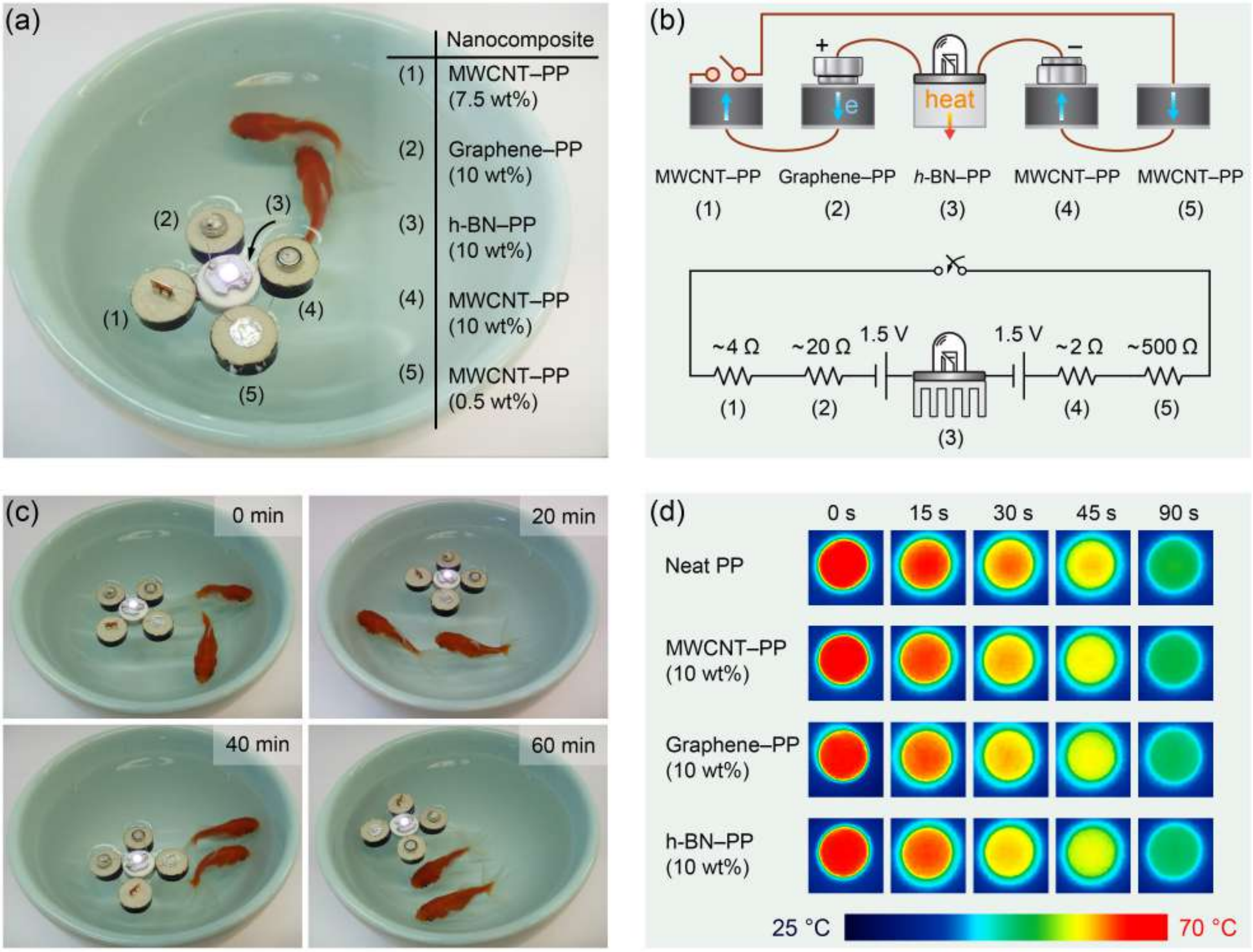

**FIG S7.** Conceptual demonstration of the lightweight, electrically, and thermally conductive polymer composites in practice. (a) The fabricated nanocomposites as part of a simple electrical circuit to light a white LED lamp so that the whole set is floating on water. (b) Schematic of the circuit along with the electrical resistance values of the nanocomposites. (c) Stability of the LED light intensity over a period of one hour, which indicates the good thermal performance of the h-BN–PP nanocomposite as a heatsink. (d) Infrared thermal images of the temperature distribution at the top surface of four preheated samples during natural cooling in ambient air at specific time intervals. The h-BN–PP nanocomposite sample shows the best thermal performance and thus has the highest thermal conductivity.

The concept of this figure is simple and straightforward. The density of neat polypropylene is $\rho_{PP} = 0.9$ g cm$^{-3}$ and slightly lower than the density of water at room temperature, i.e. $\rho_{Water} = 1.0$ g cm$^{-3}$. Therefore, neat polypropylene, which is an electrical and thermal insulator, floats naturally on water. By adding filler particles to polypropylene, the effective conductivity of the resulting nanocomposite increases accordingly, but at the same time, the effective density of the nanocomposite also increases due to the higher density of the filler phase ($\rho_{\mathrm{MWCNT}} \approx \rho_{\mathrm{Graphene}} \approx \rho_{\mathrm{h\text{-}BN}} \approx 2.1$ g cm$^{-3}$) compared to polypropylene. If the amount of filler loading in a polypropylene nanocomposite is greater than $\phi = (\rho_{\mathrm{Water}} - \rho_{\mathrm{PP}})/(\rho_{\mathrm{Filler}} - \rho_{\mathrm{PP}}) \approx 8.3$ vol%, then the nanocomposite will no longer float on water and will sink. Thus, floating on the water of our nanocomposites, while they also carry two batteries, an LED lamp, and a switch, visually demonstrates that the filler loading in our composites is far less than 8.3 vol%. In fact, the highest filler loading in our composites is about half of this value and is equal to 10 wt% ≈ 4.5 vol%.

In addition, as can be seen from Fig. S7a, we have used our nanocomposites with carbon fillers (MWCNT and graphene) as part of the wiring of the electrical circuit to light an LED lamp, and the high brightness of the LED lamp visually confirms that despite low filler loading in the nanocomposites, they have high electrical conductivity. The circuit schematic and the electrical resistance values of the nanocomposites are shown in Fig. S7b.

Moreover, the entire circuit was left floating on the water for more than an hour (Fig. S7c). The stability of the LED luminous intensity throughout this period proves that the nanocomposite with the h-BN filler has good thermal conductivity as a heat sink to transfer the heat generated by the LED to the water.

The significant increase in the thermal conductivity of the nanocomposites with 10 wt% filler loading compared to neat PP is also visually verified by the thermography technique in Fig. S7d. Taking t = 30 s as an example, the IR images show that the h-BN–PP nanocomposite has the highest thermal conductivity, while the thermal conductivity of the MWCNT–PP and Graphene–PP nanocomposites is almost equal but still considerably larger than the Neat PP thermal conductivity. The exact values of the thermal conductivity of the nanocomposites, from direct measurement by the transient plane heat source method (ISO 22007-2:2015), are as follows: 0.15 W m$^{-1}$ K$^{-1}$ for Neat PP, 0.55 W m$^{-1}$ K$^{-1}$ for 10 wt% MWCNT–PP (267% enhancement), 0.52 W m$^{-1}$ K$^{-1}$ for 10 wt% Graphene–PP (245% enhancement), and 0.76 W m$^{-1}$ K$^{-1}$ for 10 wt% h-BN–PP (407% enhancement).

## S9.2 Surface analysis of a typical 4.5 vol% (10 wt%) MWCNT-coated PP microparticle

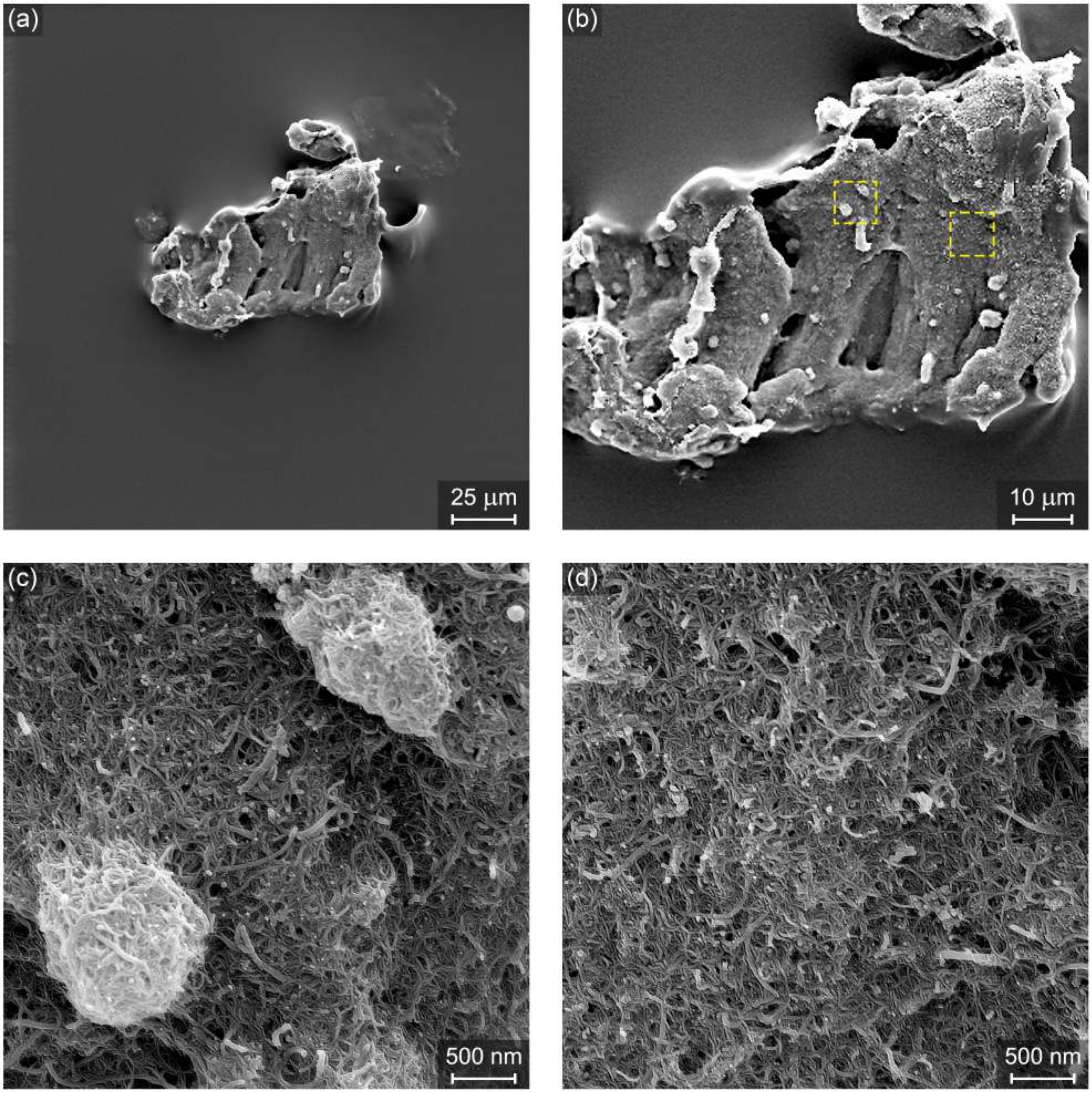

**FIG S8.** SEM images from the surface of a typical PP particle coated with $\phi = 4.5$ vol% (10 wt%) MWCNTs. (a) The PP powder microparticle with a coating of $\phi = 4.5$ vol% nanotubes, (b) A close-up view of the surface of the particle and its nanotube coating (note the two representative areas that are delineated in yellow and enlarged in the next two panels), (c) A number of nanotube bundles are formed on the surface, (d) Nevertheless, the nanotube coating on the majority of the surface is uniform.

## S10 Application of the TAGP Equation for Selected Thermal Percolation Reports

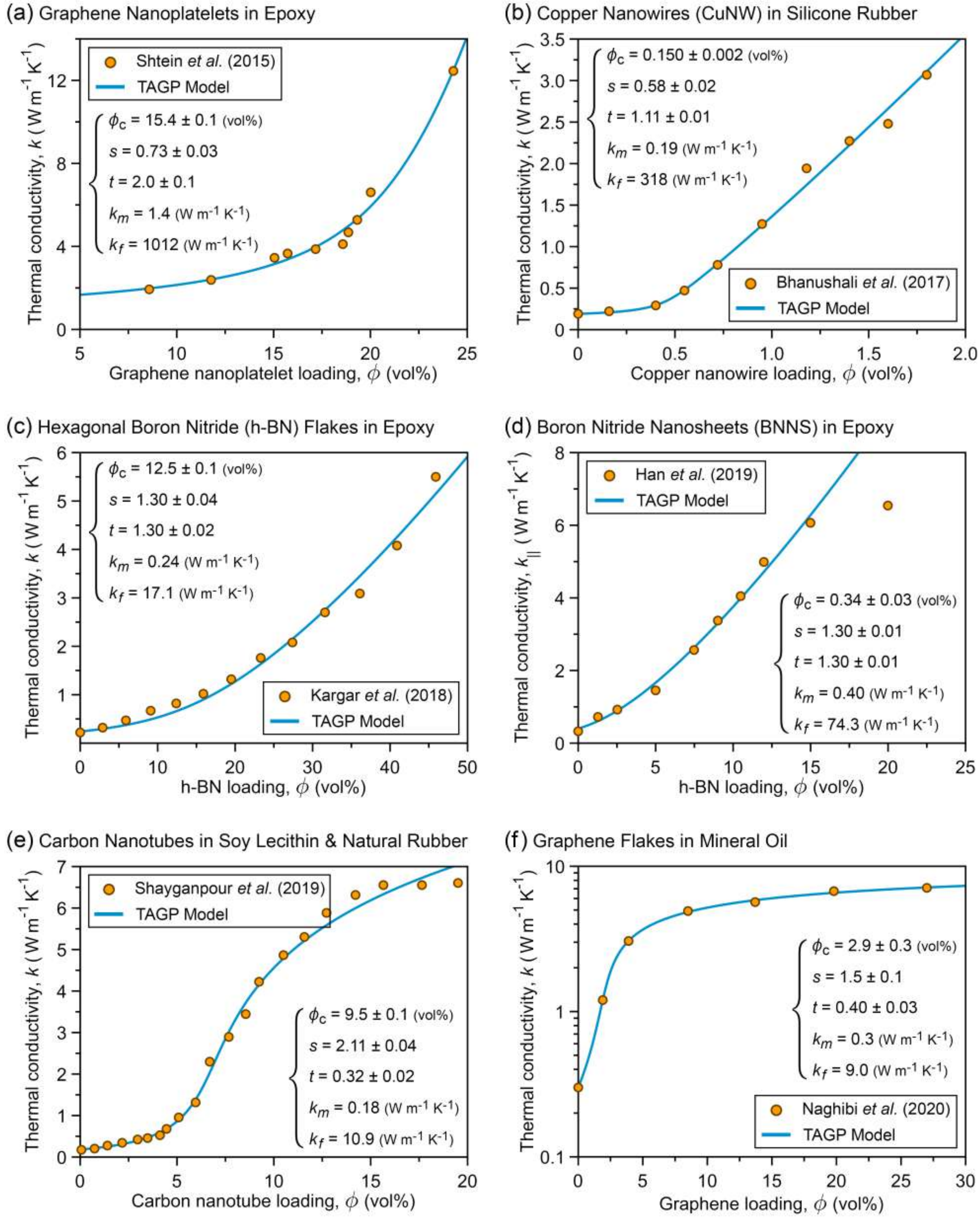

Continued on the next page

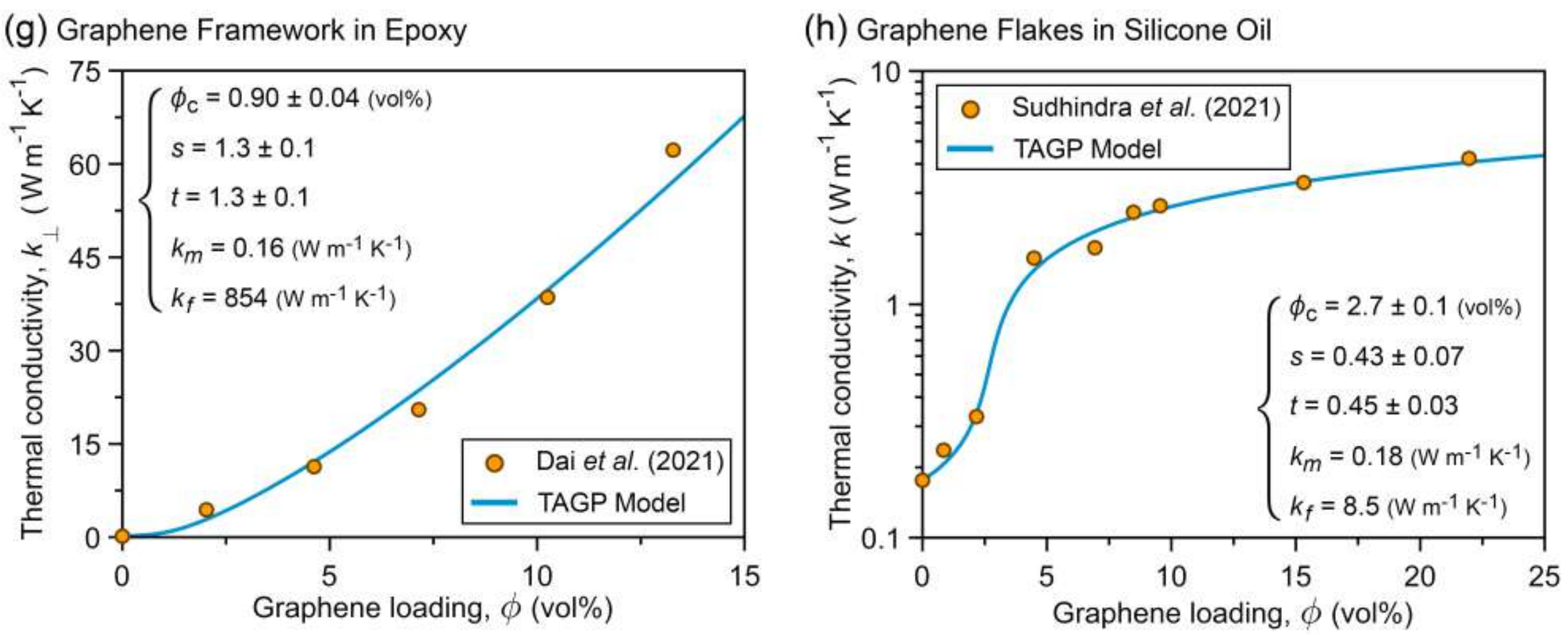

**FIG. S9.** Fitting the experimental data of thermal conductivity versus filler loading from eight selected and recent reports in the field of thermal percolation with our TAGP equation and the best-fit parameters obtained. Experimental data are from (a) Ref. 29, (b) Ref. 33, (c) Ref. 36, (d) Ref. 102, (e) Ref. 44, (f) Ref. 52, (g) Ref. 55, and (h) Ref. 103.

**Poor Conducting Phase (Matrix):** Epoxy
**Good Conducting Phase (Filler):** Graphene nanoplatelets
**Fabrication Method:** Planetary centrifugal ball milling & Mixing

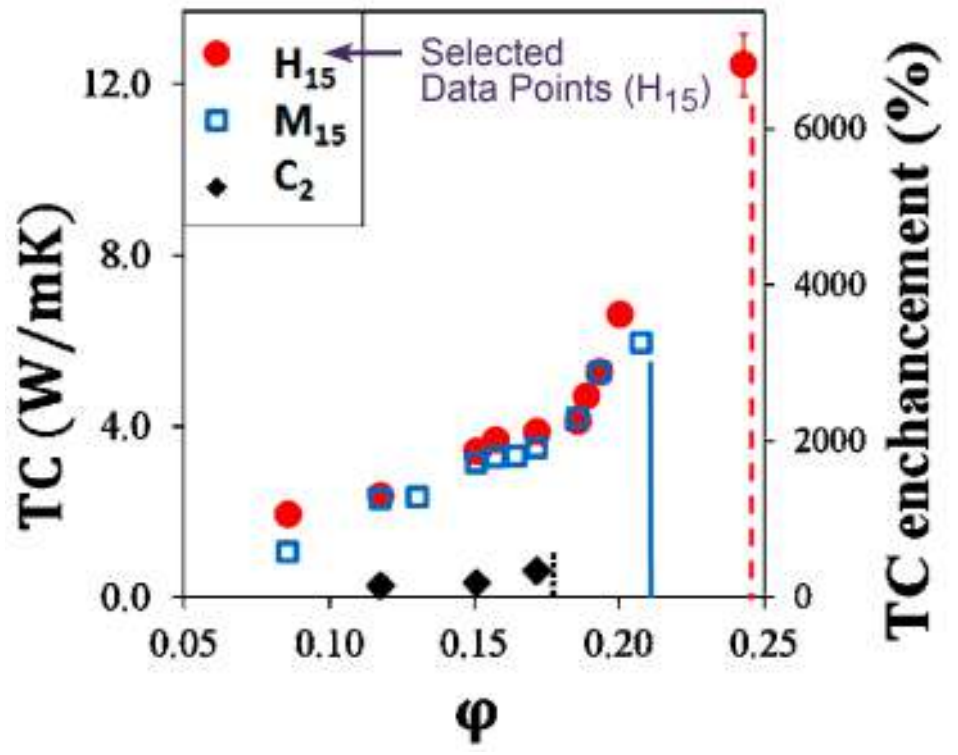

**Selected Figure:** Fig. 3(b) of the source article (Ref. [29])

**Extracted Data Points**

| # | $\phi$ (vol%) | $k$ [W/(m.K)] |
|---|---|---|
| 1 | 8.6 | 1.9 |
| 2 | 11.8 | 2.4 |
| 3 | 15.0 | 3.5 |
| 4 | 15.7 | 3.7 |
| 5 | 17.2 | 3.9 |
| 6 | 18.6 | 4.1 |
| 7 | 18.9 | 4.7 |
| 8 | 19.3 | 5.3 |
| 9 | 20.0 | 6.6 |
| 10 | 24.3 | 12.5 |

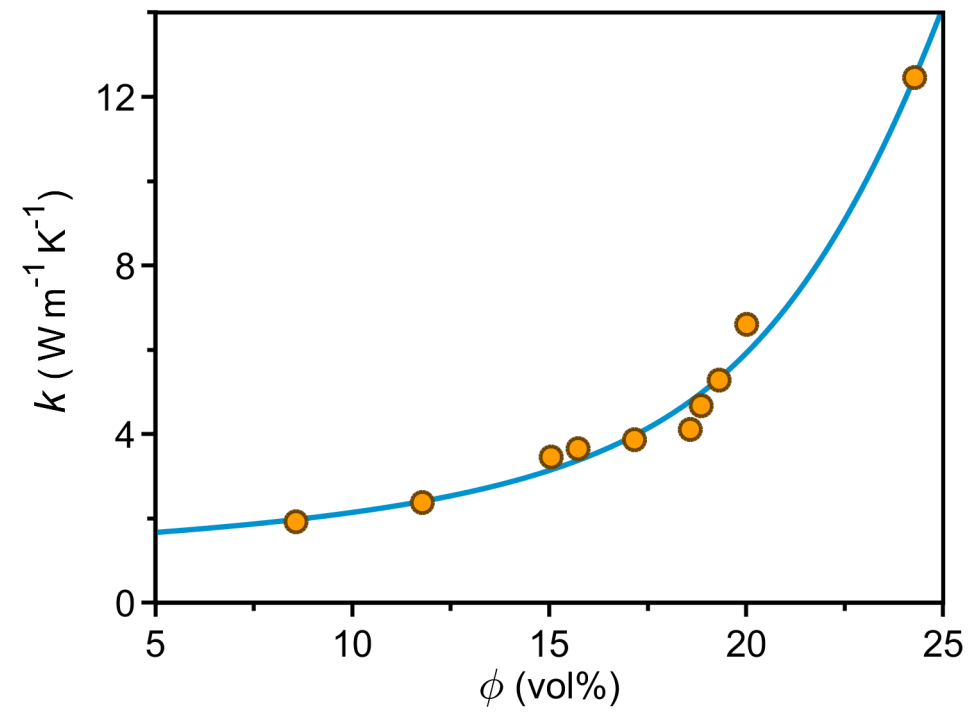

| | |
|---|---|
| $\phi_c$ (vol%) | 15.4 |
| $s$ | 0.73 |
| $t$ | 2.0 |
| $k_m$ (W $m^{-1}K^{-1}$) | 1.4 |
| $k_{f^*}$ (W $m^{-1}K^{-1}$) | 33.5 |
| $k_f$ (W $m^{-1}K^{-1}$) | 1012 |
| $h = k_m/k_{f^*}$ | $4.2 \times 10^{-2}$ |
| $R^2$ | 0.9855 |

**Fit Results:** Fitting data with the TAGP equation and the best-fit parameters

## Remarks:

- ✓ In this article, which is one of the first and most cited reports on the observation of thermal percolation in polymer composites, the authors describe their data with the classical power-law percolation model.
- ✓ The classical percolation model gives $\phi_c = 16.6$ vol%, which is close to our obtained value of $\phi_c = 15.4$ vol%.
- ✓ Interestingly, the classical percolation model gives $t = 0.84$, which is a non-universal value, while the TAGP model describes the experimental data in this work with perfectly universal critical exponents (i.e., $s = 0.73$ and $t = 2.0$).
- ✓ The difference between $k_m$ obtained from the TAGP equation ($k_m = 1.4$ W m$^{-1}$K$^{-1}$) and the experimentally reported thermal conductivity of the epoxy matrix in this work (0.19 W m$^{-1}$K$^{-1}$) implies that for this dataset, percolation is not the dominant physics of conduction in the entire range of $0 \leq \phi \leq 2\phi_c$, and $\phi = 0$ lies outside the range of the percolation regime. Repeating the data analysis with the more general version of the TAGP equation [Eq. (8) in the main text] leads to $\phi_c = 15.4$, $s = 0.73$, $t = 2.0$, $k_{poor} = 1.5$ W m$^{-1}$K$^{-1}$, $k_{good} = 28.5$ W m$^{-1}$K$^{-1}$, and $R = 0.92$. Compared to the model parameters presented in the above table [Fit to Eq. (9) of the main text], the main percolation parameters ($\phi_c$, $s$, and $t$) remain intact while the boundaries of the percolation region are determined $\phi_R^- = 1.2$ vol% and $\phi_R^+ = 29.6$ vol% with $k_{poor} = k_{eff}(\phi_R^-)$ and $k_{good} = k_{eff}(\phi_R^+)$.

**Source Article [29]:** Reproduced with permission from Chem. Mater. **27**, 2100-2106 (2015).

**Poor Conducting Phase (Matrix):** Silicone rubber
**Good Conducting Phase (Filler):** Copper nanowires (CuNW)
**Fabrication Method:** CuNW sponge filling

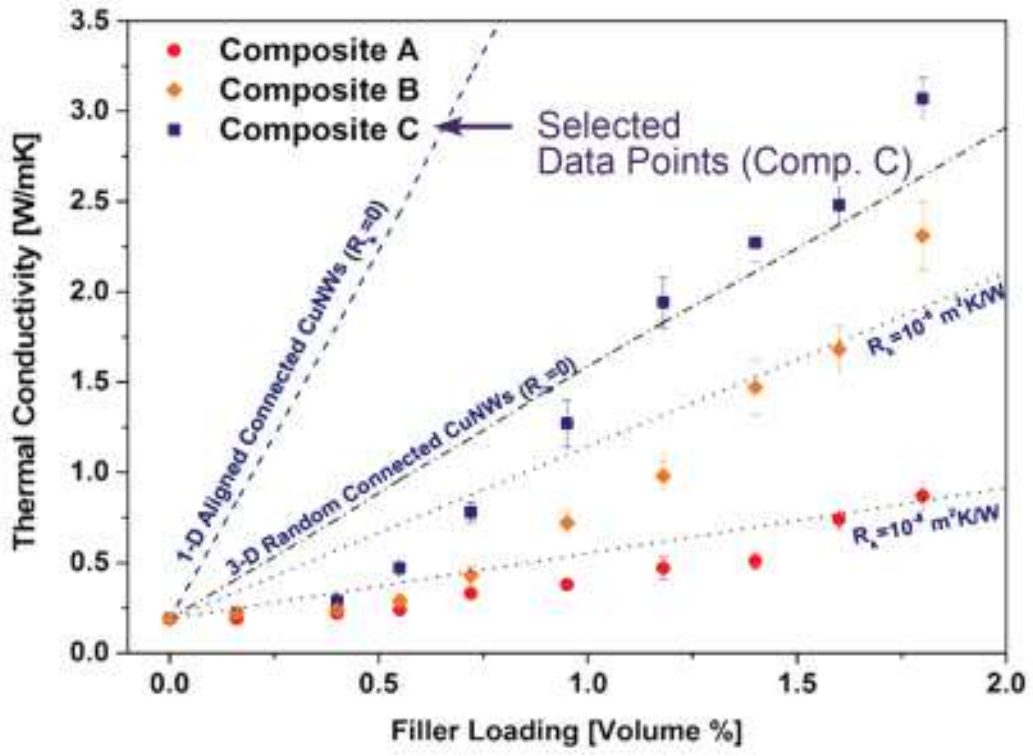

**Selected Figure:** Fig. 3(a) of the source article (Ref. [33])

**Extracted Data Points**

| # | $\phi$ (vol%) | $k$ [W/(m.K)] |
|---|---|---|
| 1 | 0.0 | 0.19 |
| 2 | 0.2 | 0.22 |
| 3 | 0.4 | 0.29 |
| 4 | 0.5 | 0.47 |
| 5 | 0.7 | 0.78 |
| 6 | 0.9 | 1.27 |
| 7 | 1.2 | 1.94 |
| 8 | 1.4 | 2.27 |
| 9 | 1.6 | 2.48 |
| 10 | 1.8 | 3.07 |

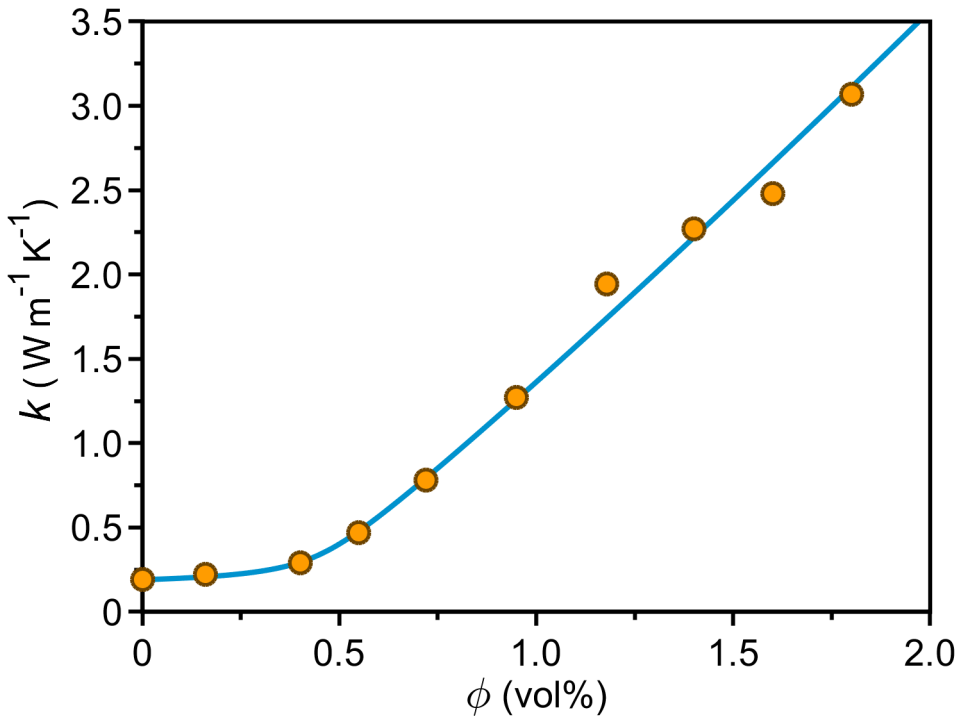

| | |
|---|---|
| $\phi_c$ (vol%) | 0.15 |
| $s$ | 0.58 |
| $t$ | 1.1 |
| $k_m$ (W $m^{-1}K^{-1}$) | 0.19 |
| $k_{f^*}$ (W $m^{-1}K^{-1}$) | 0.24 |
| $k_f$ (W $m^{-1}K^{-1}$) | 318 |
| $h = k_m/k_{f^*}$ | $7.9 \times 10^{-1}$ |
| $R^2$ | 0.9997 |

**Fit Results:** Fitting data with the TAGP equation and the best-fit parameters

## Remarks:

✓ The fillers used are metal (CuNW), heat transfer is mainly by electrons (not phonons), and there is a direct relationship between thermal and electrical conductivity (Wiedemann–Franz law).

✓ The authors found severe limitations in the classical power-law percolation model for describing their experimental data. But the TAGP model, as a generalized percolation model, can well describe the data reported in this work.

✓ The non-universality of the critical exponents, especially $t \approx 1$, indicates that after $2\phi_c$, the transition from the superlinear percolation regime to the linear EMT regime is rapid.

**Poor Conducting Phase (Matrix):** Epoxy
**Good Conducting Phase (Filler):** Hexagonal boron nitride (h-BN) flakes
**Fabrication Method:** High shear speed and mechanical mixing

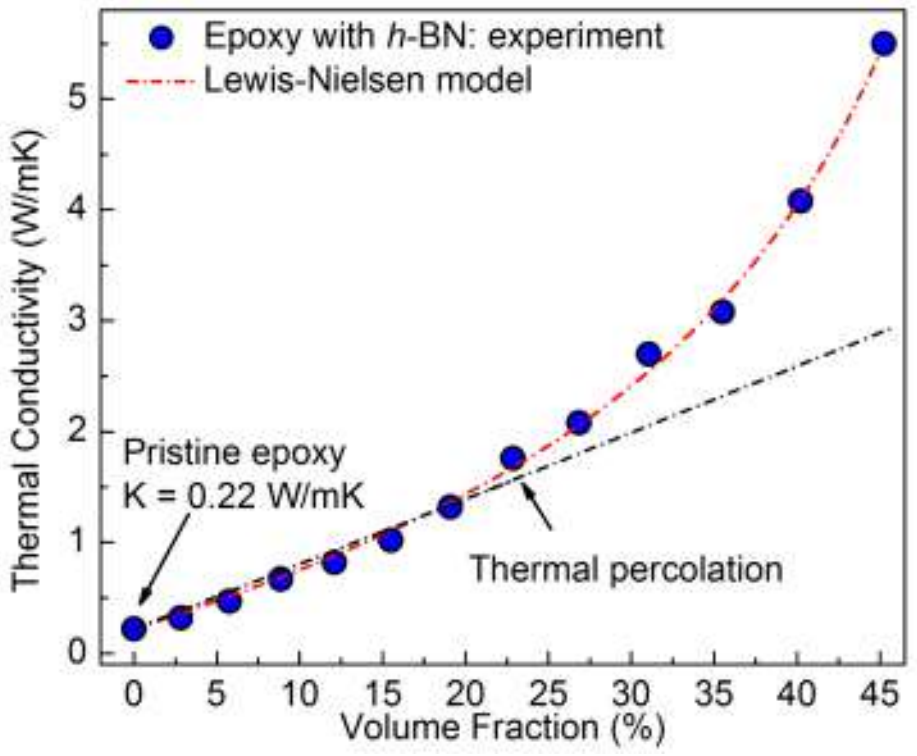

**Selected Figure:** Fig. 2(b) of the source article (Ref. [36])

**Extracted Data Points**

| # | $\phi$ (vol%) | $k$ [W/(m.K)] |
|---|---|---|
| 1 | 0.0 | 0.22 |
| 2 | 2.9 | 0.32 |
| 3 | 5.9 | 0.47 |
| 4 | 9.1 | 0.67 |
| 5 | 12.4 | 0.82 |
| 6 | 15.9 | 1.02 |
| 7 | 19.5 | 1.32 |
| 8 | 23.3 | 1.76 |
| 9 | 27.4 | 2.08 |
| 10 | 31.6 | 2.70 |
| 11 | 36.1 | 3.09 |
| 12 | 40.9 | 4.08 |
| 13 | 45.9 | 5.50 |

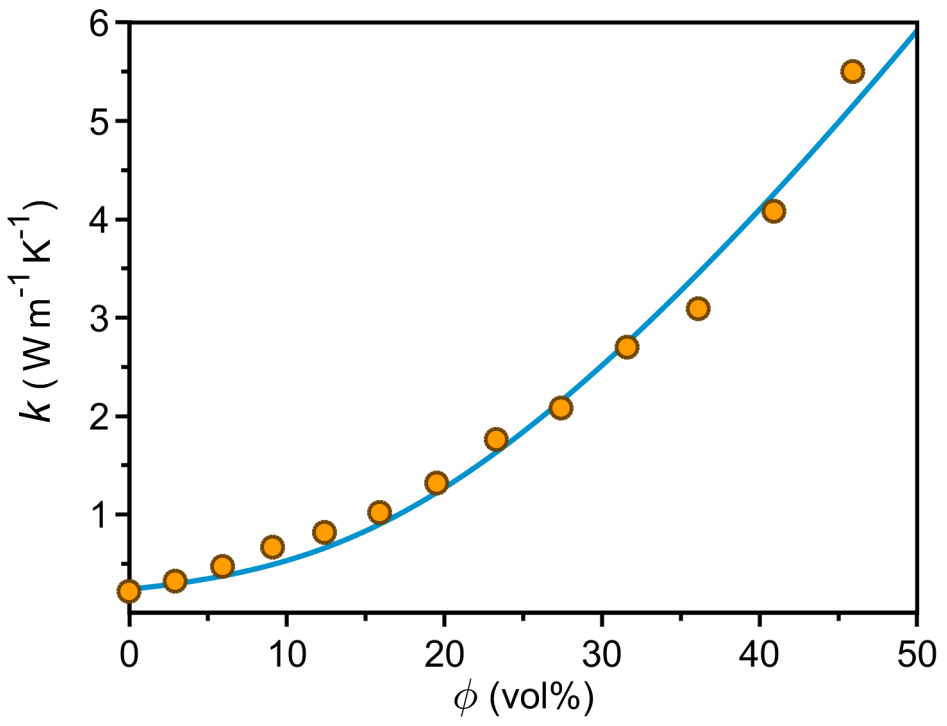

| | |
|---|---|
| $\phi_c$ (vol%) | 12.5 |
| $s$ | 1.3 |
| $t$ | 1.3 |
| $k_m$ (W $m^{-1}K^{-1}$) | 0.24 |
| $k_{f^*}$ (W $m^{-1}K^{-1}$) | 1.83 |
| $k_f$ (W $m^{-1}K^{-1}$) | 17.1 |
| $h = k_m/k_{f^*}$ | $1.3 \times 10^{-1}$ |
| $R^2$ | 0.9871 |

**Fit Results:** Fitting data with the TAGP equation and the best-fit parameters

**Remarks:**

- ✓ In this article, which is one of the most cited reports on the observation of thermal percolation in polymer composites, the authors describe their data with the semi-empirical Lewis–Nielsen model.
- ✓ The Lewis–Nielsen model gives $k_f = 16$ W m$^{-1}$K$^{-1}$ for the filler bulk thermal conductivity, which is in good agreement with the $k_f = 17.1$ W m$^{-1}$K$^{-1}$ obtained from the TAGP model.
- ✓ The Lewis–Nielsen model is not able to predict the percolation threshold. A simple linear fitting results $\phi_c =$ 23 vol%, while the TAGP model gives a percolation threshold of almost half this value and $\phi_c = 12.5$ vol%.
- ✓ The values obtained for critical exponents from TAGP, although universal, indicate reduced dimensionality in the system due to the formation of a 2D subnetwork of conductive filler particles within the 3D composite (see Sec. IV.B in the main text).

**Source Article [36]:** Reproduced with permission from ACS Appl. Mater. Interfaces **10**, 37555-37565 (2018).

**Poor Conducting Phase (Matrix):** Epoxy
**Good Conducting Phase (Filler):** Boron nitride nanosheets (BNNS)
**Fabrication Method:** Epoxy resin infiltration into the BNNS aerogel

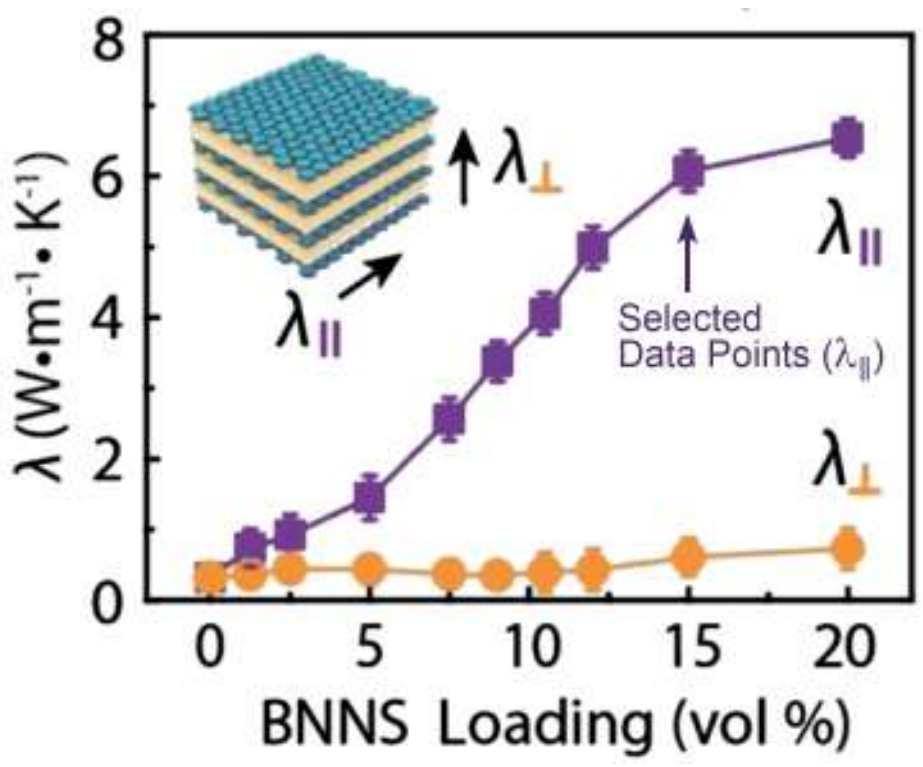

**Selected Figure:** Fig. 1(d) of the source article (Ref. [102])

**Extracted Data Points**

| # | $\phi$ (vol%) | $k$ [W/(m.K)] |
|---|---|---|
| 1 | 0.0 | 0.33 |
| 2 | 1.3 | 0.72 |
| 3 | 2.5 | 0.92 |
| 4 | 5.0 | 1.45 |
| 5 | 7.5 | 2.56 |
| 6 | 9.0 | 3.38 |
| 7 | 10.5 | 4.05 |
| 8 | 12.0 | 4.99 |
| 9 | 15.0 | 6.07 |
| 10 | 20.0 | 6.54 |

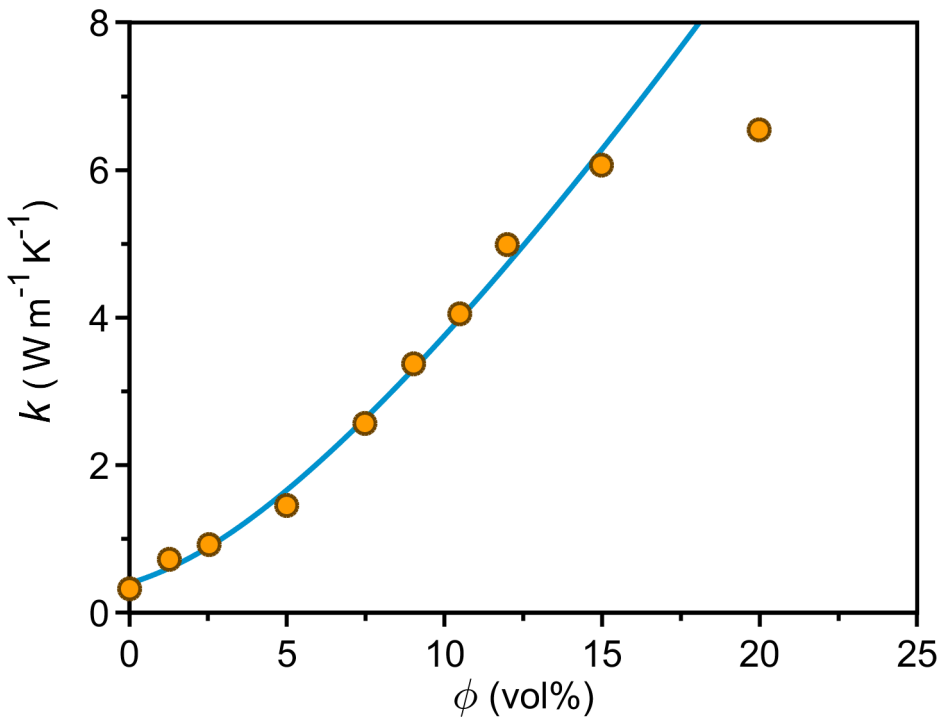

| | |
|---|---|
| $\phi_c$ (vol%) | 0.34 |
| $s$ | 1.3 |
| $t$ | 1.3 |
| $k_m$ (W $m^{-1}K^{-1}$) | 0.40 |
| $k_{f^*}$ (W $m^{-1}K^{-1}$) | 0.50 |
| $k_f$ (W $m^{-1}K^{-1}$) | 74.3 |
| $h = k_m/k_{f^*}$ | $8.0 \times 10^{-1}$ |
| $R^2$ | 0.9940 |

**Fit Results:** Fitting data with the TAGP equation and the best-fit parameters

**Remarks:**

- ✓ Thermal percolation in this anisotropic lamellar structure is observed only in the direction parallel to the layers. In the direction perpendicular to the layers, the increase in thermal conductivity with filler loading is linear and not significant.
- ✓ The values obtained for critical exponents from TAGP, although universal, indicate reduced dimensionality in the system, which is consistent with the lamellar and quasi-2D structure of these composites (see Sec. IV.B in the main text).
- ✓ The deviation of the final data point from the TAGP model can be interpreted through the transition from the percolation regime to the EMT regime at high loading levels ($\phi \gg 2\phi_c$).

**Source Article [102]:** Reproduced with permission from Adv. Funct. Mater. **29**, 1900412 (2019). Copyright 2019 Wiley-VCH.

**Poor Conducting Phase (Matrix):** Soy lecithin and natural rubber
**Good Conducting Phase (Filler):** Stacked-cup carbon nanotubes (SCCNT)
**Fabrication Method:** High-pressure homogenization and blending

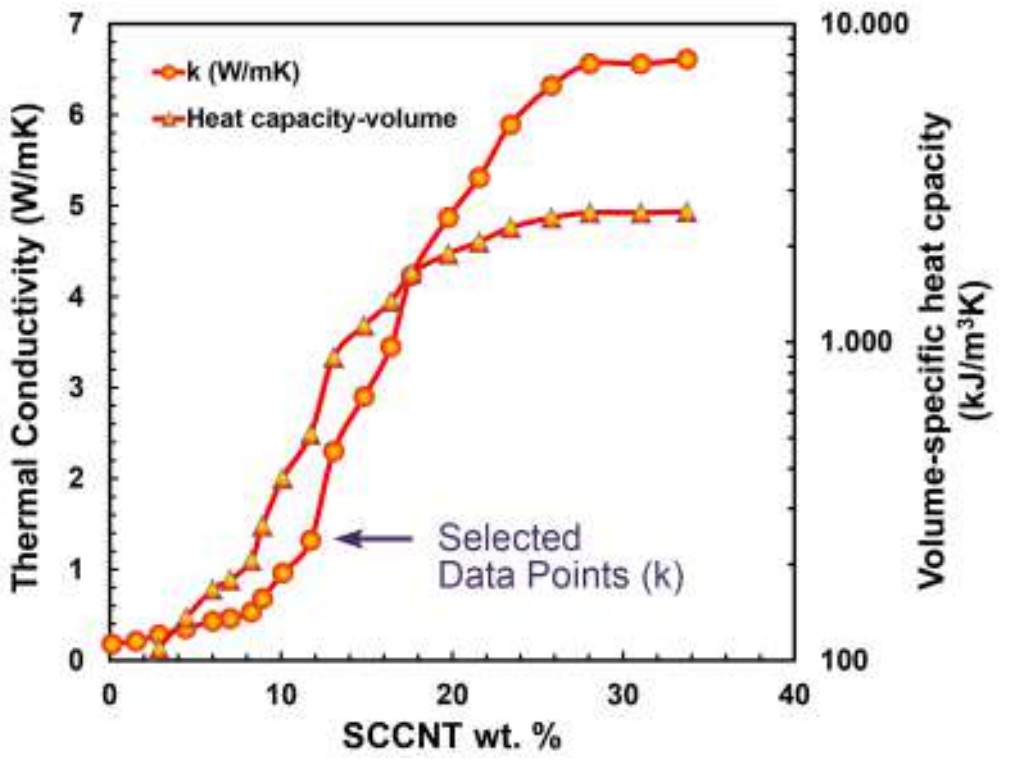

**Selected Figure:** Fig. 10(b) of the source article (Ref. [44])

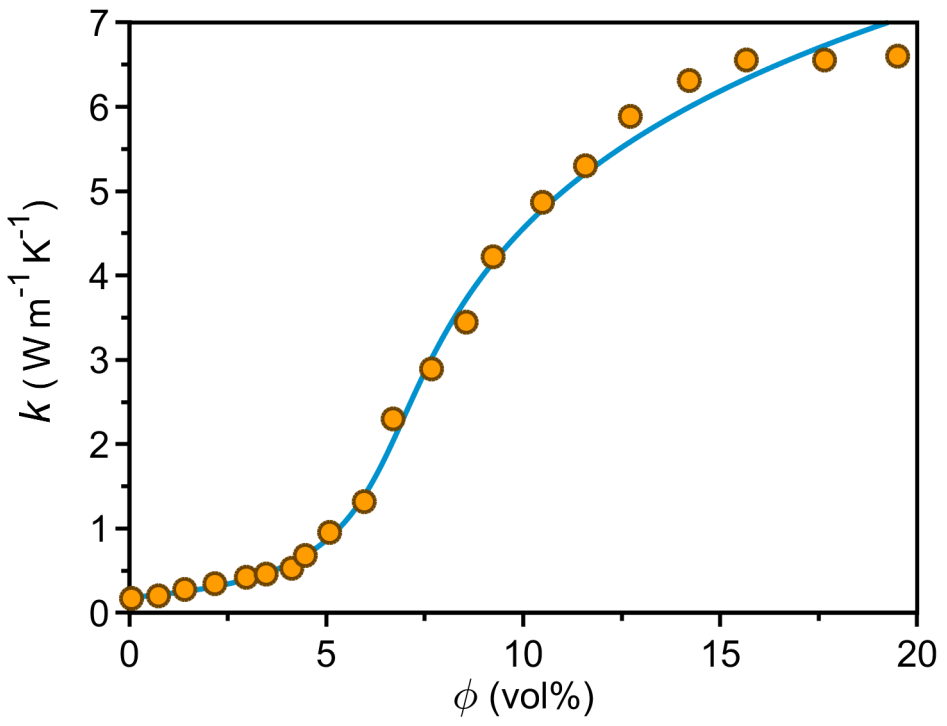

| | |
|---|---|
| $\phi_c$ (vol%) | 9.5 |
| $s$ | 2.1 |
| $t$ | 0.32 |
| $k_m$ (W $m^{-1}K^{-1}$) | 0.18 |
| $k_{f^*}$ (W $m^{-1}K^{-1}$) | 7.0 |
| $k_f$ (W $m^{-1}K^{-1}$) | 10.9 |
| $h = k_m/k_{f^*}$ | $2.6 \times 10^{-2}$ |
| $R^2$ | 0.9999 |

**Fit Results:** Fitting data with the TAGP equation and the best-fit parameters

**Extracted Data Points**

| # | $\phi$ (wt%) | $\phi$ (vol%) | $k$ [W/(m.K)] |
|---|---|---|---|
| 1 | 0.1 | 0.1 | 0.17 |
| 2 | 1.5 | 0.7 | 0.20 |
| 3 | 2.9 | 1.4 | 0.28 |
| 4 | 4.4 | 2.2 | 0.34 |
| 5 | 6.0 | 3.0 | 0.42 |
| 6 | 7.0 | 3.5 | 0.46 |
| 7 | 8.3 | 4.1 | 0.53 |
| 8 | 8.9 | 4.5 | 0.68 |
| 9 | 10.1 | 5.1 | 0.95 |
| 10 | 11.7 | 6.0 | 1.32 |
| 11 | 13.1 | 6.7 | 2.30 |
| 12 | 14.9 | 7.7 | 2.89 |
| 13 | 16.4 | 8.5 | 3.44 |
| 14 | 17.6 | 9.2 | 4.22 |
| 15 | 19.8 | 10.5 | 4.86 |
| 16 | 21.6 | 11.6 | 5.30 |
| 17 | 23.4 | 12.7 | 5.88 |
| 18 | 25.8 | 14.2 | 6.31 |
| 19 | 28.0 | 15.7 | 6.55 |
| 20 | 31.0 | 17.6 | 6.55 |
| 21 | 33.7 | 19.5 | 6.60 |

**Remarks:**

- ✓ The authors did not analyze their experimental data with any theoretical models, but as can be seen, the TAGP model can well explain the data reported in this work.
- ✓ The obtained $t = 0.32 < 1$ from the TAGP model implies severely constrained structural growth of the conductive network due to the initial ink-like and semi-fluid nature of these composites.
- ✓ The obtained $k_f = 10.9$ W m$^{-1}$K$^{-1}$ is close to the out-of-plane ($c$-axis) thermal conductivity of crystalline graphite and indicates that the heat conducting network consists mainly of stacked and aligned graphene nanosheets.

**Source Article [44]:** A. Shayganpour *et al.*, Nanomaterials **9**, 824, 2019; licensed under a Creative Commons Attribution (CC BY) license.

**Poor Conducting Phase (Matrix):** Mineral oil
**Good Conducting Phase (Filler):** Graphene flakes
**Fabrication Method:** Shear mixing

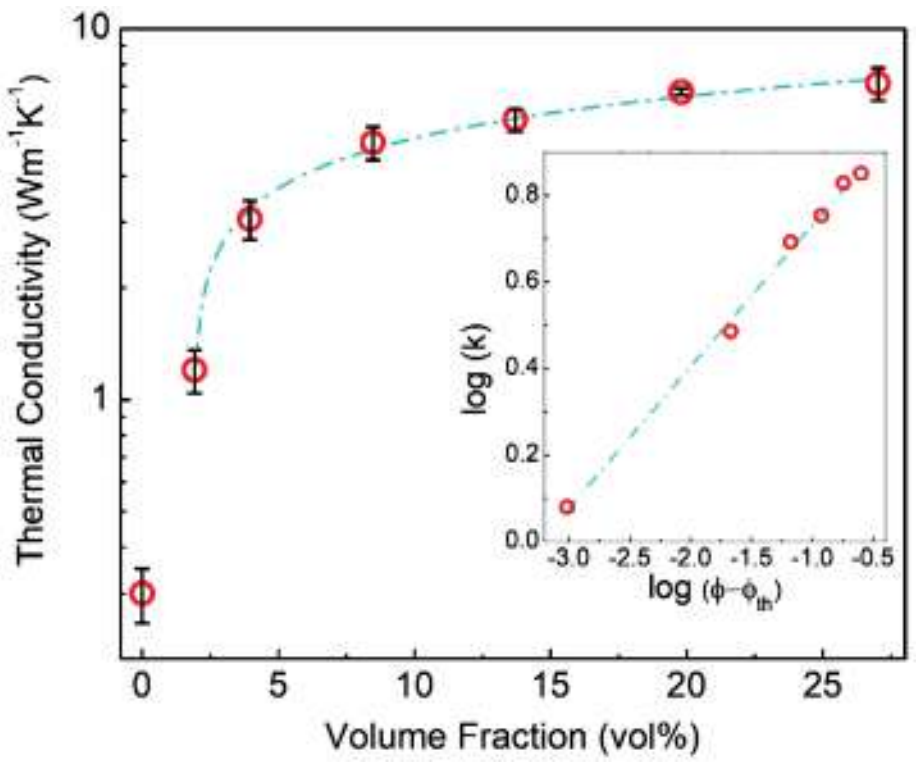

**Selected Figure:** Fig. 3 of the source article (Ref. [52])

**Extracted Data Points**

| # | $\phi$ (wt%) | $\phi$ (vol%) | $k$ [W/(m.K)] |
|---|---|---|---|
| 1 | 0 | 0.0 | 0.30 |
| 2 | 5 | 1.9 | 1.20 |
| 3 | 10 | 3.9 | 3.06 |
| 4 | 20 | 8.5 | 4.92 |
| 5 | 30 | 13.7 | 5.67 |
| 6 | 40 | 19.8 | 6.74 |
| 7 | 50 | 27.0 | 7.10 |

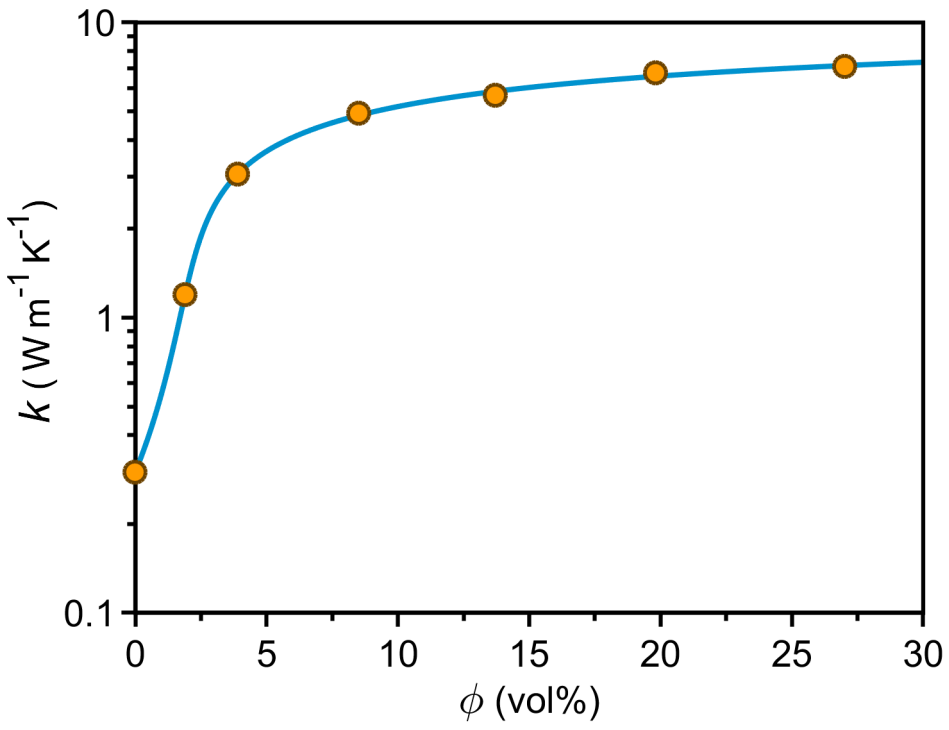

| | |
|---|---|
| $\phi_c$ (vol%) | 2.9 |
| $s$ | 1.50 |
| $t$ | 0.40 |
| $k_m$ (W $m^{-1}K^{-1}$) | 0.3 |
| $k_{f^*}$ (W $m^{-1}K^{-1}$) | 4.0 |
| $k_f$ (W $m^{-1}K^{-1}$) | 9.0 |
| $h = k_m/k_{f^*}$ | $7.4 \times 10^{-2}$ |
| $R^2$ | 1.0000 |

**Fit Results:** Fitting data with the TAGP equation and the best-fit parameters

## Remarks:

- ✓ In this article, which reports observing thermal percolation in graphene thermal pastes, the authors describe their data with the classical power-law percolation model.
- ✓ The classical percolation model gives $\phi_c = 1.9$ vol% and $t = 0.32$, which are close to the values of $\phi_c = 2.9$ vol% and $t = 0.40$, obtained from the TAGP model.
- ✓ Unlike the classical power-law percolation model, the TAGP model is capable of describing the pre-percolation region well.
- ✓ The obtained $t = 0.40 < 1$ from the TAGP model implies severely constrained structural growth of the conductive network due to the pasty and semi-fluid nature of these composites.
- ✓ The obtained $k_f = 9.0$ W $m^{-1}K^{-1}$ is close to the out-of-plane ($c$-axis) thermal conductivity of crystalline graphite and indicates that the heat conducting network consists mainly of stacked and aligned graphene nanosheets.

**Source Article [52]:** Reproduced with permission from Adv. Electron. Mater. **6**, 1901303 (2020). Copyright 2020 Wiley-VCH.

**Poor Conducting Phase (Matrix):** Epoxy
**Good Conducting Phase (Filler):** Dual assembled graphene framework (DAGF)
**Fabrication Method:** Epoxy resin infiltration into the DAGF

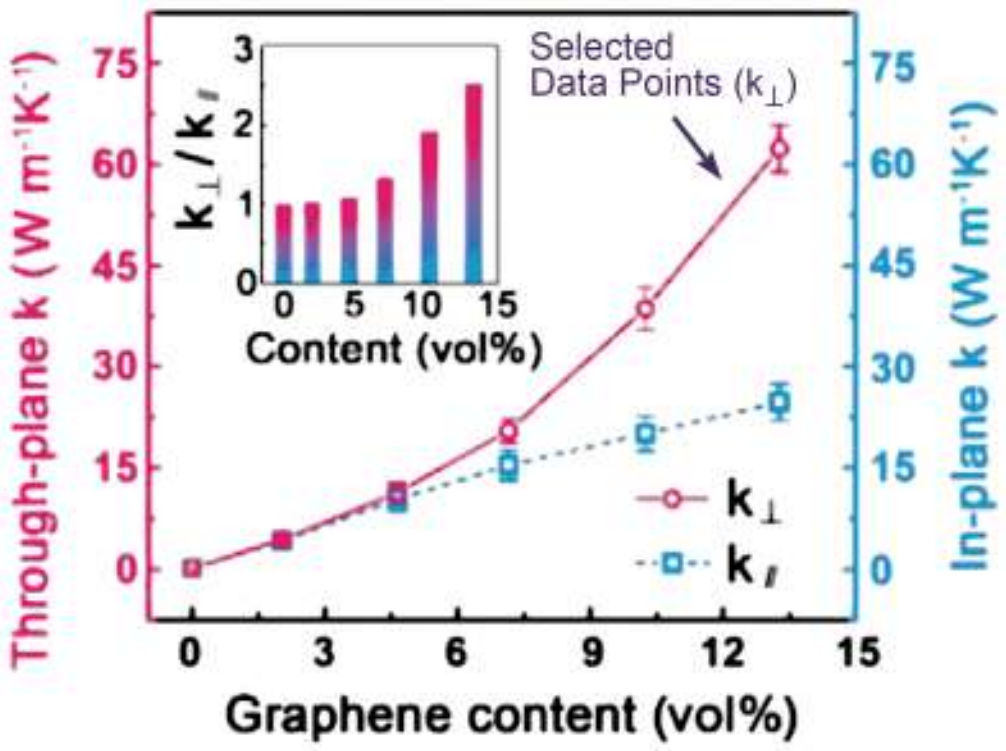

Selected Figure: Fig. 4(d) of the source article (Ref. [55])

**Extracted Data Points**

| # | $\phi$ (vol%) | $k$ [W/(m.K)] |
|---|---|---|
| 1 | 0.0 | 0.16 |
| 2 | 2.0 | 4.40 |
| 3 | 4.6 | 11.3 |
| 4 | 7.1 | 20.5 |
| 5 | 10.3 | 38.6 |
| 6 | 13.3 | 62.2 |

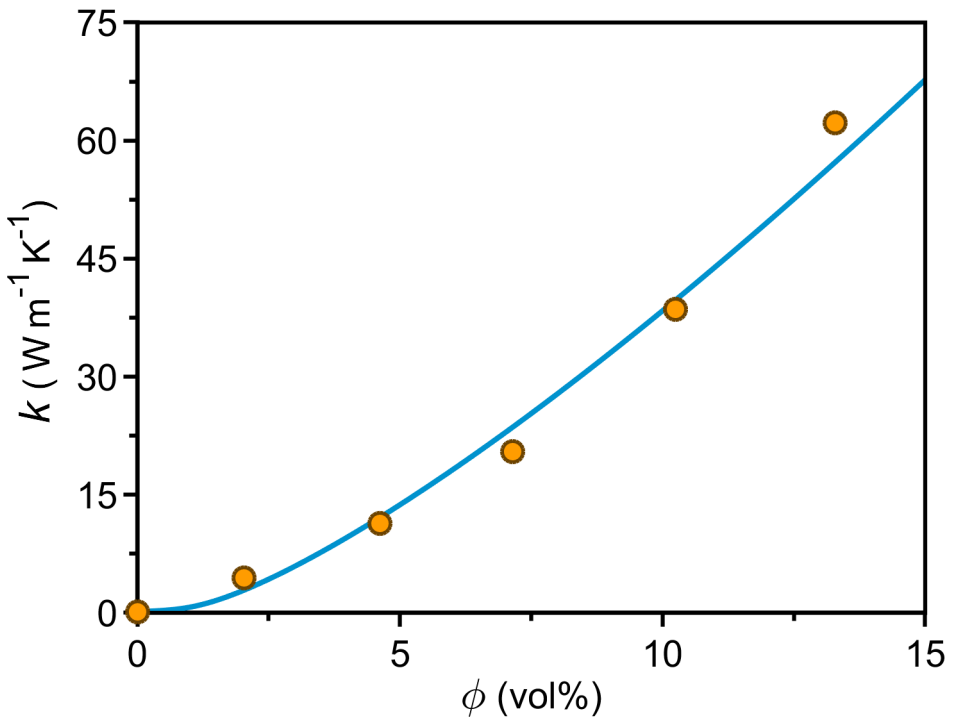

| | |
|---|---|
| $\phi_c$ (vol%) | 0.9 |
| $s$ | 1.3 |
| $t$ | 1.3 |
| $k_m$ (W $m^{-1}K^{-1}$) | 0.16 |
| $k_{f^*}$ (W $m^{-1}K^{-1}$) | 2.3 |
| $k_f$ (W $m^{-1}K^{-1}$) | 854 |
| $h = k_m/k_{f^*}$ | $7.1 \times 10^{-2}$ |
| $R^2$ | 1.0000 |

**Fit Results:** Fitting data with the TAGP equation and the best-fit parameters

**Remarks:**

✓ Thermal percolation in this anisotropic rolled-up structure is observed only in the through-plane direction. In the in-plane direction, the increase in thermal conductivity with filler loading is linear.

✓ The values obtained for critical exponents from TAGP, although universal, indicate reduced dimensionality in the system, which is consistent with the rolled-up and quasi-2D structure of these composites (see Sec. IV.B in the main text).

**Source Article [55]:** W. Dai *et al.*, Adv. Sci. **8**, 2003734, 2021; licensed under a Creative Commons Attribution (CC BY) license.

**Poor Conducting Phase (Matrix):** Silicone oil
**Good Conducting Phase (Filler):** Graphene flakes
**Fabrication Method:** Shear mixing

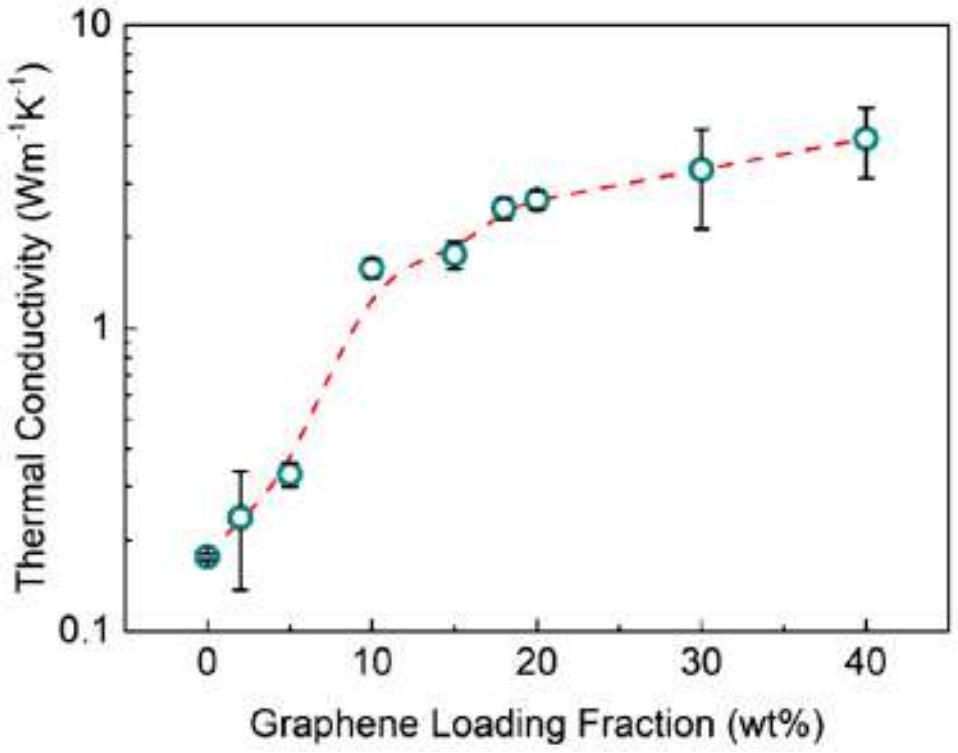

**Selected Figure:** Fig. 4 of the source article (Ref. [103])

**Extracted Data Points**

| # | $\phi$ (wt%) | $\phi$ (vol%) | $k$ [W/(m.K)] |
|---|---|---|---|
| 1 | 0 | 0.00 | 0.176 |
| 2 | 2 | 0.85 | 0.237 |
| 3 | 5 | 2.17 | 0.330 |
| 4 | 10 | 4.48 | 1.578 |
| 5 | 15 | 6.93 | 1.748 |
| 6 | 18 | 8.48 | 2.483 |
| 7 | 20 | 9.55 | 2.645 |
| 8 | 30 | 15.32 | 3.326 |
| 9 | 40 | 21.96 | 4.221 |

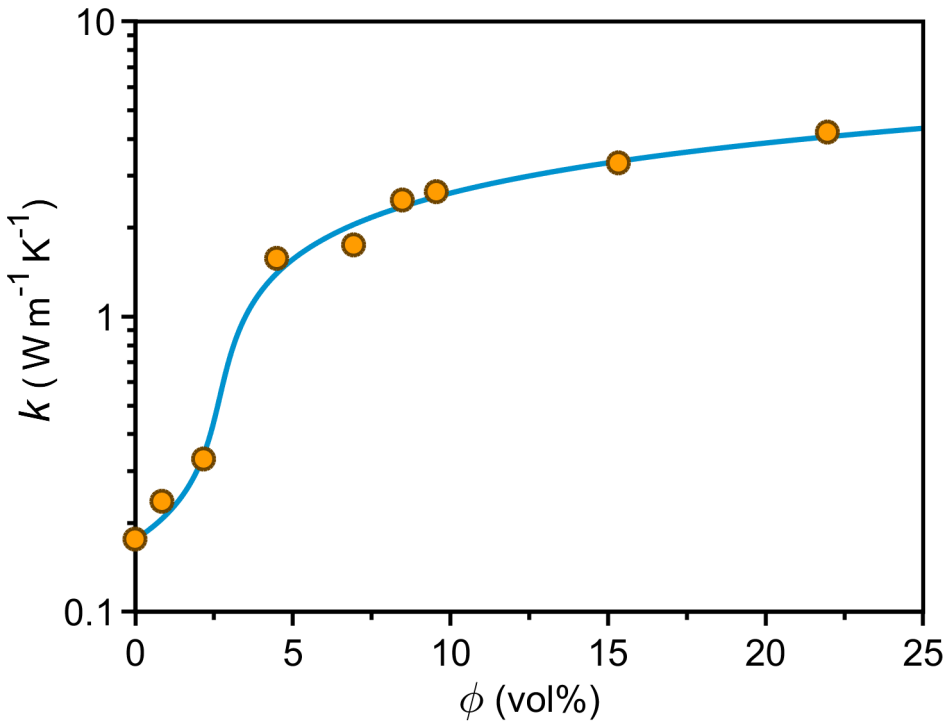

| | |
|---|---|
| $\phi_c$ (vol%) | 2.7 |
| $s$ | 0.43 |
| $t$ | 0.45 |
| $k_m$ (W $m^{-1}K^{-1}$) | 0.18 |
| $k_{f^*}$ (W $m^{-1}K^{-1}$) | 1.7 |
| $k_f$ (W $m^{-1}K^{-1}$) | 8.5 |
| $h = k_m/k_{f^*}$ | $1.1 \times 10^{-1}$ |
| $R^2$ | 0.9996 |

**Fit Results:** Fitting data with the TAGP equation and the best-fit parameters

**Remarks:**

✓ The authors did not analyze their experimental data with any theoretical models, but as can be seen, the TAGP model can well explain the data reported in this work.

✓ The obtained $t = 0.45 < 1$ from the TAGP model implies severely constrained structural growth of the conductive network due to the pasty and semi-fluid nature of these composites.

✓ The obtained $k_f = 8.5$ W $m^{-1}K^{-1}$ is close to the out-of-plane ($c$-axis) thermal conductivity of crystalline graphite and indicates that the heat conducting network consists mainly of stacked and aligned graphene nanosheets.

**Source Article [103]:** S. Sudhindra *et al.*, Nanomaterials **11**, 1699, 2021; licensed under a Creative Commons Attribution (CC BY) license.

## S11 Details of the Data Collapse

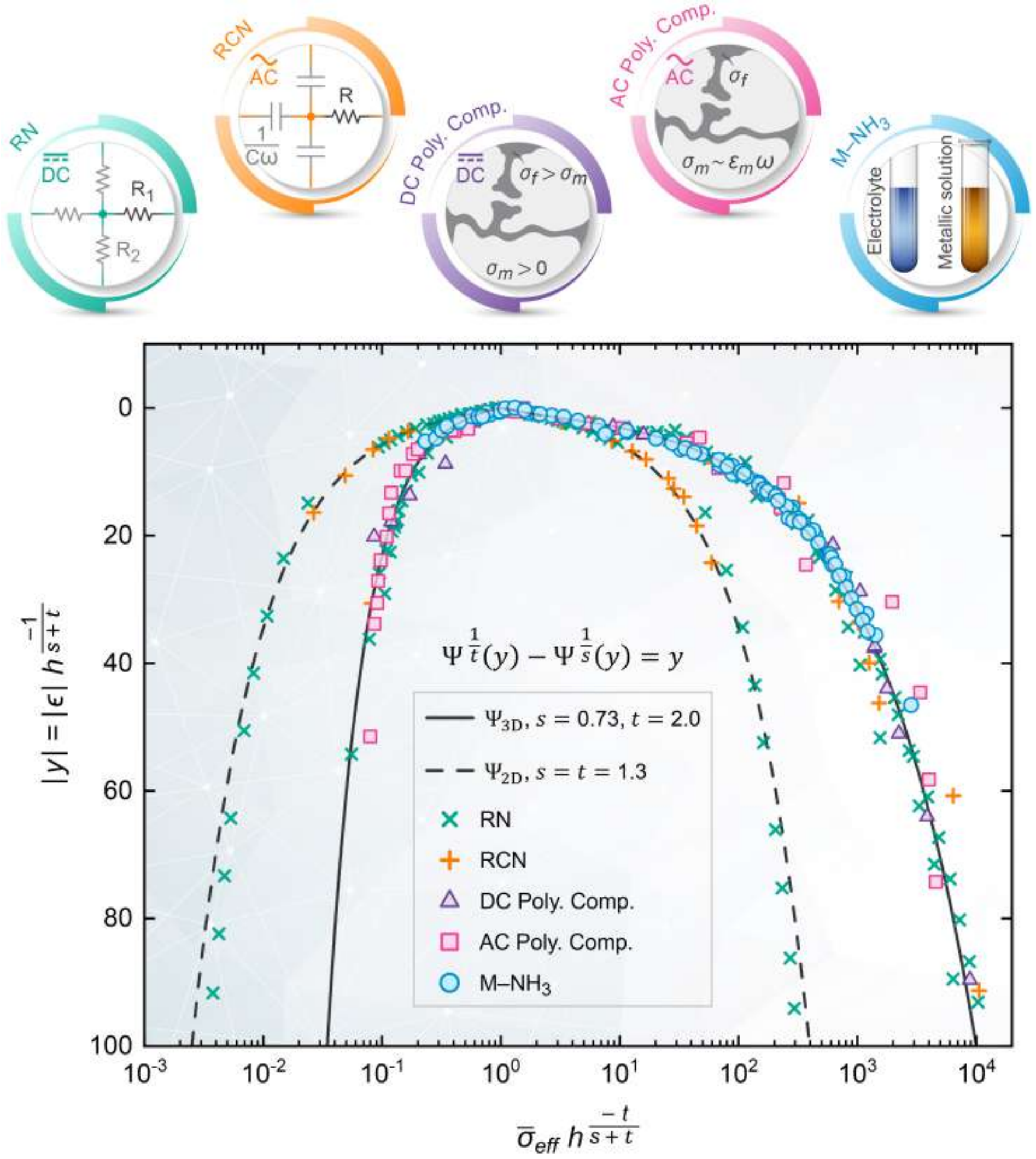

**System:** 2D square lattice (random resistor network)
**Poor Conducting Elements:** Resistors with conductivity $\sigma = 10^{-3}$ (a.u.)
**Good Conducting Elements:** Resistors with conductivity $\sigma = 1$ (a.u.)

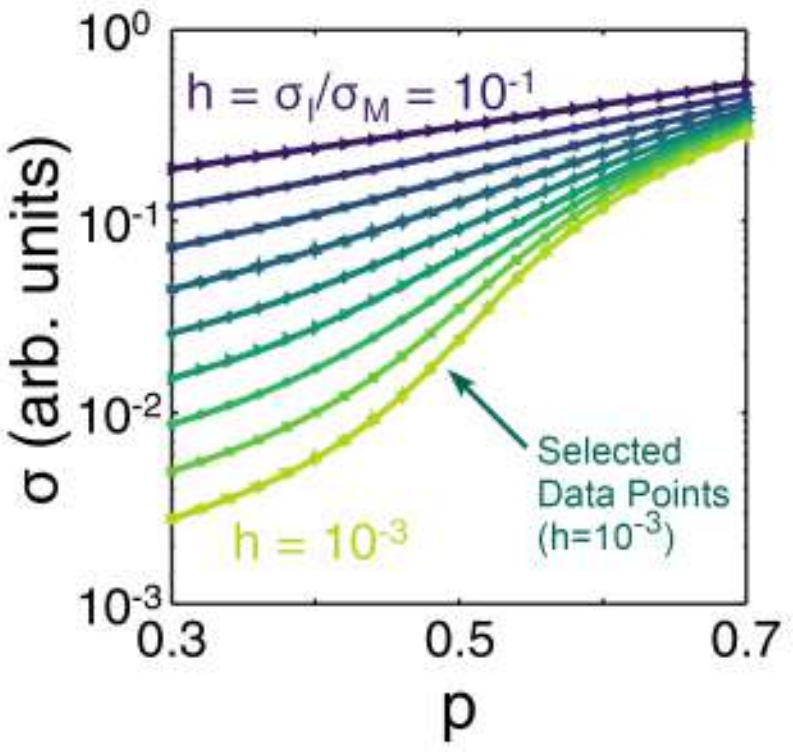

**Selected Figure:** Fig. 3(d) of the source article (Ref. [104])

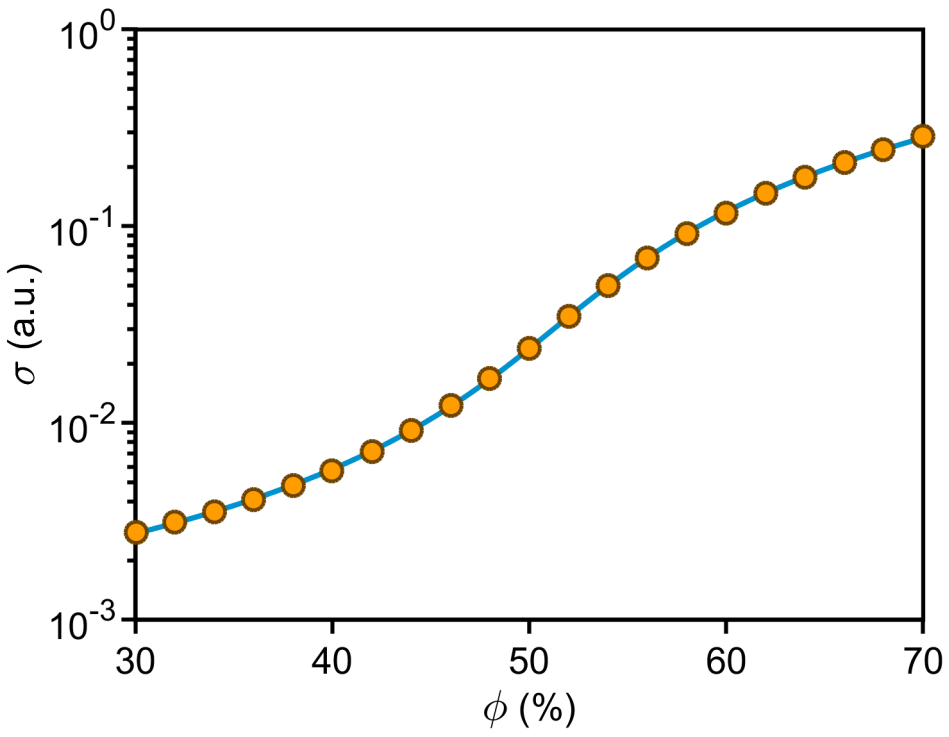

| | |
|---|---|
| $\phi_c$ (%) | 51.2 |
| $s$ | 1.30 |
| $t$ | 1.30 |
| $\sigma_m$ (a.u.) | $9.0 \times 10^{-4}$ |
| $\sigma_{f^*}$ (a.u.) | 1 |
| $\sigma_f$ (a.u.) | $9.4 \times 10^{-1}$ |
| $h = \sigma_m/\sigma_{f^*}$ | $9.0 \times 10^{-4}$ |
| $R^2$ | 0.999999 |

**Fit Results:** Fitting data with the TAGP equation and the best-fit parameters

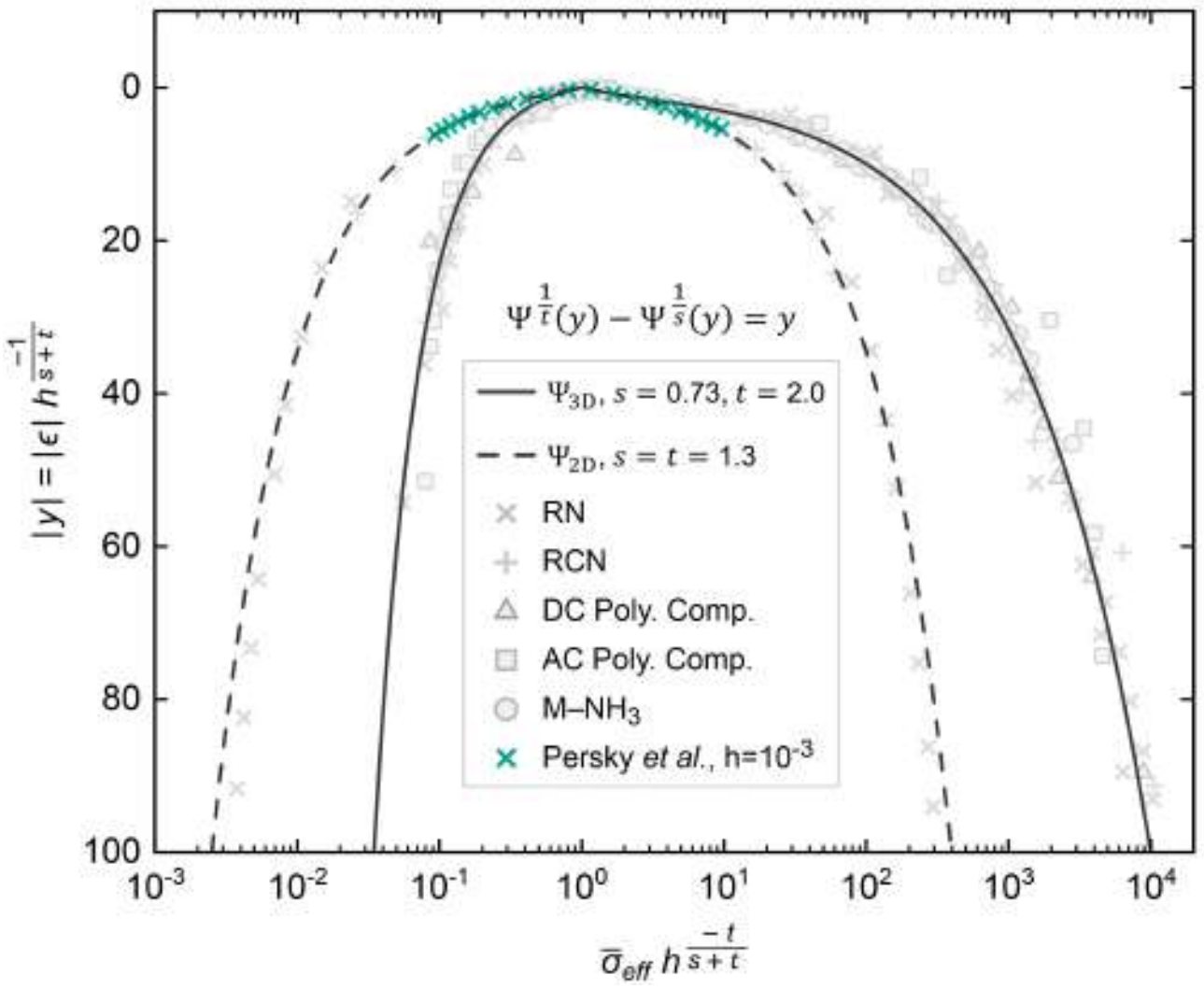

**Data Collapse:** Collapse of the rescaled data onto the proposed scaling function, $\Psi(y)$

**Extracted Data Points**

| # | $\phi$ (%) | $\sigma$ (a.u.) |
|---|---|---|
| 1 | 30.0 | 2.79E-03 |
| 2 | 32.0 | 3.14E-03 |
| 3 | 34.0 | 3.54E-03 |
| 4 | 36.0 | 4.08E-03 |
| 5 | 38.0 | 4.80E-03 |
| 6 | 40.0 | 5.74E-03 |
| 7 | 42.0 | 7.14E-03 |
| 8 | 44.0 | 9.14E-03 |
| 9 | 46.0 | 1.23E-02 |
| 10 | 48.0 | 1.69E-02 |
| 11 | 50.0 | 2.41E-02 |
| 12 | 52.0 | 3.48E-02 |
| 13 | 54.0 | 5.01E-02 |
| 14 | 56.0 | 6.89E-02 |
| 15 | 58.0 | 9.15E-02 |
| 16 | 60.0 | 1.17E-01 |
| 17 | 62.0 | 1.47E-01 |
| 18 | 64.0 | 1.78E-01 |
| 19 | 66.0 | 2.11E-01 |
| 20 | 68.0 | 2.46E-01 |
| 21 | 70.0 | 2.88E-01 |

**Source Article [104]:** E. Persky *et al.*, Nat. Commun. **12**, 3311, 2021; licensed under a Creative Commons Attribution (CC BY) license.

**System:** 2D square lattice (random resistor network)
**Poor Conducting Elements:** Resistors with conductivity $\sigma = 10^{-3}$ (a.u.)
**Good Conducting Elements:** Resistors with conductivity $\sigma = 1$ (a.u.)

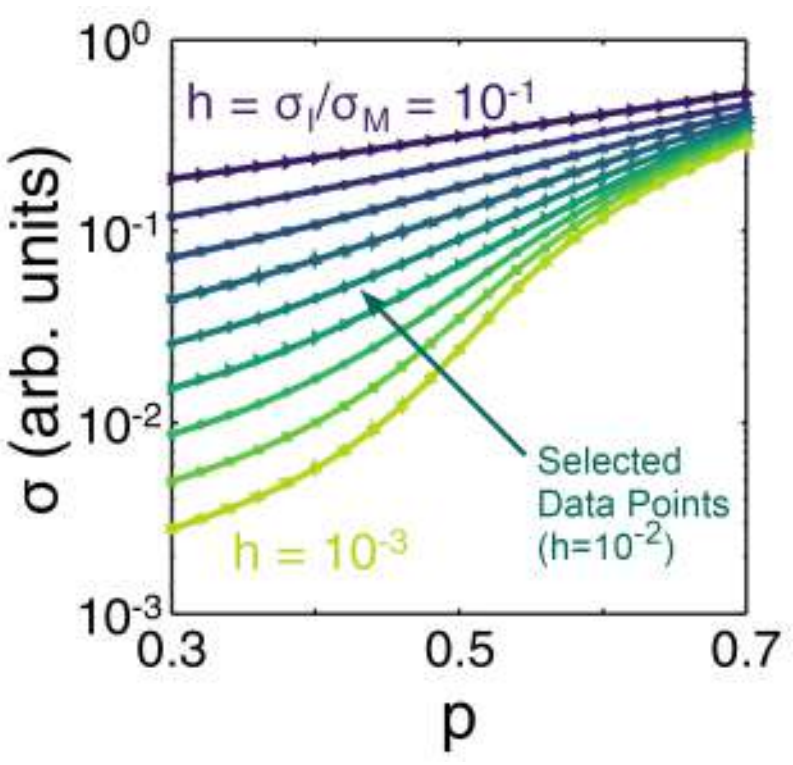

**Selected Figure:** Fig. 3(d) of the source article (Ref. [104])

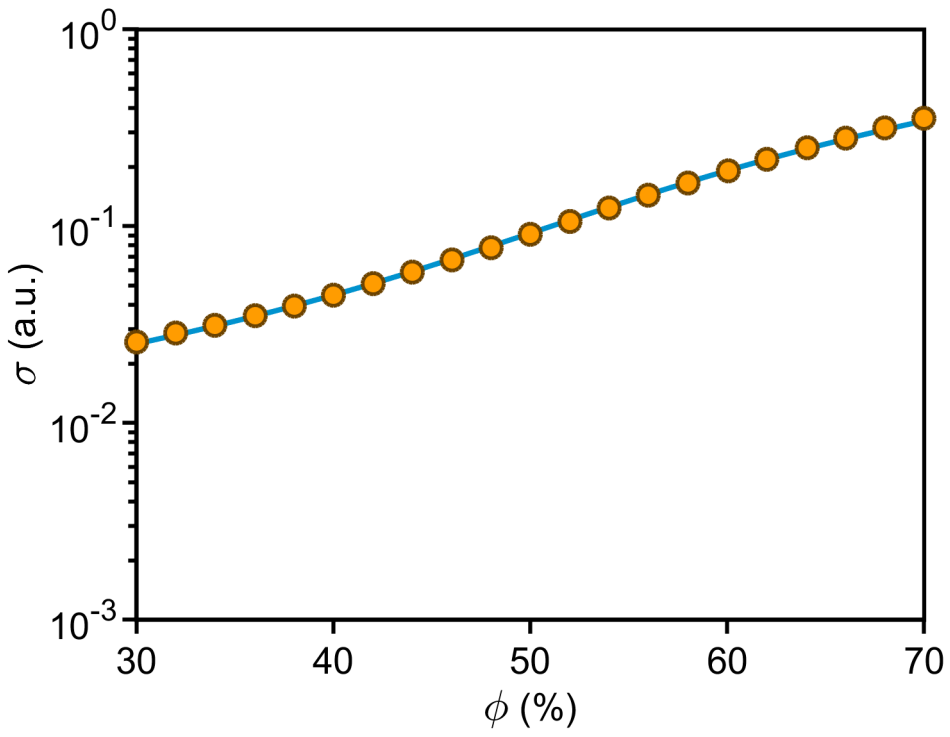

| | |
|---|---|
| $\phi_c$ (%) | 50.4 |
| $s$ | 1.30 |
| $t$ | 1.30 |
| $\sigma_m$ (a.u.) | $9.0 \times 10^{-3}$ |
| $\sigma_{f^*}$ (a.u.) | 1 |
| $\sigma_f$ (a.u.) | $9.8 \times 10^{-1}$ |
| $h = \sigma_m/\sigma_{f^*}$ | $9.0 \times 10^{-3}$ |
| $R^2$ | 0.999937 |

**Fit Results:** Fitting data with the TAGP equation and the best-fit parameters

**Extracted Data Points**

| # | $\phi$ (%) | $\sigma$ (a.u.) |
|---|---|---|
| 1 | 30.0 | 2.58E-02 |
| 2 | 32.0 | 2.86E-02 |
| 3 | 34.0 | 3.15E-02 |
| 4 | 36.0 | 3.51E-02 |
| 5 | 38.0 | 3.93E-02 |
| 6 | 40.0 | 4.45E-02 |
| 7 | 42.0 | 5.09E-02 |
| 8 | 44.0 | 5.85E-02 |
| 9 | 46.0 | 6.73E-02 |
| 10 | 48.0 | 7.79E-02 |
| 11 | 50.0 | 9.09E-02 |
| 12 | 52.0 | 1.06E-01 |
| 13 | 54.0 | 1.24E-01 |
| 14 | 56.0 | 1.44E-01 |
| 15 | 58.0 | 1.66E-01 |
| 16 | 60.1 | 1.92E-01 |
| 17 | 62.0 | 2.20E-01 |
| 18 | 64.1 | 2.51E-01 |
| 19 | 66.0 | 2.81E-01 |
| 20 | 68.0 | 3.16E-01 |
| 21 | 70.0 | 3.54E-01 |

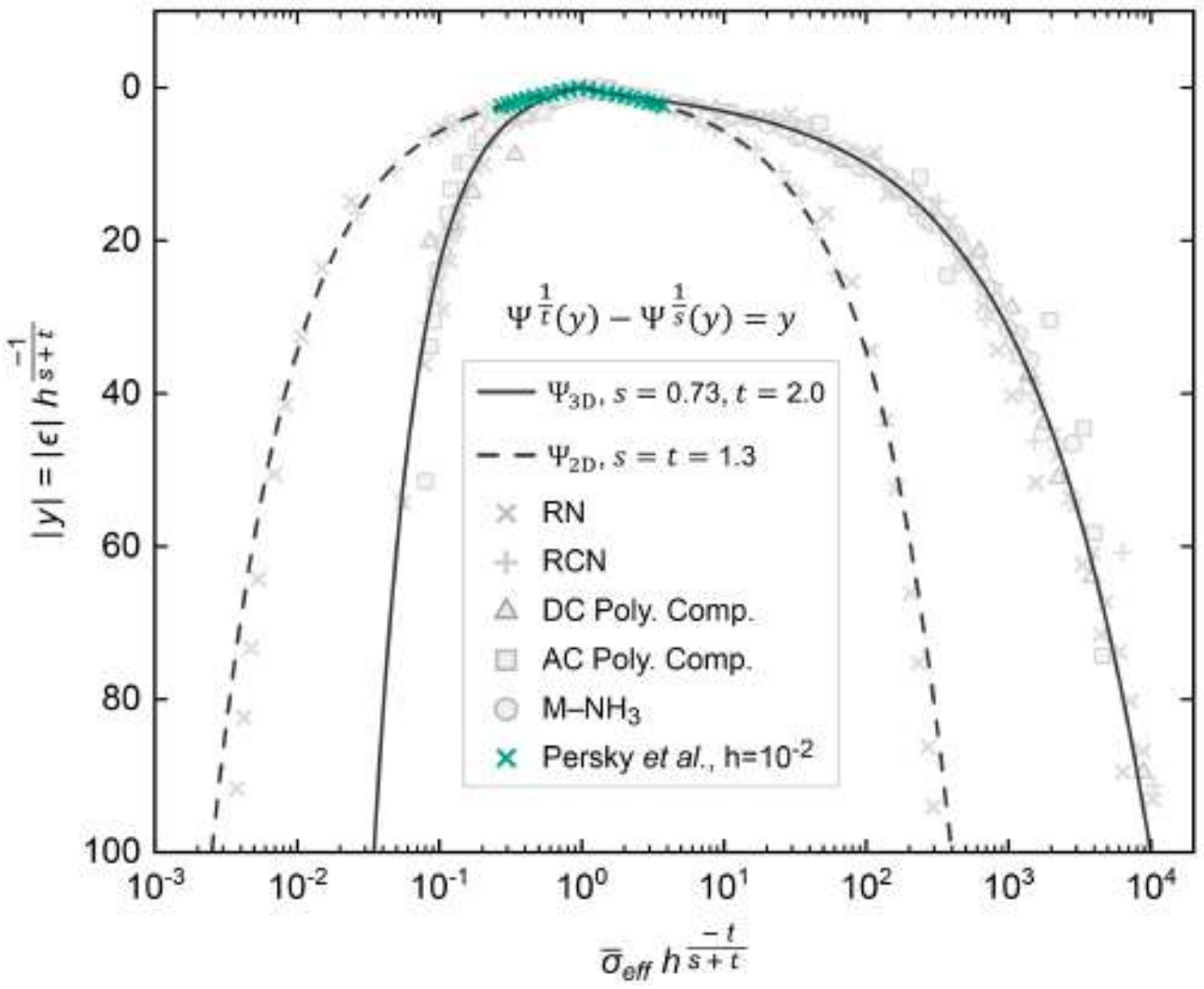

**Data Collapse:** Collapse of the rescaled data onto the proposed scaling function, $\Psi(y)$

**Source Article [104]:** E. Persky *et al.*, Nat. Commun. **12**, 3311, 2021; licensed under a Creative Commons Attribution (CC BY) license.

## Dataset 03, Resistor Network (RN) Models

**System:** 2D square lattice (random resistor network)
**Poor Conducting Elements:** Resistors with conductivity $\sigma = 10^{-3}$ (a.u.)
**Good Conducting Elements:** Resistors with conductivity $\sigma = 100$ (a.u.)

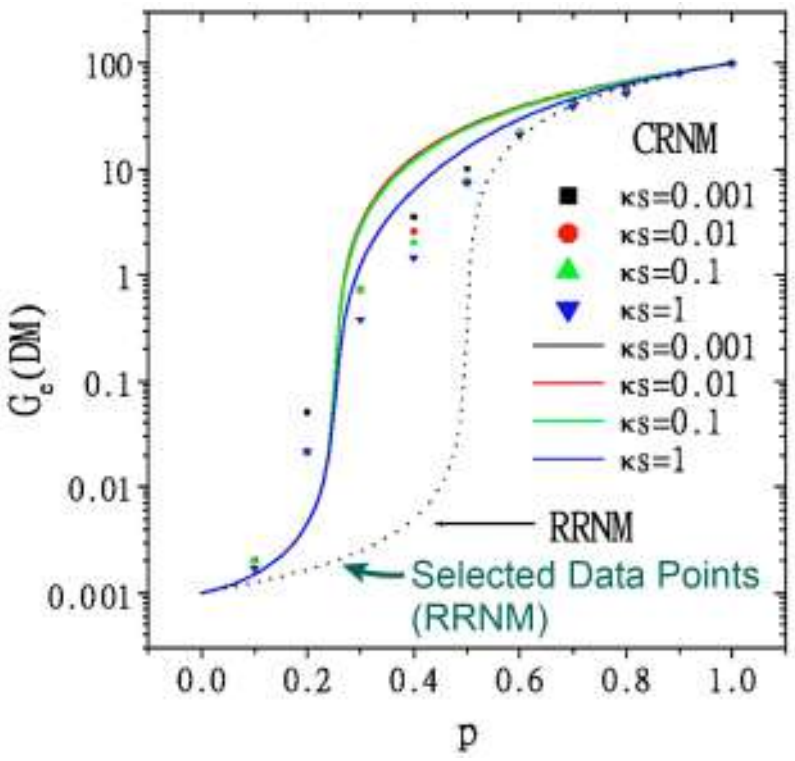

**Selected Figure:** Fig. 3 of the source article (Ref. [105])

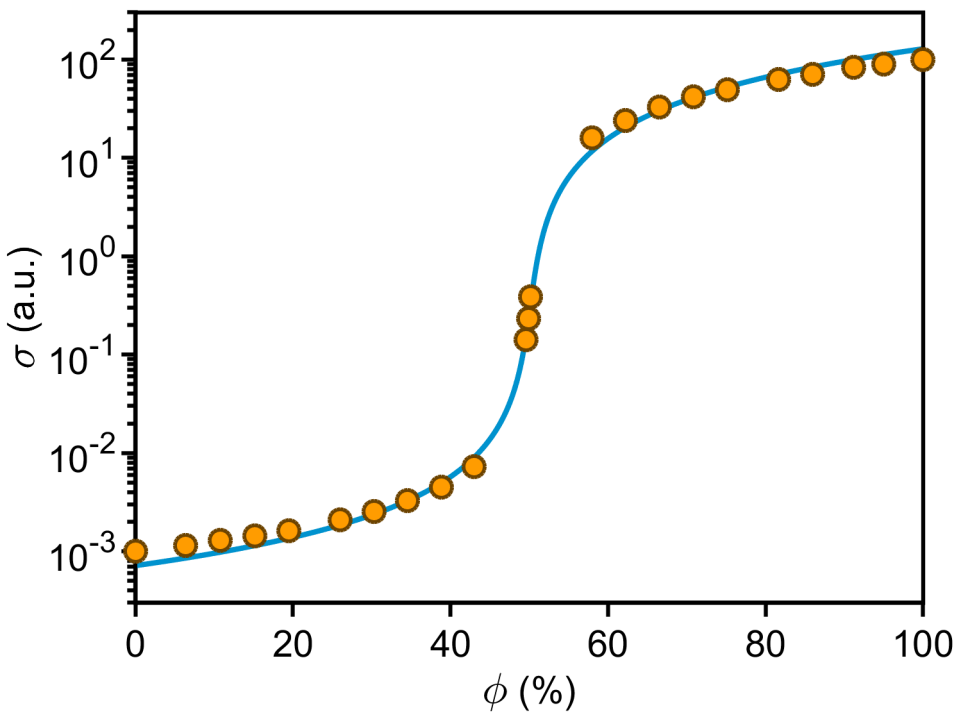

| | |
|---|---|
| $\phi_c$ (%) | 50.1 |
| $s$ | 1.30 |
| $t$ | 1.30 |
| $\sigma_m$ (a.u.) | $7.2 \times 10^{-4}$ |
| $\sigma_{f^*}$ (a.u.) | $1.3 \times 10^{2}$ |
| $\sigma_f$ (a.u.) | $1.3 \times 10^{2}$ |
| $h = \sigma_m/\sigma_{f^*}$ | $5.5 \times 10^{-6}$ |
| $R^2$ | 1.000000 |

**Fit Results:** Fitting data with the TAGP equation and the best-fit parameters

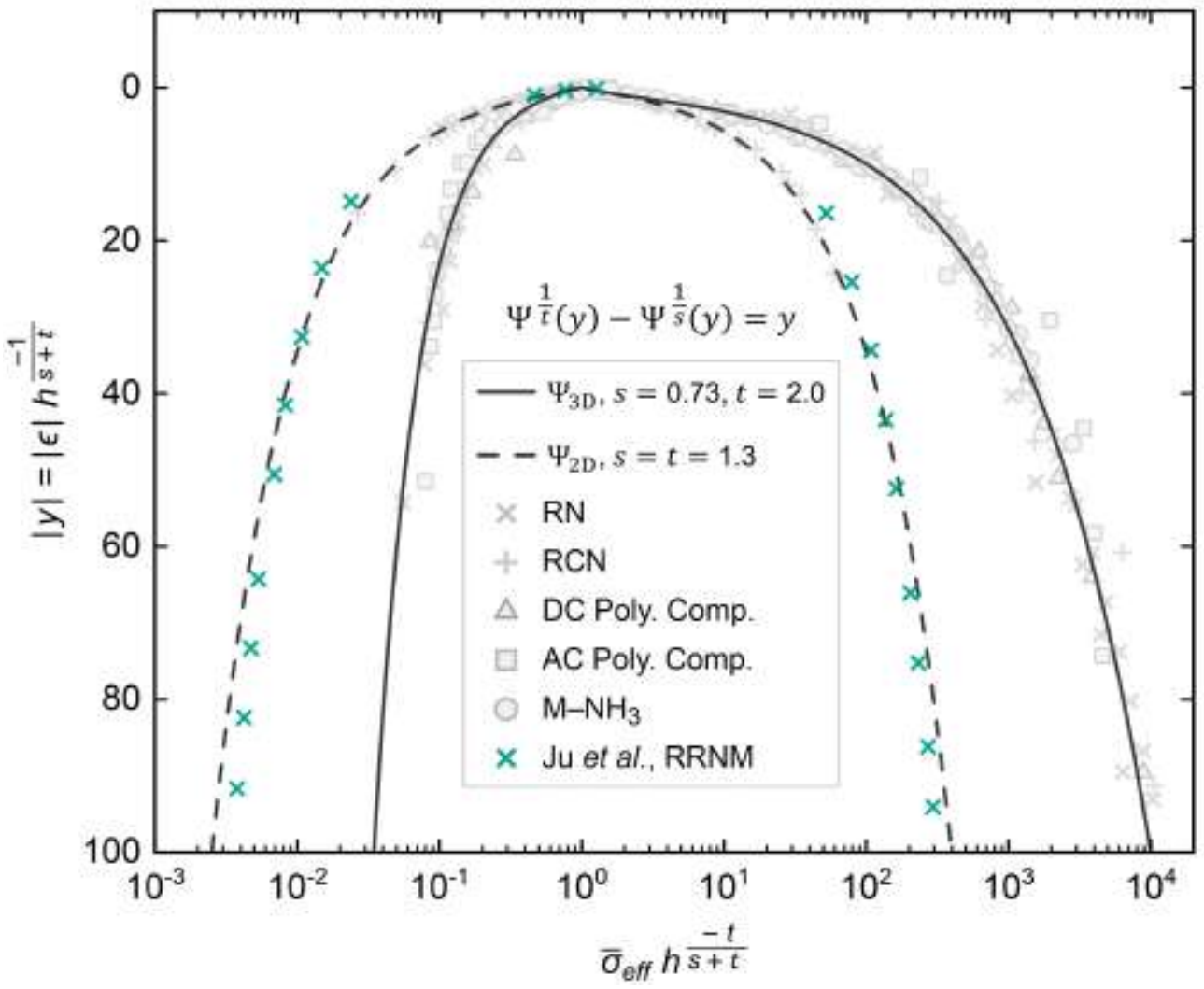

**Data Collapse:** Collapse of the rescaled data onto the proposed scaling function, $\Psi(y)$

**Extracted Data Points**

| # | $\phi$ (%) | $\sigma$ (a.u.) |
|---|---|---|
| 1 | 0.0 | 1.00E-03 |
| 2 | 6.4 | 1.15E-03 |
| 3 | 10.8 | 1.29E-03 |
| 4 | 15.1 | 1.43E-03 |
| 5 | 19.4 | 1.63E-03 |
| 6 | 26.0 | 2.10E-03 |
| 7 | 30.3 | 2.53E-03 |
| 8 | 34.6 | 3.29E-03 |
| 9 | 38.9 | 4.51E-03 |
| 10 | 43.0 | 7.24E-03 |
| 11 | 49.6 | 1.42E-01 |
| 12 | 49.9 | 2.33E-01 |
| 13 | 50.2 | 3.86E-01 |
| 14 | 57.9 | 1.59E+01 |
| 15 | 62.2 | 2.41E+01 |
| 16 | 66.5 | 3.28E+01 |
| 17 | 70.9 | 4.17E+01 |
| 18 | 75.1 | 4.93E+01 |
| 19 | 81.6 | 6.22E+01 |
| 20 | 86.0 | 7.07E+01 |
| 21 | 91.2 | 8.30E+01 |
| 22 | 95.0 | 8.97E+01 |
| 23 | 100.0 | 9.99E+01 |

**System:** 3D simple-cubic lattice (nearest-neighbor bond correlated resistor network)
**Poor Conducting Elements:** Resistors with conductivity $\sigma = 10^{-5}$ (a.u.)
**Good Conducting Elements:** Resistors with conductivity $\sigma = 1$ (a.u.)

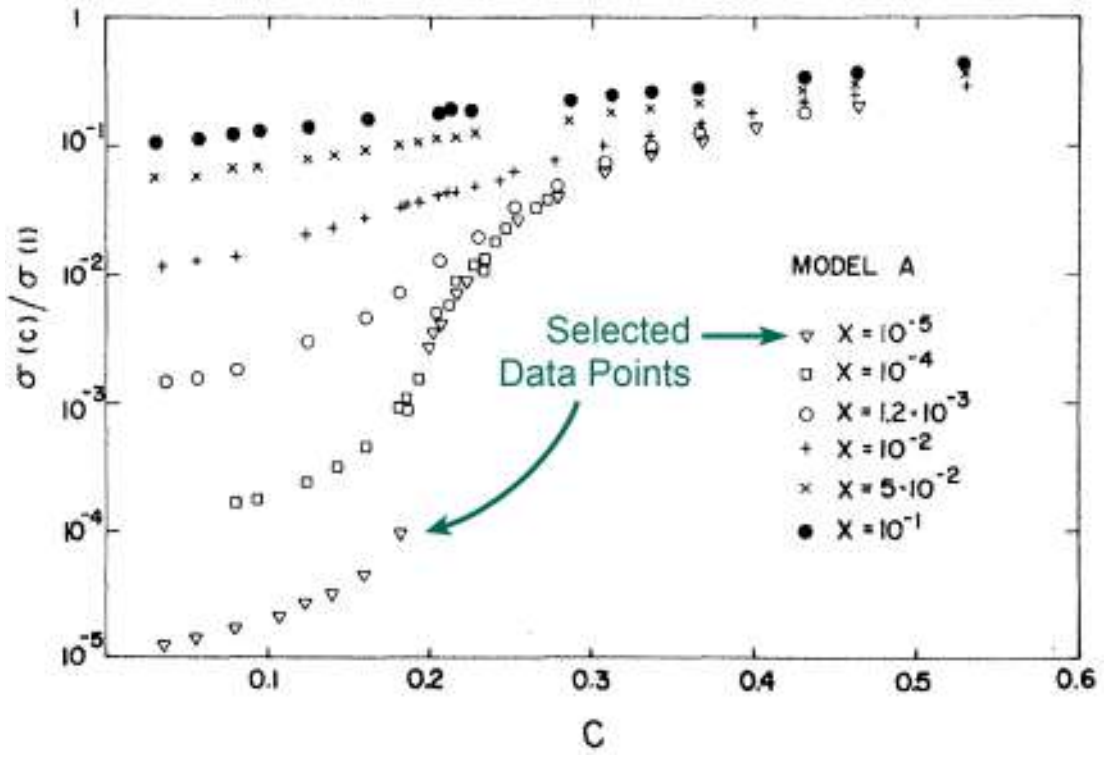

**Selected Figure:** Fig. 4 of the source article (Ref. [106])

**Extracted Data Points**

| # | $\phi$ (%) | $\sigma$ (norm.) |
|---|---|---|
| 1 | 3.6 | 1.24E-05 |
| 2 | 5.6 | 1.41E-05 |
| 3 | 8.0 | 1.69E-05 |
| 4 | 10.8 | 2.09E-05 |
| 5 | 12.3 | 2.66E-05 |
| 6 | 14.0 | 3.15E-05 |
| 7 | 16.0 | 4.41E-05 |
| 8 | 18.3 | 9.41E-05 |
| 9 | 20.0 | 2.68E-03 |
| 10 | 20.3 | 3.48E-03 |
| 11 | 20.8 | 4.10E-03 |
| 12 | 21.8 | 7.03E-03 |
| 13 | 22.4 | 8.80E-03 |
| 14 | 25.5 | 2.62E-02 |
| 15 | 28.0 | 4.04E-02 |
| 16 | 30.9 | 6.25E-02 |
| 17 | 33.7 | 8.53E-02 |
| 18 | 36.9 | 1.09E-01 |
| 19 | 40.2 | 1.38E-01 |
| 20 | 46.5 | 2.02E-01 |

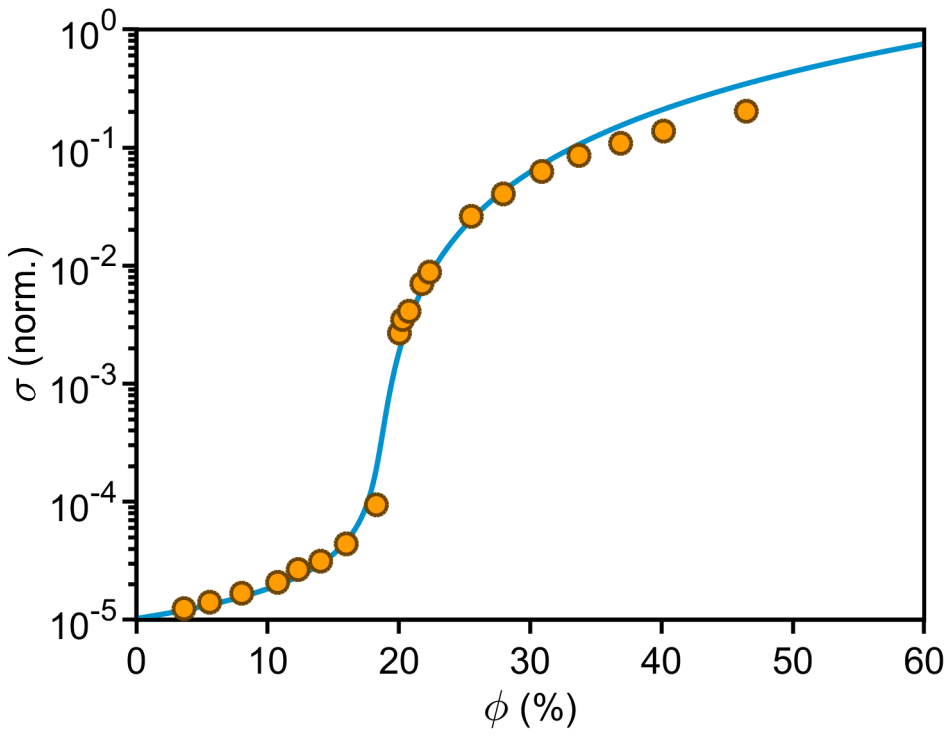

| | |
|---|---|
| $\phi_c$ (%) | 17.9 |
| $s$ | 0.73 |
| $t$ | 2.0 |
| $\sigma_m$ (norm.) | $1.0 \times 10^{-5}$ |
| $\sigma_{f^*}$ (norm.) | $1.4 \times 10^{-1}$ |
| $\sigma_f$ (norm.) | 2.9 |
| $h = \sigma_m/\sigma_{f^*}$ | $7.5 \times 10^{-5}$ |
| $R^2$ | 1.000000 |

**Fit Results:** Fitting data with the TAGP equation and the best-fit parameters

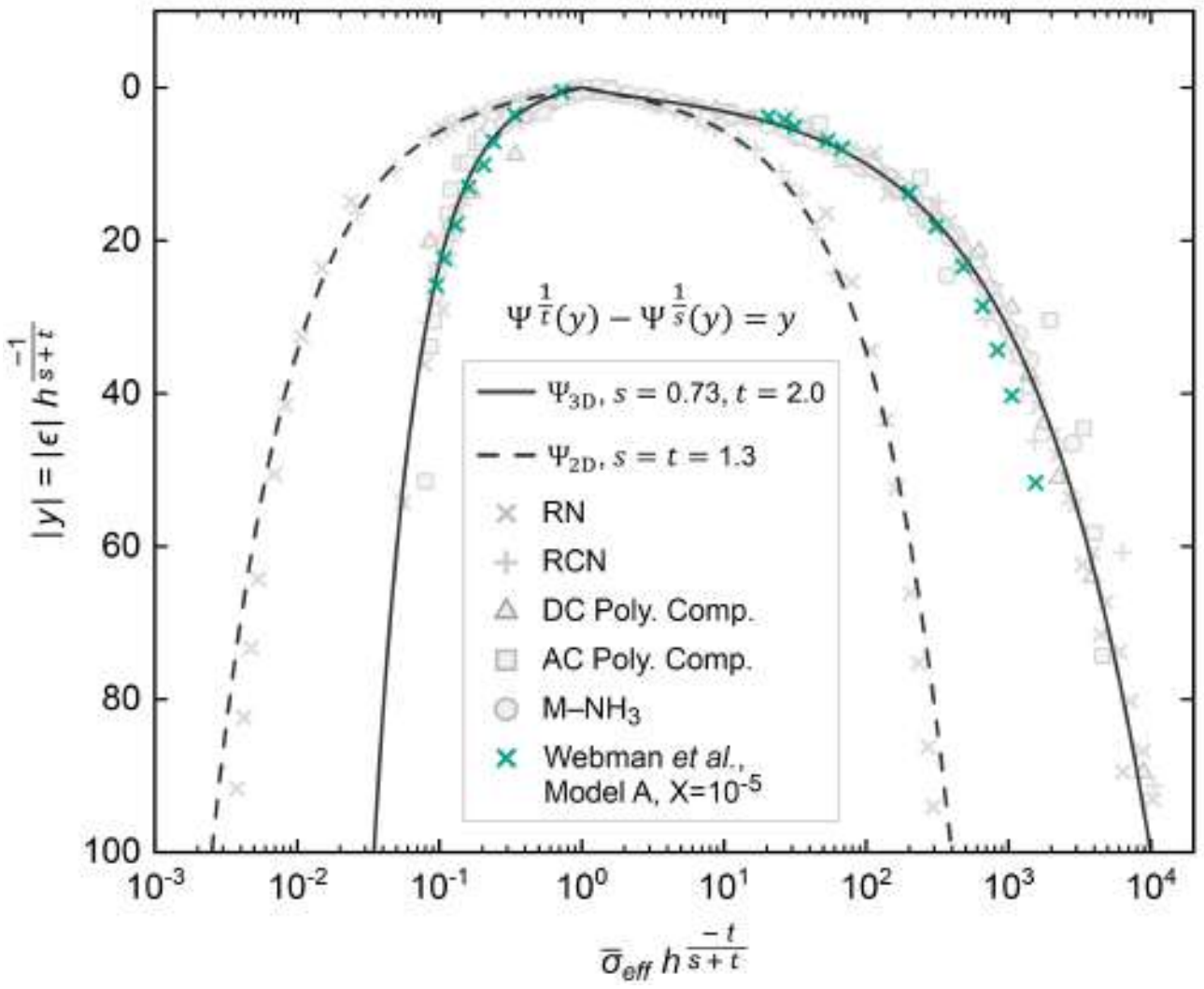

**Data Collapse:** Collapse of the rescaled data onto the proposed scaling function, $\Psi(y)$

**Source Article [106]:** Reproduced with permission from Phys. Rev. B **11**, 2885–2892 (1975). Copyright 1975 American Physical Society.

**System:** 3D simple-cubic lattice (nearest-neighbor bond correlated resistor network)
**Poor Conducting Elements:** Resistors with conductivity $\sigma = 1.2 \times 10^{-3}$ (a.u.)
**Good Conducting Elements:** Resistors with conductivity $\sigma = 1$ (a.u.)

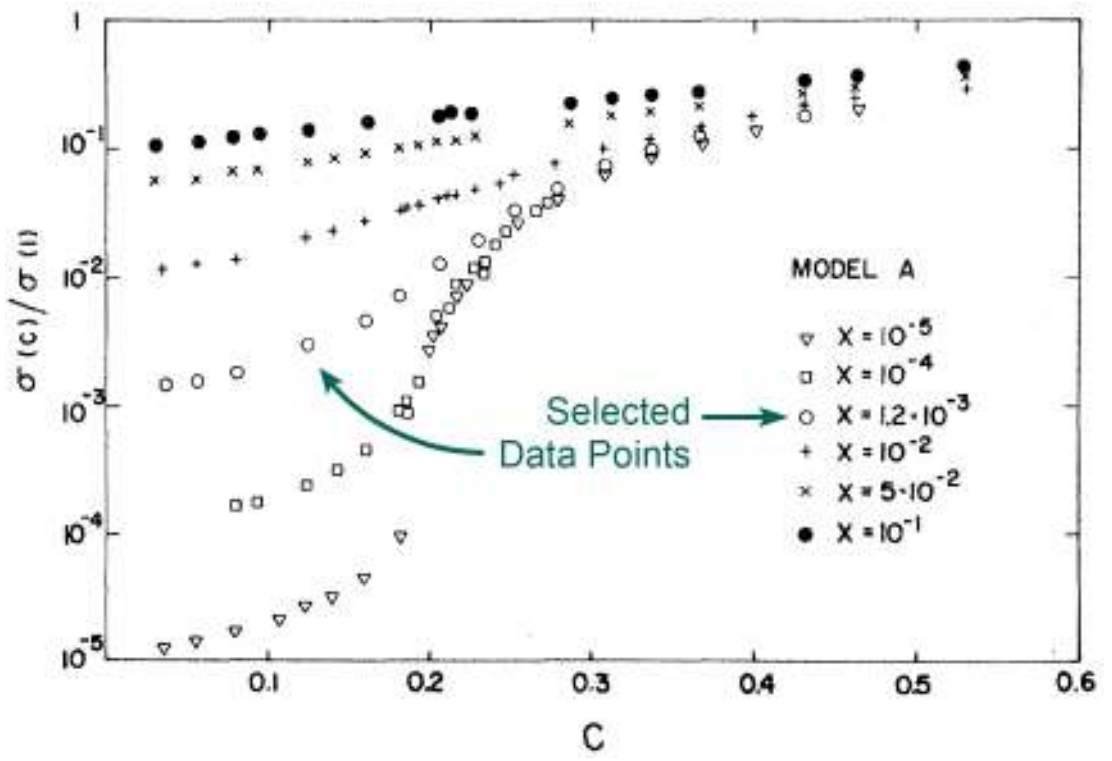

**Selected Figure:** Fig. 4 of the source article (Ref. [106])

### Extracted Data Points

| # | $\phi$ (%) | $\sigma$ (norm.) |
|---|---|---|
| 1 | 3.7 | 1.44E-03 |
| 2 | 5.7 | 1.52E-03 |
| 3 | 8.2 | 1.78E-03 |
| 4 | 12.5 | 2.95E-03 |
| 5 | 16.1 | 4.53E-03 |
| 6 | 18.2 | 7.14E-03 |
| 7 | 20.7 | 1.26E-02 |
| 8 | 23.1 | 1.92E-02 |
| 9 | 25.3 | 3.29E-02 |
| 10 | 27.9 | 4.90E-02 |
| 11 | 30.9 | 7.33E-02 |
| 12 | 33.7 | 9.73E-02 |
| 13 | 36.7 | 1.26E-01 |
| 14 | 43.1 | 1.78E-01 |

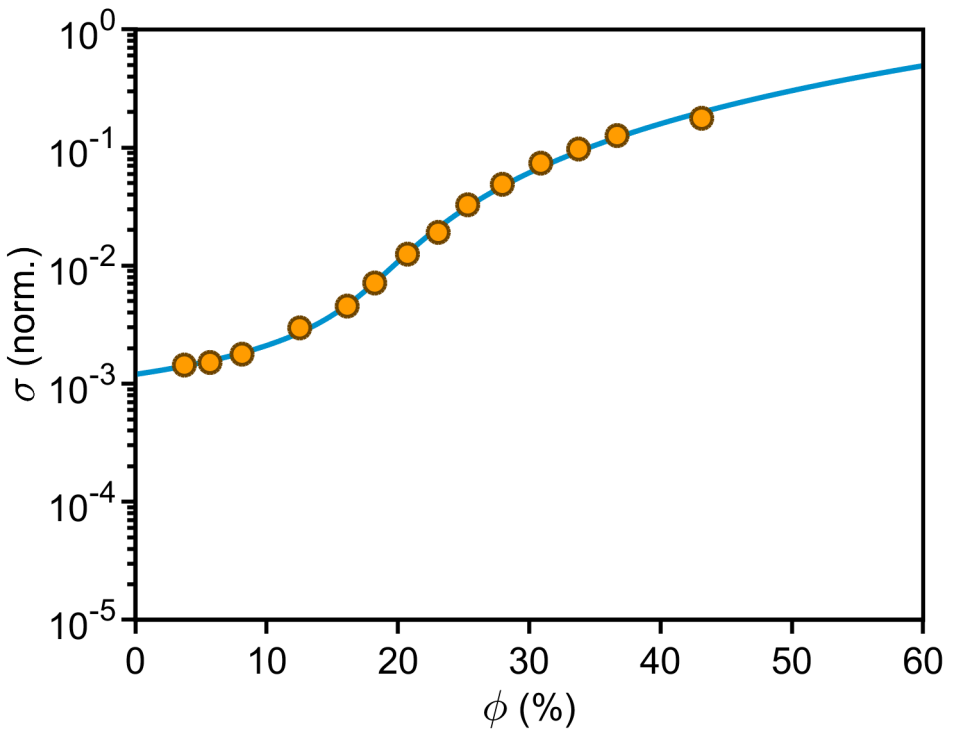

| | |
|---|---|
| $\phi_c$ (%) | 13.7 |
| $s$ | 0.73 |
| $t$ | 2.0 |
| $\sigma_m$ (norm.) | $1.2 \times 10^{-3}$ |
| $\sigma_{f^*}$ (norm.) | $4.3 \times 10^{-2}$ |
| $\sigma_f$ (norm.) | 1.7 |
| $h = \sigma_m/\sigma_{f^*}$ | $2.8 \times 10^{-2}$ |
| $R^2$ | 0.999982 |

**Fit Results:** Fitting data with the TAGP equation and the best-fit parameters

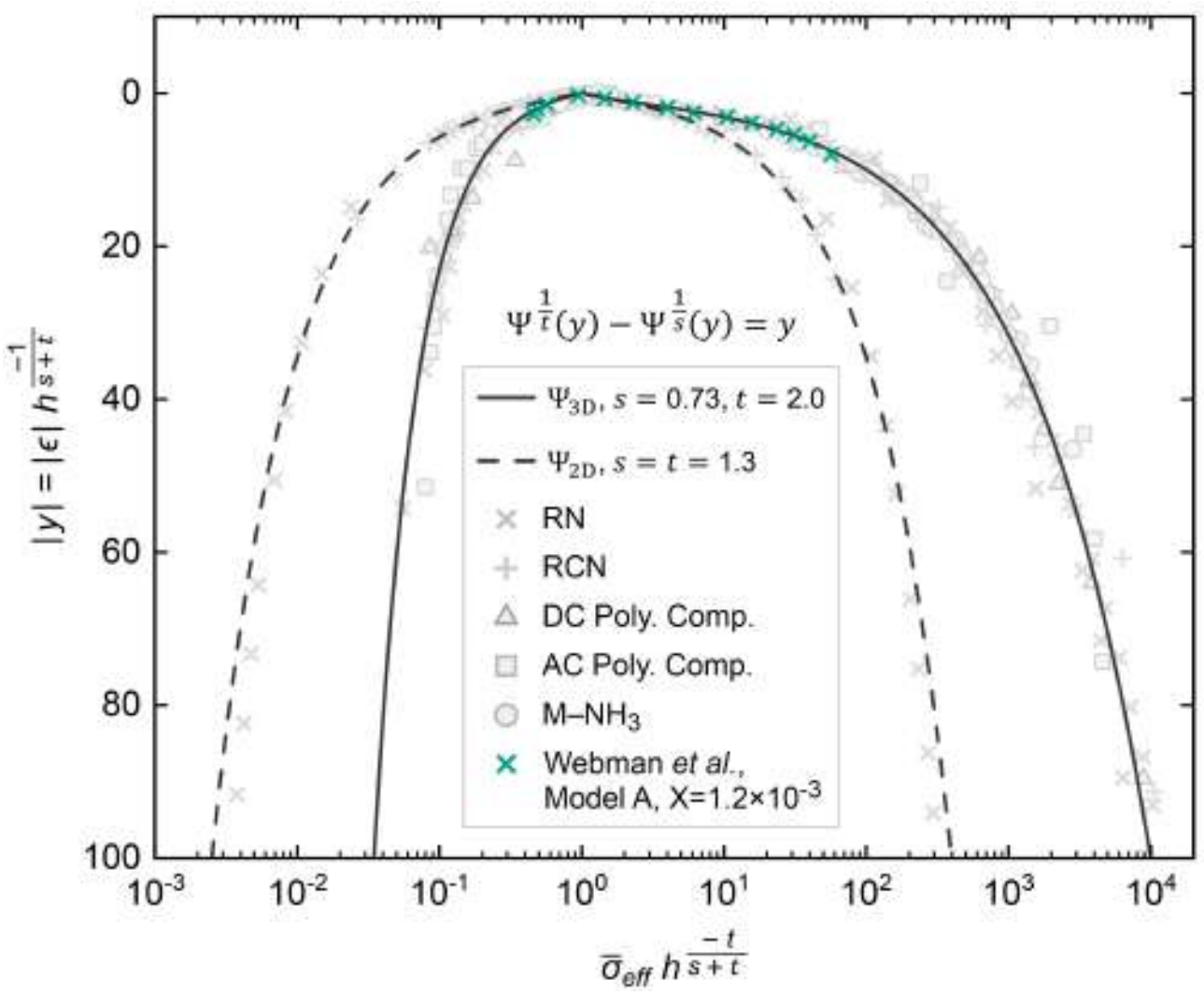

**Data Collapse:** Collapse of the rescaled data onto the proposed scaling function, $\Psi(y)$

**Source Article [106]:** Reproduced with permission from Phys. Rev. B **11**, 2885–2892 (1975). Copyright 1975 American Physical Society.

**System:** 3D simple-cubic lattice (second-order bond correlated resistor network)
**Poor Conducting Elements:** Resistors with conductivity $\sigma = 10^{-5}$ (a.u.)
**Good Conducting Elements:** Resistors with conductivity $\sigma = 1$ (a.u.)

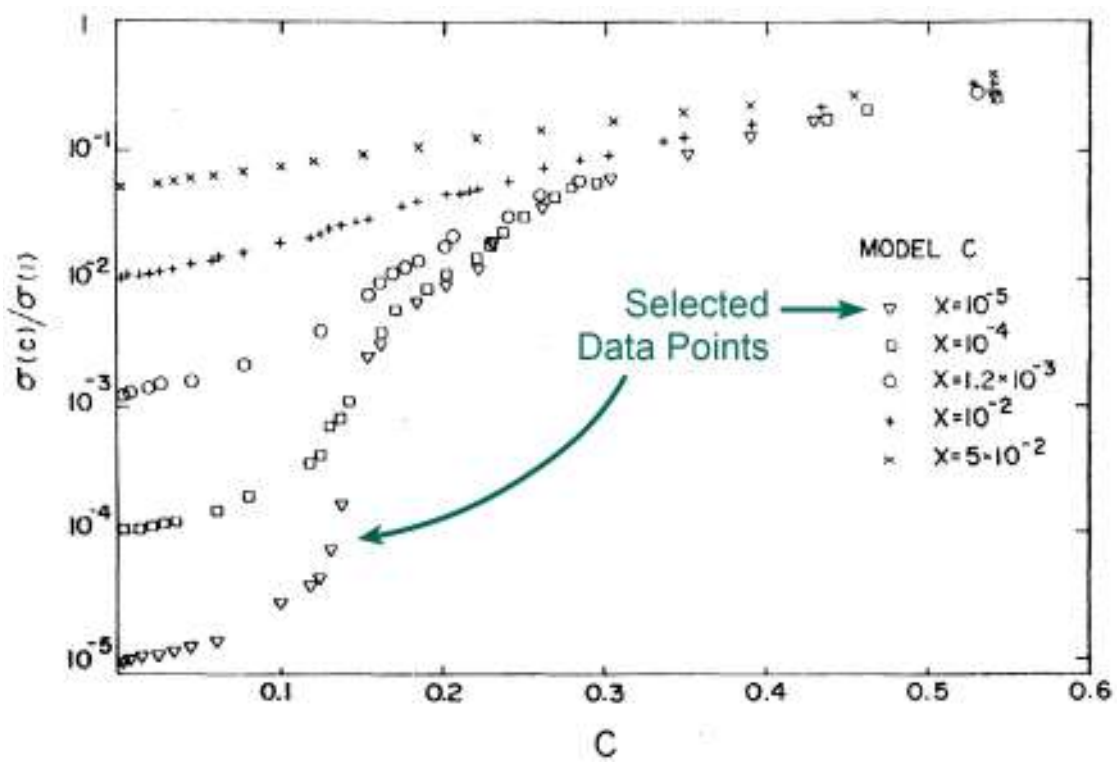

**Selected Figure:** Fig. 6 of the source article (Ref. [106])

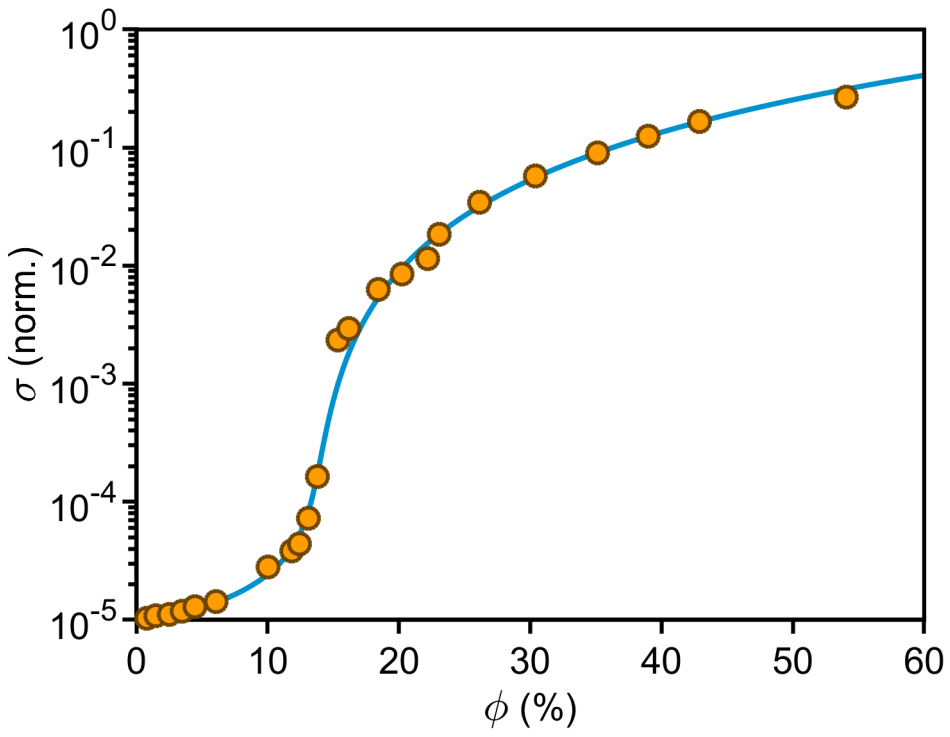

| | |
|---|---|
| $\phi_c$ (%) | 13.1 |
| $s$ | 0.73 |
| $t$ | 2.0 |
| $\sigma_m$ (norm.) | $9.1 \times 10^{-6}$ |
| $\sigma_{f^*}$ (norm.) | $3.2 \times 10^{-2}$ |
| $\sigma_f$ (norm.) | 1.4 |
| $h = \sigma_m/\sigma_{f^*}$ | $2.9 \times 10^{-4}$ |
| $R^2$ | 1.000000 |

**Fit Results:** Fitting data with the TAGP equation and the best-fit parameters

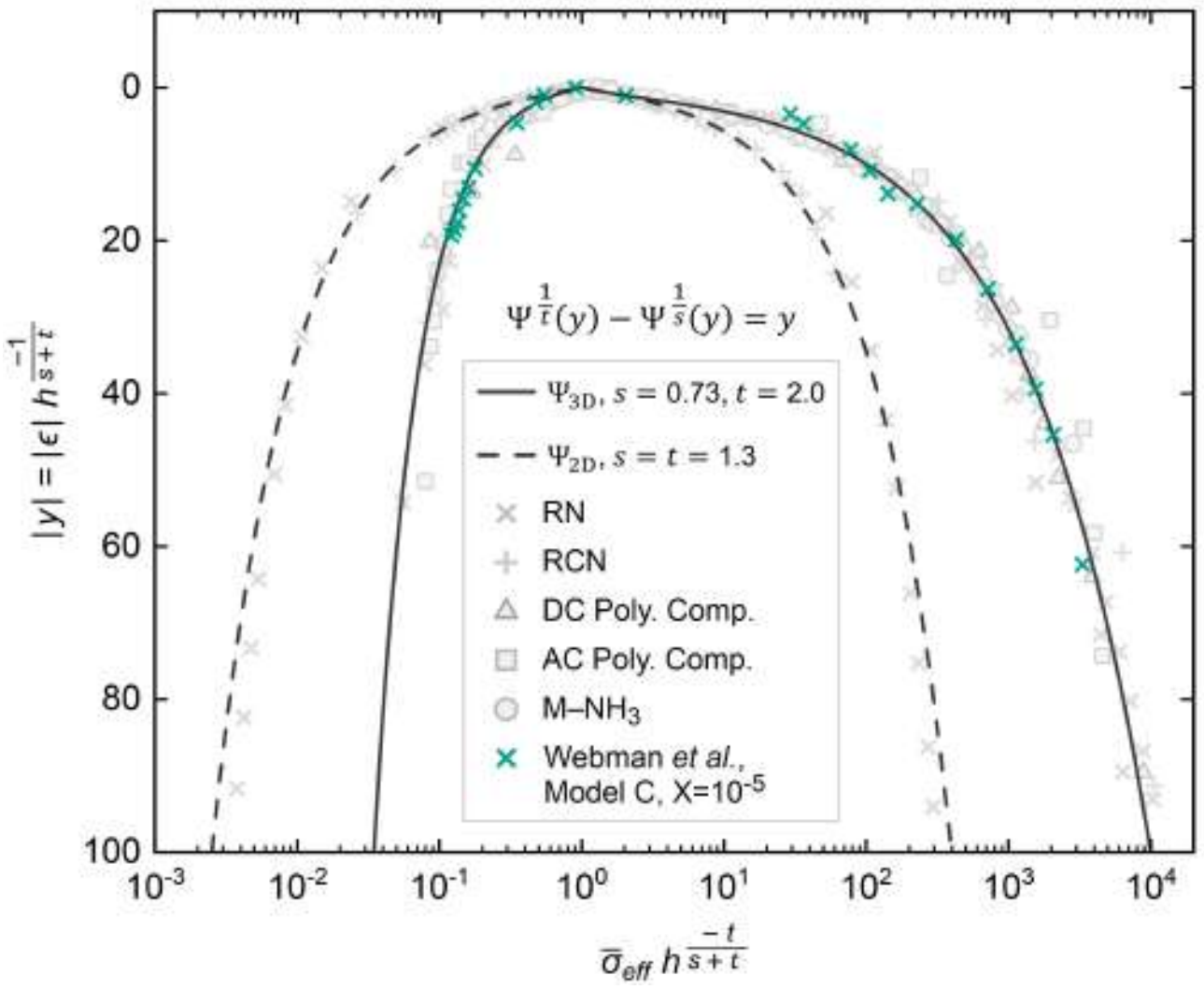

**Data Collapse:** Collapse of the rescaled data onto the proposed scaling function, $\Psi(y)$

### Extracted Data Points

| # | $\phi$ (%) | $\sigma$ (norm.) |
|---|---|---|
| 1 | 0.3 | 9.76E-06 |
| 2 | 0.8 | 1.03E-05 |
| 3 | 1.5 | 1.09E-05 |
| 4 | 2.5 | 1.11E-05 |
| 5 | 3.5 | 1.18E-05 |
| 6 | 4.5 | 1.29E-05 |
| 7 | 6.1 | 1.42E-05 |
| 8 | 10.0 | 2.81E-05 |
| 9 | 11.8 | 3.83E-05 |
| 10 | 12.4 | 4.40E-05 |
| 11 | 13.1 | 7.26E-05 |
| 12 | 13.8 | 1.64E-04 |
| 13 | 15.4 | 2.35E-03 |
| 14 | 16.2 | 2.92E-03 |
| 15 | 18.4 | 6.25E-03 |
| 16 | 20.2 | 8.50E-03 |
| 17 | 22.2 | 1.15E-02 |
| 18 | 23.1 | 1.84E-02 |
| 19 | 26.1 | 3.45E-02 |
| 20 | 30.4 | 5.78E-02 |
| 21 | 35.1 | 9.07E-02 |
| 22 | 39.0 | 1.25E-01 |
| 23 | 42.9 | 1.65E-01 |
| 24 | 54.1 | 2.66E-01 |

**Source Article [106]:**

**System:** 3D simple-cubic lattice (second-order bond correlated resistor network)
**Poor Conducting Elements:** Resistors with conductivity $\sigma = 1.2 \times 10^{-3}$ (a.u.)
**Good Conducting Elements:** Resistors with conductivity $\sigma = 1$ (a.u.)

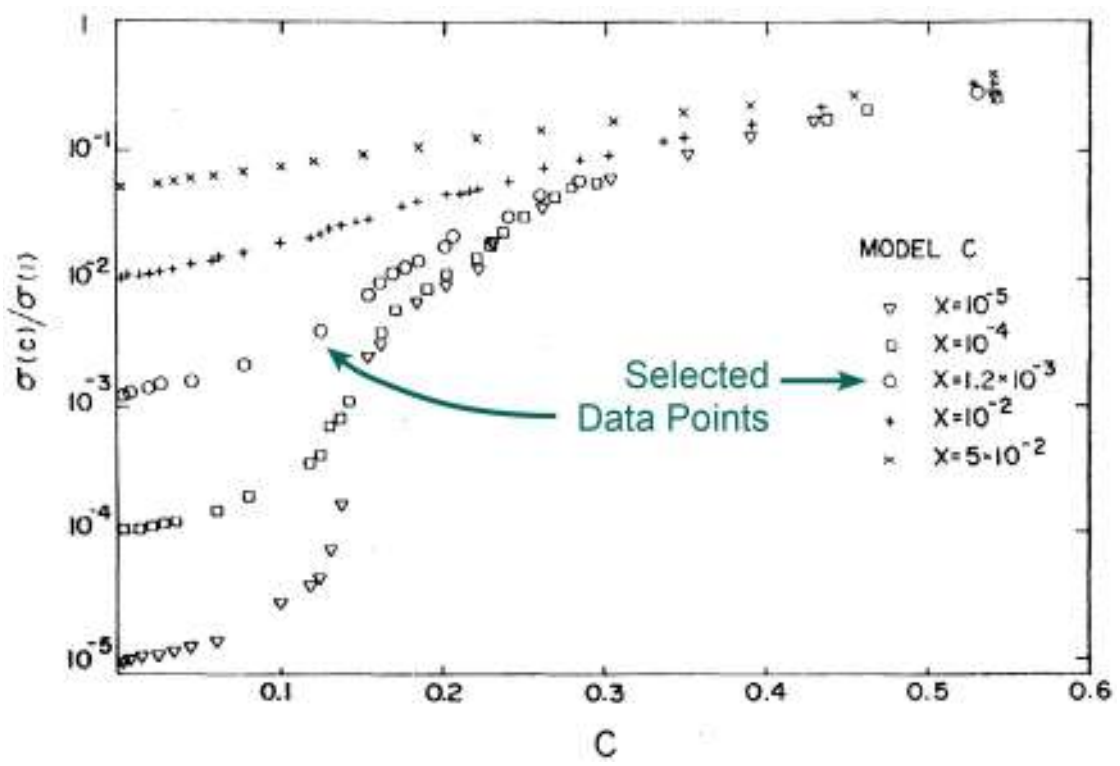

**Selected Figure:** Fig. 6 of the source article (Ref. [106])

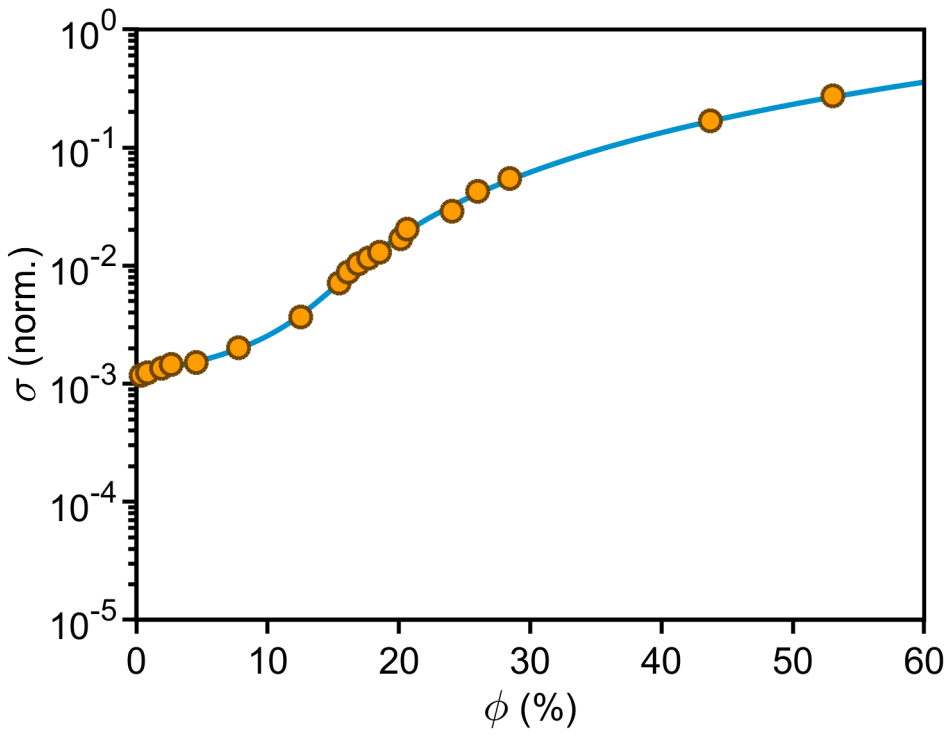

| | |
|---|---|
| $\phi_c$ (%) | 8.4 |
| $s$ | 0.73 |
| $t$ | 2.0 |
| $\sigma_m$ (norm.) | $1.2 \times 10^{-3}$ |
| $\sigma_{f^*}$ (norm.) | $9.6 \times 10^{-3}$ |
| $\sigma_f$ (norm.) | 1.1 |
| $h = \sigma_m/\sigma_{f^*}$ | $1.3 \times 10^{-1}$ |
| $R^2$ | 0.999989 |

**Fit Results:** Fitting data with the TAGP equation and the best-fit parameters

**Extracted Data Points**

| # | $\phi$ (%) | $\sigma$ (norm.) |
|---|---|---|
| 1 | 0.4 | 1.18E-03 |
| 2 | 0.8 | 1.24E-03 |
| 3 | 1.9 | 1.35E-03 |
| 4 | 2.7 | 1.45E-03 |
| 5 | 4.6 | 1.51E-03 |
| 6 | 7.8 | 2.01E-03 |
| 7 | 12.5 | 3.70E-03 |
| 8 | 15.4 | 7.09E-03 |
| 9 | 16.1 | 8.80E-03 |
| 10 | 16.9 | 1.04E-02 |
| 11 | 17.7 | 1.16E-02 |
| 12 | 18.5 | 1.30E-02 |
| 13 | 20.1 | 1.68E-02 |
| 14 | 20.6 | 2.04E-02 |
| 15 | 24.1 | 2.87E-02 |
| 16 | 26.0 | 4.26E-02 |
| 17 | 28.4 | 5.50E-02 |
| 18 | 43.7 | 1.69E-01 |
| 19 | 53.1 | 2.74E-01 |

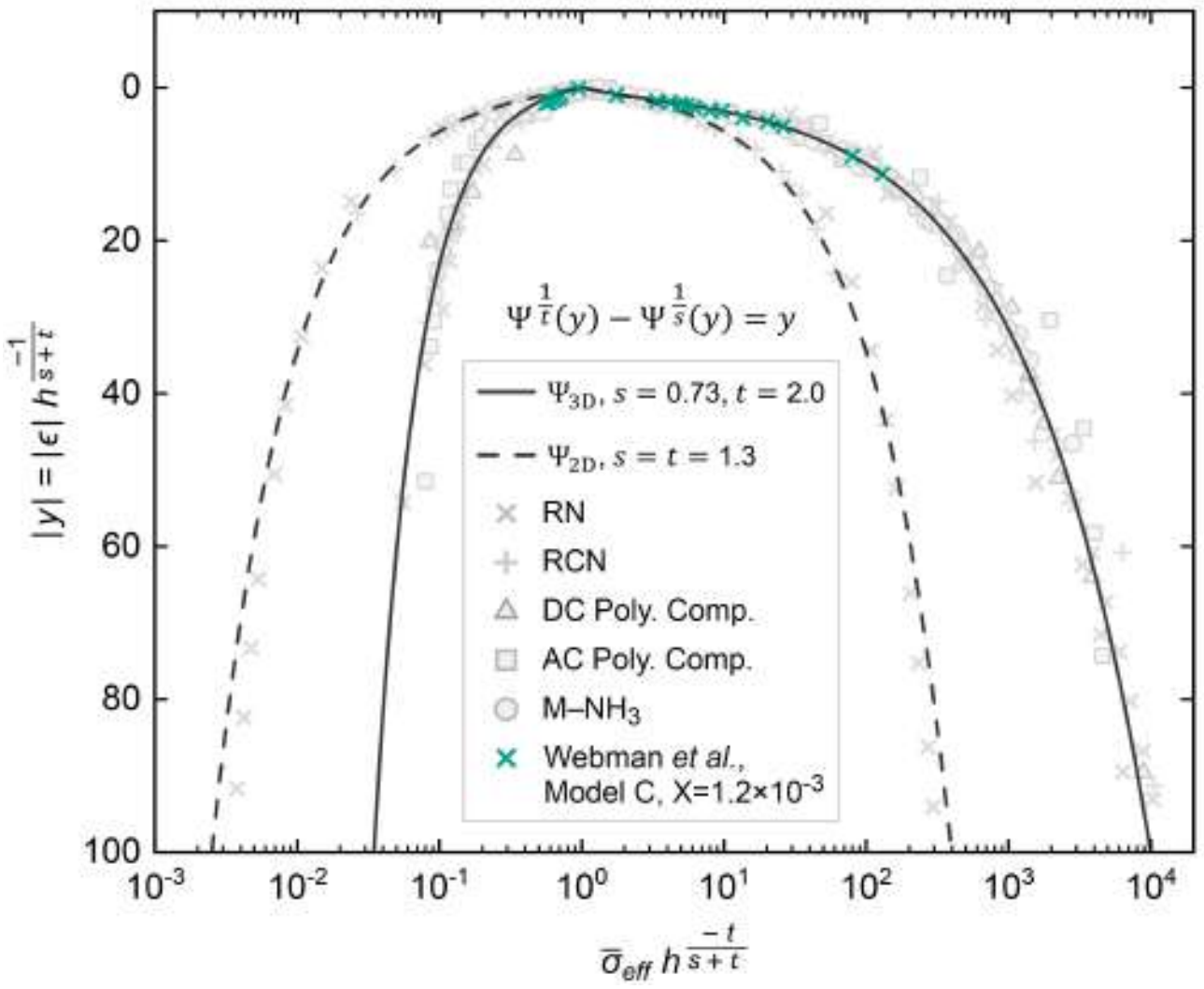

**Data Collapse:** Collapse of the rescaled data onto the proposed scaling function, $\Psi(y)$

**Source Article [106]:** Reproduced with permission from Phys. Rev. B **11**, 2885–2892 (1975). Copyright 1975 American Physical Society.

**System:** 3D simple-cubic lattice (random resistor network)
**Poor Conducting Elements:** Resistors with conductivity $\sigma = 0.1$ (S/cm)
**Good Conducting Elements:** Resistors with conductivity $\sigma = 2 \times 10^4$ (S/cm)

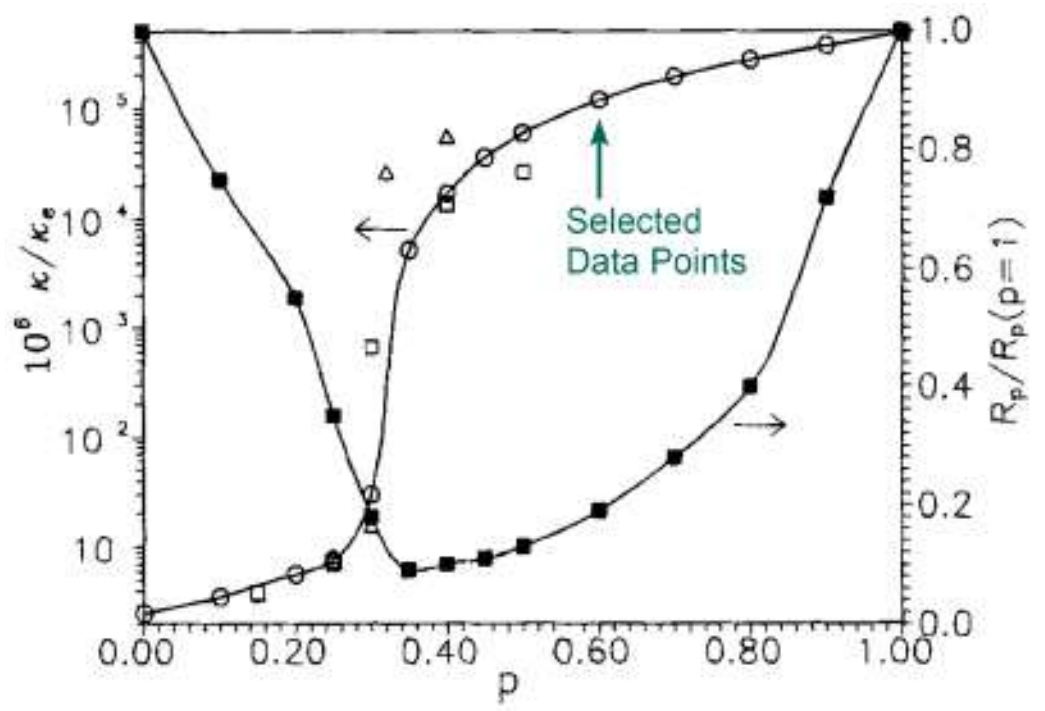

**Selected Figure:** Fig. 2 of the source article (Ref. [107])

**Extracted Data Points**

| # | $\phi$ (%) | $\sigma$ (norm.) |
|---|---|---|
| 1 | 0.0 | 2.47E+00 |
| 2 | 10.1 | 3.50E+00 |
| 3 | 20.0 | 5.63E+00 |
| 4 | 29.9 | 3.05E+01 |
| 5 | 34.9 | 5.12E+03 |
| 6 | 39.9 | 1.72E+04 |
| 7 | 44.9 | 3.62E+04 |
| 8 | 49.9 | 6.09E+04 |
| 9 | 60.0 | 1.22E+05 |
| 10 | 69.9 | 1.99E+05 |
| 11 | 79.9 | 2.87E+05 |
| 12 | 89.8 | 3.92E+05 |
| 13 | 99.9 | 5.22E+05 |

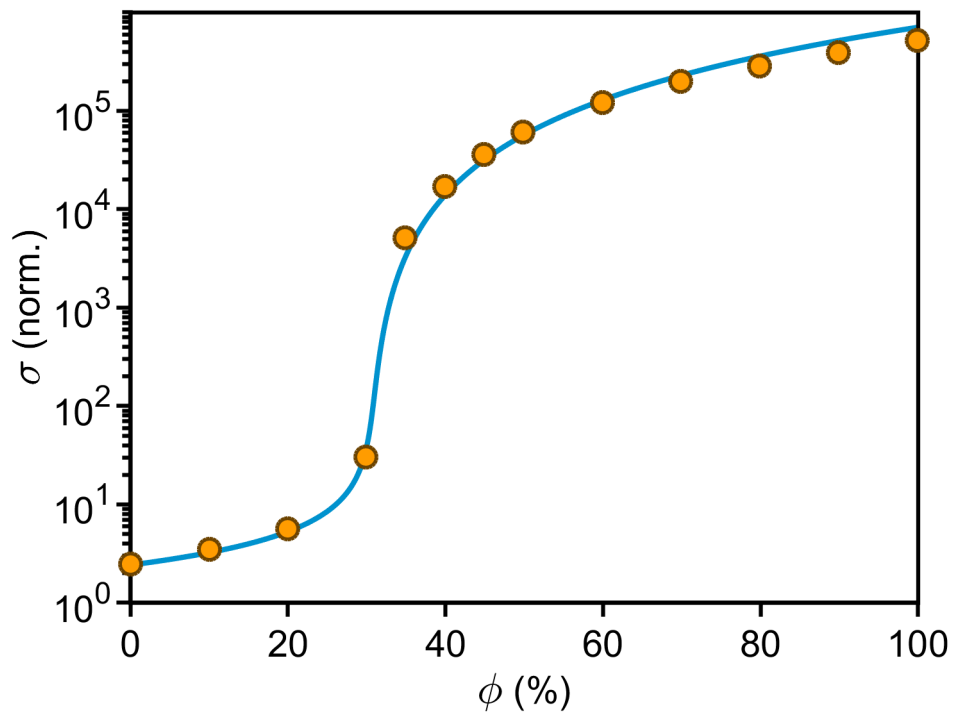

| | |
|---|---|
| $\phi_c$ (%) | 30.2 |
| $s$ | 0.73 |
| $t$ | 2.0 |
| $\sigma_m$ (norm.) | 2.4 |
| $\sigma_{f^*}$ (norm.) | $1.3 \times 10^5$ |
| $\sigma_f$ (norm.) | $7.1 \times 10^5$ |
| $h = \sigma_m/\sigma_{f^*}$ | $1.8 \times 10^{-5}$ |
| $R^2$ | 1.000000 |

**Fit Results:** Fitting data with the TAGP equation and the best-fit parameters

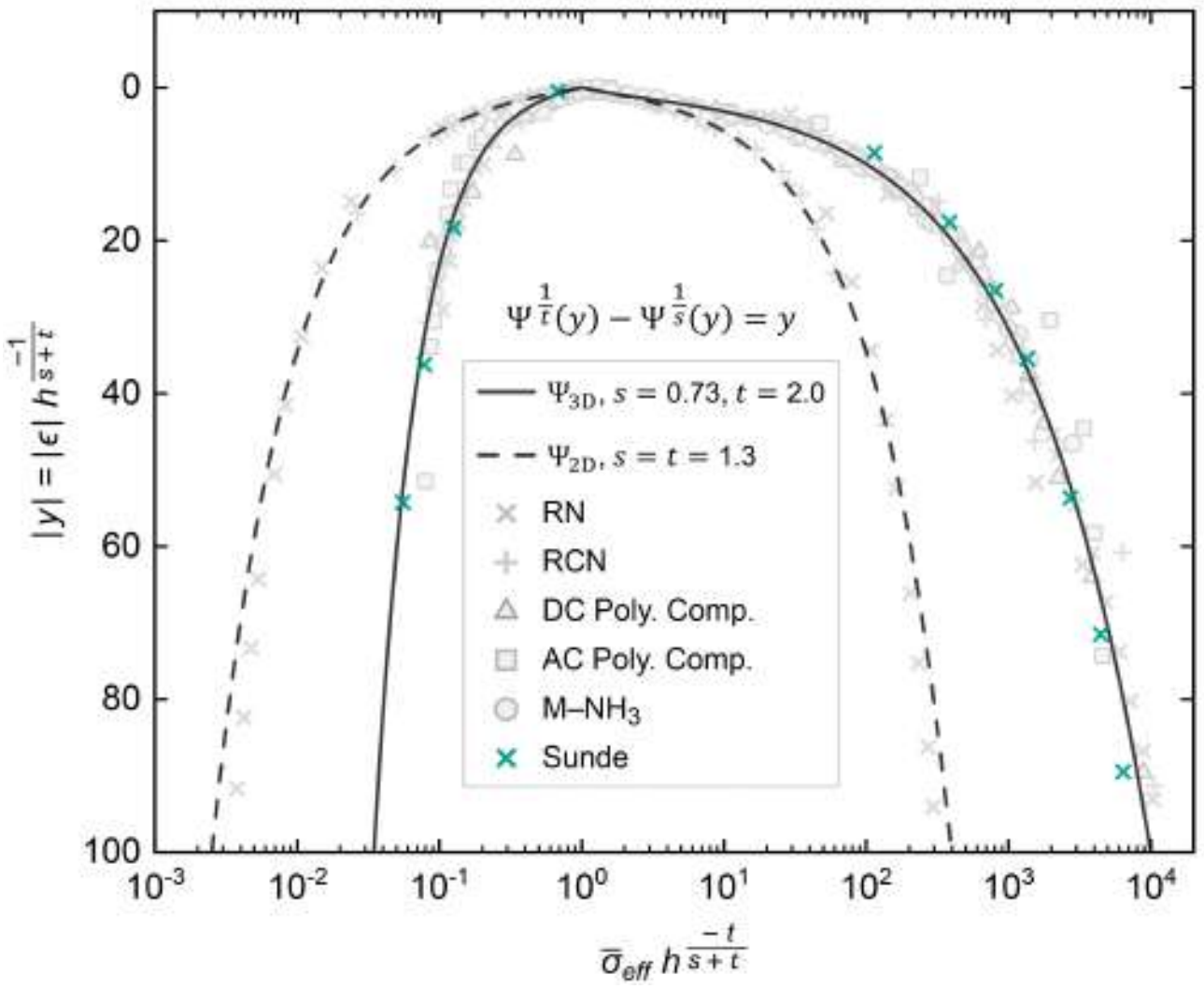

**Data Collapse:** Collapse of the rescaled data onto the proposed scaling function, $\Psi(y)$

**Source Article [107]:** Reproduced with permission from J. Electrochem. Soc. **142**, L50–L52 (1995). Copyright 1995 IOP Publishing Ltd.

**System:** 3D simple-cubic lattice (isotropically correlated resistor network)
**Poor Conducting Elements:** Resistors with conductivity $\sigma = 10^{-5}$ (S/cm)
**Good Conducting Elements:** Resistors with conductivity $\sigma = 1$ (S/cm)

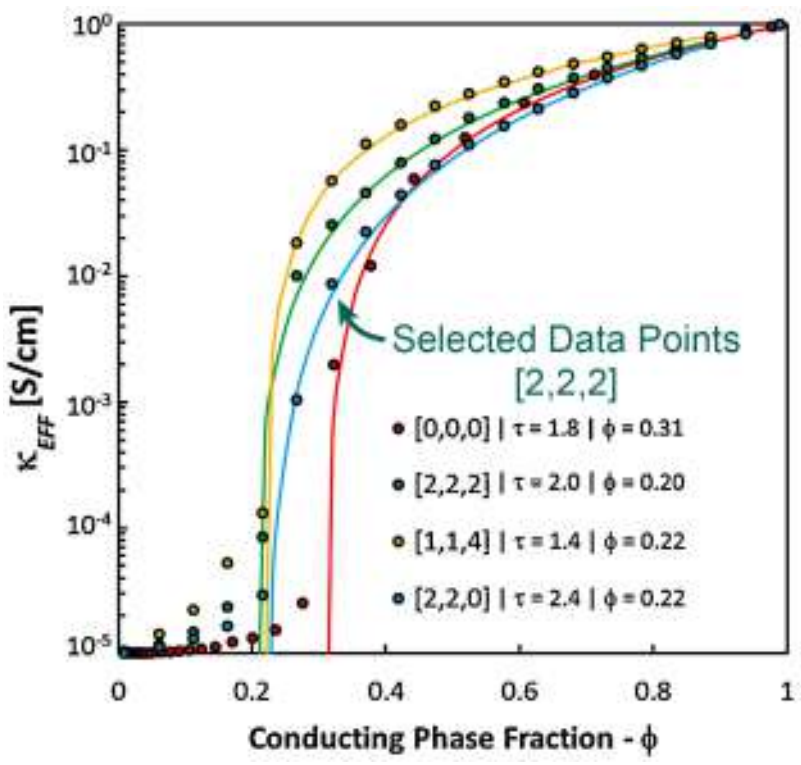

**Selected Figure:** Fig. 9 of the source article (Ref. [108])

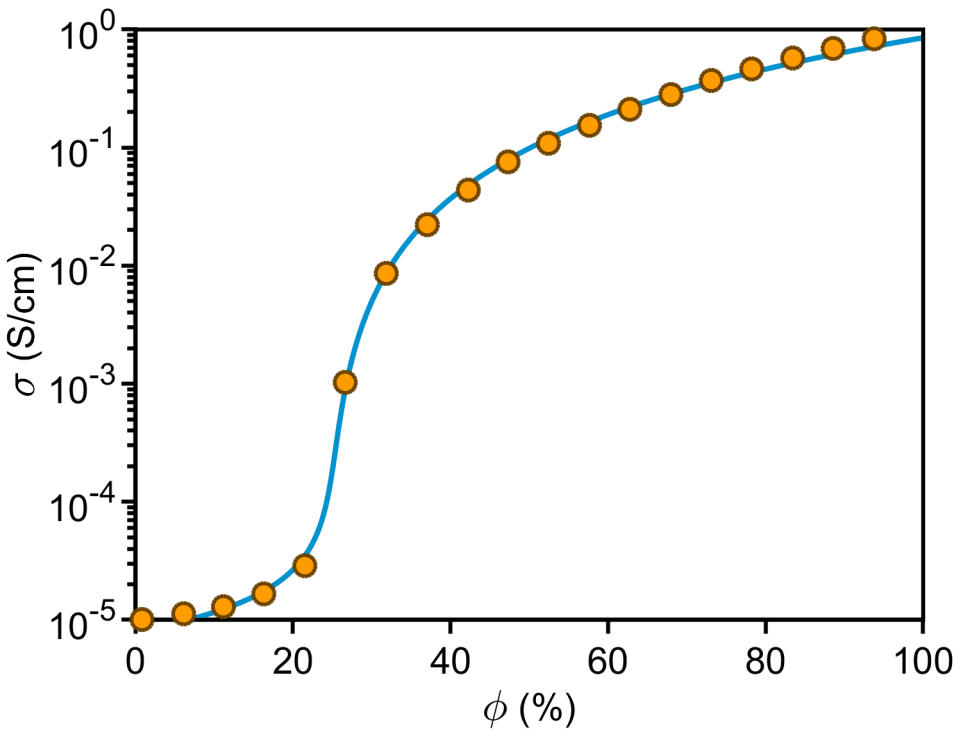

| | |
|---|---|
| $\phi_c$ (%) | 24.2 |
| $s$ | 0.73 |
| $t$ | 2.0 |
| $\sigma_m$ (S/cm) | $7.9 \times 10^{-6}$ |
| $\sigma_{f^*}$ (S/cm) | $8.7 \times 10^{-2}$ |
| $\sigma_f$ (S/cm) | $8.5 \times 10^{-1}$ |
| $h = \sigma_m/\sigma_{f^*}$ | $9.1 \times 10^{-5}$ |
| $R^2$ | 1.000000 |

**Fit Results:** Fitting data with the TAGP equation and the best-fit parameters

**Extracted Data Points**

| # | $\phi$ (%) | $\sigma$ (S/cm) |
|---|---|---|
| 1 | 0.9 | 1.01E-05 |
| 2 | 6.1 | 1.12E-05 |
| 3 | 11.2 | 1.29E-05 |
| 4 | 16.3 | 1.65E-05 |
| 5 | 21.6 | 2.89E-05 |
| 6 | 26.6 | 1.03E-03 |
| 7 | 31.9 | 8.59E-03 |
| 8 | 37.0 | 2.22E-02 |
| 9 | 42.2 | 4.38E-02 |
| 10 | 47.4 | 7.56E-02 |
| 11 | 52.4 | 1.09E-01 |
| 12 | 57.7 | 1.54E-01 |
| 13 | 62.8 | 2.10E-01 |
| 14 | 68.0 | 2.83E-01 |
| 15 | 73.1 | 3.72E-01 |
| 16 | 78.2 | 4.61E-01 |
| 17 | 83.4 | 5.76E-01 |
| 18 | 88.6 | 6.92E-01 |
| 19 | 93.8 | 8.33E-01 |
| 20 | 98.9 | 9.86E-01 |

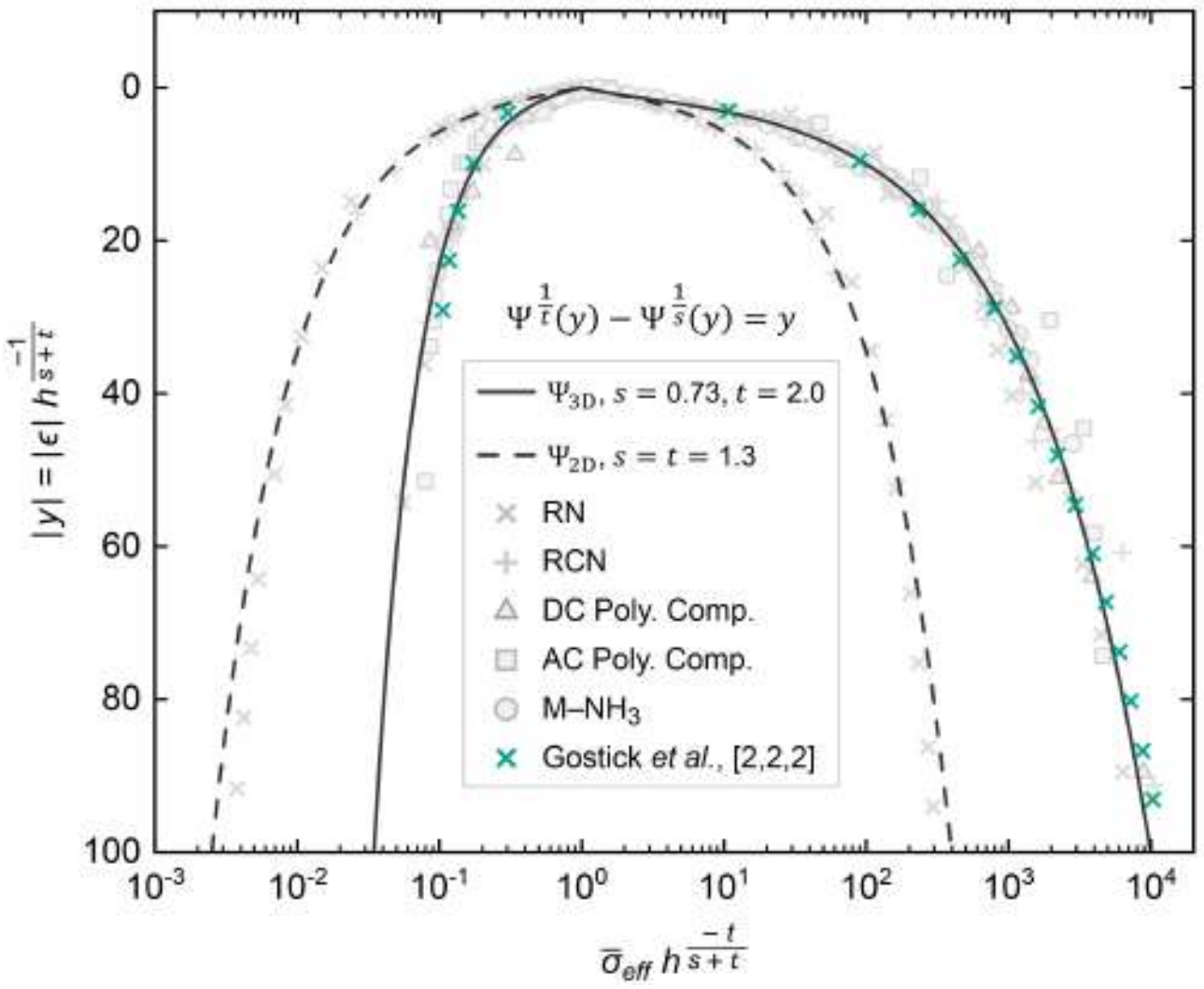

**Data Collapse:** Collapse of the rescaled data onto the proposed scaling function, $\Psi(y)$

**Source Article [108]:** Reproduced with permission from Electrochim. Acta **179**, 137–145 (2015). Copyright 2015 Elsevier.

**System:** 2D square lattice (random resistor–capacitor network)
**Poor Conducting Elements:** Capacitors with complex admittance $\sigma = j\mathrm{C}\omega$ (a.u.), $\mathrm{C} = 1$ (a.u.)
**Good Conducting Elements:** Resistors with conductivity $\sigma = \mathrm{R}^{-1}$ (a.u.), $\mathrm{R} = 1$ (a.u.)

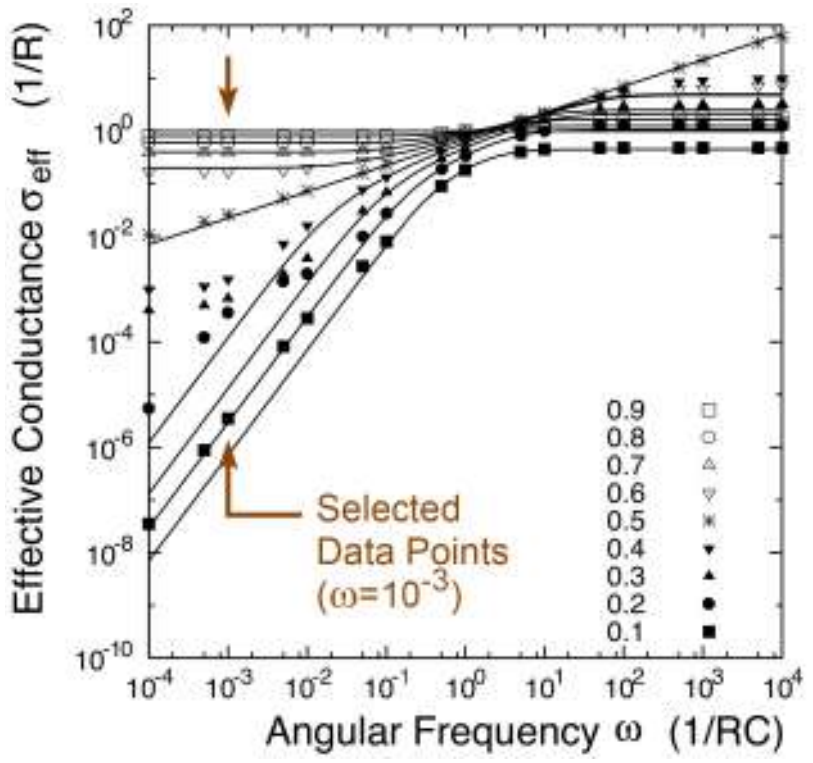

**Selected Figure:** Fig. 4(a) of the source article (Ref. [109])

**Extracted Data Points**

| # | $\phi$ (%) | $\sigma$ (a.u.) |
|---|---|---|
| 1 | 10.0 | 3.51E-06 |
| 2 | 20.0 | 3.56E-04 |
| 3 | 30.0 | 6.54E-04 |
| 4 | 40.0 | 1.52E-03 |
| 5 | 50.0 | 2.58E-02 |
| 6 | 60.0 | 1.71E-01 |
| 7 | 70.0 | 3.80E-01 |
| 8 | 80.0 | 5.94E-01 |
| 9 | 90.0 | 7.89E-01 |

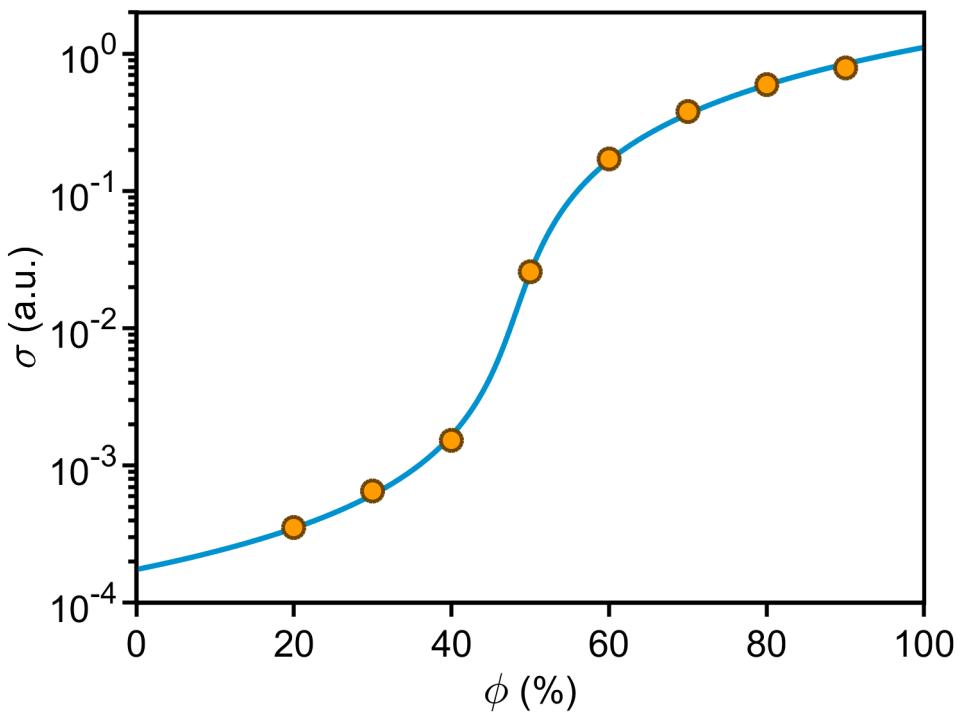

| | |
|---|---|
| $\phi_c$ (%) | 48.2 |
| $s$ | 1.30 |
| $t$ | 1.30 |
| $\sigma_m$ (a.u.) | $1.8 \times 10^{-4}$ |
| $\sigma_{f^*}$ (a.u.) | 1.0 |
| $\sigma_f$ (a.u.) | 1.1 |
| $h = \sigma_m/\sigma_{f^*}$ | $1.7 \times 10^{-4}$ |
| $R^2$ | 1.000000 |

**Fit Results:** Fitting data with the TAGP equation and the best-fit parameters

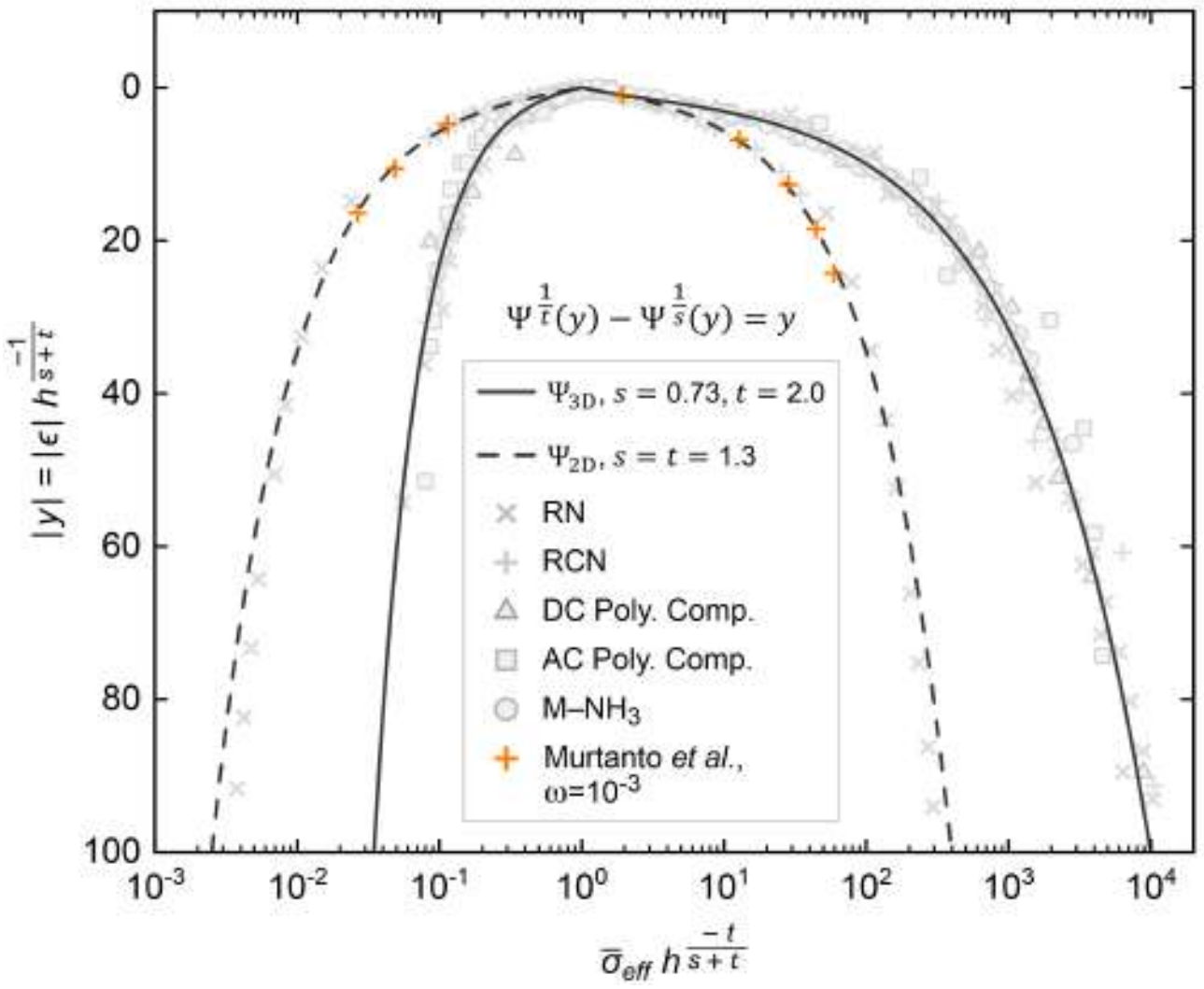

**Data Collapse:** Collapse of the rescaled data onto the proposed scaling function, $\Psi(y)$

**Source Article [109]:** Reproduced with permission from Phys. Rev. B **74**, 115206 (2006).

**System:** 2D square lattice (random resistor–capacitor network)

**Poor Conducting Elements:** Capacitors with complex admittance $\sigma = jC\omega$ (a.u.), $C = 1$ (a.u.)

**Good Conducting Elements:** Resistors with conductivity $\sigma = R^{-1}$ (a.u.), $R = 1$ (a.u.)

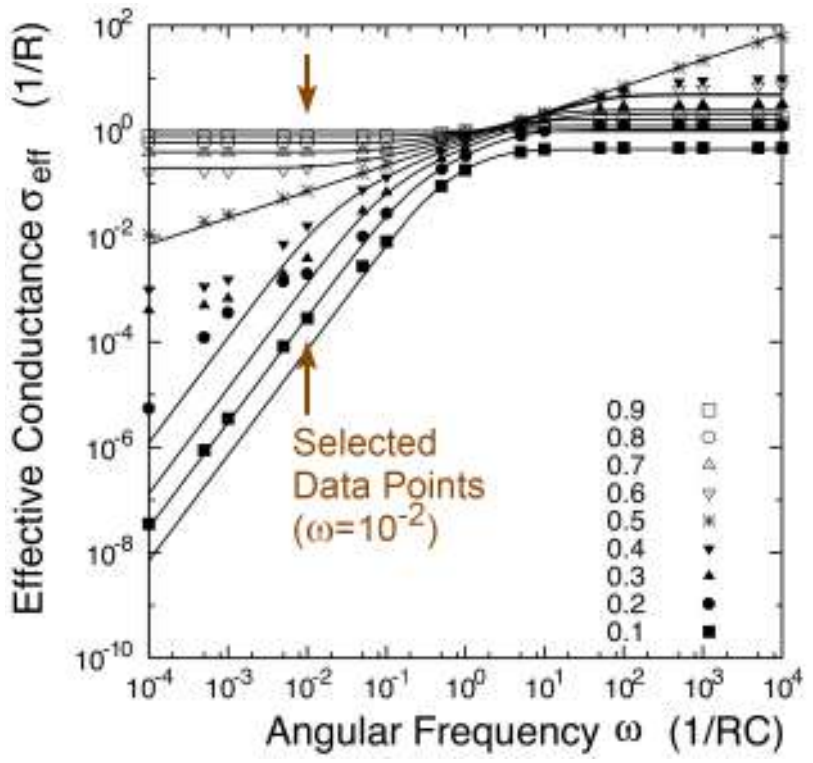

**Selected Figure:** Fig. 4(a) of the source article (Ref. [109])

**Extracted Data Points**

| # | $\phi$ (%) | $\sigma$ (a.u.) |
|---|---|---|
| 1 | 10.0 | 2.79E-04 |
| 2 | 20.0 | 1.95E-03 |
| 3 | 30.0 | 3.83E-03 |
| 4 | 40.0 | 1.60E-02 |
| 5 | 50.0 | 7.38E-02 |
| 6 | 60.0 | 1.98E-01 |
| 7 | 70.0 | 3.85E-01 |
| 8 | 80.0 | 5.94E-01 |
| 9 | 90.0 | 8.00E-01 |

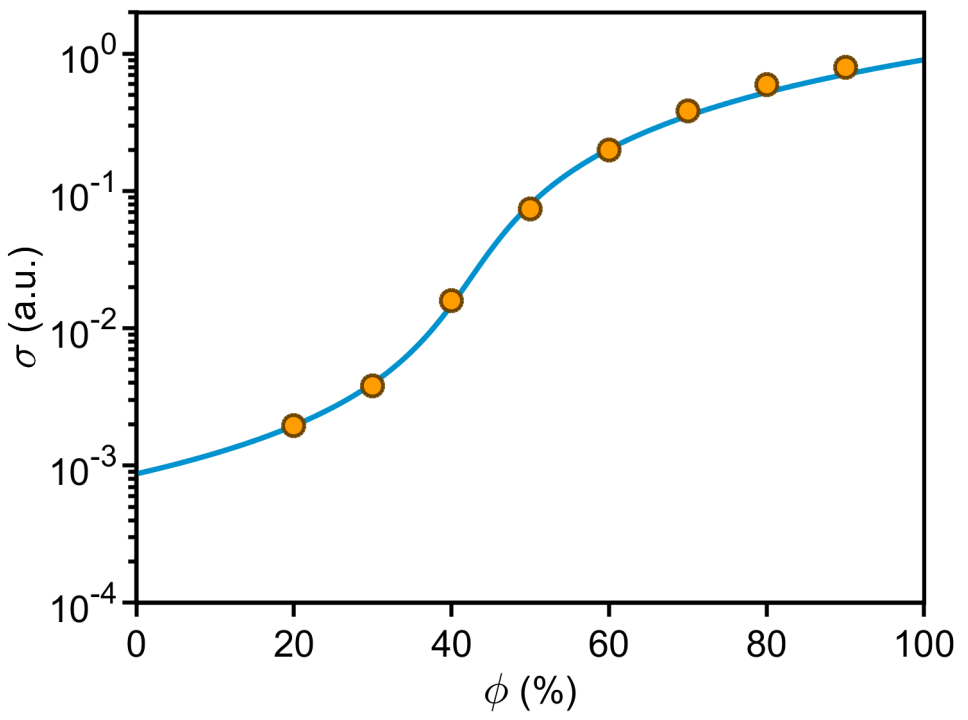

| | |
|---|---|
| $\phi_c$ (%) | 42.4 |
| $s$ | 1.30 |
| $t$ | 1.30 |
| $\sigma_m$ (a.u.) | $8.7 \times 10^{-4}$ |
| $\sigma_{f^*}$ (a.u.) | $6.1 \times 10^{-1}$ |
| $\sigma_f$ (a.u.) | $9.1 \times 10^{-1}$ |
| $h = \sigma_m/\sigma_{f^*}$ | $1.4 \times 10^{-3}$ |
| $R^2$ | 0.999983 |

**Fit Results:** Fitting data with the TAGP equation and the best-fit parameters

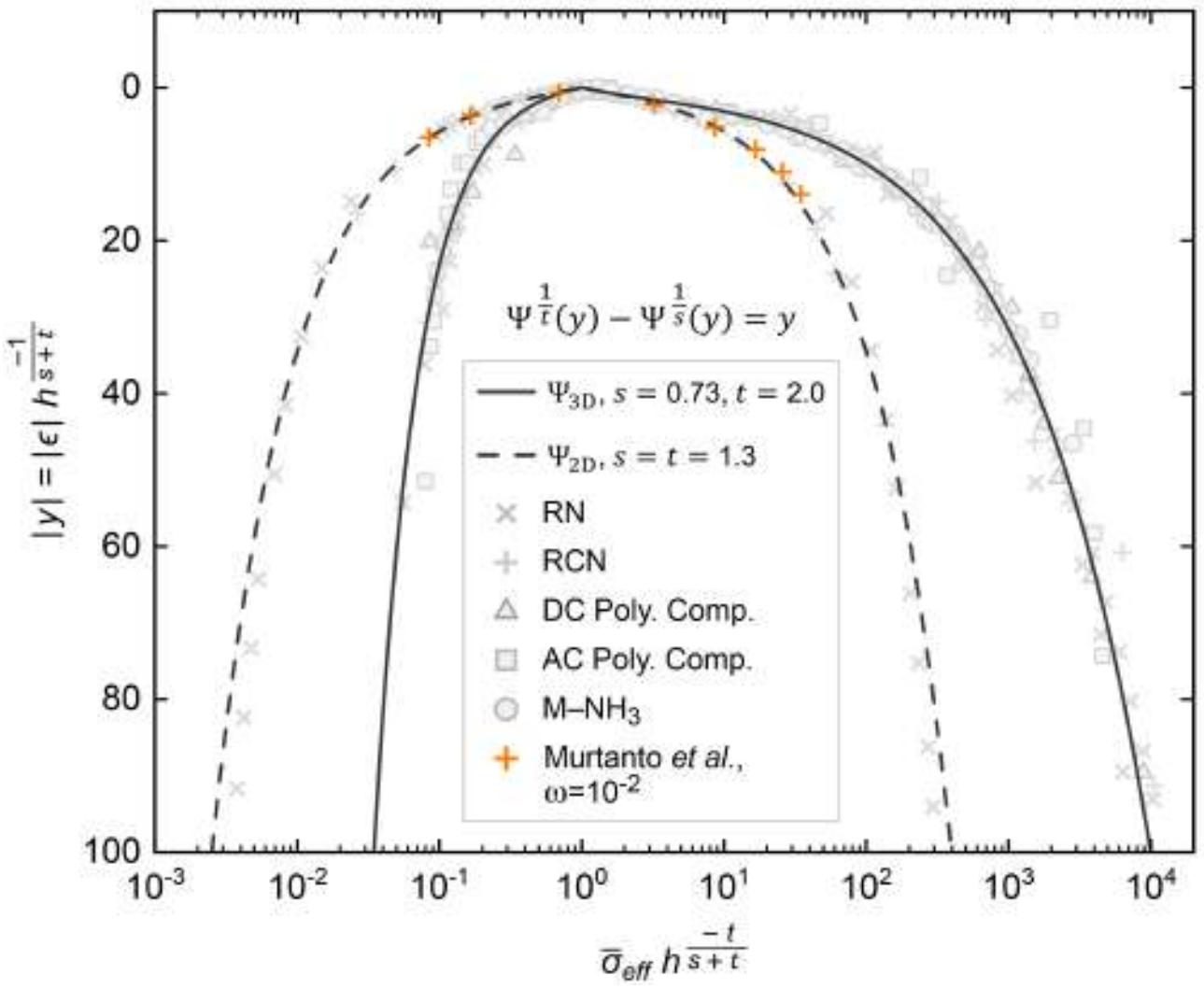

**Data Collapse:** Collapse of the rescaled data onto the proposed scaling function, $\Psi(y)$

**Source Article [109]:** Reproduced with permission from Phys. Rev. B **74**, 115206 (2006). Copyright 2006 American Physical Society.

**System:** 3D simple-cubic lattice (random resistor–capacitor network)
**Poor Conducting Elements:** Capacitors with complex admittance $\sigma = jC\omega$ (a.u.), C = 1 (nF)
**Good Conducting Elements:** Resistors with conductivity $\sigma = R^{-1}$ (a.u.), R = 1 (kΩ)

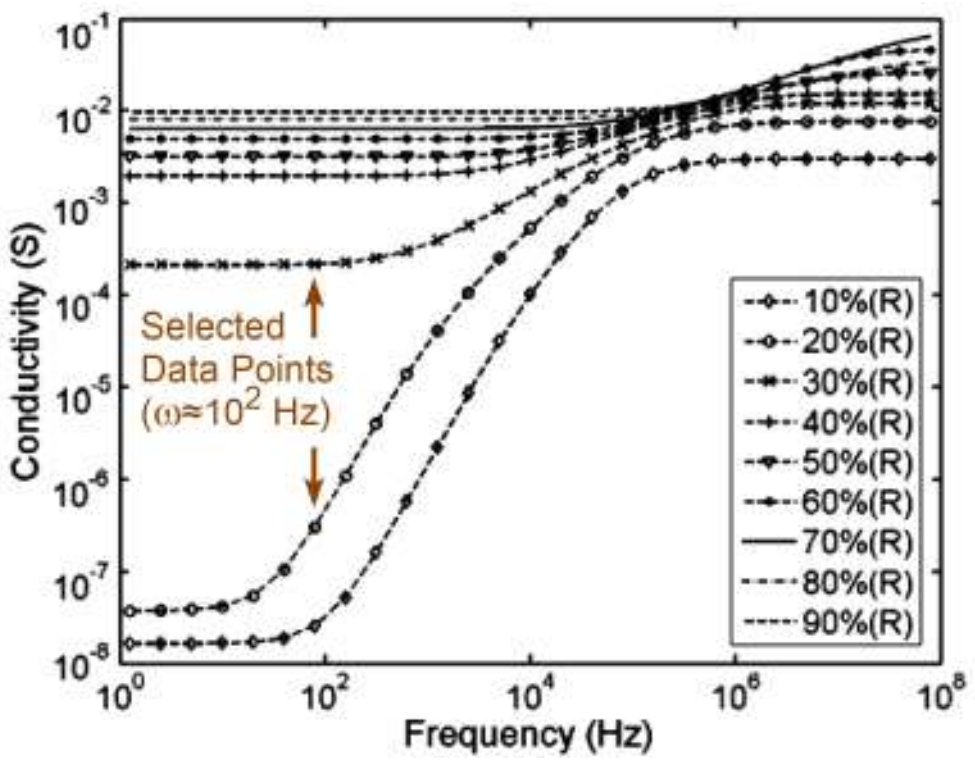

**Selected Figure:** Fig. 5 of the source article (Ref. [110])

**Extracted Data Points**

| # | $\phi$ (%) | $\sigma$ (S) |
|---|---|---|
| 1 | 10.0 | 2.61E-08 |
| 2 | 20.0 | 3.04E-07 |
| 3 | 30.0 | 2.22E-04 |
| 4 | 40.0 | 2.04E-03 |
| 5 | 50.0 | 3.37E-03 |
| 6 | 60.0 | 5.17E-03 |
| 7 | 70.0 | 6.74E-03 |
| 8 | 80.0 | 8.37E-03 |
| 9 | 90.0 | 1.01E-02 |

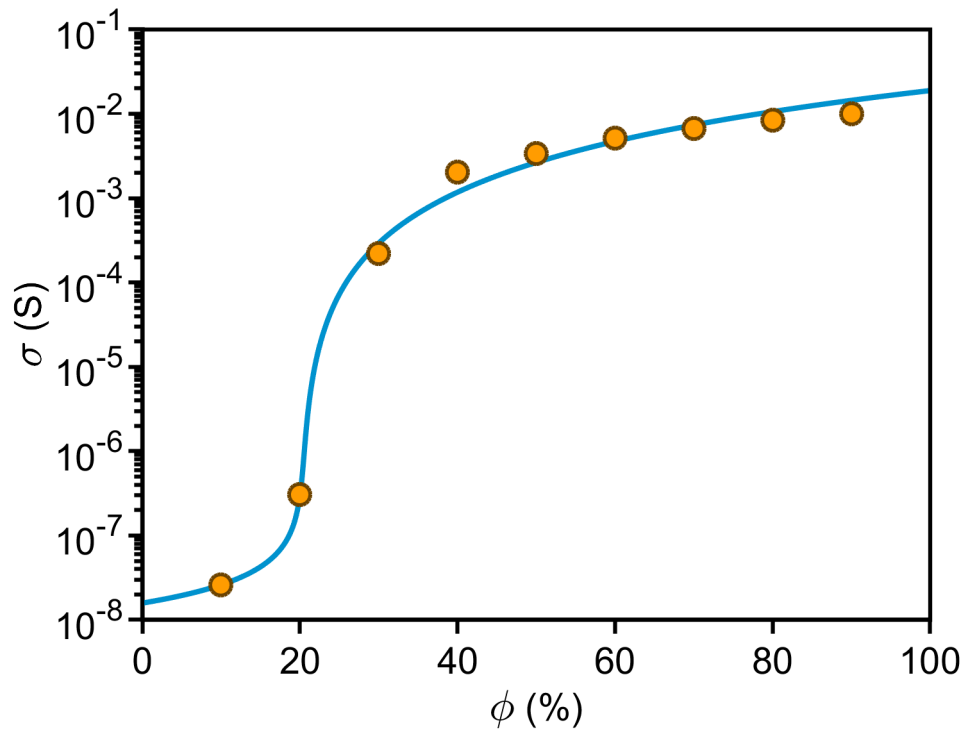

| | |
|---|---|
| $\phi_c$ (%) | 20.1 |
| $s$ | 0.73 |
| $t$ | 2.0 |
| $\sigma_m$ (S) | $1.6 \times 10^{-8}$ |
| $\sigma_{f^*}$ (S) | $1.2 \times 10^{-3}$ |
| $\sigma_f$ (S) | $1.9 \times 10^{-2}$ |
| $h = \sigma_m/\sigma_{f^*}$ | $1.3 \times 10^{-5}$ |
| $R^2$ | 1.000000 |

**Fit Results:** Fitting data with the TAGP equation and the best-fit parameters

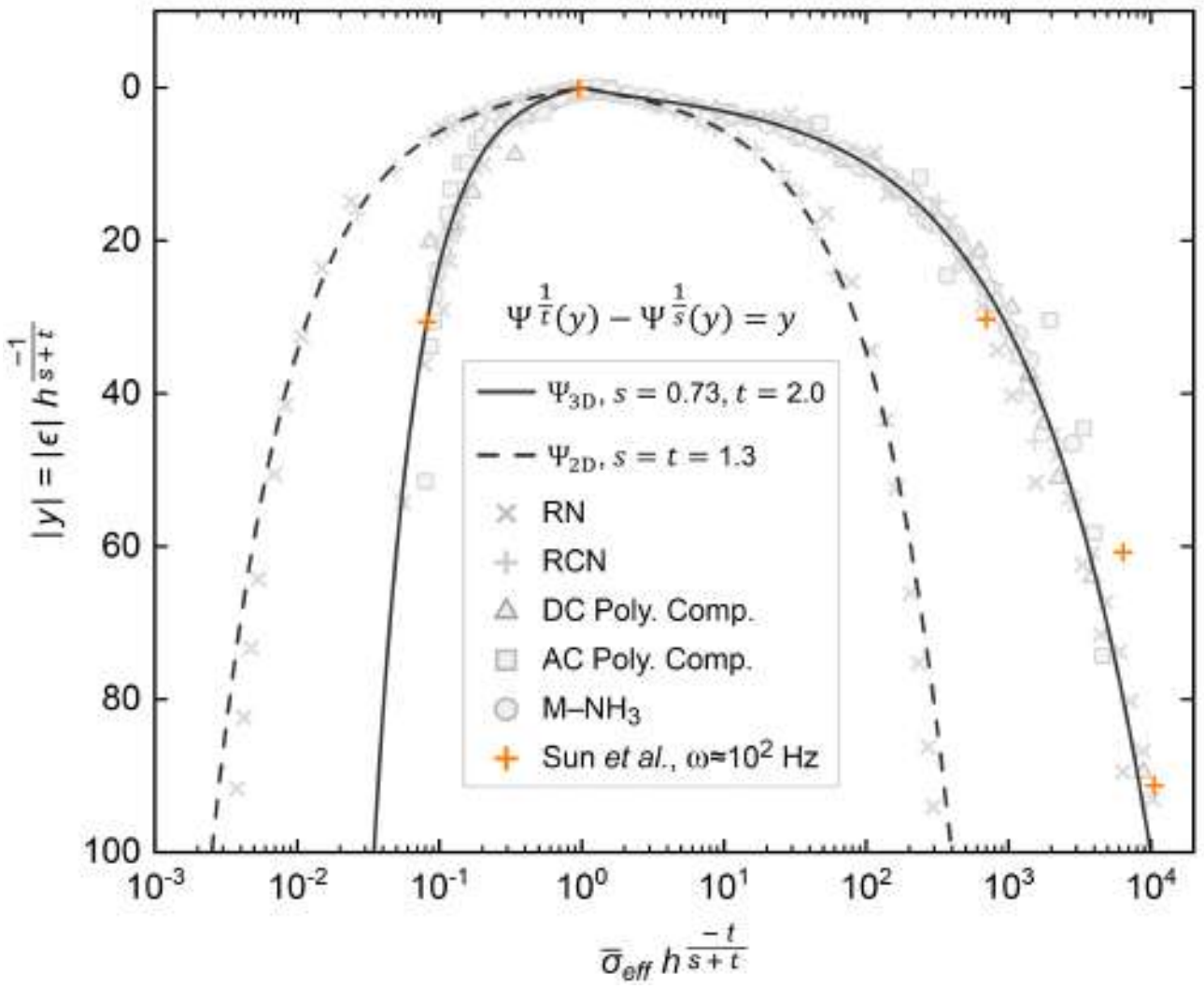

**Data Collapse:** Collapse of the rescaled data onto the proposed scaling function, $\Psi(y)$

**Source Article [110]:** Reproduced with permission from *Proceedings of ISAPE2012*, Xi'an, China, 2012 (IEEE, Piscataway, NJ), pp. 1214–1218. Copyright 2012 IEEE.

**System:** 3D simple-cubic lattice (random resistor–capacitor network)
**Poor Conducting Elements:** Capacitors with complex admittance $\sigma = jC\omega$ (a.u.), C = 1 (nF)
**Good Conducting Elements:** Resistors with conductivity $\sigma = R^{-1}$ (a.u.), R = 1 (kΩ)

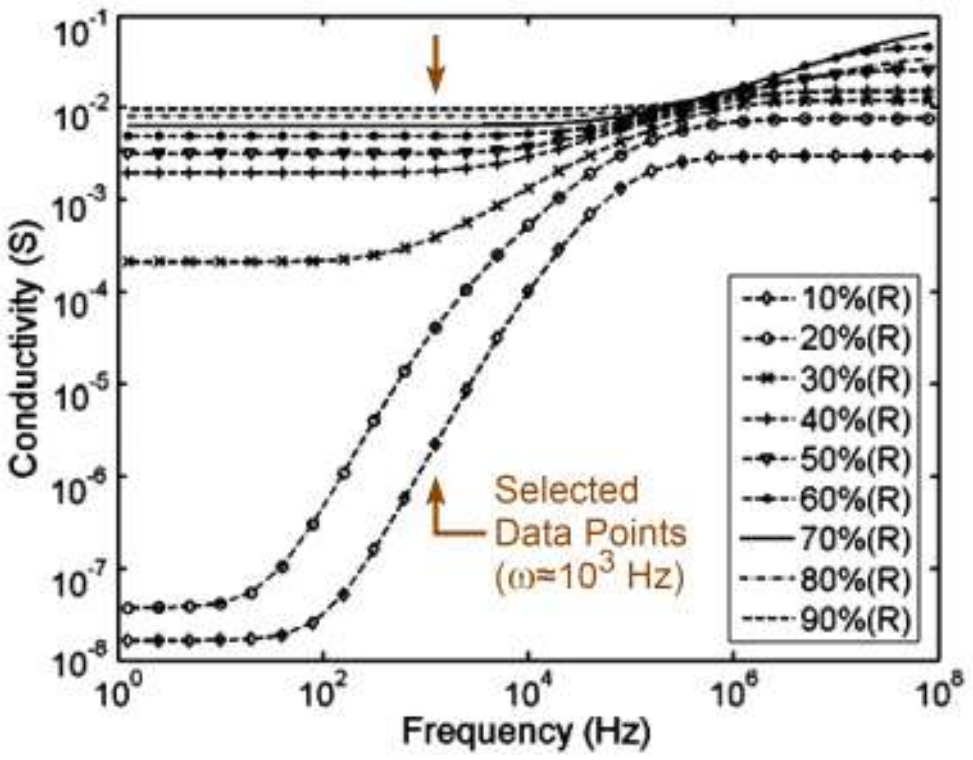

**Selected Figure:** Fig. 5 of the source article (Ref. [110])

**Extracted Data Points**

| # | $\phi$ (%) | $\sigma$ (S) |
|---|---|---|
| 1 | 10.0 | 2.25E-06 |
| 2 | 20.0 | 4.14E-05 |
| 3 | 30.0 | 4.09E-04 |
| 4 | 40.0 | 2.14E-03 |
| 5 | 50.0 | 3.38E-03 |
| 6 | 60.0 | 5.17E-03 |
| 7 | 70.0 | 6.74E-03 |
| 8 | 80.0 | 8.37E-03 |
| 9 | 90.0 | 1.01E-02 |

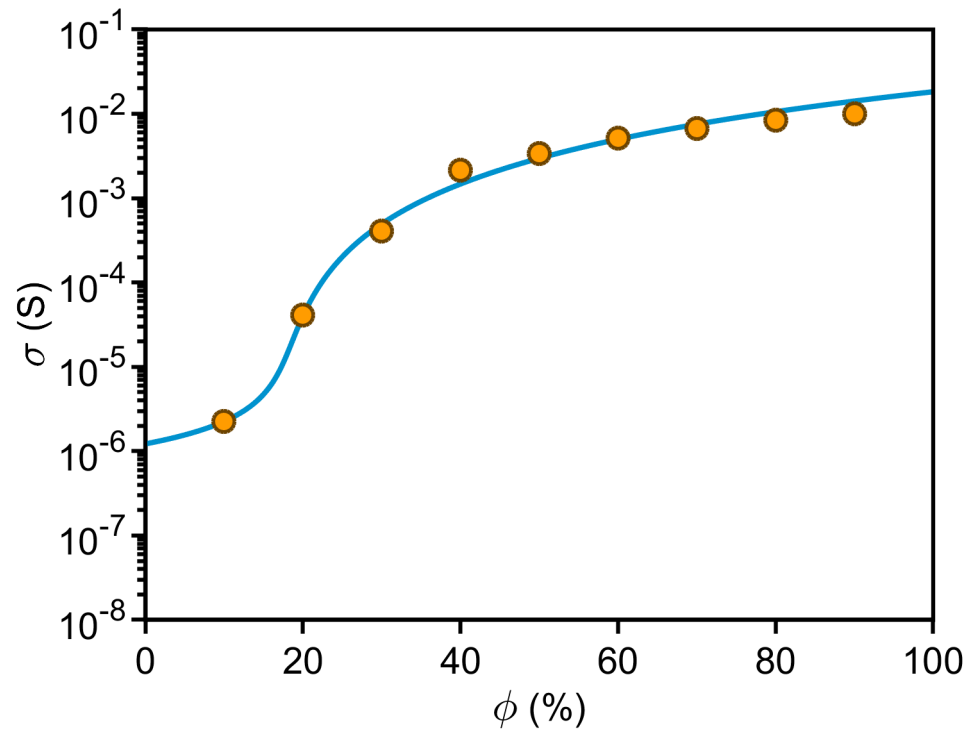

| | |
|---|---|
| $\phi_c$ (%) | 16.2 |
| $s$ | 0.73 |
| $t$ | 2.0 |
| $\sigma_m$ (S) | $1.2 \times 10^{-6}$ |
| $\sigma_{f^*}$ (S) | $6.8 \times 10^{-4}$ |
| $\sigma_f$ (S) | $1.8 \times 10^{-2}$ |
| $h = \sigma_m/\sigma_{f^*}$ | $1.8 \times 10^{-3}$ |
| $R^2$ | 1.000000 |

**Fit Results:** Fitting data with the TAGP equation and the best-fit parameters

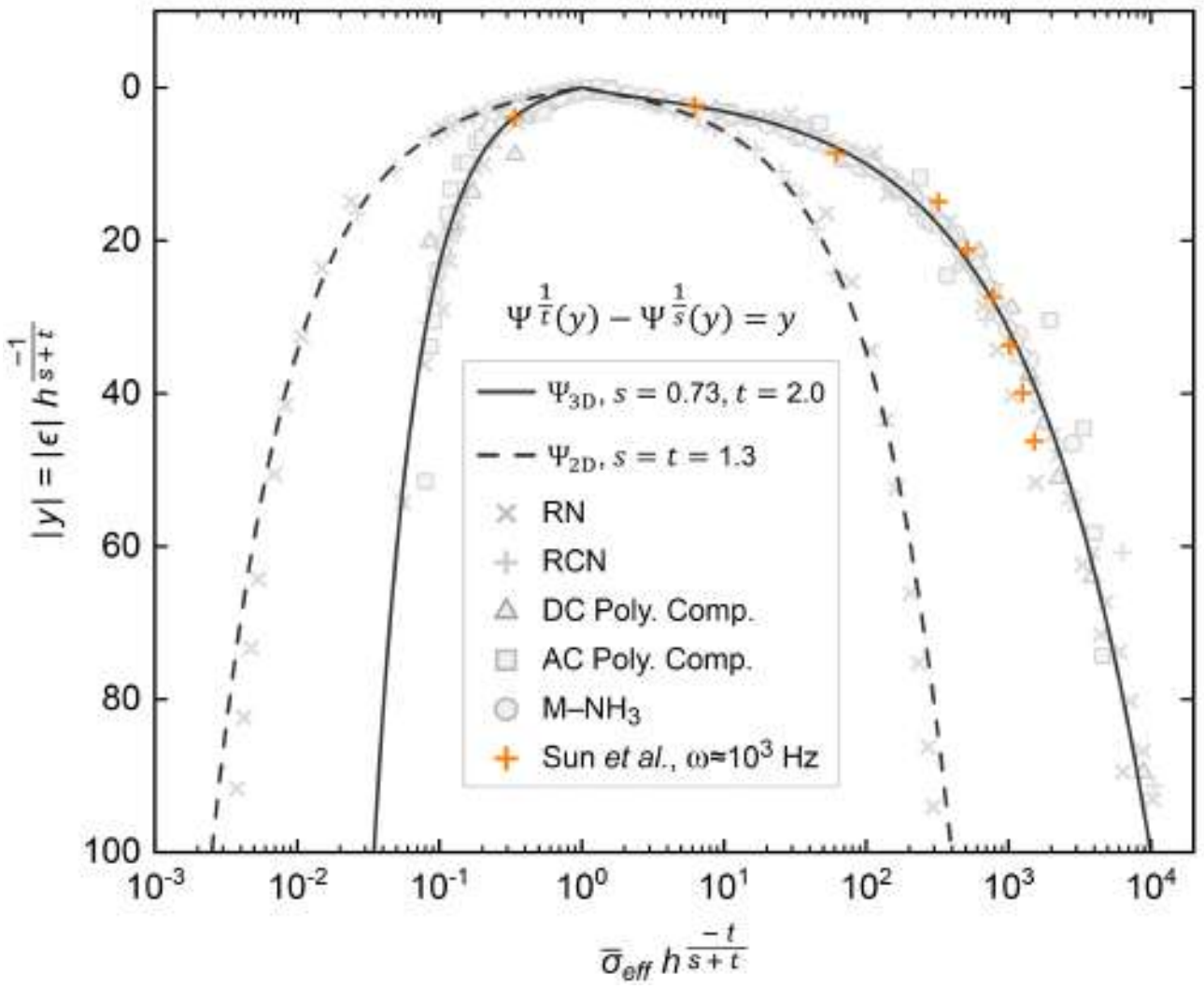

**Data Collapse:** Collapse of the rescaled data onto the proposed scaling function, $\Psi(y)$

**Source Article [110]:** Reproduced with permission from *Proceedings of ISAPE2012*, Xi'an, China, 2012 (IEEE, Piscataway, NJ), pp. 1214–1218.

**System:** Polymer composite
**Poor Conducting Phase (Matrix):** Conducting polymer poly(3-hexylthiophene) (P3HT)
**Good Conducting Phase (Filler):** Single-walled carbon nanotubes (SWCNT)

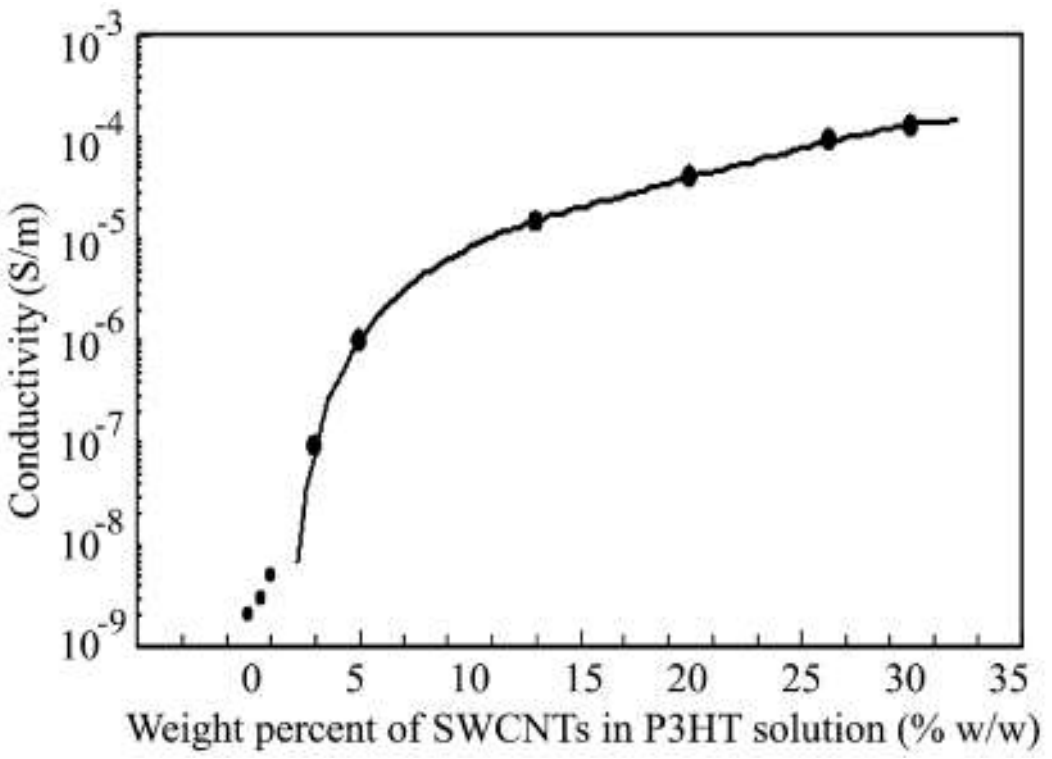

**Selected Figure:** Fig. 5 of the source article (Ref. [111])

**Extracted Data Points**

| # | $\phi$ (wt%) | $\phi$ (vol%) | $\sigma$ (S/m) |
|---|---|---|---|
| 1 | 0.0 | 0.0 | 2.09E-09 |
| 2 | 0.6 | 0.3 | 3.01E-09 |
| 3 | 1.0 | 0.5 | 5.05E-09 |
| 4 | 2.9 | 1.5 | 9.32E-08 |
| 5 | 4.9 | 2.6 | 1.00E-06 |
| 6 | 12.9 | 7.2 | 1.47E-05 |
| 7 | 19.9 | 11.5 | 4.10E-05 |
| 8 | 26.2 | 15.7 | 9.37E-05 |
| 9 | 30.0 | 18.3 | 1.28E-04 |

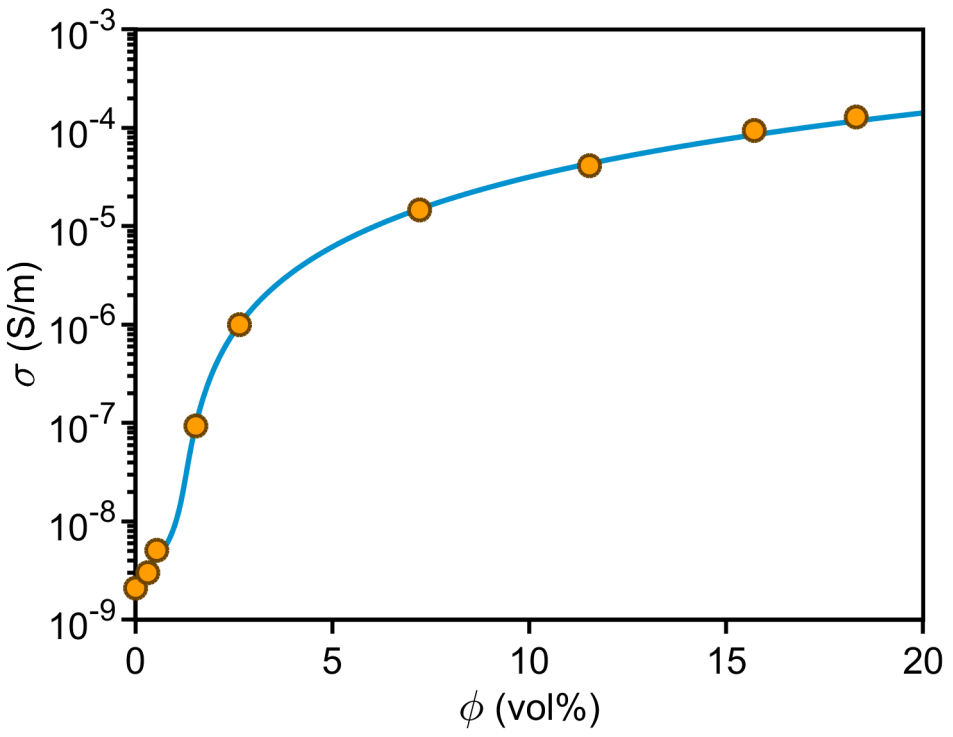

| | |
|---|---|
| $\phi_c$ (vol%) | 1.1 |
| $s$ | 0.73 |
| $t$ | 2.0 |
| $\sigma_m$ (S/m) | $2.7 \times 10^{-9}$ |
| $\sigma_{f^*}$ (S/m) | $4.4 \times 10^{-7}$ |
| $\sigma_f$ (S/m) | $3.9 \times 10^{-3}$ |
| $h = \sigma_m/\sigma_{f^*}$ | $6.3 \times 10^{-3}$ |
| $R^2$ | 1.000000 |

**Fit Results:** Fitting data with the TAGP equation and the best-fit parameters

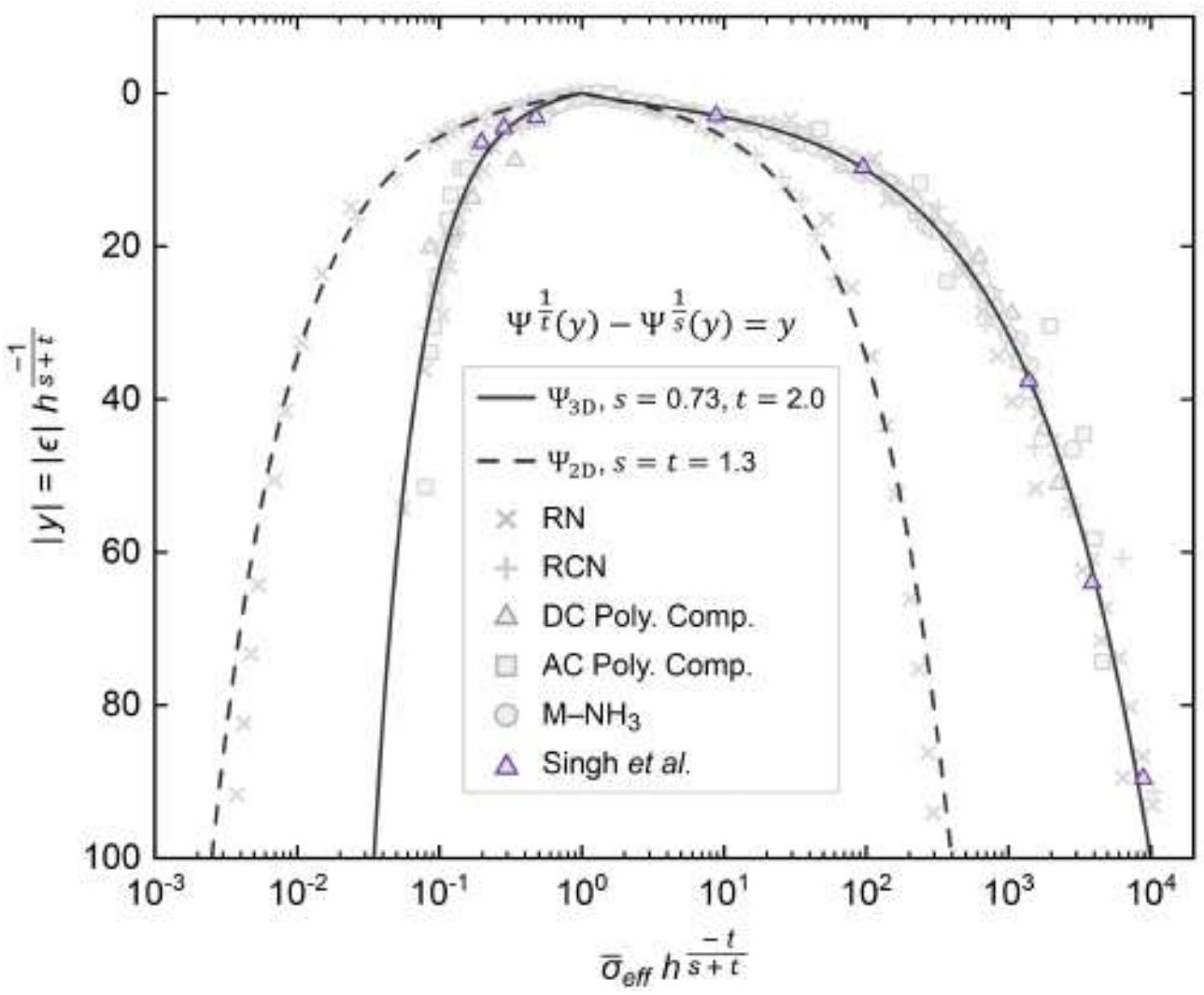

**Data Collapse:** Collapse of the rescaled data onto the proposed scaling function, $\Psi(y)$

**Source Article [111]:** Reproduced with permission from Carbon **46**, 1141–1144 (2008). Copyright 2008 Elsevier.

**System:** Polymer composite
**Poor Conducting Phase (Matrix):** Conducting polymer poly(3-octylthiophene-2,5-diyl) (P3OT)
**Good Conducting Phase (Filler):** Single-walled carbon nanotubes (SWCNT)

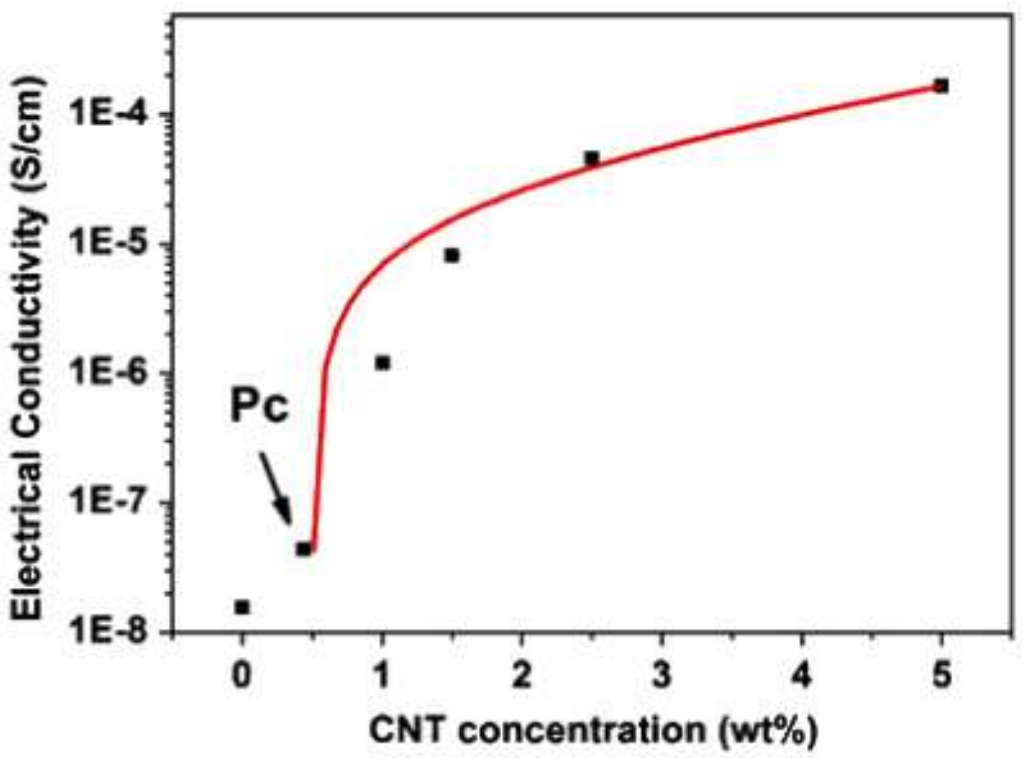

Selected Figure: Fig. 8 of the source article (Ref. [112])

**Extracted Data Points**

| # | $\phi$ (wt%) | $\phi$ (vol%) | $\sigma$ (S/cm) |
|---|---|---|---|
| 1 | 0.0 | 0.0 | 1.53E-08 |
| 2 | 0.4 | 0.2 | 4.38E-08 |
| 3 | 1.0 | 0.5 | 1.18E-06 |
| 4 | 1.5 | 0.8 | 8.11E-06 |
| 5 | 2.5 | 1.3 | 4.62E-05 |
| 6 | 5.0 | 2.7 | 1.66E-04 |

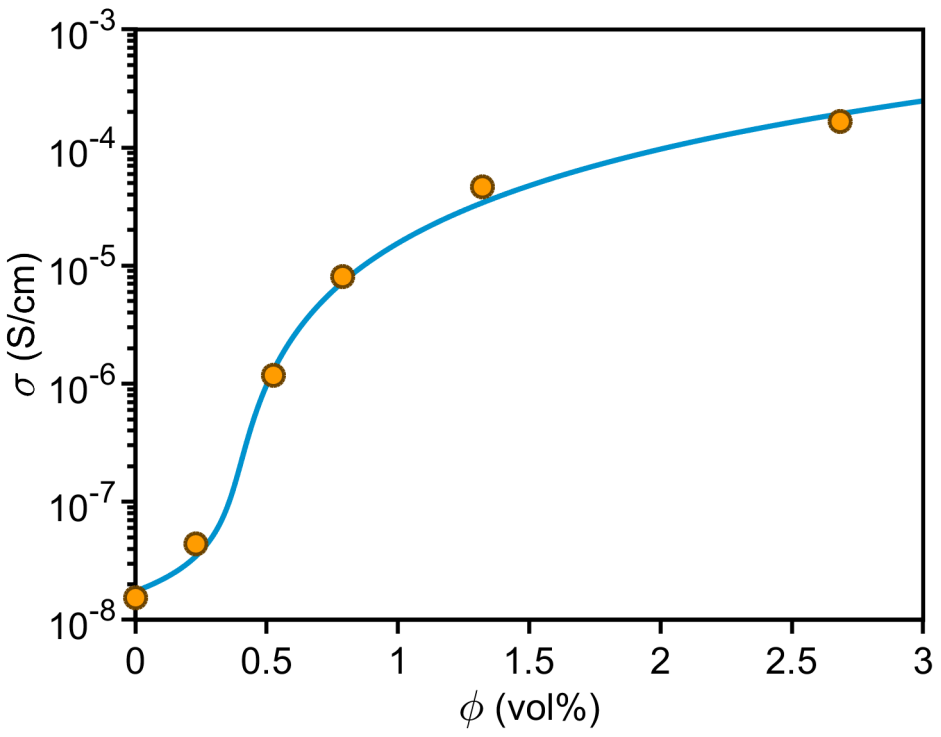

| | |
|---|---|
| $\phi_c$ (vol%) | 0.33 |
| $s$ | 0.73 |
| $t$ | 2.0 |
| $\sigma_m$ (S/cm) | $1.7 \times 10^{-8}$ |
| $\sigma_{f^*}$ (S/cm) | $3.9 \times 10^{-6}$ |
| $\sigma_f$ (S/cm) | $3.5 \times 10^{-1}$ |
| $h = \sigma_m/\sigma_{f^*}$ | $4.5 \times 10^{-3}$ |
| $R^2$ | 1.000000 |

**Fit Results:** Fitting data with the TAGP equation and the best-fit parameters

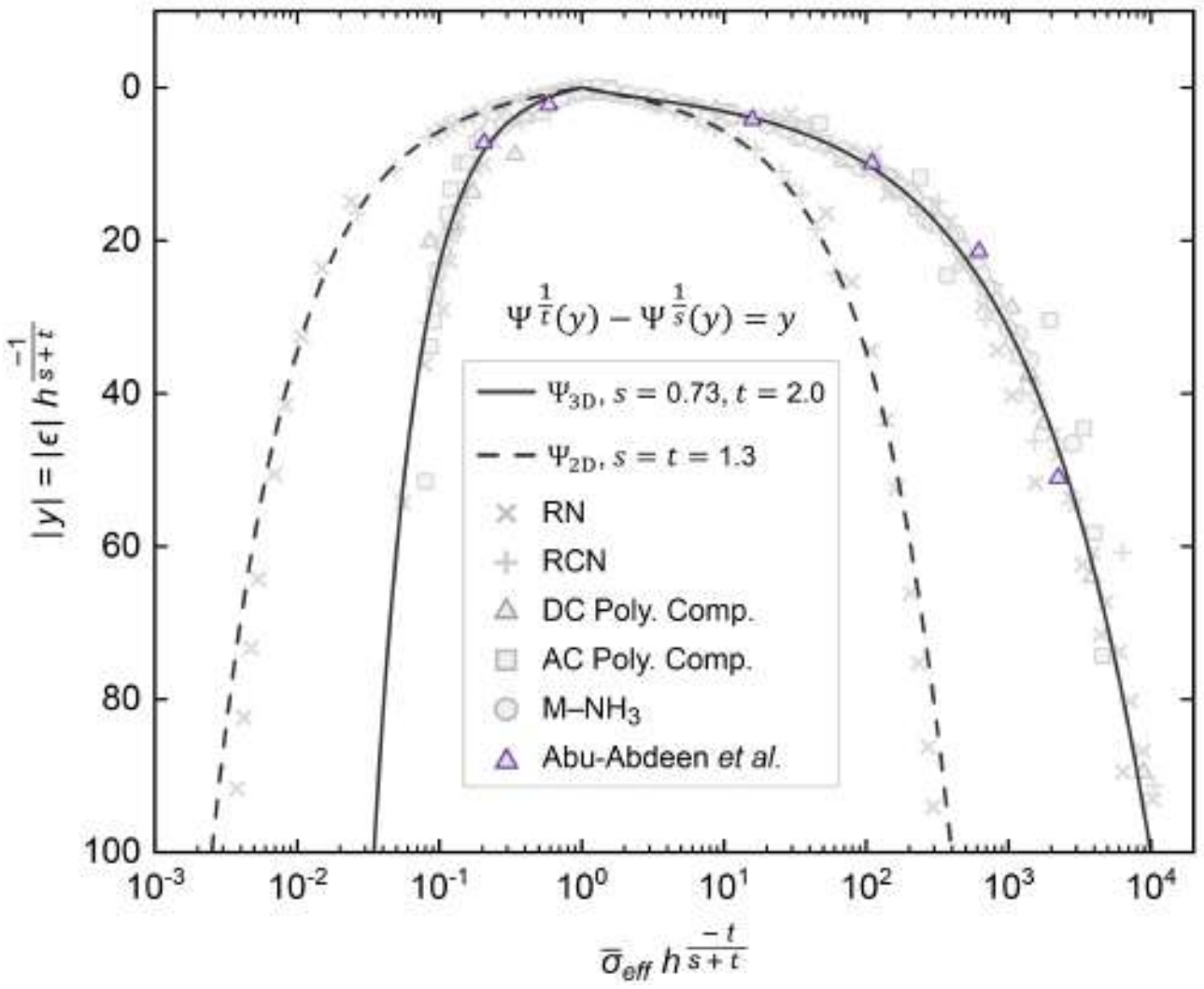

**Data Collapse:** Collapse of the rescaled data onto the proposed scaling function, $\Psi(y)$

**Source Article [112]:** Reproduced with permission from J. Polym. Res. **19**, 9839 (2012). Copyright 2012 Springer Nature.

**System:** Polymer composite
**Poor Conducting Phase (Matrix):** Conducting polymer poly(3-octylthiophene) (P3OT)
**Good Conducting Phase (Filler):** Single-walled carbon nanotubes (SWCNT)

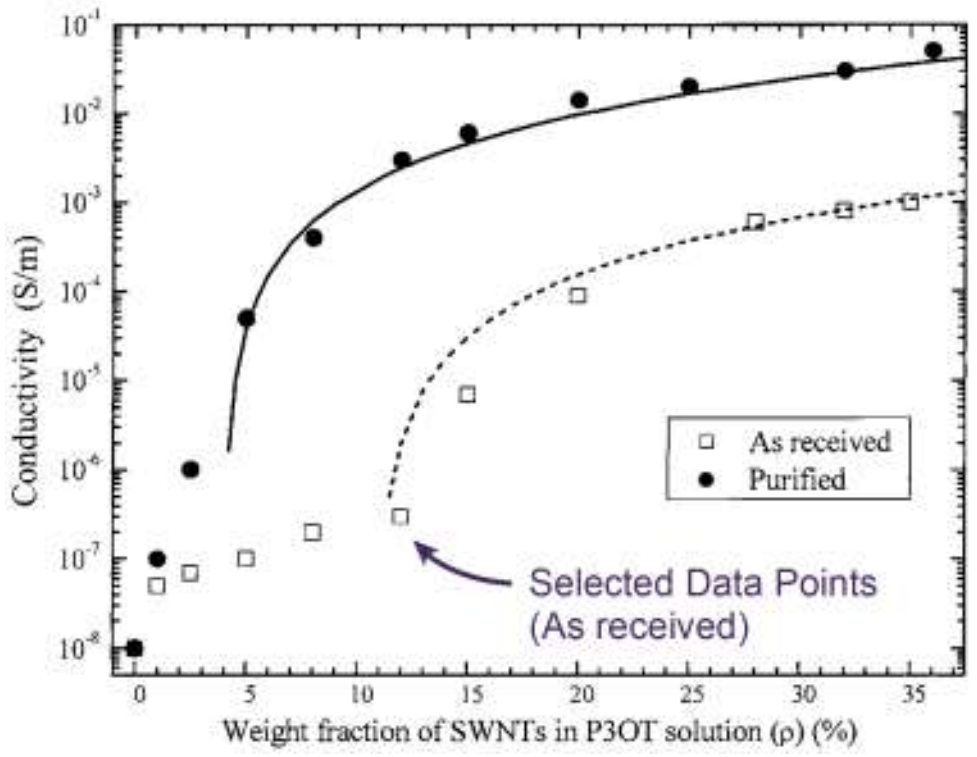

**Selected Figure:** Fig. 3 of the source article (Ref. [113])

**Extracted Data Points**

| # | $\phi$ (wt%) | $\phi$ (vol%) | $\sigma$ (S/m) |
|---|---|---|---|
| 1 | 1.0 | 0.5 | 5.03E-08 |
| 2 | 2.5 | 1.3 | 6.98E-08 |
| 3 | 5.0 | 2.7 | 1.00E-07 |
| 4 | 8.0 | 4.4 | 2.00E-07 |
| 5 | 12.0 | 6.7 | 3.00E-07 |
| 6 | 15.0 | 8.5 | 7.05E-06 |
| 7 | 20.0 | 11.6 | 9.07E-05 |
| 8 | 28.1 | 17.0 | 6.17E-04 |
| 9 | 32.1 | 19.8 | 8.20E-04 |
| 10 | 35.1 | 22.0 | 1.04E-03 |

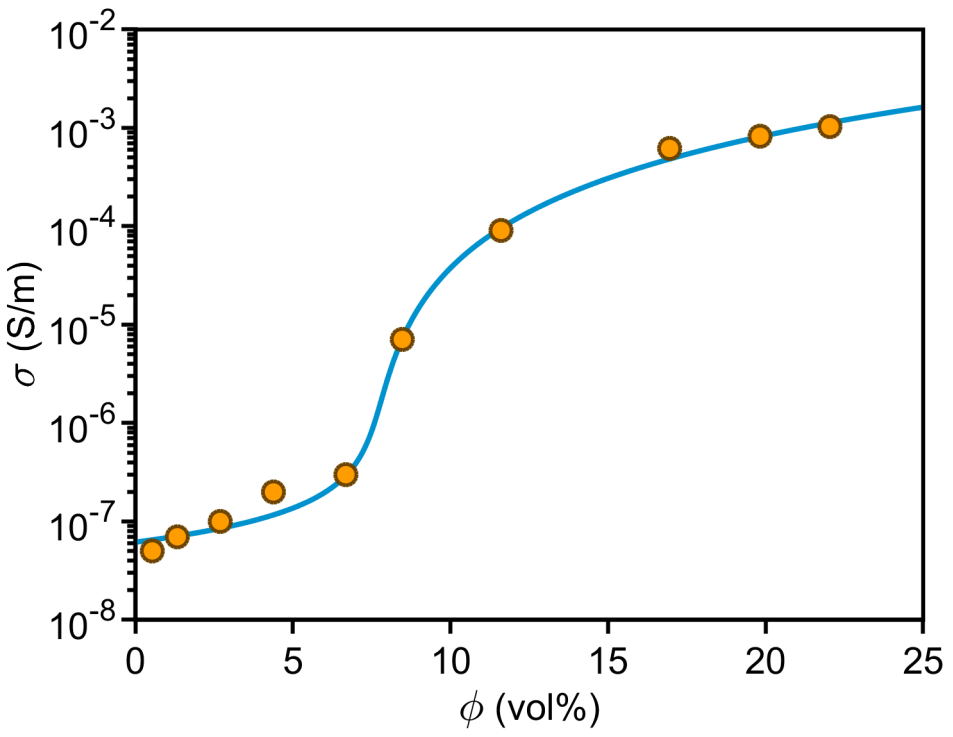

| | |
|---|---|
| $\phi_c$ (vol%) | 7.3 |
| $s$ | 0.73 |
| $t$ | 2.0 |
| $\sigma_m$ (S/m) | $6.2 \times 10^{-8}$ |
| $\sigma_{f^*}$ (S/m) | $2.8 \times 10^{-4}$ |
| $\sigma_f$ (S/m) | $4.5 \times 10^{-2}$ |
| $h = \sigma_m/\sigma_{f^*}$ | $2.2 \times 10^{-4}$ |
| $R^2$ | 1.000000 |

**Fit Results:** Fitting data with the TAGP equation and the best-fit parameters

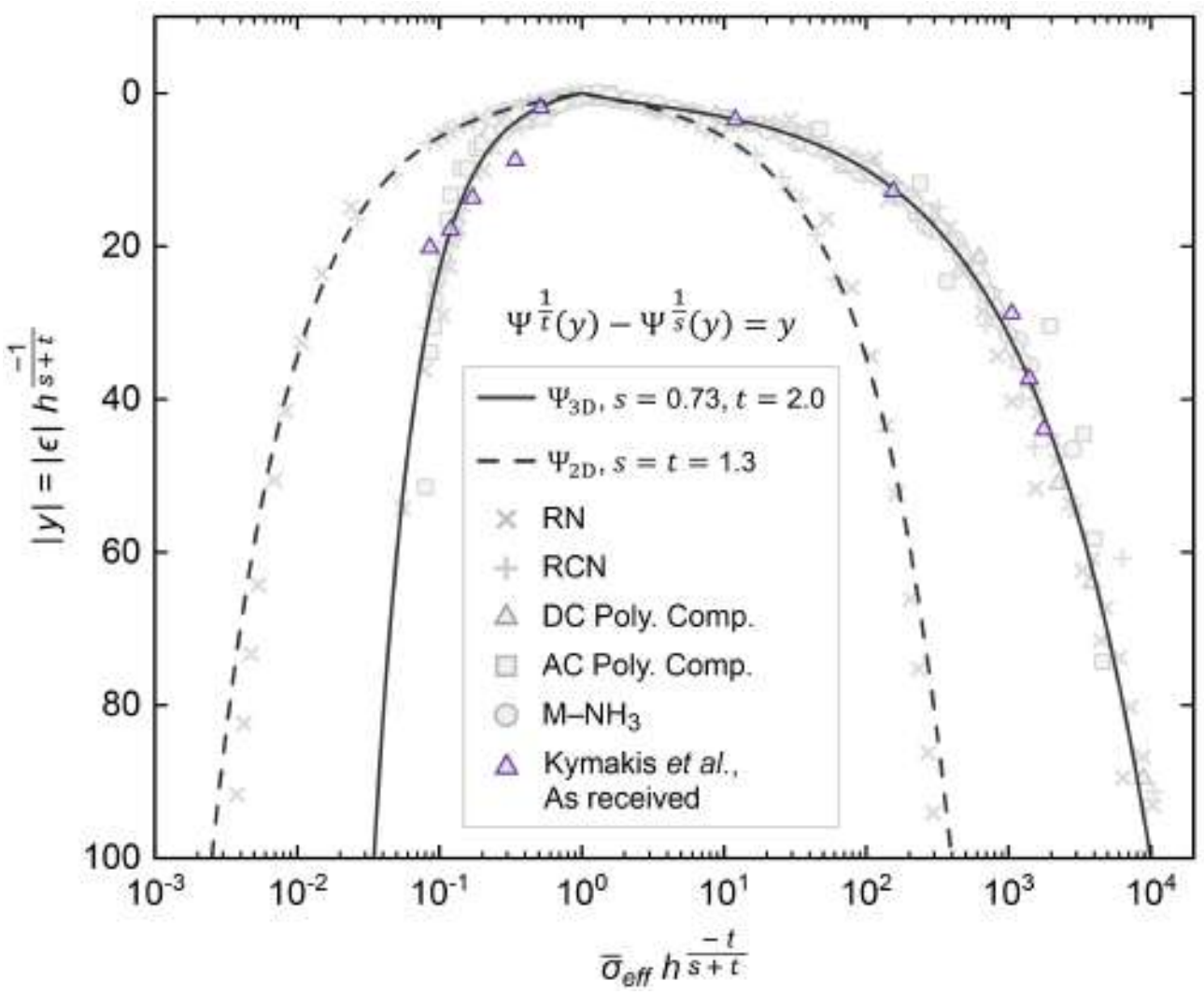

**Data Collapse:** Collapse of the rescaled data onto the proposed scaling function, $\Psi(y)$

**Source Article [113]:** Reproduced from E. Kymakis *et al.*, "Electrical properties of single-wall carbon nanotube-polymer composite films," *J. Appl. Phys.* **99**, 084302 (2006), with the permission of AIP Publishing.

**System:** Polymer composite
**Poor Conducting Phase (Matrix):** Polyvinyl alcohol (PVA)
**Good Conducting Phase (Filler):** Multiwalled carbon nanotubes (MWCNT)

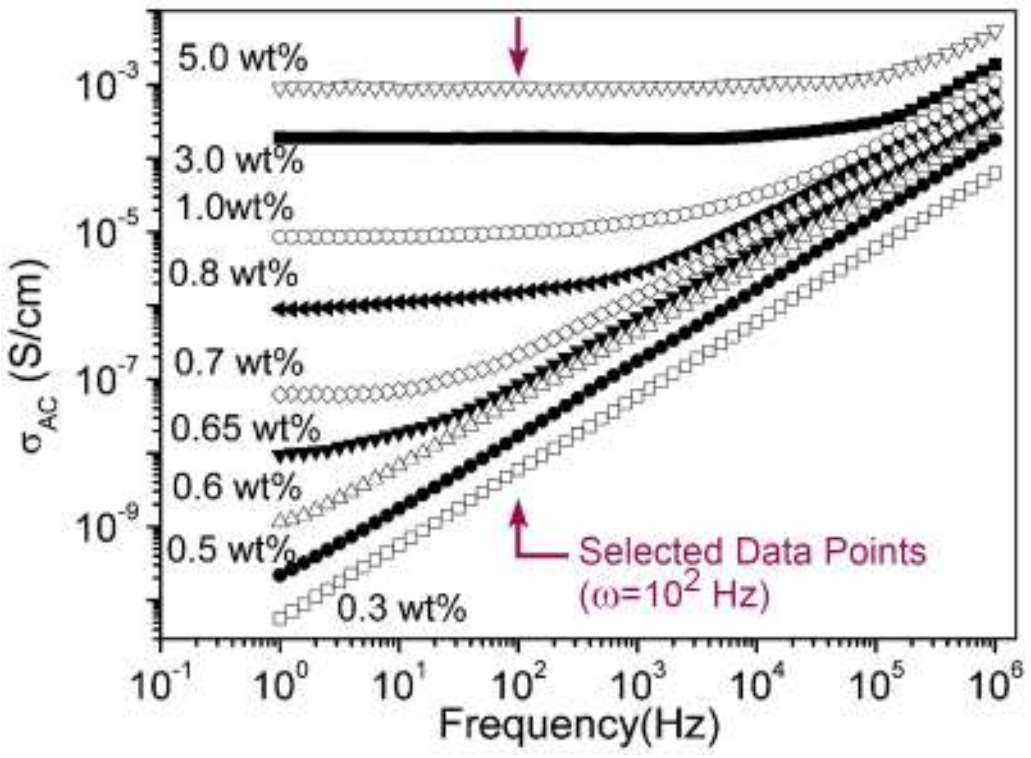

**Selected Figure:** Fig. 3 of the source article (Ref. [114])

**Extracted Data Points**

| # | $\phi$ (wt%) | $\phi$ (vol%) | $\sigma$ (S/cm) |
|---|---|---|---|
| 1 | 0.30 | 0.19 | 5.92E-09 |
| 2 | 0.50 | 0.31 | 1.69E-08 |
| 3 | 0.60 | 0.37 | 5.52E-08 |
| 4 | 0.65 | 0.40 | 8.38E-08 |
| 5 | 0.70 | 0.43 | 2.12E-07 |
| 6 | 0.80 | 0.50 | 1.51E-06 |
| 7 | 1.00 | 0.62 | 9.88E-06 |
| 8 | 3.00 | 1.88 | 1.89E-04 |
| 9 | 5.00 | 3.16 | 8.69E-04 |

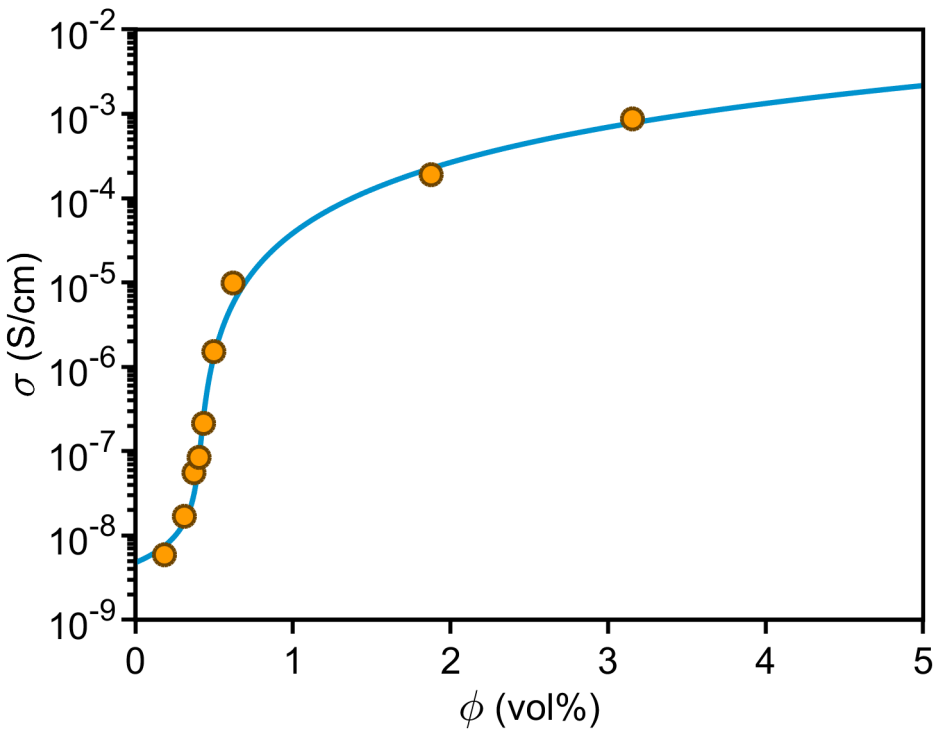

| | |
|---|---|
| $\phi_c$ (vol%) | 0.38 |
| $s$ | 0.73 |
| $t$ | 2.0 |
| $\sigma_m$ (S/cm) | $4.8 \times 10^{-9}$ |
| $\sigma_{f^*}$ (S/cm) | $1.5 \times 10^{-5}$ |
| $\sigma_f$ (S/cm) | 1.0 |
| $h = \sigma_m/\sigma_{f^*}$ | $3.2 \times 10^{-4}$ |
| $R^2$ | 1.000000 |

**Fit Results:** Fitting data with the TAGP equation and the best-fit parameters

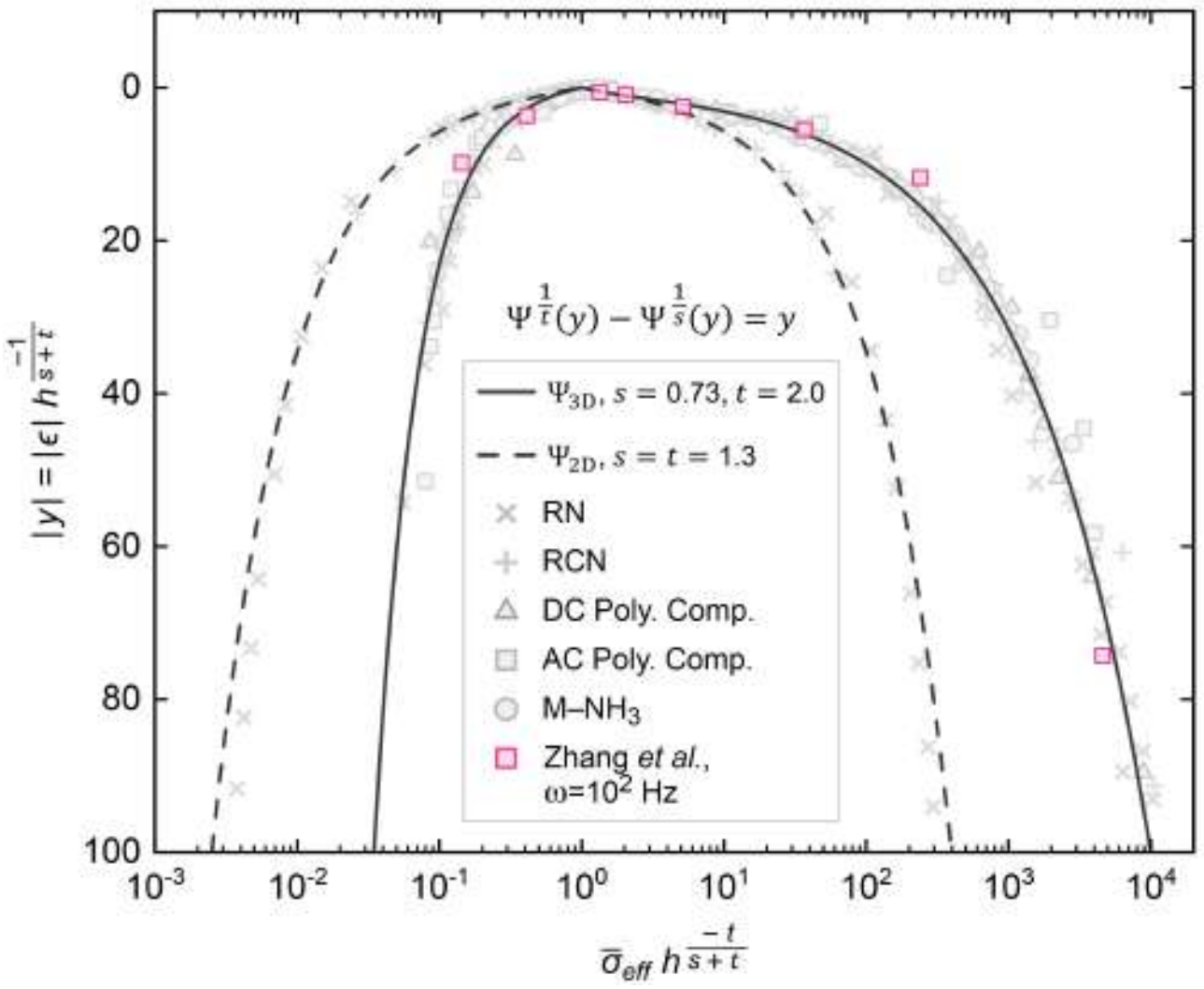

**Data Collapse:** Collapse of the rescaled data onto the proposed scaling function, $\Psi(y)$

**Source Article [114]:** Reproduced with permission from Carbon **47**, 1311–1320 (2009). Copyright 2009 Elsevier.

**System:** Polymer composite
**Poor Conducting Phase (Matrix):** Epoxy
**Good Conducting Phase (Filler):** Carbon fibers

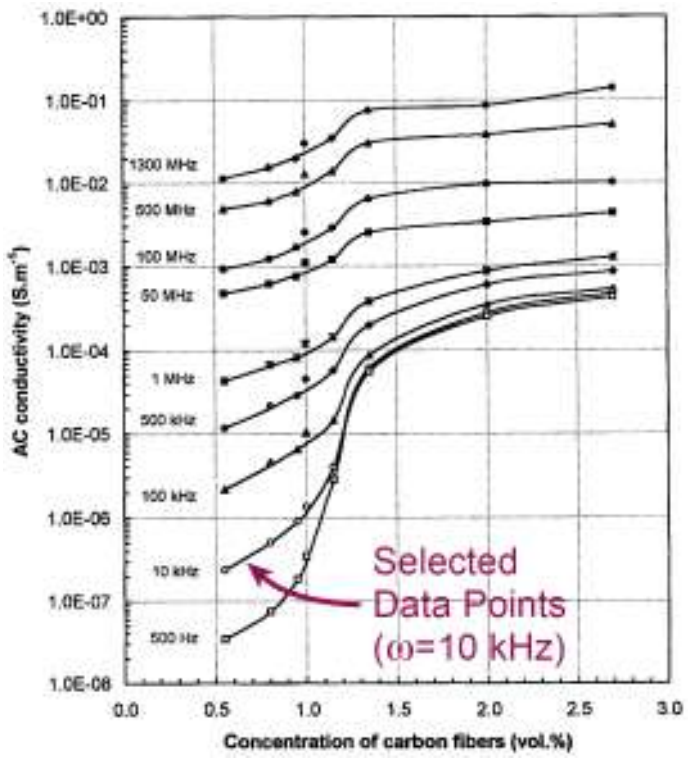

**Selected Figure:** Fig. 4 of the source article (Ref. [115])

**Extracted Data Points**

| # | $\phi$ (wt%) | $\sigma$ (S/m) |
|---|---|---|
| 1 | 0.55 | 2.32E-07 |
| 2 | 0.80 | 4.89E-07 |
| 3 | 0.95 | 8.77E-07 |
| 4 | 1.00 | 1.29E-06 |
| 5 | 1.15 | 3.96E-06 |
| 6 | 1.36 | 5.99E-05 |
| 7 | 2.01 | 2.76E-04 |
| 8 | 2.70 | 4.71E-04 |

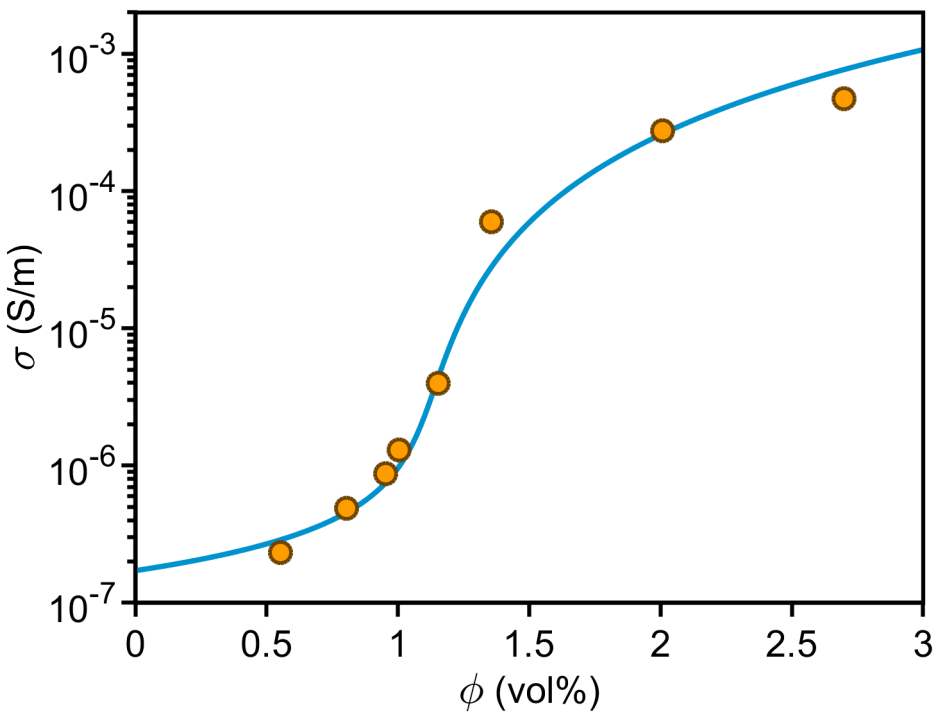

| | |
|---|---|
| $\phi_c$ (vol%) | 1.04 |
| $s$ | 0.73 |
| $t$ | 2.0 |
| $\sigma_m$ (S/m) | $1.7 \times 10^{-7}$ |
| $\sigma_{f^*}$ (S/m) | $3.0 \times 10^{-4}$ |
| $\sigma_f$ (S/m) | 2.7 |
| $h = \sigma_m/\sigma_{f^*}$ | $5.7 \times 10^{-4}$ |
| $R^2$ | 0.999997 |

**Fit Results:** Fitting data with the TAGP equation and the best-fit parameters

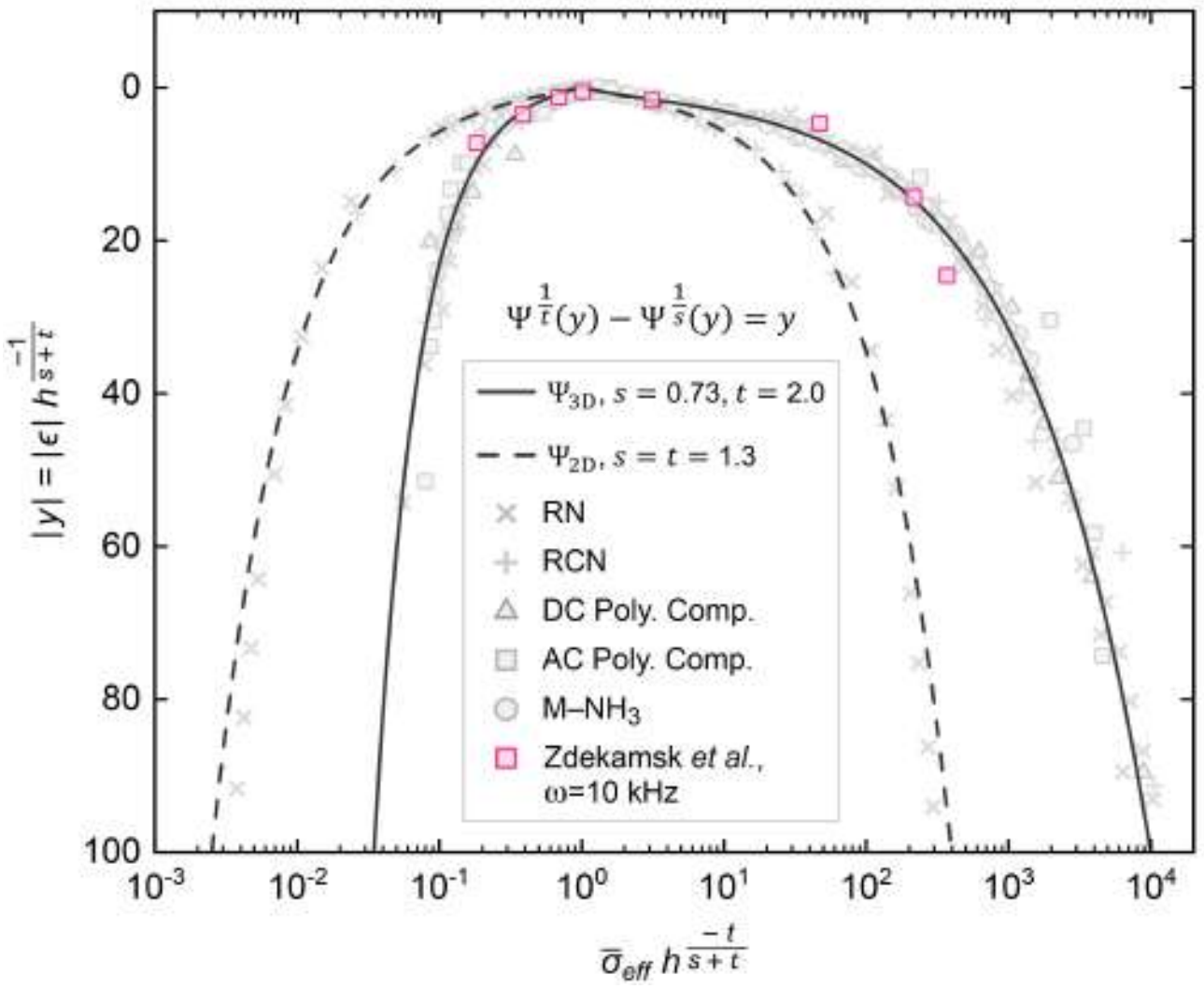

**Data Collapse:** Collapse of the rescaled data onto the proposed scaling function, $\Psi(y)$

**Source Article [115]:** Reproduced with permission from Polym. Compos. **23**, 95–103 (2002). Copyright 2002 Wiley-VCH.

**System:** Polymer composite
**Poor Conducting Phase (Matrix):** High-density polyethylene (HDPE)
**Good Conducting Phase (Filler):** Carbon black

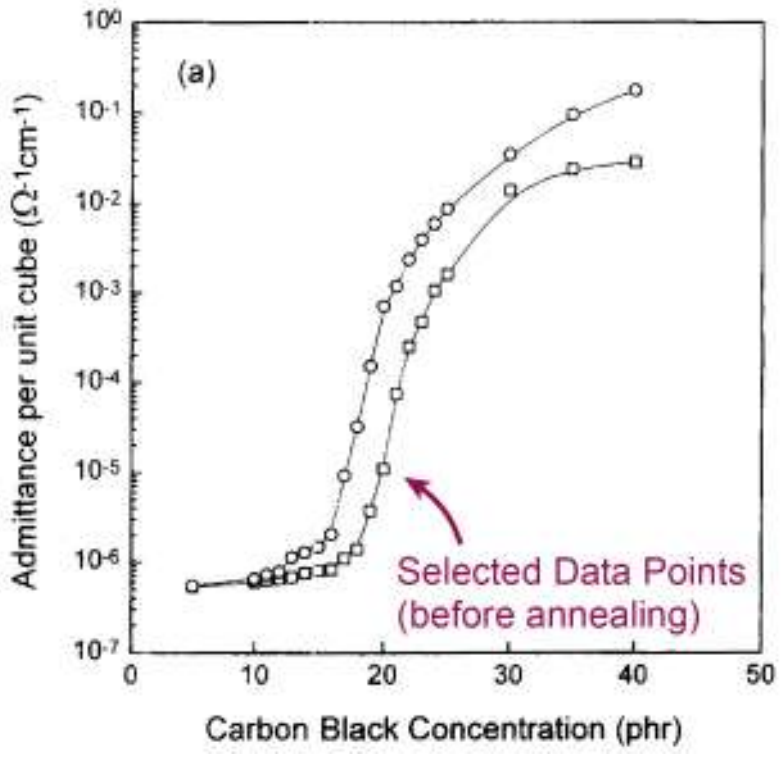

**Selected Figure:** Fig. 2 of the source article (Ref. [116])

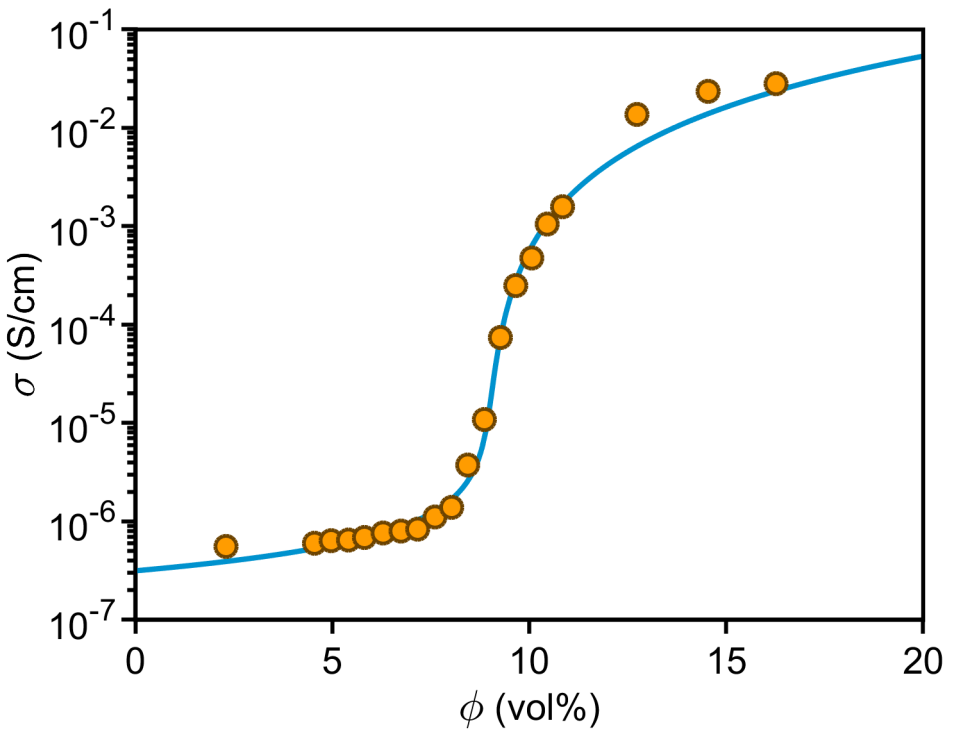

| | |
|---|---|
| $\phi_c$ (vol%) | 8.9 |
| $s$ | 0.73 |
| $t$ | 2.0 |
| $\sigma_m$ (S/cm) | $3.2 \times 10^{-7}$ |
| $\sigma_{f^*}$ (S/cm) | $3.4 \times 10^{-2}$ |
| $\sigma_f$ (S/cm) | 3.6 |
| $h = \sigma_m/\sigma_{f^*}$ | $9.3 \times 10^{-6}$ |
| $R^2$ | 1.000000 |

**Fit Results:** Fitting data with the TAGP equation and the best-fit parameters

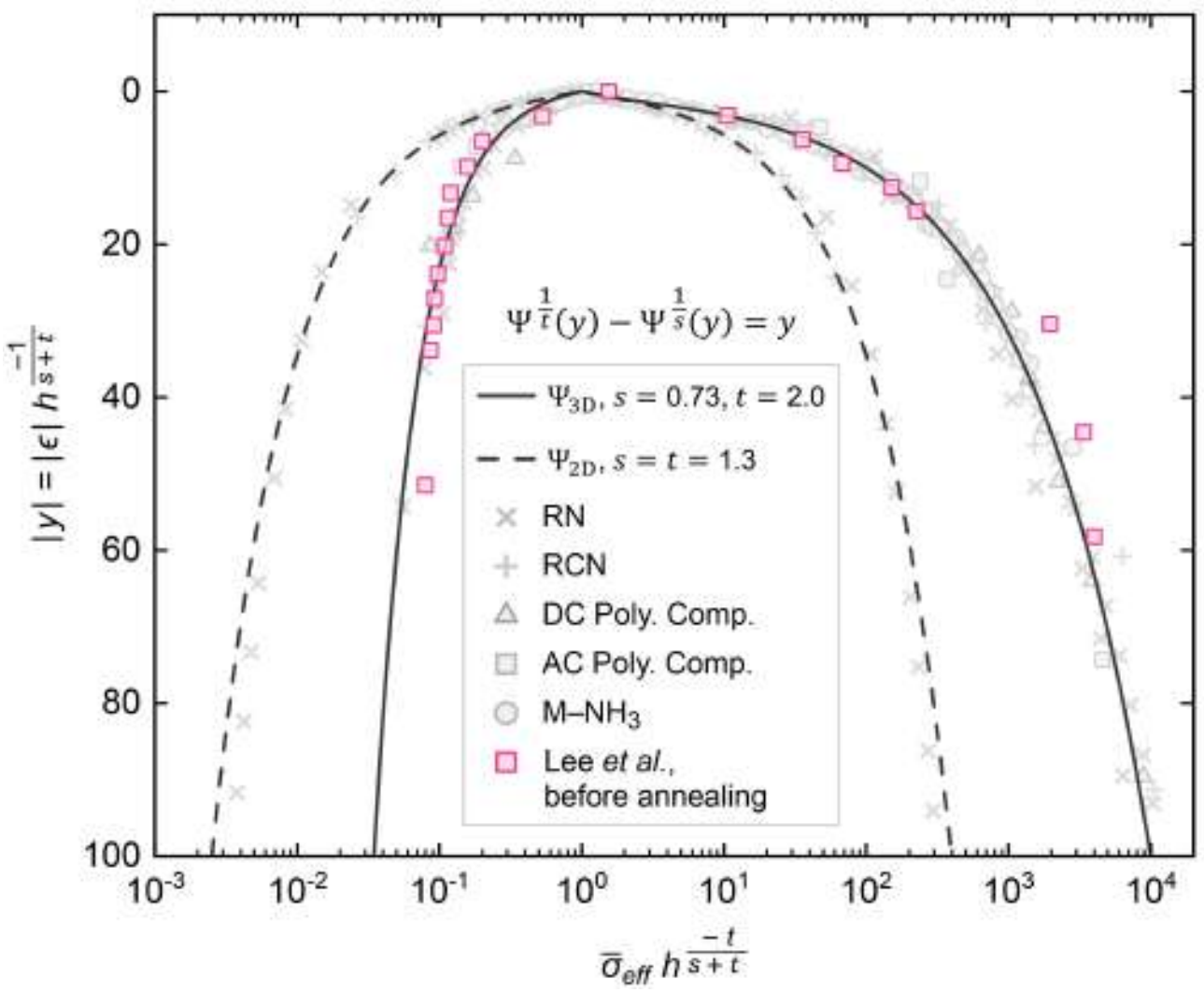

**Data Collapse:** Collapse of the rescaled data onto the proposed scaling function, $\Psi(y)$

**Extracted Data Points**

| # | $\phi$ (phr) | $\phi$ (vol%) | $\sigma$ (S/cm) |
|---|---|---|---|
| 1 | 4.8 | 2.3 | 5.56E-07 |
| 2 | 9.8 | 4.6 | 6.04E-07 |
| 3 | 10.7 | 5.0 | 6.36E-07 |
| 4 | 11.7 | 5.4 | 6.49E-07 |
| 5 | 12.7 | 5.8 | 6.83E-07 |
| 6 | 13.8 | 6.3 | 7.67E-07 |
| 7 | 14.9 | 6.7 | 8.04E-07 |
| 8 | 15.8 | 7.2 | 8.37E-07 |
| 9 | 16.9 | 7.6 | 1.10E-06 |
| 10 | 17.9 | 8.0 | 1.40E-06 |
| 11 | 18.9 | 8.4 | 3.72E-06 |
| 12 | 20.0 | 8.9 | 1.09E-05 |
| 13 | 21.0 | 9.3 | 7.41E-05 |
| 14 | 21.9 | 9.7 | 2.49E-04 |
| 15 | 23.0 | 10.1 | 4.73E-04 |
| 16 | 24.0 | 10.5 | 1.06E-03 |
| 17 | 25.0 | 10.9 | 1.59E-03 |
| 18 | 29.9 | 12.7 | 1.37E-02 |
| 19 | 34.9 | 14.5 | 2.36E-02 |
| 20 | 39.9 | 16.3 | 2.80E-02 |

**Source Article [116]:** Reproduced with permission from Polym. Eng. Sci. **38**, 471–477 (1998). Copyright 1998 Wiley-VCH.

**System:** Metal–Ammonia solution
**Poor Conducting Phase (Solvent):** Ammonia ($NH_3$)
**Good Conducting Phase (Solute):** Lithium (Li)

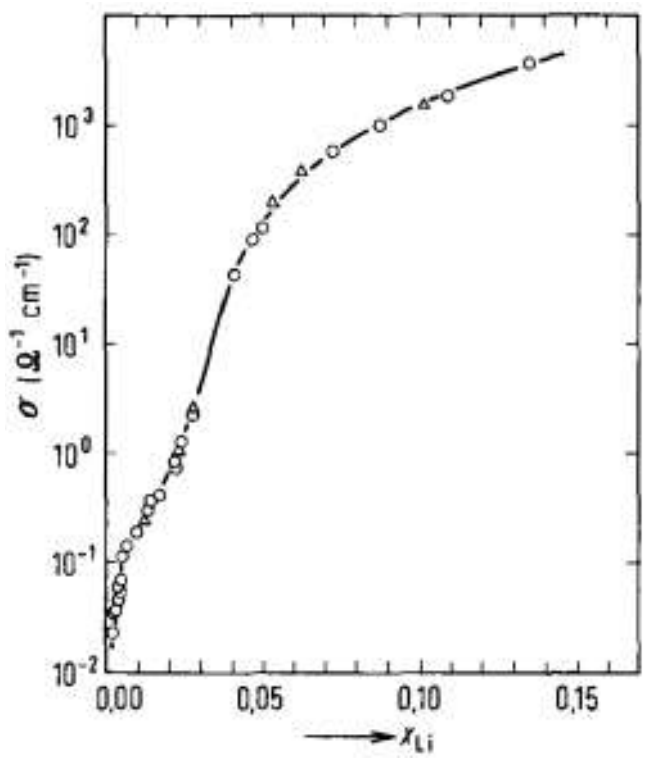

**Selected Figure:** Fig. 1 of the source article (Ref. [117])

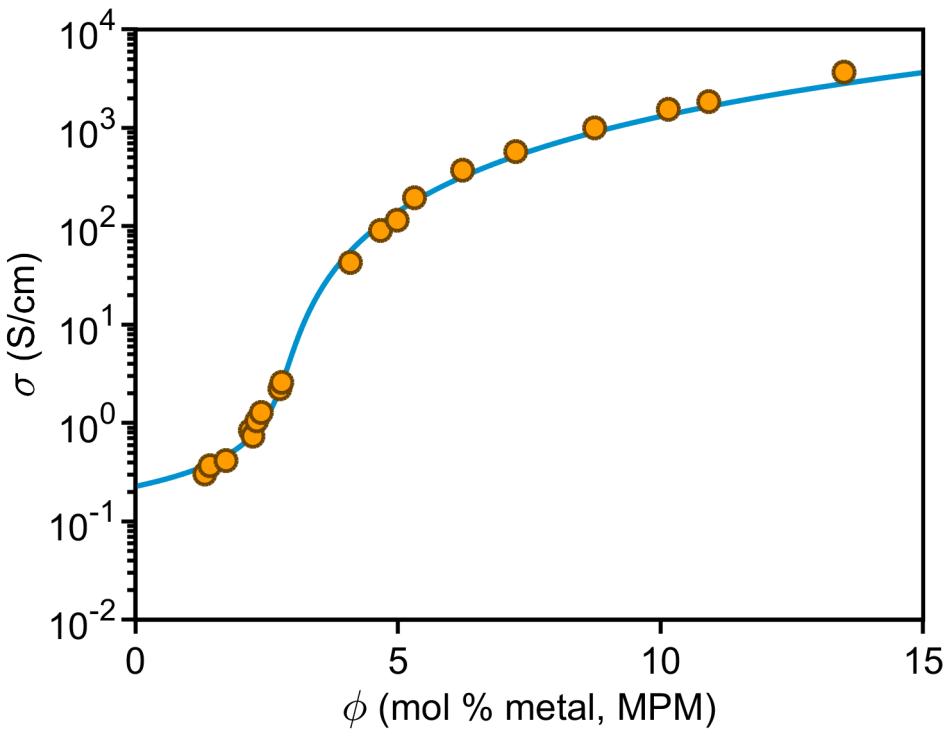

| | |
|---|---|
| $\phi_c$ (MPM) | 2.6 |
| $s$ | 0.73 |
| $t$ | 2.0 |
| $\sigma_m$ (S/cm) | $2.3 \times 10^{-1}$ |
| $\sigma_{f^*}$ (S/cm) | $1.5 \times 10^{2}$ |
| $\sigma_f$ (S/cm) | $2.2 \times 10^{5}$ |
| $h = \sigma_m/\sigma_{f^*}$ | $1.5 \times 10^{-3}$ |
| $R^2$ | 1.000000 |

**Fit Results:** Fitting data with the TAGP equation and the best-fit parameters

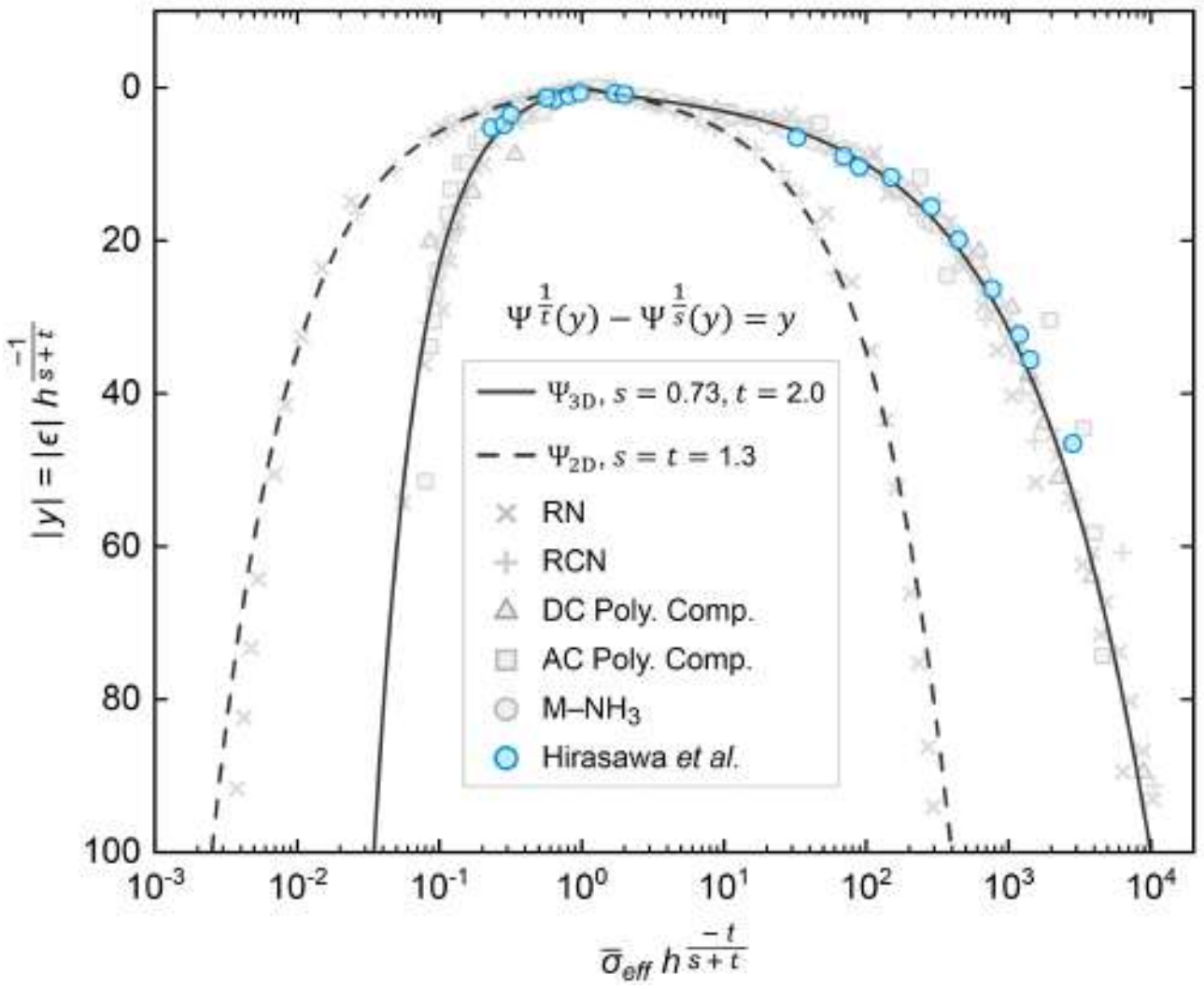

**Data Collapse:** Collapse of the rescaled data onto the proposed scaling function, $\Psi(y)$

**Extracted Data Points**

| # | $\phi$ (MPM) | $\sigma$ (S/cm) |
|---|---|---|
| 1 | 0.2 | 2.31E-02 |
| 2 | 0.2 | 3.49E-02 |
| 3 | 0.3 | 3.71E-02 |
| 4 | 0.4 | 6.06E-02 |
| 5 | 0.4 | 4.59E-02 |
| 6 | 0.5 | 5.44E-02 |
| 7 | 0.5 | 7.02E-02 |
| 8 | 0.5 | 1.15E-01 |
| 9 | 0.7 | 1.43E-01 |
| 10 | 1.0 | 1.92E-01 |
| 11 | 1.2 | 2.44E-01 |
| 12 | 1.3 | 3.01E-01 |
| 13 | 1.4 | 3.72E-01 |
| 14 | 1.7 | 4.15E-01 |
| 15 | 2.2 | 8.46E-01 |
| 16 | 2.2 | 7.33E-01 |
| 17 | 2.3 | 1.06E+00 |
| 18 | 2.4 | 1.27E+00 |
| 19 | 2.7 | 2.23E+00 |
| 20 | 2.8 | 2.61E+00 |
| 21 | 4.1 | 4.26E+01 |
| 22 | 4.7 | 9.03E+01 |
| 23 | 5.0 | 1.16E+02 |
| 24 | 5.3 | 1.95E+02 |
| 25 | 6.2 | 3.73E+02 |
| 26 | 7.2 | 5.78E+02 |
| 27 | 8.7 | 9.94E+02 |
| 28 | 10.1 | 1.55E+03 |
| 29 | 10.9 | 1.85E+03 |
| 30 | 13.5 | 3.69E+03 |

**Source Article [117]:** Reproduced with permission from *Ber. Bunsen-Ges./PCCP* **82**, 815–818 (1978). Copyright 1978 Wiley-VCH.

**System:** Metal–Ammonia solution
**Poor Conducting Phase (Solvent):** Ammonia ($NH_3$)
**Good Conducting Phase (Solute):** Sodium (Na)

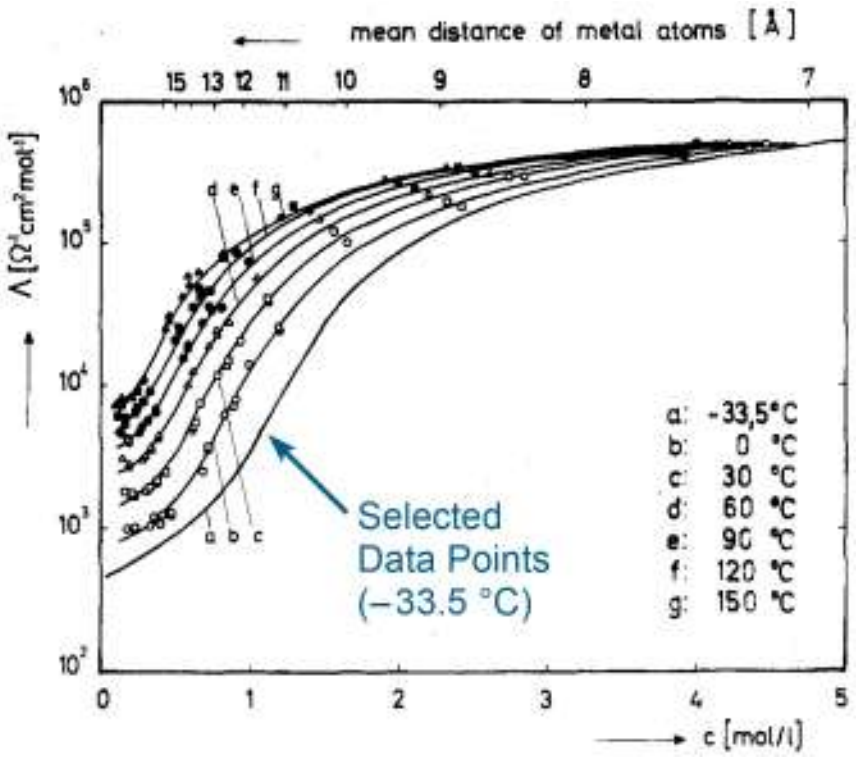

**Selected Figure:** Fig. 6 of the source article (Ref. [118])

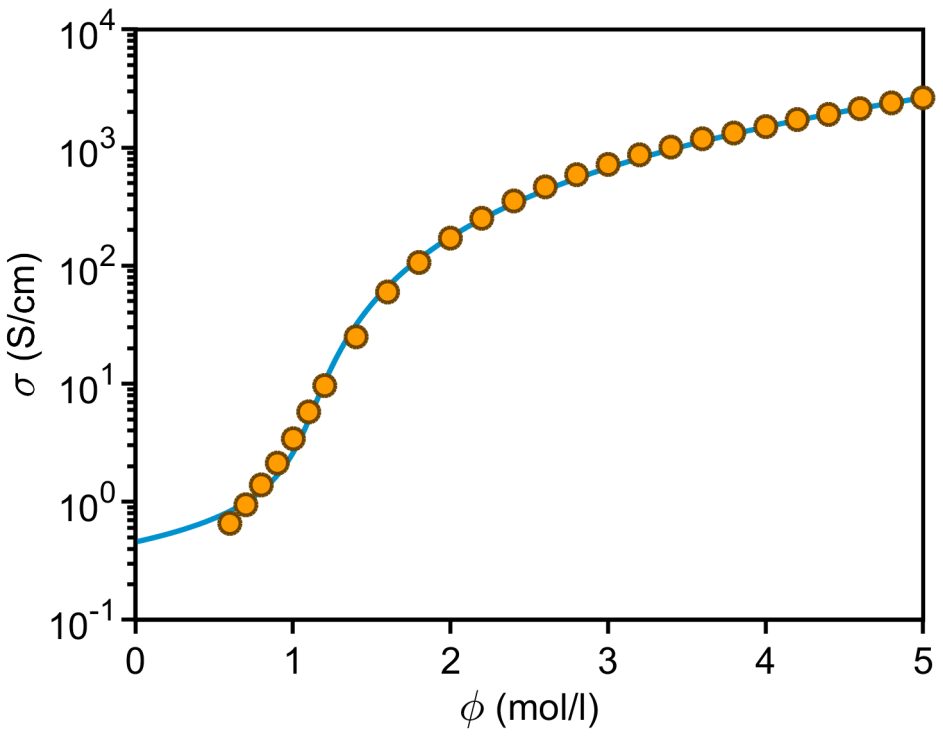

| | |
|---|---|
| $\phi_c$ (mol/l) | 0.96 |
| $s$ | 0.73 |
| $t$ | 2.0 |
| $\sigma_m$ (S/cm) | $4.6 \times 10^{-1}$ |
| $\sigma_{f^*}$ (S/cm) | $1.5 \times 10^{2}$ |
| $\sigma_f$ (S/cm) | $1.6 \times 10^{6}$ |
| $h = \sigma_m/\sigma_{f^*}$ | $3.0 \times 10^{-3}$ |
| $R^2$ | 1.000000 |

**Fit Results:** Fitting data with the TAGP equation and the best-fit parameters

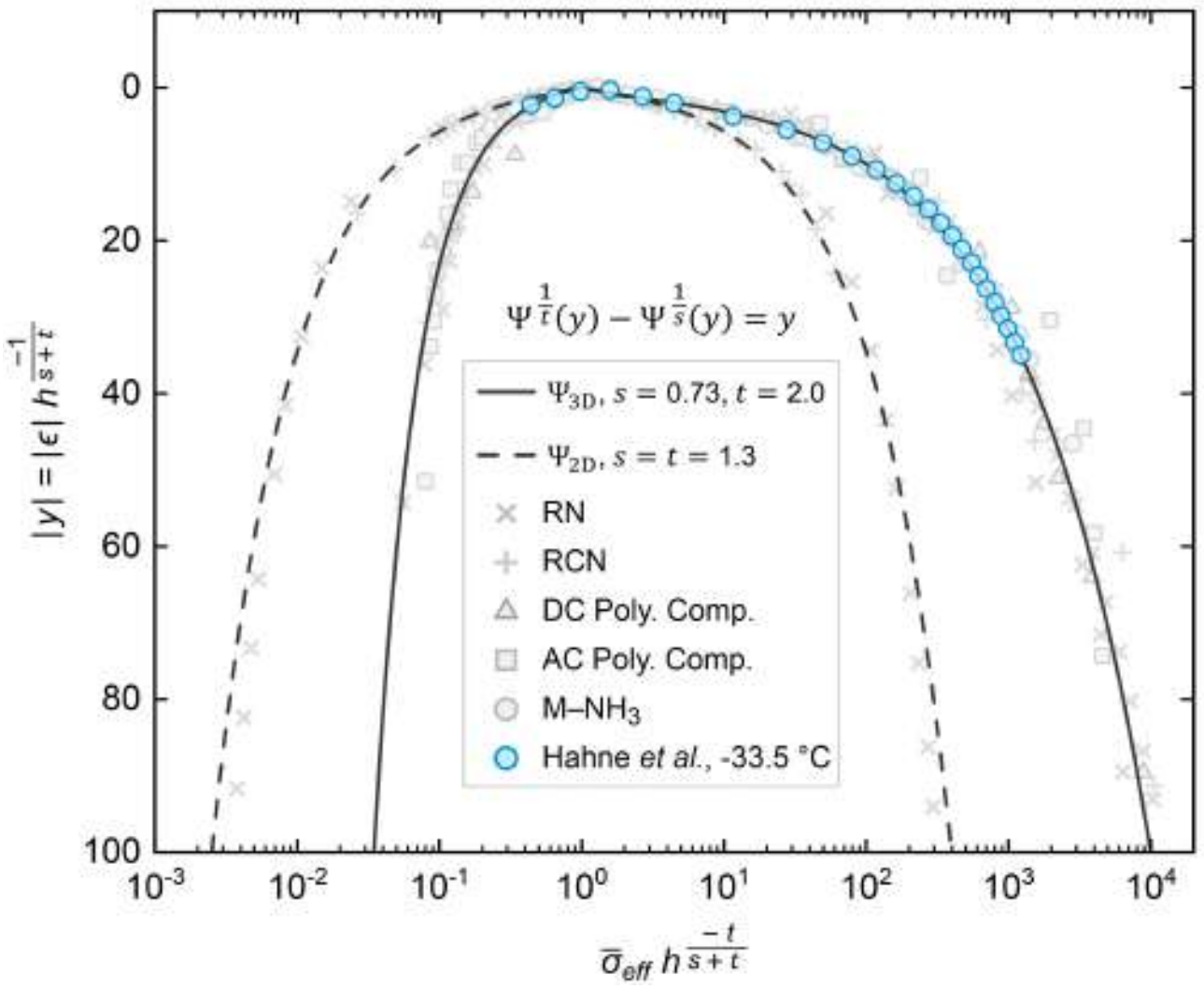

**Data Collapse:** Collapse of the rescaled data onto the proposed scaling function, $\Psi(y)$

**Extracted Data Points**

| # | $\phi$ (mol/l) | $\sigma$ (Scm²/mol) | $\sigma$ (S/cm) |
|---|---|---|---|
| 1 | 0.2 | 5.72E+02 | 1.15E-01 |
| 2 | 0.4 | 7.68E+02 | 3.08E-01 |
| 3 | 0.6 | 1.09E+03 | 6.57E-01 |
| 4 | 0.7 | 1.35E+03 | 9.48E-01 |
| 5 | 0.8 | 1.73E+03 | 1.39E+00 |
| 6 | 0.9 | 2.34E+03 | 2.11E+00 |
| 7 | 1.0 | 3.40E+03 | 3.40E+00 |
| 8 | 1.1 | 5.26E+03 | 5.79E+00 |
| 9 | 1.2 | 8.03E+03 | 9.64E+00 |
| 10 | 1.4 | 1.78E+04 | 2.50E+01 |
| 11 | 1.6 | 3.71E+04 | 5.95E+01 |
| 12 | 1.8 | 5.93E+04 | 1.07E+02 |
| 13 | 2.0 | 8.56E+04 | 1.71E+02 |
| 14 | 2.2 | 1.15E+05 | 2.53E+02 |
| 15 | 2.4 | 1.46E+05 | 3.51E+02 |
| 16 | 2.6 | 1.79E+05 | 4.65E+02 |
| 17 | 2.8 | 2.10E+05 | 5.89E+02 |
| 18 | 3.0 | 2.41E+05 | 7.22E+02 |
| 19 | 3.2 | 2.71E+05 | 8.68E+02 |
| 20 | 3.4 | 2.98E+05 | 1.01E+03 |
| 21 | 3.6 | 3.29E+05 | 1.18E+03 |
| 22 | 3.8 | 3.50E+05 | 1.33E+03 |
| 23 | 4.0 | 3.78E+05 | 1.51E+03 |
| 24 | 4.2 | 4.11E+05 | 1.73E+03 |
| 25 | 4.4 | 4.35E+05 | 1.92E+03 |
| 26 | 4.6 | 4.63E+05 | 2.13E+03 |
| 27 | 4.8 | 4.97E+05 | 2.38E+03 |
| 28 | 5.0 | 5.26E+05 | 2.63E+03 |

**Source Article [118]:** Reproduced with permission from J. Phys. Chem. **79**, 2922–2928 (1975).

**System:** Metal–Ammonia solution
**Poor Conducting Phase (Solvent):** Ammonia ($NH_3$)
**Good Conducting Phase (Solute):** Lithium (Li)

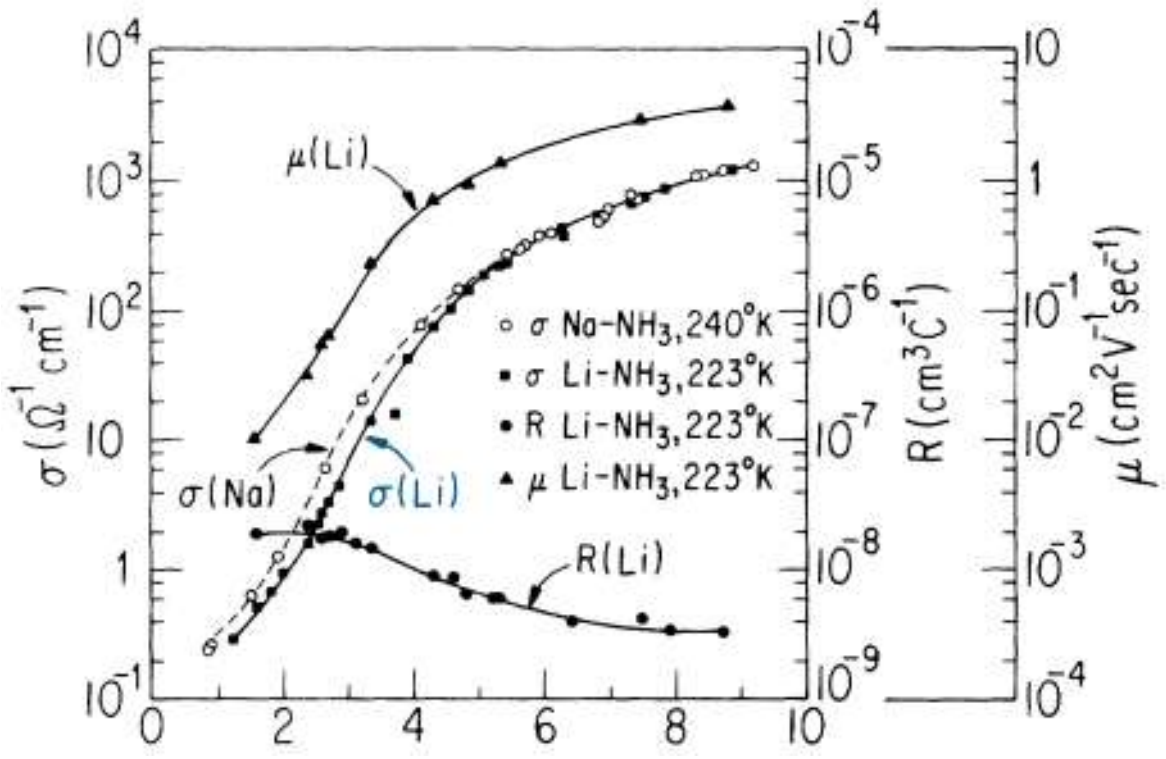

**Selected Figure:** Fig. 1 of the source article (Ref. [119])

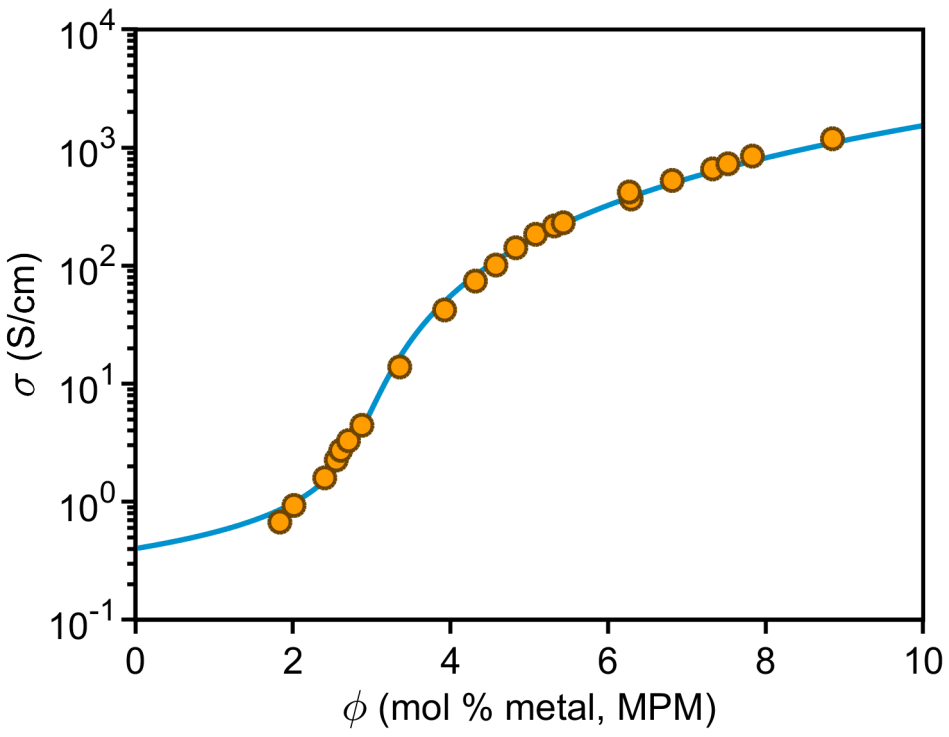

| | |
|---|---|
| $\phi_c$ (MPM) | 2.6 |
| $s$ | 0.73 |
| $t$ | 2.0 |
| $\sigma_m$ (S/cm) | $4.0 \times 10^{-1}$ |
| $\sigma_{f^*}$ (S/cm) | $1.9 \times 10^{2}$ |
| $\sigma_f$ (S/cm) | $2.6 \times 10^{5}$ |
| $h = \sigma_m/\sigma_{f^*}$ | $2.1 \times 10^{-3}$ |
| $R^2$ | 0.999999 |

**Fit Results:** Fitting data with the TAGP equation and the best-fit parameters

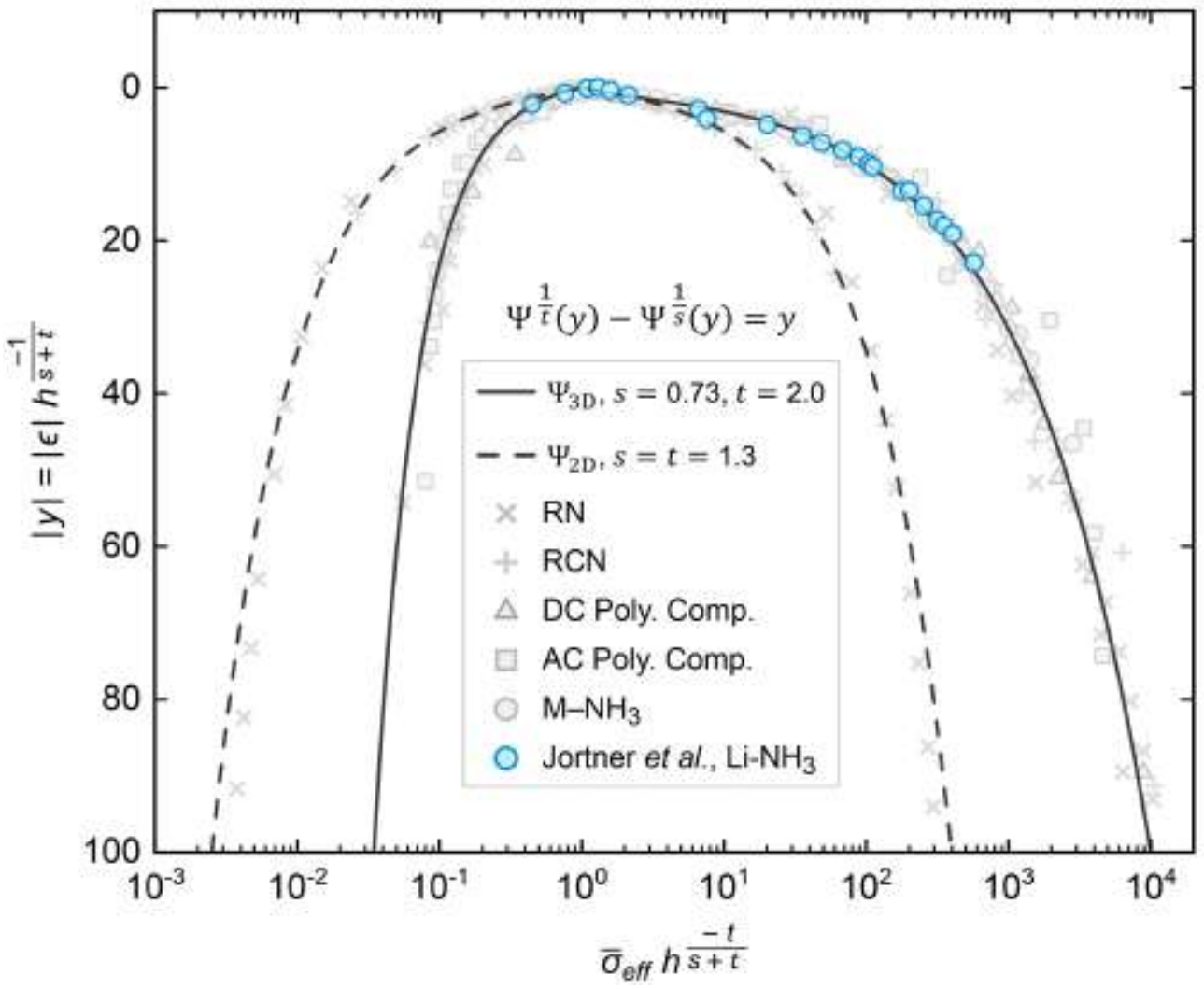

**Data Collapse:** Collapse of the rescaled data onto the proposed scaling function, $\Psi(y)$

**Extracted Data Points**

| # | $\phi$ (MPM) | $\sigma$ (S/cm) |
|---|---|---|
| 1 | 1.3 | 2.91E-01 |
| 2 | 1.6 | 5.12E-01 |
| 3 | 1.8 | 6.72E-01 |
| 4 | 2.0 | 9.35E-01 |
| 5 | 2.4 | 1.58E+00 |
| 6 | 2.6 | 2.27E+00 |
| 7 | 2.6 | 2.73E+00 |
| 8 | 2.7 | 3.31E+00 |
| 9 | 2.9 | 4.44E+00 |
| 10 | 3.4 | 1.39E+01 |
| 11 | 3.7 | 1.58E+01 |
| 12 | 3.9 | 4.22E+01 |
| 13 | 4.3 | 7.42E+01 |
| 14 | 4.6 | 1.01E+02 |
| 15 | 4.8 | 1.41E+02 |
| 16 | 5.1 | 1.85E+02 |
| 17 | 5.3 | 2.18E+02 |
| 18 | 5.4 | 2.31E+02 |
| 19 | 6.3 | 3.68E+02 |
| 20 | 6.3 | 4.21E+02 |
| 21 | 6.8 | 5.29E+02 |
| 22 | 7.3 | 6.56E+02 |
| 23 | 7.5 | 7.31E+02 |
| 24 | 7.8 | 8.46E+02 |
| 25 | 8.9 | 1.18E+03 |

**Source Article [119]:** Reproduced with permission from Phys. Rev. B **13**, 1548–1568 (1976). Copyright 1976 American Physical Society.

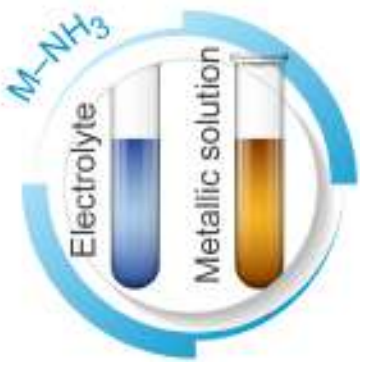

**System:** Metal–Ammonia solution
**Poor Conducting Phase (Solvent):** Ammonia ($NH_3$)
**Good Conducting Phase (Solute):** Sodium (Na)

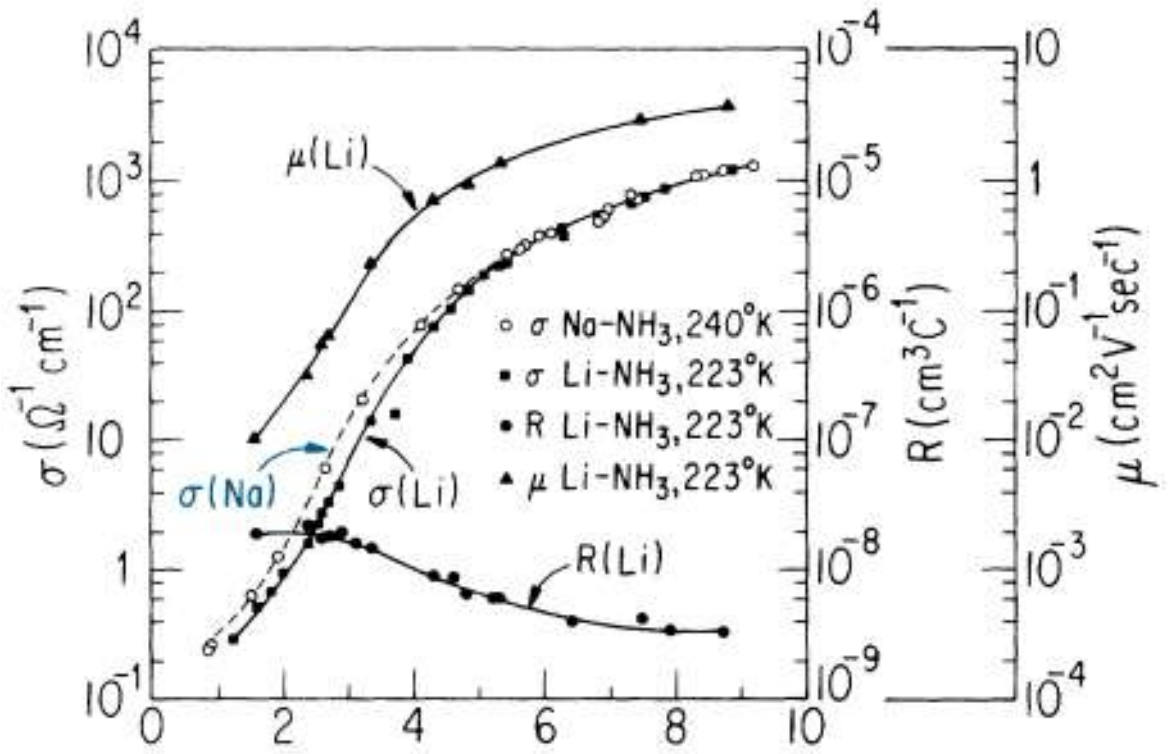

**Selected Figure:** Fig. 1 of the source article (Ref. [119])

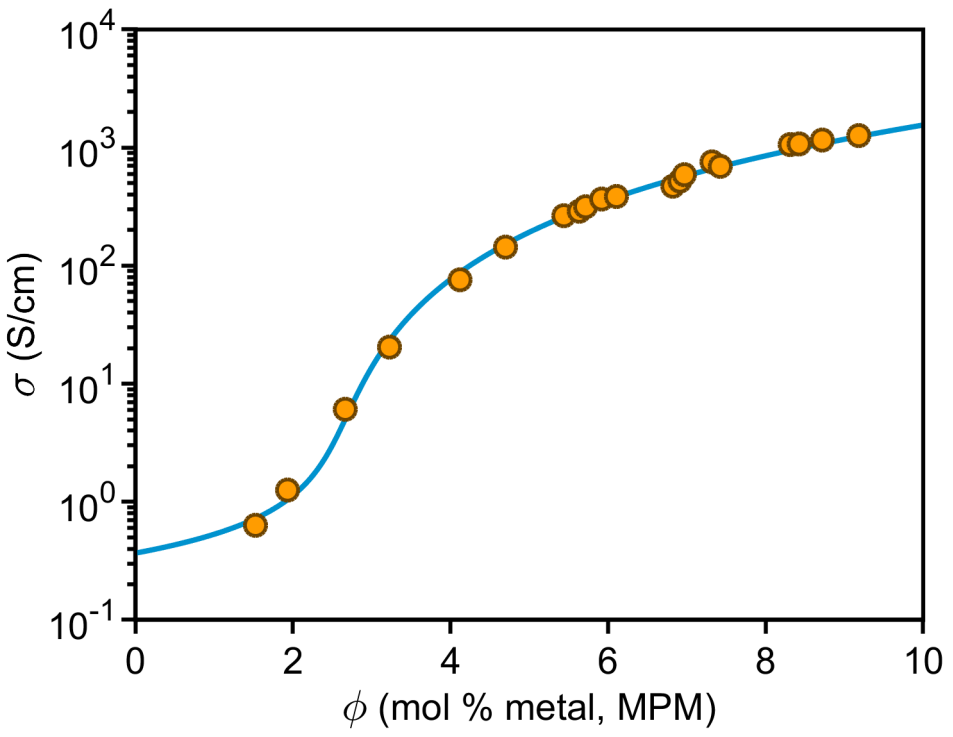

| | |
|---|---|
| $\phi_c$ (MPM) | 2.3 |
| $s$ | 0.73 |
| $t$ | 2.0 |
| $\sigma_m$ (S/cm) | $3.7 \times 10^{-1}$ |
| $\sigma_{f^*}$ (S/cm) | $1.4 \times 10^{2}$ |
| $\sigma_f$ (S/cm) | $2.5 \times 10^{5}$ |
| $h = \sigma_m/\sigma_{f^*}$ | $2.7 \times 10^{-3}$ |
| $R^2$ | 1.000000 |

**Fit Results:** Fitting data with the TAGP equation and the best-fit parameters

**Extracted Data Points**

| # | $\phi$ (MPM) | $\sigma$ (S/cm) |
|---|---|---|
| 1 | 0.9 | 2.36E-01 |
| 2 | 0.9 | 2.61E-01 |
| 3 | 1.5 | 6.29E-01 |
| 4 | 1.9 | 1.26E+00 |
| 5 | 2.7 | 6.04E+00 |
| 6 | 3.2 | 2.03E+01 |
| 7 | 4.1 | 7.62E+01 |
| 8 | 4.7 | 1.44E+02 |
| 9 | 5.4 | 2.65E+02 |
| 10 | 5.6 | 2.90E+02 |
| 11 | 5.7 | 3.14E+02 |
| 12 | 5.9 | 3.68E+02 |
| 13 | 6.1 | 3.86E+02 |
| 14 | 6.8 | 4.73E+02 |
| 15 | 6.9 | 5.19E+02 |
| 16 | 7.0 | 5.89E+02 |
| 17 | 7.3 | 7.59E+02 |
| 18 | 7.4 | 6.97E+02 |
| 19 | 8.3 | 1.06E+03 |
| 20 | 8.4 | 1.08E+03 |
| 21 | 8.7 | 1.16E+03 |
| 22 | 9.2 | 1.27E+03 |

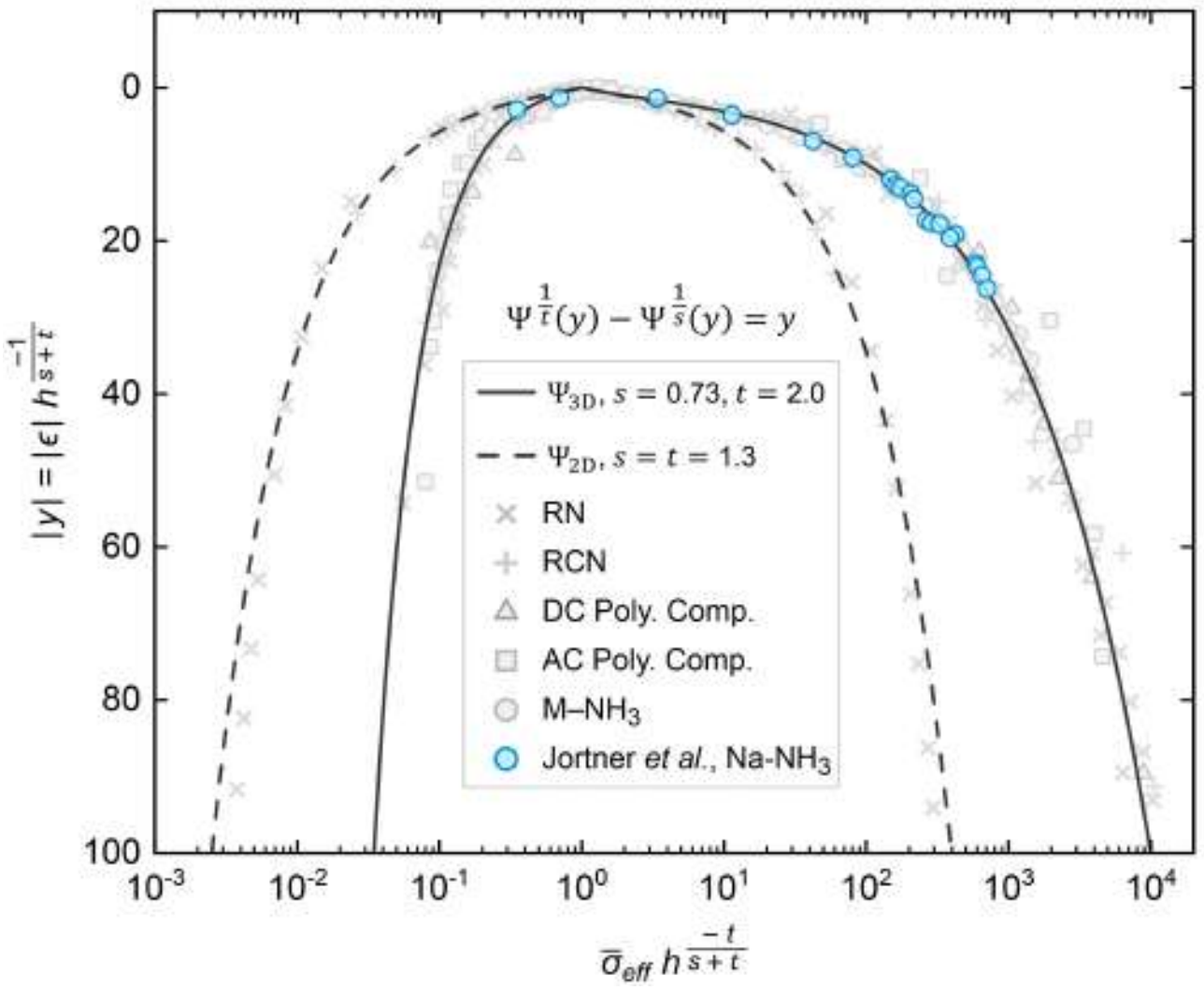

**Data Collapse:** Collapse of the rescaled data onto the proposed scaling function, $\Psi(y)$

**Source Article [119]:** Reproduced with permission from Phys. Rev. B **13**, 1548–1568 (1976).

## REFERENCES

1 S. D. Bergin *et al.*, "Multicomponent solubility parameters for single-walled carbon nanotube-solvent mixtures," *ACS Nano* **3**, 2340-2350 (2009).

2 B. Ellis and R. Smith, *Polymers: A property database*. (CRC Press/Taylor & Francis Group, 2008).

3 H. Pang, L. Xu, D. X. Yan, and Z. M. Li, "Conductive polymer composites with segregated structures," *Prog. Polym. Sci.* **39**, 1908-1933 (2014).

4 J. Shen *et al.*, "Liquid phase exfoliation of two-dimensional materials by directly probing and matching surface tension components," *Nano Lett.* **15**, 5449-5454 (2015).

5 J. Shen *et al.*, "Surface tension components based selection of cosolvents for efficient liquid phase exfoliation of 2D materials," *Small* **12**, 2741-2749 (2016).

6 W. Lei *et al.*, "Boron nitride colloidal solutions, ultralight aerogels and freestanding membranes through one-step exfoliation and functionalization," *Nat. Commun.* **6** (2015).

7 D. Lee *et al.*, "Scalable exfoliation process for highly soluble boron nitride nanoplatelets by hydroxide-assisted ball milling," *Nano Lett.* **15**, 1238-1244 (2015).

8 L. Fu *et al.*, "An ultrathin high-performance heat spreader fabricated with hydroxylated boron nitride nanosheets," *2D Mater.* **4** (2017).

9 J. R. Taylor, *Introduction to error analysis: The study of uncertainties in physical measurements*. (University Science Books, 1997).

10 P. Fornasini, *The uncertainty in physical measurements: An introduction to data analysis in the physics laboratory*. (Springer New York, 2008).

11 I. Hughes and T. Hase, *Measurements and their uncertainties: A practical guide to modern error analysis*. (OUP Oxford, 2010).

12 A. Bjorck, *Numerical methods for least squares problems*. (Society for Industrial and Applied Mathematics (SIAM, 3600 Market Street, Floor 6, Philadelphia, PA 19104), 1996).

13 Y. P. Mamunya, V. V. Davydenko, P. Pissis, and E. V. Lebedev, "Electrical and thermal conductivity of polymers filled with metal powders," *Eur. Polym. J.* **38**, 1887-1897 (2002).

14 M. J. Biercuk *et al.*, "Carbon nanotube composites for thermal management," *Appl. Phys. Lett.* **80**, 2767-2769 (2002).

15 F. H. Gojny *et al.*, "Evaluation and identification of electrical and thermal conduction mechanisms in carbon nanotube/epoxy composites," *Polymer* **47**, 2036-2045 (2006).

16 A. Moisala, Q. Li, I. A. Kinloch, and A. H. Windle, "Thermal and electrical conductivity of single- and multi-walled carbon nanotube-epoxy composites," *Compos. Sci. Technol.* **66**, 1285-1288 (2006).

17 A. Yu, M. E. Itkis, E. Bekyarova, and R. C. Haddon, "Effect of single-walled carbon nanotube purity on the thermal conductivity of carbon nanotube-based composites," *Appl. Phys. Lett.* **89** (2006).

18 P. Bonnet, D. Sireude, B. Garnier, and O. Chauvet, "Thermal properties and percolation in carbon nanotube-polymer composites," *Appl. Phys. Lett.* **91** (2007).

19 R. Haggenmueller *et al.*, "Single wall carbon nanotube/polyethylene nanocomposites: Thermal and electrical conductivity," *Macromolecules* **40**, 2417-2421 (2007).

20 Y. Mamunya *et al.*, "Electrical and thermophysical behaviour of pvc-MWCNT nanocomposites," *Compos. Sci. Technol.* **68**, 1981-1988 (2008).

21 Y. Yang, M. C. Gupta, J. N. Zalameda, and W. P. Winfree, "Dispersion behaviour, thermal and electrical conductivities of carbon nanotube-polystyrene nanocomposites," *Micro Nano Lett.* **3**, 35-40 (2008).

22 A. Yu *et al.*, "Enhanced thermal conductivity in a hybrid graphite nanoplatelet - carbon nanotube filler for epoxy composites," *Adv. Mater.* **20**, 4740-4744 (2008).

23 K. M. F. Shahil and A. A. Balandin, "Graphene-multilayer graphene nanocomposites as highly efficient thermal interface materials," *Nano Lett.* **12**, 861-867 (2012).

24 R. Gulotty *et al.*, "Effects of functionalization on thermal properties of single-wall and multi-wall carbon nanotube-polymer nanocomposites," *ACS Nano* **7**, 5114-5121 (2013).

25 J. Bouchard, A. Cayla, E. Devaux, and C. Campagne, "Electrical and thermal conductivities of multiwalled carbon nanotubes-reinforced high performance polymer nanocomposites," *Compos. Sci. Technol.* **86**, 177-184 (2013).

26 B. W. Kim, S. H. Park, R. S. Kapadia, and P. R. Bandaru, "Evidence of percolation related power law behavior in the thermal conductivity of nanotube/polymer composites," *Appl. Phys. Lett.* **102** (2013).

27 S. Y. Kwon *et al.*, "A large increase in the thermal conductivity of carbon nanotube/polymer composites produced by percolation phenomena," *Carbon* **55**, 285-290 (2013).

28 B. W. Kim, S. H. Park, and P. R. Bandaru, "Anomalous decrease of the specific heat capacity at the electrical and thermal conductivity percolation threshold in nanocomposites," *Appl. Phys. Lett.* **105** (2014).

29 M. Shtein *et al.*, "Thermally conductive graphene-polymer composites: Size, percolation, and synergy effects," *Chem. Mater.* **27**, 2100-2106 (2015).

30 M. Shtein, R. Nadiv, M. Buzaglo, and O. Regev, "Graphene-based hybrid composites for efficient thermal management of electronic devices," *ACS Appl. Mater. Interfaces* **7**, 23725-23730 (2015).

31 J. Huang *et al.*, "Massive enhancement in the thermal conductivity of polymer composites by trapping graphene at the interface of a polymer blend," *Compos. Sci. Technol.* **129**, 160-165 (2016).

32 H. S. Kim *et al.*, "Volume control of expanded graphite based on inductively coupled plasma and enhanced thermal conductivity of epoxy composite by formation of the filler network," *Carbon* **119**, 40-46 (2017).

33 S. Bhanushali, P. C. Ghosh, G. P. Simon, and W. Cheng, "Copper nanowire-filled soft elastomer composites for applications as thermal interface materials," *Adv. Mater. Interfaces* **4** (2017).

34 X. Gao *et al.*, "Topological design of inorganic–organic thermoelectric nanocomposites based on "electron–percolation phonon–insulator" concept," *ACS Appl. Energy Mater.* **1**, 2927-2933 (2018).

35 B. Shi *et al.*, "Thermal percolation in composite materials with electrically conductive fillers," *Appl. Phys. Lett.* **113** (2018).

36 F. Kargar *et al.*, "Thermal percolation threshold and thermal properties of composites with high loading of graphene and boron nitride fillers," *ACS Appl. Mater. Interfaces* **10**, 37555-37565 (2018).

37 O. Maruzhenko *et al.*, "Improving the thermal and electrical properties of polymer composites by ordered distribution of carbon micro- and nanofillers," *Int. J. Heat Mass Transfer* **138**, 75-84 (2019).

38 B. Krause, P. Rzeczkowski, and P. Pötschke, "Thermal conductivity and electrical resistivity of melt-mixed polypropylene composites containing mixtures of carbon-based fillers," *Polymers* **11** (2019).

39 J. U. Jang *et al.*, "Electrically and thermally conductive carbon fibre fabric reinforced polymer composites based on nanocarbons and an in-situ polymerizable cyclic oligoester," *Sci. Rep.* **8** (2018).

40 J. Chen, J. Han, and D. Xu, "Thermal and electrical properties of the epoxy nanocomposites reinforced with purified carbon nanotubes," *Mater. Lett.* **246**, 20-23 (2019).

41 D. Suh *et al.*, "Significantly enhanced phonon mean free path and thermal conductivity by percolation of silver nanoflowers," *Phys. Chem. Chem. Phys.* **21**, 2453-2462 (2019).

42 D. An *et al.*, "A polymer-based thermal management material with enhanced thermal conductivity by introducing three-dimensional networks and covalent bond connections," *Carbon* **155**, 258-267 (2019).

43 Z. Wu *et al.*, "Synergistic effect of aligned graphene nanosheets in graphene foam for high-performance thermally conductive composites," *Adv. Mater.* **31**, e1900199 (2019).

44 A. Shayganpour *et al.*, "Stacked-cup carbon nanotube flexible paper based on soy lecithin and natural rubber," *Nanomaterials* **9** (2019).

45 F. Kargar *et al.*, "Dual-functional graphene composites for electromagnetic shielding and thermal management," *Adv. Electron. Mater.* **5** (2019).

46 G. Zhao *et al.*, "Flame synthesis of carbon nanotubes on glass fibre fabrics and their enhancement in electrical and thermal properties of glass fibre/epoxy composites," *Composites, Part B* **198** (2020).

47 A. I. Misiura, Y. P. Mamunya, and M. P. Kulish, "Metal-filled epoxy composites: Mechanical properties and electrical/thermal conductivity," *J. Macromol. Sci., Part B: Phys.* **59**, 121-136 (2020).

48 O. V. Lebedev, O. I. Bogdanova, G. P. Goncharuk, and A. N. Ozerin, "Tribological and percolation properties of polypropylene/nanodiamond soot composites," *Polym. Polym. Compos.* **28**, 369-377 (2020).

49 A. Gurijala *et al.*, "Castable and printable dielectric composites exhibiting high thermal conductivity via percolation-enabled phonon transport," *Matter* **2**, 1015-1024 (2020).

50 Z. Barani *et al.*, "Multifunctional graphene composites for electromagnetic shielding and thermal management at elevated temperatures," *Adv. Electron. Mater.* **6** (2020).

51 M. C. Vu *et al.*, "High thermal conductivity enhancement of polymer composites with vertically aligned silicon carbide sheet scaffolds," *ACS Appl. Mater. Interfaces* **12**, 23388-23398 (2020).

52 S. Naghibi *et al.*, "Noncuring graphene thermal interface materials for advanced electronics," *Adv. Electron. Mater.* **6** (2020).

53 C. Muhammed Ajmal *et al.*, "In-situ reduced non-oxidized copper nanoparticles in nanocomposites with extraordinary high electrical and thermal conductivity," *Mater. Today* **48**, 59-71 (2021).

54 M. Cierpisz, J. McPhedran, Y. He, and A. Edrisy, "Characterization of graphene-filled fluoropolymer coatings for condensing heat exchangers," *J. Compos. Mater.* **55**, 4305-4320 (2021).

55 W. Dai *et al.*, "Multiscale structural modulation of anisotropic graphene framework for polymer composites achieving highly efficient thermal energy management," *Adv. Sci.* **8**, 2003734 (2021).

56 H. Bark, M. W. M. Tan, G. Thangavel, and P. S. Lee, "Deformable high loading liquid metal nanoparticles composites for thermal energy management," *Adv. Energy Mater.* **11** (2021).

57 S. H. Ryu *et al.*, "Quasi-isotropic thermal conduction in percolation networks: Using the pore-filling effect to enhance thermal conductivity in polymer nanocomposites," *ACS Appl. Polym. Mater.* **3**, 1293-1305 (2021).

58 G. Shachar-Michaely *et al.*, "Mixed dimensionality: Highly robust and multifunctional carbon-based composites," *Carbon* **176**, 339-348 (2021).

59 I. Y. Forero-Sandoval *et al.*, "Percolation threshold of the thermal, electrical and optical properties of carbonyl-iron microcomposites," *Appl. Compos. Mater.* **28**, 447-463 (2021).

60 K. M. Burzynski *et al.*, "Graphite nanocomposite substrates for improved performance of flexible, high-power algan/gan electronic devices," *ACS Appl. Electron. Mater.* **3**, 1228-1235 (2021).

61 S. Yang *et al.*, “The fabrication of polyethylene/graphite nanoplatelets composites for thermal management and electromagnetic interference shielding application,” *J. Mater. Sci.* **57**, 1084–1097 (2022).

62 A. Mirabedini *et al.*, “Scalable production and thermoelectrical modeling of infusible functional graphene/epoxy nanomaterials for engineering applications,” *Industrial & Engineering Chemistry Research* **61**, 5141-5157 (2022).

63 J.-u. Jang *et al.*, “Thermal percolation behavior in thermal conductivity of polymer nanocomposite with lateral size of graphene nanoplatelet,” *Polymers* **14**, 323 (2022).

64 B. S. Chang *et al.*, “Thermal percolation in well-defined nanocomposite thin films,” *ACS Appl. Mater. Interfaces* **14**, 14579-14587 (2022).

65 J. U. Jang *et al.*, “Enhanced thermal conductivity of graphene nanoplatelet filled polymer composite based on thermal percolation behavior,” *Compos. Commun.* **31** (2022).

66 S. Shi *et al.*, “3D printed polylactic acid/graphene nanocomposites with tailored multifunctionality towards superior thermal management and high-efficient electromagnetic interference shielding,” *Chem. Eng. J.* **450** (2022).

67 D. J. Bergman and D. Stroud, "Physical properties of macroscopically inhomogeneous media", in *Solid state physics - advances in research and applications*, edited by Henry Ehrenreich and David Turnbull (Academic Press, 1992), Vol. 46, pp. 147-269.

68 C. W. Nan, “Physics of inhomogeneous inorganic materials,” *Prog. Mater Sci.* **37**, 1-116 (1993).

69 A. A. Snarskii *et al.*, *Transport processes in macroscopically disordered media: From mean field theory to percolation*. (Springer New York, 2016).

70 M. Sahimi, *Applications of percolation theory*. (Taylor & Francis, 2003).

71 J. P. Clerc, G. Giraud, J. M. Laugier, and J. M. Luck, “The electrical conductivity of binary disordered systems, percolation clusters, fractals and related models,” *Adv. Phys.* **39**, 191-309 (1990).

72 D. Stauffer, “Scaling theory of percolation clusters,” *Phys. Rep.* **54**, 1-74 (1979).

73 W. Lin, "Modeling of thermal conductivity of polymer nanocomposites", in *Modeling and prediction of polymer nanocomposite properties* (Wiley, 2013), pp. 169-200.

74 R. Pal, *Electromagnetic, mechanical, and transport properties of composite materials*. (Taylor & Francis, 2014).

75 J. C. Maxwell, *A treatise on electricity and magnetism (vol 1)*. (Dover Publications, 1954).

76 T. C. Choy, *Effective medium theory: Principles and applications*. (OUP Oxford, 2015).

77 V. A. Markel, “Introduction to the Maxwell Garnett approximation: Tutorial,” *J. Opt. Soc. Am. A* **33**, 1244-1256 (2016).

78 D. P. H. Hasselman and L. F. Johnson, “Effective thermal conductivity of composites with interfacial thermal barrier resistance,” *J. Compos. Mater.* **21**, 508-515 (1987).

79 Y. Benveniste, “Effective thermal conductivity of composites with a thermal contact resistance between the constituents: Nondilute case,” *J. Appl. Phys.* **61**, 2840-2843 (1987).

80 H. Fricke, “A mathematical treatment of the electric conductivity and capacity of disperse systems i. The electric conductivity of a suspension of homogeneous spheroids,” *Phys. Rev.* **24**, 575-587 (1924).

81 H. Hatta and M. Taya, “Effective thermal conductivity of a misoriented short fiber composite,” *J. Appl. Phys.* **58**, 2478-2486 (1985).

82 C. W. Nan, R. Birringer, D. R. Clarke, and H. Gleiter, “Effective thermal conductivity of particulate composites with interfacial thermal resistance,” *J. Appl. Phys.* **81**, 6692-6699 (1997).

83 Z. Hashin and S. Shtrikman, "A variational approach to the theory of the effective magnetic permeability of multiphase materials," *J. Appl. Phys.* **33**, 3125-3131 (1962).

84 P. Cosenza *et al.*, "Effective medium theories for modelling the relationships between electromagnetic properties and hydrological variable in geomaterials: A review," *Near Surface Geophysics* **7**, 563-578 (2009).

85 D. A. G. Bruggeman, "Berechnung verschiedener physikalischer konstanten von heterogenen substanzen. I. Dielektrizitätskonstanten und leitfähigkeiten der mischkörper aus isotropen substanzen," *Ann Phys Leipzig* **416**, 636-664 (1935).

86 D. S. McLachlan, "Evaluating the microstructure of conductor-insulator composites using effective media and percolation theories," *MRS Proceedings* **411**, 309 (1995).

87 D. S. McLachlan, M. Blaszkiewicz, and R. E. Newnham, "Electrical resistivity of composites," *J. Am. Ceram. Soc.* **73**, 2187-2203 (1990).

88 E. T. Swartz and R. O. Pohl, "Thermal boundary resistance," *Rev. Mod. Phys.* **61**, 605-668 (1989).

89 Y. Cengel, *Heat and mass transfer: Fundamentals and applications*. (McGraw-Hill, 2014).

90 L. D. Landau *et al.*, *Electrodynamics of continuous media*. (Elsevier Science, 2013).

91 R. Landauer, "Electrical conductivity in inhomogeneous media," **40**, 2-45 (1978).

92 A. N. Volkov and L. V. Zhigilei, "Scaling laws and mesoscopic modeling of thermal conductivity in carbon nanotube materials," *Phys. Rev. Lett.* **104** (2010).

93 A. N. Volkov and L. V. Zhigilei, "Heat conduction in carbon nanotube materials: Strong effect of intrinsic thermal conductivity of carbon nanotubes," *Appl. Phys. Lett.* **101** (2012).

94 A. M. Marconnet, M. A. Panzer, and K. E. Goodson, "Thermal conduction phenomena in carbon nanotubes and related nanostructured materials," *Rev. Mod. Phys.* **85**, 1295-1326 (2013).

95 A. N. Volkov and L. V. Zhigilei, "Thermal conductivity of two-dimensional disordered fibrous materials defined by interfiber thermal contact conductance and intrinsic conductivity of fibers," *J. Appl. Phys.* **127**, 065102 (2020).

96 X. Zhao *et al.*, "Thermal conductivity model for nanofiber networks," *J. Appl. Phys.* **123** (2018).

97 C. Huang, X. Qian, and R. Yang, "Thermal conductivity of polymers and polymer nanocomposites," *Mater. Sci. Eng., R* **132**, 1-22 (2018).

98 M. Fujii *et al.*, "Measuring the thermal conductivity of a single carbon nanotube," *Phys. Rev. Lett.* **95**, 065502 (2005).

99 Q. Li, C. Liu, X. Wang, and S. Fan, "Measuring the thermal conductivity of individual carbon nanotubes by the raman shift method," *Nanotechnology* **20**, 145702 (2009).

100 Y. Xie *et al.*, "19-fold thermal conductivity increase of carbon nanotube bundles toward high-end thermal design applications," *Carbon* **139**, 445-458 (2018).

101 J. Yang *et al.*, "Phonon transport through point contacts between graphitic nanomaterials," *Phys. Rev. Lett.* **112**, 205901 (2014).

102 J. Han, G. Du, W. Gao, and H. Bai, "An anisotropically high thermal conductive boron nitride/epoxy composite based on nacre-mimetic 3D network," *Adv. Funct. Mater.* (2019).

103 S. Sudhindra, F. Kargar, and A. A. Balandin, "Noncured graphene thermal interface materials for high-power electronics: Minimizing the thermal contact resistance," *Nanomaterials* **11** (2021).

104 E. Persky *et al.*, "Non-universal current flow near the metal-insulator transition in an oxide interface," *Nat. Commun.* **12**, 3311 (2021).

105 S. Ju, T. Y. Cai, and Z. Y. Li, "Percolative magnetotransport and enhanced intergranular magnetoresistance in a correlated resistor network," *Phys. Rev. B* **72** (2005).

106 I. Webman, J. Jortner, and M. H. Cohen, "Numerical simulation of electrical conductivity in microscopically inhomogeneous materials," *Phys. Rev. B* **11**, 2885–2892 (1975).

107 S. Sunde, "Calculation of conductivity and polarization resistance of composite SOFC electrodes from random resistor networks," *J. Electrochem. Soc.* **142**, L50–L52 (1995).

108 J. T. Gostick and A. Z. Weber, "Resistor-network modeling of ionic conduction in polymer electrolytes," *Electrochim. Acta* **179**, 137–145 (2015).

109 T. B. Murtanto, S. Natori, J. Nakamura, and A. Natori, "AC conductivity and dielectric constant of conductor-insulator composites," *Phys. Rev. B* **74**, 115206 (2006).

110 J. Sun *et al.*, "Parallel algorithm for the effective electromagnetic properties of heterogeneous materials on 3D RC network model," in *Proceedings of the 10th International Symposium on Antennas, Propagation, and EM Theory, ISAPE2012*, Xi'an, China, 2012 (IEEE, Piscataway, NJ), pp. 1214–1218.

111 I. Singh *et al.*, "Optical and electrical characterization of conducting polymer-single walled carbon nanotube composite films," *Carbon* **46**, 1141–1144 (2008).

112 M. Abu-Abdeen, A. S. Ayesh, and A. A. Al Jaafari, "Physical characterizations of semi-conducting conjugated polymer-CNTs nanocomposites," *J. Polym. Res.* **19**, 9839 (2012).

113 E. Kymakis and G. A. J. Amaratunga, "Electrical properties of single-wall carbon nanotube-polymer composite films," *J. Appl. Phys.* **99**, 084302 (2006).

114 J. Zhang, M. Mine, D. Zhu, and M. Matsuo, "Electrical and dielectric behaviors and their origins in the three-dimensional polyvinyl alcohol/MWCNT composites with low percolation threshold," *Carbon* **47**, 1311–1320 (2009).

115 Zdekamsk, V. Kueslek, and J. Paek, "AC conductivity of carbon fiber-polymer matrix composites at the percolation threshold," *Polym. Compos.* **23**, 95–103 (2002).

116 G. J. Lee, K. D. Suh, and S. S. Im, "Study of electrical phenomena in carbon black–filled HDPE composite," *Polym. Eng. Sci.* **38**, 471–477 (1998).

117 M. Hirasawa, Y. Nakamura, and M. Shimoji, "Electrical conductivity and thermoelectric power of concentrated lithium-ammonia solutions," *Ber. Bunsen-Ges./PCCP* **82**, 815–818 (1978).

118 S. Hahne and U. Schindewolf, "Temperature and pressure dependence of the nonmetal-metal transition in sodium–ammonia solutions (electrical conductivity and pressure–volume–temperature data up to 150°C and 1000 bars)," *J. Phys. Chem.* **79**, 2922–2928 (1975).

119 J. Jortner and M. H. Cohen, "Metal-nonmetal transition in metal-ammonia solutions," *Phys. Rev. B* **13**, 1548–1568 (1976).